\documentclass[final,1p,10pt,times,amsmath,amssymb,authoryear]{elsarticle}
\usepackage{natbib}
\usepackage{geometry}
\usepackage{amssymb}

\usepackage{amsmath}
\usepackage{tensor}
\usepackage{slashed}
\usepackage{amsfonts}
\usepackage{euscript}
\usepackage{mathrsfs}
\usepackage{bm}
\usepackage{dsfont}
\usepackage{multirow}
\usepackage{ulem}
\usepackage{bm}

\usepackage{hyperref}
\hypersetup{
     colorlinks   = true,
     citecolor    = blue,
     urlcolor = blue,
     linkcolor = blue
}

\makeatletter
\newcommand{\xleftrightarrow}[2][]{\ext@arrow 3359\leftrightarrowfill@{#1}{#2}}
\makeatother

\journal{Encyclopedia of Condensed Matter Physics 2e}

\begin{document}

\begin{frontmatter}



\title{Field Theoretic Aspects of Condensed Matter Physics: An Overview\tnoteref{t1}}
\tnotetext[t1]{Work was supported in part by the US National Science Foundation through grant No. DMR 1725401 and DMR 2225920 at the University of Illinois.}


\author{Eduardo Fradkin}

\address{Department of Physics and Institute for Condensed Matter Theory,\\
University of Illinois at Urbana-Champaign, 1110 West Green St, Urbana Illinois 61801-3080, U.S.A.}

\date{\today}

\begin{abstract}
In this chapter I discuss the impact of concepts of Quantum Field Theory in modern Condensed Physics. Although the interplay between these two areas is certainly not new, the impact and mutual cross-fertilization has certainly grown enormously with time, and Quantum Field Theory has become a central conceptual tool in Condensed Matter Physics. In this chapter I cover how these ideas and tools have influenced our understanding of phase transitions, both classical and quantum, as well as topological phases of matter, and dualities.

\end{abstract}


\end{frontmatter}
\begin{itemize}
\item
Quantum Field Theory has a deep and wide impact on our understanding of modern Condensed Matter Physics.
\item 
The conceptual framework of the Renormalization Group and of the UV fixed point is the central concept in the theory and classical and quantum phase transitions
\item
The renormalization group is the foundation for universality, scale and conformal invariance at these fixed points.
\item
Here I discuss quantum criticality in the 1D transverse field Ising model and in quantum antiferromagnets.
\item
Topology is also central to our understanding of condensed matter ranging from the role of vortices and magnetic monopoles to topological terms in effective actions.
\item
Duality and duality transformations  and bosonization and fermionization transformations in matter and gauge theories provide a non-perturbative framework to our understanding of these systems.
\item
Topology has provided a sharp definition of  fractional charge, fractional statistics, braiding and fusion of quasiparticles.
\item
Topological phases of matter are physical systems with long range quantum entanglement characterized by topological invariants.
\end{itemize}



\section{Introduction}
\label{sec:intro}

Many (if not most) puzzling problems in Condensed Matter Physics involve systems with a macroscopically large number of degrees of freedom often in regimes of large fluctuations, thermal and/or quantum mechanical. The description of the physics of systems of this type requires the framework provided by Quantum Field Theory. Although quantum field theory has its origins in high-energy physics, notably in the development of Quantum Electrodynamics, it has found a nurturing home in Condensed Matter Physics. 

There is a long history of of cross-fertilization between both fields. Since the 1950's in many of the most significant developments in Condensed Matter Physics, Quantum Field Theory has played a key role if not in the original development but certainly in the eventual understanding the meaning and the further development of the discoveries. As a result, many of the discoveries and concepts developed in Condensed Matter have had a reciprocal impact in Quantum Field theory. One can already see this interplay in the development of the Bardeen-Cooper-Schrieffer theory of superconductivity \cite{schrieffer-1964} and its implications in the theory of dynamical symmetry breaking in particle physics by Nambu \cite{nambu-1961}. 

The close and vibrant relationship between both fields has continued to these days, and it is even stronger today than before. Many textbooks have been devoted to teaching these ideas and concepts to new generations of condensed matter physicists (and field theorists as well). The earlier texts focused on Green functions which are computed in perturbation theory using  Feynman diagrams \cite{abrikosov-1964} \cite{fetter-1971} \cite{doniach-1974}, while the more modern ones have a broader scope, use path integrals and attack non-perturbative problems \cite{feynman-1965,feynman-1972} \cite{faddeev-1975} \cite{fradkin-2013,fradkin-2021} \cite{altland-2010}. In two recent books I have discussed many aspects of the interrelation between condensed matter physics and quantum field theory in more depth than I can do in this chapter \cite{fradkin-2013,fradkin-2021}.

\section{Early Years: Feynman Diagrams and Correlation Functions}
\label{sec:early}

Quantum field theory played a key role in the development of the Theory of the Fermi Liquid \cite{pines-1966,baym-1991}. The theory of the Fermi liquid was first formulated by Landau using the framework of hydrodynamics and the quantum Boltzmann equation.  Landau's ideas were later given a microscopic basis using Green functions and Feynman diagrams \cite{abrikosov-1964}, including the effects of quantum fluctuations at finite temperature and non-equilibrium behavior \cite{kadanoff-1962,kamenev-2011}. Linear response theory was developed which allowed the computation of response functions (such as electrical conductivities and magnetic susceptibilities) from the computation of correlation functions for a given microscopic theory \cite{martin-1968}. In turn, correlation functions can be computed in terms of a set of Feynman diagrams. These concepts and tools borrowed many concepts from field theory including the study of the analytic structure of the generalized susceptibilities and the associated spectral functions (together with the use of dispersion relations). These developments led to the derivation of the fluctuation-dissipation theorem. These ideas were widely applied to metals \cite{baym-1991} and superconductors \cite{schrieffer-1964}, as well as to quantum magnets \cite{mattis-1965b}.

The spectrum of an interacting system has low energy excitations characterized by a set of quantum numbers associated with the symmetries of the theory. These low energy excitations are known as {\it quasiparticles}. In the case of the Landau theory of the Fermi liquid the quasiparticle is  a ``dressed'' electron: it is  a low energy excitation with the same quantum numbers (charge and spin) and an electron but with a renormalized effective mass. There are many such quasiparticles in condensed matter physics. The correlation functions (the propagators) of a physical system has a specific analytic structure. In momentum (and frequency) space, the quasiparticle spectrum is given by the poles of the correlators. 

The role of symmetries and, in particular of gauge invariance, in the structure of correlation functions  was investigated extensively. A direct consequence of symmetries is the existence of Ward identities which must be satisfied by all the correlation functions of the theory. Ward identities are exact relations that relate different correlation functions. Such identities contain a host of important results. For example, in a  theory with a globally conserved charge, the Hamiltonian (and the action) have a global U$(1)$ symmetry associated with the transformation of the local field operator $\phi(x)$ (which can be fermionic or bosonic) to a new field $\phi'(x)=e^{i\theta} \phi(x)$ (where $\theta$ is a constant phase). Theories with a global continuous symmetry have a locally conserved current (and satisfy a continuity equation). The Ward identity requires the correlators of these currents (and densities) to be be transverse (i.e. they should have vanishing divergence). In the absence of so-called quantum anomalies (which we will discuss below) global symmetries can be made local and become gauge symmetries.

 In many circumstances a global symmetry can be spontaneously broken. If the global symmetry is continuous, then the Ward identities imply the existence of gapless excitations known as Goldstone bosons. For example in the case of a superfluid, which has a spontaneously broken U$(1)$ symmetry the Goldstone boson is the gapless phase mode. Instead, a the N\'eel phase of a quantum antiferromagnet has two gapless Goldstone bosons, the magnons of the spontaneously broken SO$(3)$ global symmetry of this state of matter. Another example is the Ward identity of quantum electrodynamics, which relates the electron self-energy to the electron-photon vertex function, which also holds in non-relativistic electron fluids.  In addition, these identities implied the existence of sum rules that the spectral functions must satisfy. All of these results became part of the standard toolkit of condensed matter experimentalists in analyzing their data and for theorists to make predictions. Ward identities and sum rules also imply restrictions on the allowed approximations which are often needed to obtain predictions from a microscopic model.

\section{Critical Phenomena}  
\label{sec:critical-phenomena}

\subsection{Classical Critical Phenomena}
\label{sec:classical}

The late 1960s and particularly 1970s brought about an intense back and forth between condensed matter physics and field theory in the context of the problem of classical critical phenomena and phase transitions. This was going to become a profound revolution on the description of macroscopic physical systems with large-scale fluctuations. The problem of continuous (``second order'') phase transitions has a long history going back to the work of Landau \cite{landau-1937,landau-1959} who introduced the concept of an {\it order parameter field}. This turned out to be a powerful concept of broad applicability in many physical systems sometimes quite different from each other at the microscopic level. 

A simple example is that of a ferromagnet with uniaxial anisotropy in which the spins of the atoms in a crystal are strongly favored to be aligned (or anti-aligned) along certain directions of the crystal.  The simplest microscopic model for this problem is the Ising model, a spin system in which the individual spins are  allowed to take only two values, $\sigma=\pm 1$. The partition function of the Ising model (in any dimension) is
\begin{equation}
Z=\sum_{[\sigma]} \exp\left(-\frac{J}{T} \sum_{\langle {\bm r}, {\bm r}'\rangle} \sigma({\bm r}) \sigma({\bm r}')\right)
\label{eq:Ising}
\end{equation}
where $J$ is the exchange coupling constant and $T$ is the temperature (measured in energy units); here $[\sigma]$ denotes the sume over the $2^N$ spin configurations (for a lattice with $N$ sites), and $\langle {\bm r}, {\bm r}'\rangle$ are nearest neighbor sites of the lattice. The order parameter of the Ising model is the local magnetization $\langle \sigma({\bm r}) \langle$ which, in the case of a ferromagnet, is uniform. The partition function of the Ising model can be computed trivially in one dimension. The solution of the two-dimensional Ising model by Onsager constituted a tour-de-force in theoretical physics \cite{onsager-1944}. Its actual meaning remained obscure for some time. The work of Schultz, Mattis and Lieb \cite{schultz-1964}  evinced a deep connection between Onsager's solution and the problem if the spectrum of one-dimensional quantum spin chains \cite{lieb-1961} (specifically the one-dimensional Ising model in a transverse field \cite{pfeuty-1970}). One important result was that the Ising model was in fact a theory of (free) fermions which, crudely speaking, represented the configurations of domain walls of the magnet.  However, even this simple model cannot be solved exactly  in general dimension,  and approximate mean field theories of various sorts were devised over time to understand its physics. 

\subsection{Landau-Ginzburg Theory}
\label{sec:LG}

Landau's approach assumed that close enough to a phase transition, the important spin configurations are those for which the local magnetization varies slowly  on lattice scales. In this picture the  local magnetization, on long enough length scales, becomes an order parameter field that takes values on the real numbers, and can be positive of negative. Thus, the order parameter field is effectively the  average of local magnetizations on some scale large compared to the lattice scale, which we will denote by a real field  $\phi(x)$. The thermodynamics properties of a system of this type in $d$ dimensions can be described in terms of a free energy
\begin{equation}
F[\phi]=\int d^dx \left[\frac{\kappa}{2} \left(\bm{\bigtriangledown} \phi(x)\right)^2+a (T-T_c) \phi^2(x)+u \phi^4(x)+\ldots\right]
\label{eq:GL}
\end{equation}
which is known as the Ginzburg-Landau free energy. Here $\kappa$ is the stiffness of the order parameter field, $T_c$ is the (mean-field) critical temperature; $a$ and $u$ are two (positive) constants. This expression make sense if the transition is continuous and hence that the order parameter is small near the transition. The energy of the Ising model is invariant under the global symmetry $[\sigma] \mapsto [-\sigma]$. This is the symmetry of the group $\mathbb{Z}_2$. Likewise, the Ginzburg-Landau free energy has the global (discrete) symmetry $[\phi(x)] \mapsto [-\phi(x)]$, and also has a $\mathbb{Z}_2$ global symmetry. 

In Landau's approach, which was a mean field theory, the equilibrium state is the global minimum of this free energy. The nature of the equilibrium state depends on whether $T>T_c$ or $T<T_c$: for $T>T_c$ the global minimum is the trivial configuration, $\bar \phi(x)=0$ (this is the paramagnetic state), whereas for $T<T_c$ the equilibrium state is two fold degenerate, $\bar \phi(x)=\pm (a(T_c-T)/2)^{\beta}$, with the two degenerate states being related by the $\mathbb{Z}_2$ symmetry (this is the ferromagnetic state). In the Landau theory the critical exponent of the magnetization is $\beta=1/2$ and the critical exponent  of the correlation length is $\nu=1/2$. However, in the case of the 2D Ising model the order parameter exponent is $\beta=1/8$ \cite{yang-1952} and the correlation length exponent is $\nu=1$. These (and other) apparent discrepancies led many theorists for much of the 1960s believe that each model was different and that these behaviors reflected microscopic differences. In addition, Landau's theory was regarded as phenomenological and believed to be of questionable validity.

\subsection{The Renormalization Group}
\label{sec:RG}

This situation was to change with the development of the Renormalization Group, due primarily to the work of Leo P. Kadanoff \cite{kadanoff-1966a,kadanoff-1966b,efrati-2014} and Kenneth G. Wilson \cite{wilson-1971a,wilson-1971b,wilson-1972,wilson-1974,wilson-1975,wilson-1983}. The renormalization Group was going to have (and still has) a profound effect both in Condensed Matter Physics and in Quantum Field Theory (and beyond).

\subsubsection{Scaling}
\label{sec:scaling}

Several phenomenological theories were proposed in the 1960s to describe the singular behavior of physical observables near a continuous phase transition \cite{patashinskii-1966,fisher-1967,fisher-1974}. 
These early works argued that in order to explain the singular behavior of the observables the free energy density had to have a singular part which should be a homogeneous function of the temperature, 
magnetic field, etc. A function $f(x)$ is homogeneous if it satisfies  the property that it transforms irreducibly under dilations, i.e. $f(\lambda x)=\lambda^k f(x)$, there $\lambda$ is a real positive number (a scale) and $k$ is called the degree. These heuristic ideas then implied that the critical exponents should obey several identities.  In 1966 Kadanoff wrote and insightful paper in which he showed that the homogeneity hypothesis 
implied that in that regime these systems should obey scaling. He showed that this can be justified by performing a sequence of block-spin transformations in which the  configurations that vary rapidly at the 
lattice scale $a$ become averaged at the scale of a larger sized block of length scale $ba>a$ which resulted in a renormalization of the coupling constants from $\{ K \} $ at scale $a$ to $\{ K' \} $ at the new scale $ba$ \cite{kadanoff-1966a,efrati-2014}.  In other words, the block spin transformation amounts to a scale transformation and a renormalization of the couplings (and operators). From this condensed matter/statistical physics perspective the important physics is in the long distance (``infrared'') behavior. 

A significant consequence of these  ideas was that close enough to a critical point, if the distance $|x-y|$ between two local observables $O(x)$ and $O(y)$ is large compared to the lattice spacing $a$ but small compared to the correlation length $\xi$, their correlation function takes the form of a power law 
\begin{equation}
\langle O(x) O(y) \rangle \sim \frac{\textrm{const.}}{|x-y|^{2\Delta_O}}
\label{eq:scaling}
\end{equation}
where $\Delta_O$ is a positive real number known as the scaling dimension of the operator $O$ \cite{patashinskii-1966,kadanoff-1966a,kadanoff-1967}. These conjectures were known to be satisfied in the non-trivial case of the 2D Ising model \cite{kadanoff-1966b}, as well as in the  Landau-Ginzburg theory once the effects of Gaussian fluctuations were included \cite{kadanoff-1967}. For example, in the 2D Ising model, the scaling dimensions of the local magnetization $\sigma$ is $\Delta_\sigma=1/8$ and of the energy density $\varepsilon$ is $\Delta_\varepsilon=1$, which were sufficient to explain all the singular behaviors known at that time. 

The concept of renormalization actually originated earlier in quantum field theory as part of the development of Quantum Electrodynamics (QED). In QED the notion of renormalization was used to ``hide'' the short distance (``ultraviolet'') divergencies of the Feynman diagrams needed to compute physical processes involving electrons (and positrons) and photons, i.e. their strong, divergent, dependence of an artificially introduced short-distance cutoff or regulator.  In particular the sum of the leading diagrams that enter in the electron-photon vertex amounted to a redefinition (renormalization) of the coupling constant. It was observed by Murray Gell-Mann and Francis Low that this renormalization was equivalent to the solution of a first order differential equation that governed the infinitesimal change of the coupling, the fine structure constant  $\alpha=e^2/4\pi$, under an infinitesimal change of the UV cutoff $\Lambda$ \cite{gell_mann-1954}
\begin{equation}
\Lambda \frac{d\alpha}{d\Lambda}\equiv \beta(\alpha)=\frac{2}{3\pi} \alpha^2+O(\alpha^3)
\label{eq:gell_mann-low}
\end{equation}
where $\beta(\alpha)$ is the Gell-Mann-Low beta function. Except for the work by Nikolai Bogoliubov and coworkers \cite{bogoliubov-1959}, this reinterpretation by Gell-Mann and Low was not actively pursued, partly because it predicted that the renormalized coupling became very large at short distances, $\alpha \to \infty$, and, conversely, it vanished in the deep long distance regime, $\alpha \to 0$ (if the electron bare mass is zero). In other terms, QED is strongly coupled  in the UV and trivial in the IR. The same behavior was found in the case of the theory of a scalar field $\phi(x)$ with an $\phi^4$ interaction which is relevant in the theory of phase transitions. In addition to these puzzles, the1960s saw the experimental development of the physics of hadrons which involve strong interactions. For these reasons, for much of that decade most high-energy theorists had largely abandoned the use of quantum field theory, and explored other, phenomenologically motivated, approaches (which led to an early version of string theory.)
At  any rate the notion that the physics may depend on the scale was present as was the notion that in some regimes field theories may exhibit scale-invariance at least in an approximate form.

\subsubsection{The Operator Product Expansion}
\label{sec:ope}

The next stage of the development of these ideas was the concept of the operator product expansion (OPE). If we denote by $\{ O_j(x) \} $ the set of all possible local operators in a theory (a field theory or a statistical mechanical system near criticality), then the product of two  observables on this list closer to each other than to any other observable (and to the correlation length $\xi$) obeys the expansion
\begin{equation}
\lim_{x\to y} O_j(x) O_k(y) = \lim_{x \to y} \sum_l \frac{C_{jkl}}{|x-y|^{\Delta_j+\Delta_k-\Delta_l}} O_l \left(\frac{x+y}{2}\right)
\label{eq:OPE}
\end{equation}
where this equation should be understood as a weak identity, valid inside an expectation value. Remarkably, this concept was derived independently and simultaneously by Leo Kadanoff \cite{kadanoff-1969} (who was working in critical phenomena), by Kenneth Wilson \cite{wilson-1969} (who was interested in the short distance singularities arising in Feynman diagrams), and by Alexander Polyakov \cite{polyakov-1970,polyakov-1974} (also working in critical phenomena). In Eq.\eqref{eq:OPE} $\{ \Delta_j \} $ are  the scaling dimensions of the operators $\{ O_j \}$. The coefficients $C_{jkl}$ are (like the dimensions) universal numbers. In a follow up paper Polyakov showed that if the theory has conformal invariance, i.e. scale invariance augmented by conformal transformations which preserve angles, then, provided the operators $O_j$ are suitably normalized, the coefficients $C_{jkl}$ of the OPE are determined by a three point correlator
\begin{equation}
\langle O_j(x) O_k(y) O_l(z) \rangle=\frac{C_{jkl}} {|x-y|^{\Delta_{jk}} |y-z|^{\Delta_{kl}} |z-x|^{\Delta_{lj}}}
\label{eq:jkl}
\end{equation}
where $\Delta_{jk}=\Delta_j+\Delta_k-\Delta_l$.
These results constitute the beginnings of Conformal Field Theory. In a nontrivial check, Kadanoff and Ceva showed that the OPE holds for the local observables of the 2D Ising model \cite{kadanoff-1971}.

\subsubsection{Fixed Points}
\label{sec:FP}

The next and crucial step in the development of the renormalization Group was made by Kenneth Wilson. Wilson was a high-energy theorist who wanted to know how to properly define a quantum field theory and the physical meaning of renormalization. 

In a Lorentz invariant quantum field theory one is interested in the computation of the expectation value of time-ordered operators. In the case of a self-interacting scalar field $\phi(x)$ in $D$-dimensional Euclidean space-time, obtained by analytic continuation from Minkowski spacetime to imaginary time, the observables are computed from the functional (or path) integral by functional differentiation of the partition function  
\begin{equation}
Z=\int \mathcal{D} \phi \exp\left(-S(\phi,\partial_\mu \phi)+\int d^Dx \; J(x) \phi(x)\right)
\label{eq:path-integral}
\end{equation}
with respect to the local sources $J(x)$. For a scalar field the Euclidean action is
\begin{equation}
S=\int d^Dx \left[\frac{1}{2} (\partial_\mu \phi(x))^2+\frac{m^2}{2} \phi^2(x)+\frac{\lambda}{4!} \phi^4(x)\right]
\label{eq:phi4}
\end{equation}
which has the same form as the free energy of the Landau-Ginzburg theory of phase transitions shown in Eq.\eqref{eq:GL}. It is apparent that the Landau-Ginzburg theory is the classical limit of the theory of the scalar field whose partition function is a sum over all histories of the field. It is easy to see that en expansion of the partition function (or of a correlator) in powers of the the coupling constant $\lambda$ can be cast in the form of a sum of Feynman diagrams. To lowest order in $\lambda$ a typical Feynman diagram involves a one-loop integral in momentum space of the form
\begin{equation}
I(p)=\int \frac{d^Dq}{(2\pi)^D} \frac{1}{(q^2+m^2) ((q-p)^2+m^2)}
\label{eq:one-loop}
\end{equation}
As noted by Wilson in his Nobel Lecture \cite{wilson-1983}, this integral has large contributions from the IR region of small momenta $q \sim 0$, but for any dimension $D \geq 4$ has a much larger contribution form large the UV region of large momenta, which requires the introduction of a UV cutoff $\Lambda$ in momentum space (or a lattice spacing $a$ in real space by defining the theory on a hypercubic lattice). In quantum field theory one then has to require that somehow one takes the limit $a \to 0$ (or $\Lambda \to \infty$). To take this limit in the field theory is very much  analogous to the definition of a conventional integral in terms of a limit of a Riemann sum. The difference is that this is a {\it functional} integral. While  for a function of bounded variation in a finite interval $(a,b)$ the  limit of a partition of the interval into $N$ steps each of length $\Delta x$, such that $N \Delta x=b-a$, exists and defines the integral of the function
\begin{equation}
\lim_{\Delta x \to 0} \lim_{N \to \infty}\sum_{j=1}^N  f(x_j) \Delta x_j=\int_a^b dx \; f(x),
\end{equation}
the analogous statement does not obviously exists in general for a {\it functional integral}, i.e. an integral over a space of functions which is what is required. In fact, although thousands of integrals of a function are known to exist, there are extremely few examples for a functional integral. Moreover, in order  to take the {\it continuum limit} the lattice spacing must approach zero, $a \to 0$. This means that physical scales, such as the correlation length $\xi$, must diverge in lattice units so that they can be fixed in physical units. But to do that one has to be asymptotically close to a continuous phase transition! Hence, the problem of defining a quantum field theory is {\it equivalent} to the problem of critical phenomena at a continuous phase transition!

 Wilson gave a systematic formulation to the Renormalization Group by generalizing the earlier ideas introduced by Kadanoff and the earlier work by Gell-Mann and Low. Wilson's key contribution was the  introduction of the  concept of a {\it fixed point} of the Renormalization Group transformation \cite{wilson-1971a,wilson-1971b,wilson-1972,wilson-1974}.
As we saw, the block-spin transformation is a procedure for coarse graining the degrees of freedom of a physical system resulting in a renormalization of the coupling constants. Upon the repeated action of the RG transformation its effect can be pictured as a {\it flow} in the space of coupling constants. However, in addition of integrating-out short distance degrees of freedom one needs to restore the units of length which have changed under that process. This requires a rescaling of lengths. Once this is done,  Wilson showed that the resulting RG flows necessarily have {\it fixed points}, special values of the couplings which are invariant (fixed) under the action of the RG transformation. He then deduced that at a fixed point the theory has no scales, aside from the linear size $L$ of the system and the microscopic UV cutoff (the lattice spacing $a$ in a spin system). 

This analysis means that for length scales long compared to $a \to 0$ but short compared to $L \to \infty$ the theory acquired a new, emergent, symmetry: scale invariance. Therefore, at a fixed point the correlators of all local observables must be homogeneous functions (hence, must scale). 

\subsubsection{Universality}
\label{sec:universality}

A crucial consequence of the concept if the fixed point is that phase transitions can be classified into {\it universality classes}. Universality means that a large class of physical systems with different microscopic properties have fixed points with the same properties, i.e. the same scaling dimensions, operator product expansions and correlation functions at long distances independent on how they are defined microscopically. Although the renormalization group transformation is a transformation scheme that we define and, because of that the location in coupling constant space of the fixed point itself does depends on the  scheme we choose, its universal properties are the same. Thus, universality classes depend only on features such as the space (and spacetime) dimension and the global symmetries of the system. But the systems themselves may be quite different. This we speak if the Ising universality class in 2D, on the superfluid (or XY) transition class in 3D, etc. This concept, which originated in the theory of phase transition, has been adopted and generalized in the development of conformal field theory.

\subsubsection{RG flows}
\label{sec:RG-flows}

Combined with the condition that the correlators decay at long separations,  homogeneity  implies that the correlators must have the form of Eq.\eqref{eq:scaling}. In addition, this equation also implies that at a fixed point the operators (the local observables) have certain scaling dimensions. Let us consider a theory close to a fixed point whose action we will denote by $S^*$. Let $\{ O_j(x) \}$ be a complete set of local observables whose scaling dimensions are $\{ \Delta_j \}$. The total action of the theory close to the fixed point then can be expanded as a linear combination of the operators  with dimensionless coupling constants $\{ g_j \}$
\begin{equation}
S=S^*+\int d^Dx \sum_j g_j a^{\Delta_j-D} O_j(x)
\label{eq:S}
\end{equation}
Under a change of length scale $x \to x'=bx$, with $b>1$, the operators (which must transform homogeneously) change as $O_j(bx)=b^{-\Delta_j} O_j(x)$. Since the phase space changes as $d^Dx'=b^D d^dx$, we can keep the form of the action provided the coupling constants also change to compensate for these changes as $g'_j=b^{D-\Delta_j}g_j$. Let $b=|x'|/|x|=1+da/a$, where $da$ is an infinitesimal change of the UV cutoff $a$. Then, if we integrate-out the degrees of freedom in the range $a < |x| < a+da$, the rate of change of the coupling constants $\{ g_j \}$ under this rescaling is
\begin{equation}
a\frac{dg_j}{da}\equiv \beta(g_j)=(D-\Delta_j) g_j+\ldots
\label{beta-functions-tree-level}
\end{equation}
which we recognize as a Gell-Mann Low beta function for each coupling constant.

This result says that if the scaling dimension $\Delta_j<D$, then the renormalized coupling will {\it increase} as we increase the length scale, $g'_j>g_j$, and along this direction in coupling constant space the RG flows {\it away} from the fixed point. Conversely, if $\Delta_j>D$ the renormalized coupling flows to smaller values, $g_j'<g_j$ and the RG flows {\it into} the fixed point. We then say that an operator is {\it relevant} if its scaling dimension satisfies $\Delta_j <D$, and that it is {\it irrelevant}  if $\Delta_j>D$. If $\Delta_j=D$ then we say that operator is {\it marginal}.

To go beyond this simple dimensional analysis one has to include the effects of fluctuations. To lowest orders in the couplings one finds \cite{cardy-1996}
\begin{equation}
a \frac{dg_j}{da}=(D-\Delta_j) g_j+\sum_{k,l} C_{jkl} \; g_k g_l+\ldots
\label{eq:one-loop-beta-functions}
\end{equation}
where $\{C_{jkl} \}$ are the coefficients of the OPE shown in Eq.\eqref{eq:jkl}. This expression is the general form of a perturbative renormalization group and it is valid close enough to a fixed point.

Wilson and Fisher \cite{wilson-1972} used a similar approach to analyze how fluctuations alter the results of the Landau-Ginzburg theory. They considered the partition function of Eq.\eqref{eq:path-integral} with the action of Eq.\eqref{eq:phi4}. Instead of working in real space they considered the problem  in momentum space and partitioned the field configurations into slow and fast modes
\begin{equation}
\phi(x)=\phi_<(x)+\phi_>(x)
\label{eq:fast-slow}
\end{equation}
where $\phi_>(x)$ are configurations whose Fourier components have momenta in the range $b\Lambda < |p| <\Lambda$, where $\Lambda$ is a UV momentum cutoff and $b<1$. Hence, if we choose $b\to 1$, the fast modes $\phi>(x)$ have components in a thin momentum shell near the UV cutoff $\Lambda$. Conversely, the slow modes $\phi_<(x)$ have momenta in the range $0\leq |p|<b\Lambda$. One can then use Feynman diagrams to integrate out the fast modes and derive an effective low-energy action with renormalized couplings. In the case of a free field theory (with $\lambda=0$)  the scaling dimension of the $\phi^4$ operator is $\Delta_4=2(D-2)$ whereas the $\phi^2$ operator has dimension $\Delta_2=D-2$. Upon defining a dimensionless  mass and  coupling constant by $m^2=t\Lambda^2$ and 
$\lambda=g \Lambda^{4-D}$, the beta functions are found to be \cite{fradkin-2021}
\begin{align}
\beta(t)=&-\Lambda \frac{dt}{d\Lambda}=2t+\frac{g}{2}-\frac{gt}{2}+\ldots\\
\beta(g)=&-\Lambda \frac{dg}{d\Lambda}=(4-D)g-\frac{3}{2} g^2 +\ldots
\label{eq:phi4-RG}
\end{align}
(where we absorbed an uninteresting factor if the definition of the coupling constant).

The RG flows of Eq.\eqref{eq:phi4-RG} show that the free-field (Gaussian) fixed point at $g=0$ is stable for $D>4$ and the asymptotic IR behavior is the same as predicted by the Landau-Ginzburg theory. However, for $D<4$, the free-field fixed point becomes unstable and a new fixed point arises at $g^*=\frac{2}{3} \epsilon +O(\epsilon^2)$, where we have set $\epsilon=4-D$. This is the Wilson-Fisher fixed point. At this fixed point the correlation length diverges with an exponent $\nu=\frac{1}{2}+\frac{\epsilon}{12}+O(\epsilon^2)$, which deviates from the predictions of the Landau-Ginzburg theory.  The small parameter of this expansion is $\epsilon$, and this is known as the $\epsilon$ expansion.

The Wilson-Fisher (WF) fixed point is an example of a non-trivial fixed point at which the correlation length is divergent. It has only one relevant operator, the mass term, which in the IR flows into the symmetric phase for $t>0$ and flows to the broken symmetry for $t<0$. Conversely, in the UV it flows into the WF fixed point. For these reasons condensed matter physicists say that this is an IR unstable (or critical) fixed point while high-energy physicists say that it is the UV fixed point. At this fixed point a non-trivial field theory can be defined with non-trivial interactions. UV fixed points also define examples of what in high-energy physics are called renormalizable field theories and can be used to define a continuum field theory.

The $D=4$-dimensional theory is special in that the $\phi^4$ operator is marginal. As can be seen in Eq.\eqref{eq:phi4-RG}, at $D=4$ the beta function for the dimensionless coupling constant $g$ does not have a linear term and is quadratic in $g$. In this case the operator is marginally irrelevant, and its beta function  has the same behavior as the beta function of Gell-Mann and Low for QED. Such theories are said to have a ``triviality problem'' since, up to logarithmic corrections to scaling, there are no interactions in the IR and, conversely, become large in the UV.

There are also fixed points at which the correlation length $\xi \to 0$. These fixed points are IR stable (and in a sense trivial). These stable fixed points are  sinks of the IR RG flows. Such fixed points define stable phases of matter, e.g. the broken symmetry state, the symmetric (or unbroken state), etc. However in the UV they are unstable and in high-energy physics such fixed points correspond to non-renormalizable field theories.

More sophisticated methods are needed to go beyond the lowest order beta functions of Eq.\eqref{eq:one-loop-beta-functions}, and the computation of critical exponents beyond the leading non-trivial order. Possibly the record high-precision calculations have been done for $\phi^4$ theory for which the beta function is know to $O(\epsilon^5)$. This has been achieved using the method of dimensional regularization \cite{thooft-1972,bollini-1972,amit-book} (with minimal subtraction). Special resummation methods  (Borel-Pad\'e) have been used to do these calculations in $D=3$ dimensions \cite{zinn-book}. Remarkably, these results are so precise that in the case of the superfluid transition, which is well described by a $\phi^4$ theory with a complex field, the results could only be tested in the microgravity environment of the International Space Station!

\subsubsection{Asymptotic Freedom}
\label{sec:asymptotic-freedom}

There are several physical systems systems of great interest whose beta function has the form
\begin{equation}
\beta(g)=A g^2+ \ldots
\label{eq:asymptotic-freedom}
\end{equation}
The coupling constant has a different interpretation in each theory and the constant $A>0$, opposite to the sign of the beta function found in QED and $\phi^4$ theory in $D=4$ dimensions. This beta function means that while the associated operator is marginal, with this sign is actually marginally relevant. This also means that the fixed point is unstable in the IR but the departure from the fixed point is logarithmically small. Conversely, the in the UV the RG flows into the fixed point and the effective constant is weak at short distances. This is the origin of the term asymptotic freedom \cite{gross-1973}.
The paradigmatic examples of theories with a beta function of this form are the Kondo problem, the 2D non-linear sigma model, and the $D=4$ dimensional Yang-Mills non-abelian gauge theory. 

The Kondo problem is the theory of a localized spin-1/2 degree of freedom in a metal. the electrons of the metal couple to this quantum impurity through an  exchange interaction of the impurity and the magnetic moment density of the mobile electrons in the metal with coupling constant $J$. This problem is actually one dimensional since only the s wave channel of the mobile (conduction band) electrons  actually couple to the localized impurity. In 1970 Philip Anderson developed a  theory of the Kondo problem in terms  of the renormalization of the Kondo coupling constant $g$ as a function of the energy scale \cite{anderson-1970b}. Anderson used perturbation theory in $J$ to progressively integrated out the modes of the conduction electrons close to an effective  bandwidth $E_c$ and found that the beta function has the form of Eq.\eqref{eq:asymptotic-freedom}
\begin{equation}
E_c \frac{dJ}{dE_c}=-\rho J^2+\ldots
\label{eq:kondo}
\end{equation}
where  $\rho$ is the density of states at the Fermi energy of the conduction electrons. 
 This work implied that the free-impurity fixed point is IR unstable and that the effective coupling constant $J$ increases as the energy cutoff $E_c$ is lowered. He  argued that at some energy scale, the Kondo scale, perturbation theory breaks down and that there is a crossover to a strong coupling regime which is not accessible in perturbation theory.  

Shortly thereafter, in 1973  Wilson developed a numerical renormalization group approach which showed that the Kondo problem is indeed a crossover from the free impurity fixed point to the ``renormalized'' Fermi liquid  \cite{wilson-1975}. In addition, Wilson use the numerical renormalization group to examine the approach to the strong coupling fixed point and showed that it is characterized by a Wilson ratio, a universal number obtained from the low temperature specific heat and the impurity magnetic susceptibility (in suitable units). Wilson's numerical RG predicted a number close to $2\pi$ for the this ratio. In 1980 N. Andrei and P. Wiegmann showed (independently)  that the Kondo problem is an example of an integrable field theory that can be solved by the Bethe ansatz \cite{andrei-1980,wiegmann-1980}. Their exact result was consistent with Wilson's RG, including the numerical value of the Wilson ratio. 

In 1972 Gerard 't Hooft and Martinus Veltman showed that Yang-Mills gauge theory is renormalizable \cite{thooft-1972}. This groundbreaking result opened the door to use quantum field theory to develop the theory of strong interactions in particle physics known as Quantum Chromodynamics (QCD). In 1973 David Gross and Frank Wilczek \cite{gross-1973} and, independently, David Politzer \cite{politzer-1973} computed the renormalization group beta function of Yang-Mills theory with gauge group $G$ and found it to be of the same form as Eq.\eqref{eq:asymptotic-freedom},
\begin{equation}
\Lambda \frac{dg}{d\Lambda}=-\frac{g^3}{16\pi^2}\frac{11}{3} C_2(G)+\ldots
\label{eq:YM}
\end{equation}
where $g$ is the Yang-Mills coupling constant and $\Lambda$ is a UV momentum scale. Here $C_2(G)$ is the quadratic Casimir for a gauge group $G$. For $SU(3)$, the case of physical interest, $C_2(SU(3))=3$. This result implies that under the RG at large momenta (short distances)  the Yang-Mills coupling constant flows to zero  (up to logarithmic corrections). This result holds in the presence of quarks provided the number of quark flavors is less than a critical value. Hence at short distances the effective coupling is weak. Gross and Wilczek called this phenomenon asymptotic freedom. This behavior was consistent with the  observation of weakly coupled quarks in deep inelastic scattering experiments. However, the flip side of asymptotic freedom is that at low energies (long distances) the coupling constant grows without limit, which implies that at low energies perturbation theory is not applicable. This strong infrared behavior suggested that in QCD quarks are permanently confined in color neutral bound states (hadrons). However, unlike the Kondo problem we just discussed, QCD is not an integrable theory (so far as we know) and to show that it confines has required the development of Lattice Gauge Theory \cite{wilson-1974b,kogut-1975}. To this date the best evidence for quark confinement has been obtained using large-scale Monte Carlo simulations in Lattice Gauge Theory \cite{creutz-1983b}.
 
We close this subsection with a discussion of an important case: the non-linear sigma models. The $O(N)$ non-linear sigma model is the continuum limit of the classical Heisenberg model for a spin with $N$ components. Historically, the non-linear sigma model is the effective field theory for pions in particle physics. We will discuss its role in the theory of quantum antiferromagnets in  subsection \ref{sec:nlsm} and especially in the case of quantum antiferromagnetic spin chains in subsection \ref{sec:quantum-spin-chains}. 

The simplest non-linear sigma model is a theory of an $N$-component scalar field ${\bm n}(x)$ which satisfies the unite length local constraint, ${\bm n}^2(x)=1$. The Euclidean Lagrangian is 
\begin{equation}
\mathcal{L}=\frac{1}{2g} (\partial_\mu {\bm n}(x))^2
\label{eq:NLSM}
\end{equation}
where $g$ is the coupling constant (the temperature in the classical Heisenberg model). At the classical level, i.e. in the broken symmetry phase, where $\langle {\bm n}\rangle \neq0$, this model describes the $N-1$ massless modes (Goldstone bosons) of the spontaneously broken $O(N)$ symmetry. Dimensional analysis shows that the coupling constant has units of $\ell^{D-2}$. Hence, we expect to find marginal behavior at $D=2$. In 1975 Alexander Polyakov used a momentum shell renormalization group in $D=2$ dimensions and showed that the beta function of this model is (here $a$ is the short-distance cutoff) \cite{polyakov-1975}
\begin{equation}
\beta(g)=a\frac{dg}{da}=\frac{N-2}{2\pi} g^2+O(g^3)
\label{eq:NLSM-RG}
\end{equation}
Hence, in $D=2$ dimensions also this theory is asymptotically free. As in the other examples we just discussed, asymptotic freedom here also implies that the coupling constant $g$ grows to large values in the low-energy (long-distance) regime. In close analogy with Yang-Mills theory in $D=4$ dimensions, Polyakov conjectured that the $O(N)$ non-linera sigma model also undergoes dynamical dimensional transmutation \cite{gross-1973}, that the global $O(N)$ symmetry is restored and that for all values of the coupling constant $g$ the theory is in a massive with a finite correlation length $\xi \sim \exp((N-2)/2\pi g)$. Extensive numerical simulations, used to construct a renormalization group using Monte Carlo simulations \cite{shenker-1980}, showed that there is indeed a smooth crossover from the weak coupling (low temperature) regime to the high temperature regime where the correlation length is finite. The non-linear sigma model is a renormalizable field theory in $D=2$ dimensions \cite{brezin-1976}. For $D>2$ dimensions it can be studied using the $2+\epsilon$ expansion \cite{brezin-1976,zinn-book}, which predicts the existence of a nontrivial UV fixed point and a phase transition from a Goldstone phase to a symmetric phase.

It turns out that there is a significant number of asymptotically free non-linear sigma models in $D=2$ dimensions, many of physical interest  \cite{friedan-1985}, in particular non-linear sigma models whose target manifold is a coset space, a quotient of a group $G$ and a subgroup $H$. The $O(N)$ non linear sigma model is an example since the broken symmetry space leaves the $O(N-1)$ subgroup unbroken (the manifold of the Goldstone bosons). In that case the quotient is $O(N)/O(N-1)$ which is isomorphic to the $N-1$ dimensional sphere $S_{N-1}$. In later sections we will discuss other examples in which more general non-linear sigma models play an important role.

Models on coset spaces arise in the theory of Anderson localization in $D=2$ dimensions. Anderson localization is the problem of a fermion (an electron) in a disordered system in which the electron experiences a random electrostatic potential. In the limit of strong disorder Philip Anderson showed that all one-particle states are exponentially localized and the diffusion constant (and the conductivity) vanishes\cite{anderson-1958}. There was still the question of when it is possible for the electron to have a finite diffusion constant (and conductivity). In $D=2$ dimensions the conductivity is a dimensionless number which suggests that this may be the critical dimension for diffusion. Abrahams, Anderson, Licciardello and Ramakrishnan used a weak disorder calculation to construct a scaling theory that implied that in $D=2$ dimensions the RG flow of the conductivity at long distances (large samples) flows to zero and all states are localized \cite{abrahams-1979}. Shortly thereafter Wegner gave strong arguments that showed that the existence of diffusion implied that there are low-energy ``diffusson'' modes which behaved as Goldstone modes of a non-linear sigma model on the quotient manifold $O(N_++N_-)/O(N_+) \times O(N_-)$ in the ``replica limit'' $N_\pm \to 0$ \cite{wegner-1979}. A field theory approach to this non-linear sigma model was developed by McKane and Stone \cite{mckane-1981} and by Hikami \cite{hikami-1981}.

\section{Quantum Criticality}
\label{sec:qcp}

Quantum criticality is the theory of a phase transition of a system (e.g. a magnetic system) at zero temperature that occurs  as a coupling constant (or parameter) is varied continuously. Although not necessarily under that name, this question has existed as a conceptual problem for a long time, In particular, already in 1973 Wilson considered the problem of the behavior of quantum filed theories blow four  spacetime dimensions and their phase transitions \cite{wilson-1973}. 

The modern interest in condensed matter physics stems from discoveries made since the late 1980s.  Since that time he behavior of condensed matter systems at a {\it quantum} critical point has emerged as a major focus in the field. There were several motivations for this problem. One was (and still is) to understand the behavior of quantum antiferromagnets in the presence of frustrating interactions. Frustrating interactions are interactions which favor incompatible types of antiferromagnetic orders. The result is the presence of intermediate non-magnetic ``valence bond'' phases that favor the formation of spin singlets between nearby spins. these phases typically either break the point group  symmetry of the lattice or are spin liquids (which will be discussed below). Another motivation is that doped quantum antiferromagnets typically harbor superconducting phases (among others) whose high-temperature behavior is a ``strange'' metal that violates the basic assumptions (and behaviors) of Fermi liquids. The most studied version of this problem is the case of the copper oxide high temperature superconductors. It was conjectured that there is a quantum critical point inside the superconducting phase at which the antiferromagnetic order (or other orders) disappears and which may be the reason for the strange metal behavior above the superconducting critical temperature. Many of these questions are discussed in depth  by Sondhi and coworkers \cite{sondhi-1997} and  in the  textbook by Sachdev \cite{sachdev-1999}.

\subsection{Dynamic Scaling}
\label{sec:dynamic-scaling}

We will consider a general quantum phase transition and assume that it is scale invariance. However, except for the case of relativistic quantum field theories, in condensed matter systems space and time do not need to scale in the same way. Let us assume that the system of interest has just one coupling constant $g$ and that the system of interest has a quantum phase transition (at zero temperature) at some critical value $g_c$ between two phases, for instance one with a spontaneously broken symmetry and a symmetric phase. If the quantum phase transition is continuous then the correlation length $\xi$ will  diverge at $g_c$ and so will the correlation time $\xi_t$. However these two scales are in general different and do not necessarily diverge at the same rate. So, in general, if  some physical quantity is measured at the quantum critical point  at some momentum ${\bm p}$ and frequency $\omega$, the  length scale of the measurement is $2\pi/|{\bm p}|$ and at a frequency is $\omega\sim |{\bm p}|^z$, where $z$ is the dynamic critical exponent. Let us say that we measure the observable ${\mathcal O}$ at momentum ${\bm p}$ and frequency $\omega$ at $g_c$. Scale invariance in both space and time means that at $g_c$ the observable $\mathcal O({\bm p},\omega)$ at momentum ${\bm p}$ and frequency $\omega$ must scale as
\begin{equation}
{\mathcal O}({\bm p}, \omega)=|{\bm p}|^{-\Delta_{\mathcal O}} {\tilde {\mathcal O}}(|{\bm p}|^z/\omega)
\label{eq:dynamic-scaling}
\end{equation}
where $\Delta_{\mathcal O}$ is the scaling dimension of the observable $\mathcal{O}$.

The situation changes at finite temperature $T$. A quantum field theory at temperature $T$ is described by a path integral on a manifold which  along the imaginary time direction $\tau$ is finite of length $2\pi/T$ and periodic for a theory bosonic fields and anti-periodic for fermionic fields \cite{fradkin-2021}. Since the imaginary time direction is finite, the behavior for correlation times $\xi_\tau <2\pi/T$ and $\xi_\tau > 2\pi/T$ must be different. Indeed, in the first regime the behavior is essentially the same as at $T=0$, while in the second it should be given by the classical theory in the same space dimension.At the quantum critical point $g_c$ there is only one time scale $\xi_\tau\sim 2\pi/T$ and only one length scale $\xi\sim (2\pi/T)^{1/z}$. 

\subsection{The Ising Model in a Transverse Field}
\label{sec:TFIM}

The prototype of the quantum phase transition is the Ising model in a transverse field. This model describes a system of spin-1/2 degrees of freedom with ferromagnetic interactions with uniaxial anisotropy in the presence of a transverse uniform magnetic field. The Hamiltonian is
\begin{equation}
H=-J\sum_{\langle {\bm r}, {\bm r}'\rangle} \sigma_3({\bm r}) \sigma_3({\bm r}')-h \sum_{{\bm r}} \sigma_1({\bm r})
\label{eq:TFIM}
\end{equation}
where $J$ and $h$ are the exchange coupling constant and the strength of the transverse field, respectively. Here $\sigma_1$ and $\sigma_3$ are the two Pauli matrices defined on the sites ${\{ \bm r} \} $ of a lattice with ferromagnetic interactions between spins on nearest neighboring sites. The Hilbert space is the tensor product of the states of the spins at each site of the lattice. At each site there are two natural bases of states: the eigenstates of $\sigma_3$, which we denote by $|\uparrow\rangle$ and $|\downarrow\rangle$ (whose eigenvalues are $\pm 1$), and the eigenstates of $\sigma_1$, which we denote by $|\pm \rangle$ (whose eigenvalues are also $\pm 1$). 

It is well known that the Ising Model in a Transverse Field on a hypercubic lattice in $D$ dimensions is equivalent to a classical Ising model in $D+1$ dimensions \cite{fradkin-1978,kogut-1979}. These two models are related through the transfer matrix. Indeed, a classical Ising model can be regarded as a path integral representation of the quantum model in one dimension less. For simplicity we will see how this work for the 2D the classical ferromagnetic Ising model of Eq.\eqref{eq:Ising}, but the construction is general. We will regard the configuration of spins on a row of the 2D lattice as a state of a quantum system, and the set of states on all rows as the  evolution of the state along the perpendicular direction that we will regard as a discretized imaginary time. The contribution from two adjacent rows to the partition function  defines the matrix element of a matrix between two arbitrary configurations. In Statistical Physics this matrix is known as the Transfer Matrix $\hat T$ and the full partition function (with periodic boundary conditions) is 
\begin{equation}
Z=\textrm{tr} {\hat T}^{N_\tau}
\label{eq:trace}
\end{equation}
where $N_\tau$ is the number of rows. For the case of the ferromagnetic Ising model (actually, for any unfrustrated model) the transfer matrix can always be constructed to be hermitian. This property holds in fact for any theory that satisfies a property known as reflection positivity which requires that all (suitably defined) correlation functions be positive. For theories of this type, and the Ising model is an example, the matrix elements of the transfer matrix can be identified with the matrix element of the evolution operator of a quantum theory for a small imaginary time step \cite{fradkin-1978}. Also, the positivity of the correlators is equivalent to the condition of positivity of the norm of states in the quantum theory. 

For the classical models that satisfy these properties, all directions of the lattice are equivalent. Moreover, asymptotically close to the critical point, the behavior of all the correlators becomes isotropic, i.e. invariant under the symmetries of Euclidean space. This means that the arbitrary choice of the direction for the transfer matrix is irrelevant. Consequently, tin the quantum model its equal-time correlators behave the same way as its correlation functions in imaginary time. In other words space and time  behave in the same way and the quantum theory is relativistically invariant. This implies at the quantum critical point the energy $\varepsilon({\bm p})$ of its massless excitations should behave as $\varepsilon({\bm p})=v \; |\bm p|$. In a relativistic theory the dynamical critical exponent must be $z=1$ and the coefficient $v$ is  the speed of the excitations (the ``speed of light''). We should note that this is not necessarily always the case. There are in fact classical systems, e.g. liquid crystals \cite{chaikin-1995}, which are spatially anisotropic and map onto quantum mechanical theories in one less dimension for which the dynamical critical exponent $z \neq 1$. One such example are the Lifshitz transitions of nematic liquid crystals in three dimensions and the associated quantum Lifshitz model in $D=2$ dimensions, for which the dynamical exponent is $z=2$ \cite{ardonne-2004}.

Just as in the classical counterpart in $D+1$ dimensions, the quantum model in $D\geq 1$ has two phases:  a broken symmetry ferromagnetic phase for $J\gg h$ and a symmetric paramagnetic phase for $h\gg J$. In the symmetric phase the ground state is unique (asymptotically is the eigenstate of $\sigma_1$ with eigenvalue $+1$), while in the broken symmetry phase the ground state is doubly degenerate (and asymptotically is an eigenstate of $\sigma_3$) and there is a non-vanishing expectation value of the local order parameter $\langle \sigma_3({\bm r})\rangle \neq 0$. In the symmetric phase the correlation function of the local order parameter decays exponentially with distance with a correlation length $\xi$, as does the connected correlation function in the broken symmetry phase. The model has a continuous quantum phase transition at a  critical value of the ratio $h/J$. For general space dimensions $D>1$ this model is not exactly solvable and much of what we know about it is due to large-scale numerical simulations. 

This problem was solved exactly  in one-dimension \cite{pfeuty-1970} using the Jordan-Wigner transformation that maps a one dimensional quantum spin system  to a theory of free fermions \cite{lieb-1961}. The fermion operators at site $j$ are
\begin{equation}
\chi_1(j)=K(j-1)  \sigma_3(j), \qquad  \chi_2(j)=i \hat K(j) \sigma_3(j)
\label{eq:majoranas}
\end{equation}
where $K(j)$ is the kink creation operator (i.e. the operator that creates a domain wall between sites $j$ and $j+1$ \cite{fradkin-1978}, and is given by
\begin{equation}
K(j)=\prod_{n\leq j} \sigma_1(n)
\label{eq:kink}
\end{equation}
The operators $\chi_1(j)$ and $\chi_2(j)$ are hermitian, $\chi^\dagger_j(n)=\chi_j(n)$, and obey the {\it anticommutation} algebra
\begin{equation}
\{ \chi_j(n), \chi_{j'}(n') \}= 2 \delta_{jj'} \delta_{nn'}
\label{eq:clifford}
\end{equation}
Hence, they are fermionic operators are hermitian, anti-commute with each other and square to the identity. Operators of this type are called {\it Majorana fermions}.

Alternatively, we can use the more conventional (Dirac) fermion operators $c(n)$ and its adjoint $c^\dagger(n)$ which are related to the Majorana fermions as
\begin{equation}
c(n)=\chi_1(n)+i \chi_2(n), \qquad c^\dagger(n)=\chi_1(n)-i\chi_2(n)
\label{eq:dirac-fermions}
\end{equation}
which obey the standard anticommutation algebra
\begin{equation}
\{ c(n), c(n') \}=\{ c^\dagger(n), c^\dagger)n') \}=0, \qquad \{ c(n), c^\dagger(n') \}=\delta-{nn'}
\label{eq:dirac-fermion-algebra}
\end{equation}
In this sense, a Majorana fermion is half of a Dirac fermion.

In terms of the Majorana operators the Hamiltonian of Eq.\eqref{eq:TFIM} becomes
\begin{equation}
H=i\sum_j \chi_1(j)\chi_2(j)+ig \sum_j\chi_2(j) \chi_1(j+1)
\label{eq:H-Majorana}
\end{equation}
where we have rescaled the Hamiltonian by a factor of $h$ and the coupling constant is $g=J/h$. Here we have not specified the boundary conditions (which depend on the fermion parity).  Qualitatively, the Majorana fermions can be identified with the domain walls of the classical models. In the Ising model the number of domain walls on each row is not conserved but their parity is. Likewise, the number of Majorana fermions $N_F$ is not conserved either but the fermion parity, $(-1)^{N_F}$, is conserved.

It is an elementary excercise to show that the spectrum of this theory has a gap $G(g)$ which vanishes at $g_c=1$ as $G(g) \sim |g-g_c|^\nu$, with an exponent $\nu=1$. Since the Hamiltonian of Eq.\eqref{eq:H-Majorana} is quadratic in the Majorana operators,  these operators obey linear equations of motion.  In the scaling regime  we take the limit of the lattice spacing $a \to 0$ and the coupling constant $g \to g_c=1$, while keeping the quantity $m=(g-g_c)/a$ fixed. In this regime, the two-component hermitian spinor field $\chi=(\chi_1,\chi_1)$, and $\chi^\dagger=\chi$, satisfies a Dirac equation
\begin{equation}
(i \slashed{\partial} -m) \chi=0
\label{eq:Majorana-Dirac}
\end{equation}
where we set the speed $\varv=1$, and where defined the $2 \times 2$ Dirac gamma-matrices $\gamma_0=\sigma_2$, $\gamma_1=i\sigma_3$, and $\gamma_5=\sigma_1$. Upon defining $\bar \chi=\chi^T \gamma_0$, we find that the Lagrangian of this field theory is
\begin{equation}
\mathcal{L}={\bar \chi} i \slashed{\partial} \chi-\frac{1}{2} m\; {\bar \chi} \chi
\label{eq:Majorana-Lagrangian}
\end{equation}
which indeed becomes massless at the quantum phase transition of the Ising spin chain. For these considerations, we say that the phase transition of the Ising model (2D classical or 1D quantum) is in the universality class of massless Majorana fermions where $m \to 0$. In Eq.\eqref{eq:Majorana-Lagrangian} we have used the standard Feynman slash notation, $\slashed{a}=\gamma_\mu a^\mu$, where $a^\mu$ is a vector.

\subsection{Quantum Antiferromagnets and Non-Linear Sigma Models}
\label{sec:nlsm}

As we noted above, the discovery of high temperature superconductors in the copper oxide compounds prompted the study of the behavior of these strongly correlated materials at low temperatures and of possible quantum phase transitions which they may host. The prototypical cuprate material  La$_2$CuO$_4$ is a quasi two-dimensional Mott insulator which exhibits long-range antiferromagnetic order below a critical temperature $T_c$. A simple microscopic model is a spin-S quantum Heisenberg antiferromagnet on the 2D square lattice of the Cu atoms, whose Hamiltonian is
\begin{equation}
H= \frac{1}{2} \sum_{ {\bm r}, {\bm r}'} J({|\bm r}-{\bm r}'|)\;  {\bm S}({\bm r}) \cdot {\bm S}({\bm r}')
\label{eq:2dHAFM}
\end{equation}
where ${\bm S}$ are the spin-S angular momentum operators.
We will consider the case where the exchange interaction for nearest neighbors  $J$ is dominant and a weaker $J' \ll J$ for next nearest neighbors. In this section we do not consider the  regime $J'\simeq J$ in which the interactions compete for incompatible ground states due to frustration effects. 

\subsubsection{Spin coherent states}
\label{sec:spin-coherent}

The simplest way to see the physics of this antiferromagnet is to construct a path-integral representation for a spin-$S$ system using spin coherent states \cite{perelomov-1986,fradkin-1988,wiegmann-1989}. For details see Ref.\cite{fradkin-2013} which we follow here. A coherent state of the ($2S+1$-dimensional) spin-$S$ representation of SU(2) is the state $|{\bm n}\rangle$, labeled by the spin polarization unit vector ${\bm n}$ 
\begin{equation}
|{\bm n}\rangle=e^{i \theta({\bm n}_0\times {\bm n}) \cdot {\bm S}}\; |S, S\rangle
\label{eq:spin-coherent-state}
\end{equation}\
where ${\bm n}^2=1$.
The states of the spin-$S$ representation are spanned by the eigenstates of $S_3$ and ${\bm S}^2$, 
\begin{equation}
S_3\; |S, M\rangle=M |S, M\rangle, \qquad {\bm S}^2 |S, M\rangle=S(S+1) |S, M\rangle
\label{eq:spin-S}
\end{equation}
and $|S, S\rangle$ is the highest weight state with eigenvalues $S$ and $S(S+1)$. In Eq.\eqref{eq:spin-coherent-state} ${\bm n}_0$ is a unit vector along the axis of quantization (the direction ${\bm e}_3$), and  $\theta$ is the colatitude, such that ${\bm n} \cdot {\bm n}_0=\cos \theta$. Two spin coherent states, $|{\bm n}_1\rangle$ and $|{\bm n}_2\rangle$, are not orthonormal,
\begin{equation}
\langle {\bm n}_1|{\bm n}_2\rangle=e^{i \Phi({\bm n}_1,{\bm n}_2,{\bm n}_0) \; S} \left(\frac{1+{\bm n}_1\cdot {\bm n}_2)}{2}\right)^S
\label{eq:ipr-spin}
\end{equation}
where $\Phi({\bm n}_1,{\bm n}_2,{\bm n}_0)$ is the area of the spherical triangle of the unit sphere spanned by the  unit vectors ${\bm n}_1$, ${\bm n}_2$ and ${\bm n}_0$. However,  there is an ambiguity in the definition of the area of the spherical triangle since the sphere is a 2-manifold without boundaries: if the ``inside'' triangle has spherical area $\Phi$, the complement (``outside'') triangle has area $4\pi-\Phi$. Thus, the ambiguity of the phase prefactor of Eq.\eqref{eq:ipr-spin} is 
\begin{equation}
e^{i 4\pi S}=1
\label{eq:ambiguity}
\end{equation}
since $S$ is an integer or a half-integer. So, the quantization of the representations of SU(2) makes the ambiguity unobservable. In addition, the spin coherent states $|{\bm n} \rangle$ satisfy the resolution of the identity
\begin{equation}
I=\int  |{\bm n}\rangle \langle {\bm n}|\; \left(\frac{2S+1}{4\pi}\right)  \delta({\bm n}^2-1)\; d^3n
\label{eq:resolution}
\end{equation}
and
\begin{equation}
\langle {\bm n}| {\bm S} |{\bm n}\rangle=S {\bm n}
\label{eq:me-spin}
\end{equation}

\subsubsection{Path integral for a spin-$S$ degree of freedom}
\label{sec:spin-S}

As an example consider problem of a spin-$S$ degree of freedom coupled to an external magnetic field ${\bm B}(t)$ that varies slowly in time. The (time-dependent) Hamiltonian is given by the Zeeman coupling
\begin{equation}
H(t)={\bm B}(t) \cdot {\bm S}
\label{eq:zeeman}
\end{equation}
As usual, the path-integral is obtained by inserting the (over-complete) set of coherent states at a large number of intermediate times. The resulting path integral is a sum of the histories of the spin polarization vector ${\bm n}(t)$
\begin{equation}
Z=\textrm{tr} \; \exp\left(i \int_0^T dt \; H(t)\right) =\int \mathcal{D} {\bm n}\; \exp\left(i\mathcal{S}[{\bm n}]\right)\; \prod_t \delta({\bm n}^2(t)-1)
\label{eq:spin-path-integral}
\end{equation}
where the action is
\begin{equation}
\mathcal{S}=S \mathcal{S}_{WZ}[{\bm n}]-S \int_0^T dt\; {\bm B}(t)\cdot {\bm n}(t)
\label{eq:spin-action1}
\end{equation}
where $\mathcal{S}_{WZ}[{\bm n}]$ is the Wess-Zumino action
\begin{align}
\mathcal{S}_{WZ}[\bm n]=&\int_0^T dt \; {\bm A}[\bm n]\cdot \partial_t {\bm n}\\
=&\int_0^1d\tau \; \int_0^T dt \; {\bm n}(t,\tau)\cdot \partial_t {\bm n}(t,\tau)\times \partial_\tau {\bm n}(t,\tau)
\label{eq:WZ}
\end{align}
where ${\bm A}[\bm n]$ is the vector potential of a Dirac magnetic monopole (of unit magnetic charge) at the center of the unit sphere. The vector potential ${\bm A}[\bm n]$ has a singularity associated with the Dirac string of the monopole. We can write an equivalent expression which is singularity-free using Stokes Theorem. We did this in the second line of Eq.\eqref{eq:WZ} which required to extend the circulation of ${\bm A}$ on the closed path described by ${\bm n}(t)$ to the flux of the vector potential through the submanifold $\Sigma$ of the unit sphere $S_2$  whose boundary is the history ${\bm n}(t)$, i.e. the area of $\Sigma$. The smooth (and arbitrary) extension of configuration ${\bm n}(t)$  to the interior of $\Sigma$ is done by defining ${\bm n}(t,\tau)$ such that ${\bm n}(t,1)={\bm n}_0$, ${\bm n}(t,0)={\bm n}(t)$, and ${\bm n}(0,\tau)={\bm n}(T,\tau)$. 
Since $\mathcal{S}_{WZ}$ is the area  of the submanifold $\Sigma$ of the unit sphere $S_2$, just as in Eq.\eqref{eq:ambiguity}, here too there is an ambiguity of $4\pi$ in the definition of the area. Here too, this ambiguity is invisible since the spin $S$ is an integer or a half-integer. 

The path integral of Eq.\eqref{eq:spin-path-integral} was derived first by Michael Berry \cite{berry-1984} (and extended by Barry Simon \cite{simon-1983}). The first term (which we called Wess-Zumino by analogy with its field theoretic versions) is called the {\it Berry Phase}. The role of this term, which is first order in time derivatives, is to govern the quantum dynamics of the spin which, in presence of a uniform magnetic field, executes a precessional motion of the (Bloch) sphere. It is also apparent from this expression that in the large-$S$ limit, the path integral can be evaluated by means of a semiclassical approximation.

The coherent-state construction shows that this problem is equivalent to the path integral of a formally  massless non-relativistic particle of unit electric charge on the surface of the unit sphere with a magnetic monopole of magnetic charge $S$ in its interior! This is not surprising since the Hilbert space of a non-relativistic particle moving on the surface of a sphere with and radial magnetic field (the field of a magnetic monopole) has a Landau level type spectrum with a degeneracy given by the flux. The condition of a massless particle means that only the lowest Landau level survives and all other levels have an infinite energy gap. 

The coherent state approach has been used to derive a path integral formulation for ferromagnets and antiferromagnets. A detailed derivation can be found in Ref. \cite{fradkin-2013}. 

\subsubsection{Quantum Ferromagnet}
\label{sec:QHF}

We will consider first the simpler case of a quantum ferromagnet and in Eq.\eqref{eq:2dHAFM} we will set $J=-|J|<0$ for nearest neighbors and zero otherwise. The action for the path-integral for the spin-$S$ quantum Heisenberg ferromagnet on a hypercubic lattice is
\begin{equation}
\mathcal{S}=S \sum_{\bm r} \mathcal{S}_{WZ}[{\bm n}({\bm r},t)]-\frac{|J|S^2}{2} \sum_{\langle {\bm r}, {\bm r}' \rangle} \int_0^T dt\; \left( {\bm n}({\bm r}, t)-{\bm n}({\bm r}',t)\right)^2
\label{eq:qfm}
\end{equation}
where we have subtracted the classical ground state energy. The oder parameter for this theory is the expectation value of the local magnetization, ${\bm n}=\langle {\bm n}({\bm r})\rangle$, which is constant in space but points in an arbitrary direction in spin space.

In the low energy regime the important configurations are slowly varying in space and we can simply approximate the action of Eq.\eqref{eq:qfm} by its continuum version in $d$ space dimensions 
\begin{equation}
\mathcal{S}=\frac{S}{a_0^d} \int d^dx\;  \; \mathcal{S}_{WZ}[{\bm n}({\bm x},t)]-\frac{|J|S^2}{2a_0^d} \int d^dx \int_0^Tdt \left( {\bm \bigtriangledown} {\bm n}({\bm x},t) \right)^2
\label{eq:qfm-continuum}
\end{equation}
where $a_0$ is the lattice spacing. As before, the path integral; is done for a field which satisfies everywhere in space-time the constraint ${\bm n}^2({\bm x},t)=1$. This action can be regarded as non-relativistic non-linear sigma model. 

It is straightforward to show that the classical equations of motion for this theory are the Landau-Lifshitz equations
\begin{equation}
\partial_t {\bm n}=|J|S a_0^2\; {\bm n} \times \bigtriangledown^2 {\bm n}
\label{eq:LLE}
\end{equation}
subject to the constraint ${\bm n}^2=1$. Due to the constraint, the Landau-Lifshitz equation is  non-linear. We will a decomposition of  the field into a longitudinal and two transverse components, $\sigma$ and ${\bm \pi}$, respectively
\begin{equation}
{\bm n}=
\begin{pmatrix}
\sigma\\
{\bm \pi}
\end{pmatrix}
\label{eq:parametrization}
\end{equation}
subject to the constraint $\sigma^2+{\bm \pi}^2=1$. The linearized Landau-Lifshitz equations become (to linear order in ${\bm \pi}$)
\begin{equation}
\partial_t \pi_1\simeq   -|J| Sa_0^2 \bigtriangledown^2\pi_2, \qquad
\partial_t  \pi_2\simeq  +|J|Sa_0^2 \bigtriangledown^2 \pi_1
\label{eq:linear-LLE}
\end{equation}
The solution to these equations are ferromagnetic spin waves (magnons or Bloch waves) which satisfy the dispersion relation
\begin{equation}
\omega({\bm p})\simeq |J|Sa_0^2 {\bm p}^2+O({\bm p}^4)
\label{eq:ferro-dispersion}
\end{equation}
which shows that the dynamic exponent for a ferromagnet is $z=2$. Notice that in this case the two transverse components are not independent (they are effectively a dynamical pair). These are the Goldstone bosons of a ferromagnet. 

\subsubsection{Quantum Antiferromagnet}
\label{sec:QHAF}

Formally, the quantum antiferromagnet has a coherent state path integral whose action is
\begin{equation}
\mathcal{S}=S \sum_{\bm r} \mathcal{S}_{WZ}[{\bm n}({\bm r},t)]-\frac{J S^2}{2} \sum_{\langle {\bm r}, {\bm r}' \rangle} \int_0^T dt\; {\bm n}({\bm r}, t) \cdot {\bm n}({\bm r}',t)
\label{eq:qafm-lattice-1}
\end{equation}
with $J>0$. For a bipartite lattice, e.g. the 1D chain, and the square and cubic lattices, the classical ground state is an antiferromagnet with a N\'eel order parameter, the staggered magnetization. Let ${\bm m}({\bm r})$ be the expectation value of the local magnetization. A bipartite lattice is the union of two interpenetrating sublattices,  and the local magnetization is staggered, i.e. it takes  values with opposite signs (with equal values) on the two sublattices. Thus, we make the change of variables, ${\bm n}({\bm r},t) \to (-1)^{\bm r} {\bm n}({\bm r},t)$ in Eq.\eqref{eq:qafm-lattice-1} and find 
\begin{equation}
\mathcal{S}=S \sum_{\bm r} (-1)^{\bm r} \mathcal{S}_{WZ}[{\bm n}({\bm r},t)]-\frac{J S^2}{2} \sum_{\langle {\bm r}, {\bm r}' \rangle} \int_0^T dt\; ({\bm n}({\bm r}, t) - {\bm n}({\bm r}',t))^2
\label{eq:qafm-lattice-2}
\end{equation}
We want to obtain the low energy effective action for the field ${\bm n}({\bm r},t)$. To this end, we decompose this field into a slowly varying part, that we will call ${\bm m}({\bm r},t)$, and a small rapidly varying part ${\bm l}({\bm r},t)$ (which represents ferromagnetic fluctuations)
\begin{equation}
{\bm n}({\bm r},t)={\bm m}({\bm r},t)+(-1)^{\bm r} a_0 {\bm l}({\bm r},t)
\label{eq:decomposition}
\end{equation}
Since ${\bm n}^2({\bm r},t)=1$, we will demand that the slowly varying part also obeys the constraint, ${\bm m}^2({\bm r},t)=1$, and require that the two components be orthogonal to each other, ${\bm m}\cdot {\bm l}=0$.

Due to the behavior of the staggered Wess-Zumino terms of Eq.\eqref{eq:qafm-lattice-2}, the resulting continuum field theory turns out to have subtle but important differences between one dimension and   higher dimensions. Here we will state the results for two and higher dimensions. We will discuss in detail the one-dimensional below when we discuss the role of topology.

It turns out that if the dimension $d>1$, the contribution of the staggered Wess-Zumino terms  for {\it smooth} field configurations is \cite{fradkin-1988,dombre-1988,haldane-1988}
\begin{equation}
\lim_{a_0 \to 0} S \sum_{{\bm r}} (-1)^{\bm r} \mathcal{S}_{WZ}[{\bm n}({\bm r},t)] =
S \int d^3x \; {\bm l}({\bm x},t)\cdot {\bm m}({\bm x},t) \times \partial_t {\bm m}({\bm x},t) 
\label{eq:cont-limit-2d-1}
\end{equation}
The continuum limit of the second term of Eq.\eqref{eq:qafm-lattice-2} in the case of a two-dimensional system is
\begin{equation}
\lim_{a_0\to 0} \frac{JS^2}{2} \sum_{\langle {\bm r},{\bm r}'\rangle} \int_0^Tdt\; \left({\bm n}({\bm r},t)-{\bm n}({\bm r}',t)\right)^2=
a_0 \frac{JS^2}{2} \int d^3x \; \left( ({\bm \bigtriangledown} {\bm m}({\bm x},t))^2+4{\bm l}^2({\bm x},t) \right)
\label{eq:cont-limit-2d-2}
\end{equation}
The massive field  ${\bm l}[{\bm x},t]$ represents ferromagnetic fluctuations. Since this  is a massive field  it can be integrated-out leading to an effective field theory for the {\it antiferromagnetic} fluctuations ${\bm m}({\bm x},t)$ whose Lagrangian is that of a non-linear sigma model
\begin{equation}
\mathcal{L}=\frac{1}{2g} \left(\frac{1}{\varv_s}(\partial_t {\bm m}({\bm x},t)-\varv_s ({\bm \bigtriangledown} {\bm m}({\bm x},t))^2\right)
\label{eq:nlsm-2d}
\end{equation}
where the coupling constant is $g=2/S$ and the spin-wave velocity is $\varv_s=4a_0JS$.  If we  to allow for a weak next-nearest-neighbor interaction $J'>0$, the coupling constant $g$ and the spin wave velocity $\varv_s$ become renormalized to $g'\simeq g/\sqrt{1-2J'/J}$ and $\varv_s'\simeq \varv_s \sqrt{1-2J'/J}$. 

We conclude that that the quantum fluctuations about a N\'eel state are well described by a  non-linear sigma model. Provided the frustration effects of  the next-nearest-neighbor interactions are weak enough, the  long-range antiferromagnetic N\'eel order should extend up to a critical value of the coupling constant $g_c$ where the RG beta function has a non-trivial zero, which signals a quantum phase transition to a strong coupling phase without long-range antiferromagnetic order. 

Motivated by the discovery of high temperature superconductivity in the strongly correlated quantum antiferromagnet La$_2$CuO$_4$ (at finite hole doping) in 1988 Chakravarty, Halperin and Nelson \cite{chakravarty-1988} utilized a quantum non-linear sigma model to analyze  this system and its quantum phase transition.  La$_2$CuO$_4$ is a quasi-two-dimensional material and so it exhibits strong quantum and thermal fluctuations. The upshot of this analysis is that while at $T=0$ the non-linear sigma model has a quantum phase transition, at $T>0$ the long range order is absent in a strictly 2D system but present in the actual material due to the weak-three-dimensional interaction. So, in the strict 2D case there is no phase transition but two different {\it crossover} regimes: a renormalized classical regime (without long range order), a quantum disordered regime and a quantum critical regime. La$_2$CuO$_4$ has long range N\'eel (antiferromagnetic) order at $T=0$ and is in the renormalized classical regime  (with long range order due to the weak 3D interaction).

The non-linear sigma model does not  describe the nature of the ground state for $g>g_c$ beyond saying that there is no long range order. The problem is that, unlike the Ising model in a transverse field, the microscopic tuning parameter is the next nearest neighbor antiferromagnetic coupling $J'$, and to reach the regime $g \simeq g_c$ one has to make $J' \simeq J$. This is the regime in which frustration effects become strong. In this regime the assumption that the important configurations are smooth and close to the classical N\'eel state is incorrect. The nature of the ground state turns out to depend on the value of $S$.

\section{Topological Excitations}
\label{sec:topological-excitations}

Topology has come to play a crucial role both in Condensed Matter Physics and in Quantum Field Theory. Topological concepts have been used to classify topological excitations such as vortices and dislocations and to provide a mechanism for phase transitions, quantum number fractionalization, tunneling processes in field theories, and nonperturbative construction of vacuum states. Here we will discuss a few representative cases of what has become a very vast subject.

\subsection{Topological Excitations: Vortices and Magnetic Monopoles}
\label{sec:topo-excitations}

In Condensed Matter Physics topological excitations play a central role in the description of topological defects and on their role in phase transitions. Here topology integers in the classification of the configuration space into equivalence classes characterized by topological invariants \cite{mermin-1979}. The most studied example are vortices. Vortices play a key role in the mixed phase of  type II superconductors in a uniform magnetic field \cite{abrikosov-1957}. Vortices also play a key role in the Statistical Mechanics of 2D superfluids and the the 2D classical XY model \cite{kosterlitz-1973,kosterlitz-1974} \cite{berezinskii-1971,berezinskii-1972} \cite{jose-1977}. Dislocations and disclinations play an analogous role in the theory of classical melting \cite{kosterlitz-1973,halperin-1978,young-1979}, and 2D and 3D classical liquid crystals \cite{toner-1981,nelson-1981,chaikin-1995}.  

A similar problem occurs in Quantum Field Theory. Theories with global symmetries, such as the two-dimensional $O(3)$ non-linear sigma model discussed above, when formulated in Euclidean space-time have {\it instantons}. Typically instantons are finite Euclidean action configurations, which are also classified into equivalence classes (associated with homotopy groups) labeled by topological invariants \cite{coleman-1985,rajaraman-1985}. Instantons play a central role in understanding the non-perturbative structure of gauge theories. Gauge theories with a compact gauge group coupled to matter fields have non-trivial vortex \cite{nielsen-1973} and monopole \cite{thooft-1976,polyakov-1977,polyakov-book} configurations, as do non-abelian Yang-Mills gauge theories \cite{callan-1976}. Instantons have also played a central role in Condensed Matter Physics as well, notably in Haldane's work  on 1D quantum antiferromagnets (discussed below), and in the problem of macroscopic quantum tunneling and coherence \cite{caldeira-1983,leggett-1987}.

\subsubsection{Vortices in two dimensions}
\label{sec:2d-vortices}

In this section I will focus on the the problem of the superfluid transition in 2D and the closely related problem of the phase transition of a magnet with an easy-plane anisotropy,  the classical XY model. 
A superfluid is described by an order parameter that is a one-component complex field $\phi({\bm x})$. If electromagnetic fluctuations are ignored, this description also applies to a superconductor. The complex field can be written in terms of an amplitude $|\phi({\bm x})|$, whose square represents the local superfluid density, and a phase $\theta({\bm x})=\textrm{arg}(\phi({\bm x}))$. Deep in the superfluid phase the amplitude is essentially constant, that we will set to be a real positive number $\phi_0$, while the phase field $\theta({\bm x})$ is periodic with period $2\pi$ and can fluctuate. Similarly, an easy-plane ferromagnet is described by a two-component real order parameter field ${\bm M}(\bm x)=(M_1({\bm x}),M_2(\bm x))=|{\bm M}({\bm x})|(\cos \theta(\bm x),\sin \theta(\bm x))$. Deep in the ferromagnetic phase the amplitude $|\bm M|$ is essentially constant but the phase field $\theta(\bm x)$ can fluctuate. 

We will assume that we are in a regime where the local superfluid density $|\phi_0|^2$ is well formed (or, equivalently that $|\bm M|$ is locally well formed) but that the phase field is fluctuating. In this regime the problem at hand is an $O(2)\simeq U(1)$ non-linear sigma model, and its partition function takes the form
\begin{equation}
Z=\int \mathcal{D}\theta \exp \left(-\int d^2x \; \frac{1}{2g}\left(\partial_\mu\theta({\bm x})\right)^2\right)
\label{eq:nlsm-U(1)}
\end{equation}
where we defined the coupling constant $g=T/J |\phi_0|^2$, where $T$ is the temperature, $J$ is an interaction strength, and $|\phi_0|^2$ is the magnitude (squared) of the amplitude of the order parameter, which we will take to be constant;  $\kappa=J|\phi_0|^2$ is the phase stiffness. 

Except for the requirement that the phase field be {\it locally} periodic, $\theta\simeq \theta +2\pi$, superficially  this  seems to be a trivial free (Gaussian) field theory. We will  see that the periodicity (or, {\it compactification}) condition makes this theory non-trivial. Indeed, configurations of the phase field that are weak enough that that do not see the periodicity condition, for all practical purposes, can regarded as being non-compact and ranging from $-\infty$ to $+\infty$. However there are many configurations for which the periodicity condition is essential. Such configurations are called {\it vortices}. 

Even in the absence of vortices, the periodic (compact) nature of the phase field is essential to the physics of this problem. In fact the only allowed observables must be invariant under local periodic shifts of the phase field. This implies that the phase field $\theta$  itself is not a physical observable but that exponentials of the phase of the form $\exp(i n \theta(\bm x))$ are physical. This operator is just the order parameter field of the $XY$ model. In Conformal Field theory  operators of this type are called vertex operators \cite{difrancesco-1997,ginsparg-1988}. We will see below that this theory has a dual field $\vartheta$, associated with vortices, and that there are vertex operators of the dual field. In String Theory the model of a compactified scalar is known as the compactified boson and represents the coordinate of a string on a compactified space, in this case a circle $S_1$ \cite{polchinski-book}.

To picture a vortex consider a large closed curve $C$ on the 2D plane. Hence, topologically a closed curve is isomorphic to a circle, $C \simeq S_1$. The phase field $\theta(\bm x)$ is equivalent to a unit circle $S_1$. Therefore the configuration space are maps of $S_1$ (the large circle) onto $S_1$ (the unit circle of the order parameter space). The configurations can be classified by the number of times the phase winds on the large circle $C$. The winding number is an integer called the topological charge of the configuration, the vorticity. Thus, a vortex is a configuration of the phase field $\theta(\bm x)$ that winds by $2\pi m$ (where $m$ is an integer):
\begin{equation}
\frac{(\Delta \theta)_C}{2\pi}=\frac{1}{2\pi} \oint_C d{\bm x}\cdot {\bm \bigtriangledown}\theta({\bm x})
					 \equiv  i \oint_0^{2\pi} \frac{d\varphi}{2\pi} \; e^{i \theta(\varphi)} \partial_\varphi e^{-i \theta(\varphi)}=m
\label{eq:vorticity}
\end{equation}
where $\varphi\in (0, 2\pi]$ is the azimuthal angle for a vector at the center of the large circle $C$. Here $n$ is the vorticity or {\it winding number} of the configuration; $m>0$ is a vortex and $m<0$ is an anti-vortex. The vorticity is a {\it topological invariant} of the field the configuration $\theta(\bm x)$ which does not change under smooth changes.

The winding number of a vortex is a topological invariant that classifies the configurations of the phase field as continuous maps of a large circle $S_1$ onto the unit circle $S_1$ defined by the phase field. In Topology such continuous maps are called homotopies. The winding number classifies these maps into a discrete set of equivalence classes, which form a {\it homotopy group} under the composition of two configurations. In this case the homotopy group is called $\Pi_1(S_1)$. Since the equivalence classes are classified by a topological invariant that takes integer values, the homotopy group $\Pi_1(S_1)$  is isomorphic to the group of integers, $\mathbb{Z}$ \cite{mermin-1979}.

The field $j_\mu(\bm x)= \partial_\mu \theta(\bm x)$ is  the superfluid current, and the {\it vorticity} $\omega(\bm x)$ is the curl of the current, i.e. 
\begin{equation}
\omega(\bm x)=\epsilon_{\mu \nu} \partial_\mu j_\nu(\bm x)=\epsilon_{\mu \nu} \partial_\mu \partial_\nu \theta(\bm x)
\label{eq:local-vorticity}
\end{equation}
which vanishes unless $\theta(\bm x)$ has a branch-cut singularity across which the phase field jumps by $2\pi n$.
Let $\omega(\bm x)$ be the vorticity field with singularities at the locations $\{ {\bm x}_j\}$ of vortices with topological charge $m_j$
\begin{equation}
\omega(\bm x)=2\pi \sum_j  m_j \delta^2({\bm x}-{\bm x}_j)
\label{eq:vortex-configs}
\end{equation}
which is satisfied by the phase field configuration
\begin{equation}
\theta(\bm x)=\sum_j 2\pi m_j \textrm{Im}\; \ln (z-z_j)
\label{eq:branch-cuts}
\end{equation}
where we have used the complex coordinates $z=x_1+ix_2$.
Away from the singularities $\{ {\bm x}_j \}$, this configuration obeys the Laplace equation. Hence, it has a Cauchy-Riemann dual field $\vartheta(\bm x)$ which satisfies the Cauchy-Riemnann equation
\begin{equation}
\partial_\mu \vartheta=\epsilon_{\mu \nu} \partial_\nu \theta
\label{eq:Cauchy-Riemann}
\end{equation}
which satisfies the Poisson equation
\begin{equation}
-\bigtriangledown^2 \vartheta(\bm x)=\omega(\bm x)
\label{eq:Poisson}
\end{equation}
whose solution is
\begin{equation}
\vartheta(\bm x)=\int d^2y \; G(|{\bm x}-{\bm y}|) \; \omega(\bm y)
\label{eq:inh-soln}
\end{equation}
where $G(|{\bm x}-{\bm y}|) $ is the Green function of the 2D Laplacian
\begin{equation}
-\bigtriangledown^2 G(|{\bm x}-{\bm y}|) =\delta^2({\bm x}-{\bm y})
\label{eq:Green-Function}
\end{equation}
In 2D this Green function is
\begin{equation}
G(|{\bm x}-{\bm y}|)=\frac{1}{2\pi}\ln \left(\frac{a}{|{\bm x}-{\bm y}|}\right)
\label{eq:GF-2D}
\end{equation}
where $a$ is a short distance cutoff (a lattice spacing). In what follows we will assume that the Green function of Eq.\eqref{eq:GF-2D} has been cutoff so that  $G(|{\bm x}-{\bm y}|)=0$ for $|{\bm x}-{\bm y}|\leq a$.

The energy of a configuration of vortices $\{ n_j \}$ with vanishing total vorticity, $\sum_j m_j=0$, is 
\begin{align}
E[\theta]=&\frac{J \phi_0^2}{2} \int d^2x \; \left(\partial_\mu \theta\right)^2 \nonumber\\
	     =&\frac{J\phi_0^2}{2} \int d^2x \int d^2y \; \omega(\bm x) \; G(|{\bm x}-{\bm y}|)\; \omega(\bm y) 
	     =2\pi J\phi_0^2 \sum_{j>k} m_j m_k \ln \left(\frac{a}{|{\bm x}_j-{\bm x}_k}\right)
   \label{eq:energy-vortices}
   \end{align}
   where we used that configurations with non-vanishing vorticity do not contribute to the partition function since they have infinite energy in the thermodynamic limit.    
   We conclude that, up to an unimportant  prefactor, that the partition function of Eq.\eqref{eq:nlsm-U(1)} is the same as the partition function a gas of charges $\{ m_j \}$ (the vortices) with total vanishing vorticity, $\sum_j m_j=0$,
\begin{equation}
Z_{2DCG}= \sum_ {[\{ m_j \}]}   \exp \left(-2\pi \frac{J\phi_0^2}{T} \sum_{j<k} m_j m_k \ln \left(\frac{|{\bm x}_j-{\bm x}_k|}{a}\right)\right)
\label{eq:2DCG}
\end{equation}
which is known as the two-dimensional (neutral) Coulomb gas \cite{kosterlitz-1973}.

Kosterlitz and Thouless argued that the neutral two dimensional Coulomb gas has a phase transition a critical temperature $T_{KT}$ between a low temperature phase dielectric phase in which vortices and anti-vortices are bound into neutral pairs and a high temperature phase in which pairs unbind, the vortex charges are screened and the vortex gas is in a plasma phase. A simple estimate of the critical temperature is found by computing the free energy of a single vortex $F_{\rm vortex}=E_{\rm vortex}-TS_{\rm vortex}$, where $E_{\rm vortex}$ of a single vortex and $S_{\rm vortex}$ is the vortex entropy. In 2D the energy of a single vortex of unit topological charge, $n=1$, is $\pi J\phi_0^2 \ln (L/a)$, where $L$ is the linear size of the system, and the vortex entropy is $S_{\rm vortex}=\ln(L/a)^2$, where $(L/a)^2$ is the number of places where a vortex can be located, i.e. the area of the system in units of the spacing $a$. Vortices unbind (proliferate) when the entropy beats the energy. Hence, the critical point occurs at a temperature $T_{KT}$ where the vortex free energy vanishes, $F_{\rm vortex}[T_{KT}]=0$. This yields the estimate 
\begin{equation}
T_{KT}= \frac{\pi}{2} \; J\phi_0^2
\label{eq:T_KT}
\end{equation}
which is the Kosterlitz-Thouless critical temperature.

There is an alternative, equivalent way to see this physics. Let is consider the theory if Eq.\eqref{eq:nlsm-U(1)} in the background of a U(1) gauge field $A_\mu(x)$ 
\begin{equation}
Z[A]=\int \mathcal{D}\theta \; \exp\left(-\frac{1}{2g}\int d^2x \left(\partial_\mu \theta-A_\mu\right)^2\right)
\label{eq:Z[A]}
\end{equation}
where the curl of the background gauge field $A_\mu$  is chosen to be equal to the local vorticity of Eq.\eqref{eq:local-vorticity}, 
\begin{equation}
\epsilon_{\mu \nu} \partial_\mu A_\nu(x)=\omega(x)
\label{eq:omega}
\end{equation}
In other words, the gauge $A_\mu$ field forces the phase field to have vortices at the locations $\{ x_j \}$.

Next we will use a Hubbard-Stratonovich transformation (a Gaussian identity) to write Eq.\eqref{eq:Z[A]} in the equivalent form
\begin{equation}
Z[A]=\int \mathcal{D} \theta \int \mathcal{D} a_\mu \; \exp\left(-\frac{g}{2} \int d^2x \; a_\mu^2+i \int d^2x \; a_\mu(x) \left(\partial_\mu \theta(x)-A_\mu(x)\right)\right)
\label{eq:mixed-A-theta}
\end{equation}
where we neglected an unimportant prefactor. In Eq.\eqref{eq:mixed-A-theta} the phase field can be integrated-out  (after an integration by parts) to yield the constraint $\partial_\mu a_\mu=0$. This constraint is solved by introducing the dual field $\vartheta(x)$ such that
\begin{equation}
a_\mu(x)=\epsilon_{\mu \nu}\partial_\nu \vartheta(x)
\label{eq:2D-duality-1}
\end{equation}
Hence, the partition function $Z[A]$ becomes
\begin{equation}
Z[A]=\int \mathcal{D} \vartheta \; \exp\left(-\frac{g}{2}\int d^2x \; (\partial_\mu \vartheta)^2+i \int d^2x \; \vartheta(x)\; \omega(x)\right)
\label{eq:2D-duality-2}
\end{equation}
where we have assumed that the total vorticity is zero, $\int d^2 x \; \omega(x)=2\pi \sum_j m_j=0$. This identity show that the quantization of the vortex topological charge requires that the dual field $\vartheta$ should also be periodic (compact) with period $1$.

It is elementary to see that after performing the gaussian integral over the field $\vartheta$ in Eq.\eqref{eq:2D-duality-2} and summing over all vortex configurations (with total vanishing vorticity) we reproduce the partition function of the 2D neutral Coulomb gas \cite{kadanoff-1978,nienhuis-1987}. Thus, the theory of Eq.\eqref{eq:Z[A]} defined in terms of the phase field $\theta$ and the theory of Eq.\eqref{eq:2D-duality-2} are equivalent. This is an example of a {\it duality transformation} which amounts to the exchange of the fields 
\begin{equation}
\theta \leftrightarrow \vartheta, \qquad \textrm{and} \qquad g \leftrightarrow 1/g
\label{eq:particle-vortex-duality-2D}
\end{equation}
A more careful analysis done on the lattice shows that the dual field $\vartheta$ is defined on the dual of the square lattice (which is also a square lattice) \cite{jose-1977,kadanoff-1978}. In addition, by differentiating both the partition function of the phase field of Eq.\eqref{eq:Z[A]} and the partition function of the dual field $\vartheta$ of Eq.\eqref{eq:2D-duality-2} with respect to the background field $A_\mu$ we obtain the duality identification of the currents
\begin{equation}
\frac{1}{g} \partial_\mu \theta \longleftrightarrow g \epsilon_{\mu \nu} \partial_\nu \vartheta
\label{eq:duality-currents}
\end{equation}
as an operator identity. This identity has the same form as the duality of forms in geometry. In this case, the dual of the current $\partial_\mu \theta$, which is a 1-form field, is another 1-form field, $\partial_\mu \vartheta$. In both cases the field $\theta$ and its dual $\vartheta$ are associated with global symmetries. Later we will encounter similar duality identifications but their content will be different in different dimensions.
Moreover, running these identities backwards it is easy to see that the insertion of the order parameter operator $\exp(i p\theta(x))$ (with $p$ and integer), which acts as a {\it source} on the phase field $\theta$, in the partition function is is equivalent to minimally couple the dual field $\vartheta$ to a singular gauge field $B_\mu(x)$ whose curl is $p$ at the location $x$. In other words, {\it charges are dual to vortices}.

We will now use the identity we derived in Eq.\eqref{eq:2D-duality-2} to compute the full partition function which is a sum over all vortex configurations. Now we will assume that the vortices have a large core energy (i.e. the energy associated with the short distance behavior of the Green function of Eq.\eqref{eq:GF-2D}). For a vortex with unit topological charge we will call this energy $u$, and for a vortex with charge $m$ the core energy will be $u m^2$. The effects of the core energy can be accounted for in terms of a vortex fugacity $z=\exp(-un^2/T)$. The total partition function now is
\begin{equation}
Z=\sum_{\{ m_j \}} \; \int \mathcal{D} \vartheta \; \exp\left(-\frac{g}{2} \int d^2x\; (\partial_\mu \vartheta)^2+i\sum_j 2\pi m_j \vartheta(x_j)-\frac{u}{T} \sum_j m_j^2\right) 
\label{eq:almost-SG}
\end{equation}
where, once again, we assumed global charge neutrality. 

In the limit in which $z=\exp(-u/T) \ll 1$, only the vortices with the lowest topological charges $m=\pm 1$ contribute and, furthermore, they  are dilute. In this limit we can perform over the  configurations with the lowest topological charge and derive the effective field theory of the dual field $\vartheta$:
\begin{equation}
Z_{\rm SG}=\int \mathcal{D} \vartheta \exp \left(- \int d^2x \left(\frac{g}{2}(\partial_\mu \vartheta)^2-\varv \cos (2\pi \vartheta) \right)\right)
\label{eq:SG}
\end{equation}
which is known as the sine-Gordon field theory. The coupling constant of the sine-Gordon theory is $\varv=2\exp(-u/T)/a^2$, where $a$ is the core size (the lattice spacing). Hence, the cosine operator of the sine-Gordon theory represents the effects of the vortices.

We can now analyze the role of vortices by looking at the renormalization group of the sine-Gordon theory. The first question is what is the scaling dimension of the cosine operator. This can be computed by looking at the vortex operator correlator in the theory with $\varv=0$:
\begin{equation}
\langle \exp(i 2\pi \vartheta(x)) \exp(-i 2\pi \vartheta (y))\rangle=\frac{\rm const.}{|x-y|^{2\pi/g}}
\label{eq:vortex-correlator}
\end{equation}
which implies that the vortex scaling dimension is 
\begin{equation}
\Delta_{\rm vortex}=\frac{\pi}{g}
\label{eq:vortex-scaling-dimension}
\end{equation}
 Eq.\eqref{beta-functions-tree-level} implies that in $D=2$ dimensions the operator is irrelevant if $\Delta>2$, relevant for  $\Delta<2$ and marginal for $\Delta=2$. Hence, vortices are marginal if $\Delta_{\rm vortex}=2$ which happens if $g=\pi/2$. This is the same as to say that the system is at the Kosterlitz-Thouless critical temperature $T=T_{KT}$. Hence, vortices are irrelevant if $T<T_{KT}$ and relevant for $T> T_{KT}$. 
 
 This RG analysis tells us that $T>T_{KT}$, when vortices proliferate, the coupling constant $\nu$ flows to strong coupling to a regime where $\vartheta$ is pinned to an integer value and the theory is in a massive phase. In this phase the connected vortex correlator decays exponentially with distance which is the same as to say that the vortex charge is screened. This is why in this phase the vortices proliferate. It can be shown that in this phase the correlator of the spins of the $XY$ model, i.e. the correlator of the vertex operator $\exp(i \theta(x))$, decays exponentially with distance and the systems is in its disordered phase. 
 
 There is still the question of the nature of the phase with $T<T_{KT}$. The Mermin-Wagner Theorem \cite{mermin-1966} (and its generalizations by Hohenberg \cite{hohenberg-1967} and Coleman \cite{coleman-1973}) states that classical statistical mechanical systems with a global continuous symmetry group cannot undergo spontaneous symmetry breaking in space dimensions $D\leq 2$ (and quantum systems with space-time dimensions $D \leq 2$). Does this theory violate this the Mermin-Wagner Theorem? The answer is no. It is easy to see that in the phase in which the vortices are irrelevant, i.e. for $T<T_{KT}$, the correlator of the order parameter operator is always a power law of the distance $|x-y|$,
 \begin{equation}
 \langle \exp(i \theta(x)) \exp(-i\theta(y))\rangle=\frac{\rm const.}{|x-y|^{g/2\pi}}
 \label{eq:spin-spjn}
 \end{equation}
with an exponent that depends on temperature and satisfies   \begin{equation}
\frac{g}{2\pi}=\frac{T}{2\pi J\phi_0^2} \leq \frac{1}{4}
\label{eq:2D-inequality}
\end{equation}
In other words, the entire low temperature phase is not an ordered phase of matter since the correlator is not constant at long distance. On the other hand, the correlators of  the order parameter $\exp(i\theta)$ and of the vortices $\exp(i \vartheta)$ have a power la behavior for $T<T_{KT}$, we conclude that in this temperature range the system is scale invariant and that it has a line of critical points. 

In summary, we succeeded in expressing the partition function as a sum over configurations of the topological excitations, the vortices. We succeeded in doing that because vortices are labeled by their coordinates on the plane. In addition, we found a non-trivial phase transition since the entropy and the energy both scale logarithmically with the linear size of the system or, equivalently, that vortices became marginally relevant at a critical temperature. This is the mechanism behind the Kosterlitz-Thouless transition.

One may ask if this construction is generic and the answer is no. As an example consider the Abelian Higgs model in $D=2$. This model has a complex scalar field minimally coupled to a Maxwell gauge field and, hence, its gauge group is U(1). In the classical spontaneously broken phase, the gauge field becomes massive. In this phase the long range coherence of the phase field of the vortices of the scalar field is screened  at the scale of the penetration depth. As a result the Euclidean action of the vortices is now finite. Furthermore, on longer scales the interaction energy between vortices becomes short ranged. Hence, instead of a 2D Coulomb gas now one has a gas of particles (vortices and anti-vortices) with short range interactions. In this case the entropy always dominates and the vortices proliferate \cite{schaposnik-1978}. This behavior is quite analogous to the restoration of symmetry by proliferation of domain walls in one-dimensional classical spin chains \cite{landau-1959} and to the analogous problem of tunneling in the path integral formulation of quantum mechanical double-well potentials \cite{coleman-1985}. As a caveat, we should note that, in spite of the obvious similarities, the 2D Abelian Higgs model does not describe correctly a 2D superconductor (i.e. a superconducting film) since the electromagnetic field is not confined to the film and this renders the electromagnetic action nonlocal.
 
\subsubsection{Magnetic monopoles in compact electrodynamics}
\label{sec:monopoles-3D}

Instantons are of great interest in Quantum Field Theory since they provide for a mechanism to understand the non-perturbative structure of these theories. For this reasons they have been used to understand the mechanisms of quark confinement and the role of quantum anomalies in non-Abelian gauge theories \cite{callan-1976,polyakov-1977,polyakov-book,thooft-1976,thooft-1976b}. Instantons in non-Abelian gauge theories are magnetic monopoles and the condensation of monopoles have long been argued to be the mechanism behind quark confinement.

The simplest non-Abelian gauge theory that has magnetic monopoles is the Georgi-Glashow \cite{georgi-1974}. This model has a three-component real field ${\bm \phi}$ and an SU(2) Yang-Mills  gauge field ${\bm A}_\mu$. In its Higgs phase the scalar field $\bm \phi$ acquires an expectation value which breaks the gauge symmetry group SU(2) down to its diagonal U(1) subgroup. Since U(1) $\subset$ SU(2), this Abelian gauge group is compact, meaning that its magnetic fluxes are quantized. Polyakov \cite{polyakov-1975b} and 't Hooft \cite{thooft-1976b} showed that in 2+1 Euclidean dimensions have non-singular instanton solutions which at long distances resemble the magnetic monopole originally proposed by Dirac in 1931 \cite{dirac-1931}
\begin{equation}
B_i(\bm x)=\frac{q}{2} \frac{x_i}{|\bm x|^2} -2\pi q \delta_{i,3} \delta(x_1) \delta(x_2) \theta(-x_3)
\label{eq:dirac-monopole}
\end{equation}
The first term in Eq.\eqref{eq:dirac-monopole} is the magnetic radial field of a monopole of magnetic charge $q$. The second term represents an infinitely long infinitesimally thin solenoid ending at the location of the monopole, ${\bm x}={\bm 0}$, that supplies the quantized magnetic flux $2\pi q$. This singular term is known as the Dirac string. The string itself (and its orientation) is physically unobservable to any electrically charged particle that obeys the Dirac quantization condition, $qe=2\pi$ (in units where $\hbar=c=1$).

In the language of a lattice gauge theory \cite{wilson-1975}, a theory with a compact (i.e. periodic) U(1) gauge fields on a $D=3$ cubic lattice, describing a compact gauge field in $2+1$ dimensions.  This theory should have instantons that resemble magnetic monopoles much in the same way as a theory with a compact global U(1) symmetry has vortices. The simplest example is Polyakov's compact electrodynamics \cite{polyakov-1977} whose partition function is
\begin{equation}
Z=\prod_{x,\mu} \int_{0}^{2\pi} \frac{dA_\mu(x)}{2\pi} \; \exp\left(\frac{1}{4e^2}\sum_{x,\mu,\nu}\cos(F_{\mu \nu}(x))\right)
\label{eq:compact-qed}
\end{equation}
where $F_{\mu \nu}(x)=\Delta_\mu A_\nu(x)-\Delta_\nu A_\mu(x) \equiv \sum_{\mu} A_\mu$ is the magnetic flux through the elementary plaquette   labeled by a site $x$ and a pair of directions, $\mu$ and $\nu$, with $\mu=1,2,3$. This theory is invariant under local gauge transformations $A_\mu(x) \to A_\mu(x)+\Delta_\mu \Phi(x)$ and it is also invariant under local periodic shifts of the gauge fields $A_\mu(x) \to A_\mu(x)+2\pi \ell_\mu(x)$, where $\ell_\mu(x) \in \mathbb{Z}$. The plaquette flux operator satisfies the lattice version of the Bianchi identity that the product of exponentials of the flux on the faces of every elementary cube of the lattice is
\begin{equation}
\prod_{\rm cube faces} e^{iF_{\mu \nu}(x)}=1
\label{eq:latt-bainchi}
\end{equation}
which says that the theory can have magnetic monopoles of integer magnetic charge.

We will analyze this theory following an approach analogous to what we used for vortices in section \ref{sec:2d-vortices}.To this end we will consider the partition function
\begin{equation}
Z[B_{\mu \nu}]=\int \mathcal{D}A_\mu \; \exp\left(-\frac{1}{4e^2} \int d^3x \; \left(F_{\mu \nu}(x)-B_{\mu \nu} (x)\right)^2\right)
\label{eq:Z[B]}
\end{equation}
where $F_{\mu \nu}(x)=\partial_\mu A_\nu -\partial_\nu A_\mu$ is the field strength of the abelian U(1) Maxwell gauge field $A_\mu$. Here $B_{\mu \nu}(x)$ is an (anti-symmetric) two-form background gauge field. The coupling constant of this theory is $e^2$. Since $A_\mu$ is a connection it has units of length$^{-1}$, and $F_{\mu \nu}^2$ is a dimension $4$ field.  Then, in $D=3$ dimensions, $e^2$ has units of length$^{-1}$.

The theory is invariant under two local transformations, namely the usual invariance under gauge transformations
\begin{equation}
A_\mu(x) \to A_\mu(x)+\partial_\mu \Phi(x), \qquad B_{\mu \nu}(x) \to B_{\mu \nu}(x)
\label{eq:U1-gauge}
\end{equation}
where $\Phi(x)$ is an arbitrary smooth function of $x$. The presence of the background two-form field $B_{\mu \nu}$ now requires  invariance under one-form gauge transformations
\begin{equation}
A_\mu(x) \to A_\mu(x)+a_\mu(x), \qquad B_{\mu \nu}(x)\to B_{\mu \nu}(x)+\partial_\mu a_\nu -\partial_\nu a_\mu
\label{eq:2-form-gauge-invariance}
\end{equation}

The two-form gauge field $B_{\mu \nu}$ essentially represents the magnetic monopoles. Let $\{ m_j \}$ be a configuration of monopoles of charges $m_j$ with coordinates $\{ x_j \}$, with total vanishing monopole charge, $\sum_j m_j=0$. Let $\mathcal{M}(x)$ be the magnetic monopole density at $x$,
\begin{equation}
\mathcal{M}(x)=2\pi \sum_j \; m_j \; \delta^3 (x-x_j)
\label{eq:monople-configuration}
\end{equation}
which can be expressed as the curl of the  two-form gauge field $B_{\mu \nu}$,
\begin{equation}
\mathcal{M}(x)=\frac{1}{2}\epsilon_{\mu \nu \lambda} \partial_\mu B_{\nu \lambda}(x)
\label{eq:curl-B}
\end{equation}

We will proceed next much in the same way as in Eq.\eqref{eq:mixed-A-theta} and rewrite the partition function of Eq.\eqref{eq:Z[B]} in terms of a two-form Hubbard-Stratonovich field $b_{\mu \nu}(x)$ such that
\begin{align}
Z[B]=&\int \mathcal{D}A_\mu \int\mathcal{D}b_{\mu \nu} \exp\left(-\frac{e^2}{4} \int d^3x\; b_{\mu \nu}^2(x)+i \int d^3x \; \frac{1}{2} b_{\mu \nu}(x) \left[F_{\mu \nu}(x)- B_{\mu \nu}(x)\right]\right)
\nonumber\\
=&\int \mathcal{D}A_\mu \int\mathcal{D}b_{\mu \nu} \exp \left(-\frac{e^2}{4} \int d^3x\; b_{\mu \nu}^2(x) +i  \int d^3x \left[A_\mu (x) \partial_\nu b_{\mu \nu}(x)  -\frac{1}{2} b_{\mu \nu} (x) B_{\mu \nu}(x)\right]\right)
\label{eq:mixed-B-A}
\end{align}
Thus, the gauge field $A_\mu$ plays the role of a Lagrange multiplier field the enforces the constraint
\begin{equation}
\partial_\nu b_{\mu \nu}(x)=0
\label{eq:2-form-constraint}
\end{equation}
which is solved in terms of a compact scalar field $\vartheta(x)$
\begin{equation}
b_{\mu \nu}(x)= \epsilon_{\mu \nu \lambda} \partial_\lambda \vartheta(x)
\label{eq:dual-3D-1}
\end{equation}
Using this identity and the definition of the monopole density $\mathcal{M}(x)$ we find that the partition function $Z[B_{\mu \nu}]$ of Eq.\eqref{eq:Z[B]} becomes
\begin{align}
Z[B]=&\int \mathcal{D}\vartheta \; \exp\left[-\int d^3x \left(\frac{e^2}{2} (\partial_\mu \vartheta(x))^2+ i \mathcal{M}(x) \vartheta(x)\right)\right]\nonumber\\
=& \int \mathcal{D}\vartheta \; \exp\left[- \frac{e^2}{2} \int d^3x (\partial_\mu \vartheta(x))^2+ 2\pi i \sum_j m_j \vartheta(x_j)  \right]
\label{eq:dual-3D-2}
\end{align}
which requires that the field $\vartheta$ obeys the compactification condition $\vartheta \to \vartheta+n$, where $n$ is an arbitrary integer. Eq.\eqref{eq:dual-3D-2} says that the magnetic monopole instantons of the compact U(1) gauge theory are dual to charges of the dual phase field $\vartheta$, which has a  compact U(1) global symmetry.

The full partition function is obtained by summing over all monopole configurations satisfying the total neutrality condition, $\sum_j m_j=0$. As in section section \ref{sec:2d-vortices}, we will weigh the configurations with a coupling $u$ and find
\begin{equation}
Z=\sum_{ \{ m_j \} } Z{\{ m_j \} } \int \mathcal{D}\vartheta\; \exp \left(-\frac{e^2}{2} \int d^3x\; (\partial_\mu \vartheta)^2+ \sum_j 2\pi i m_j \vartheta(x_j)-u \sum_j m_j^2\right)
\label{eq:almost-SG-3D}
\end{equation}
which is the same theory we found  in Eq.\eqref{eq:almost-SG} except that now we are in 3D. Moreover, summing only over dilute configurations of monopoles and anti-monopoles we find, once again the sine-Gordon theory but now in $D=3$ dimensions:
\begin{equation}
Z=\int \mathcal{D} \vartheta \exp \left(-\int d^3x \;\left(\frac{e^2}{2} (\partial_\mu \vartheta)^2-\varv \cos(2\pi \vartheta)\right)\right)
\label{eq:SG-3D}
\end{equation}
with $\varv=2\exp(-u)/a^3$.

In spite of the similarities between Eq.\eqref{eq:SG-3D} and the sine-Gordon theory in 2D, Eq.\eqref{eq:SG}, the physics is very different. It is straightforward to see that, in the limit $v=0$, the monopole operator correlator is
\begin{equation}
\langle \exp\left(2\pi i \vartheta(x)) \exp(-2\pi i \vartheta(y)\right)=  \exp\left(\frac{4\pi^2}{e^2} [G(|x-y|)-G(0)]\right)\simeq \exp\left(\frac{\pi}{2e^2}\left[\frac{1}{R}-\frac{1}{a}\right]\right)
\label{eq:monopole-correlator}
\end{equation}
where $a$ is the short-distance cutoff. Unlike the behavior of the correlator of the vortex operators in 2D found in Eq.\eqref{eq:vortex-correlator}, Eq.\eqref{eq:monopole-correlator}  does not show a power-law behavior. The reason is that at $\varv=0$  the compactified field $\vartheta$ should be regarded as the Goldstone boson of a spontaneously broken U(1) symmetry. However, the cosine operator is now always relevant and the field $\vartheta$ is  pinned and its fluctuations are  actually massive.

Looking back at the partition function of Eq.\eqref{eq:almost-SG-3D}, we could integrate out the field $\vartheta$ and obtain an expression with the same form as the Coulomb gas of Eq.\eqref{eq:2DCG} except that now this is the three-dimensional neutral Coulomb gas \cite{polyakov-1977}
\begin{equation}
Z_{3DCG}= \sum_ {[\{ m_j \}]}   \exp \left(-\frac{\pi}{2e^2} \sum_{j<k} m_j m_k \left[\frac{1}{|{\bm x}_j-{\bm x}_k|}-\frac{1}{a}\right] \right)
\label{eq:3DCG}
\end{equation}
Now the energy of an isolated monopole is finite (in the infrared) while the entropy is still logarithmically divergent. The conclusion is that the monopoles proliferate and the 3D neutral Coulomb gas is always in a plasma phase withe screened interactions. These results also imply that  the expectation value of the Wilson loop operator in the U(1) gauge theory decays exponentially with the {\it area} of the loop. The conclusion is that in 2+1 dimensions compact electrodynamics is in a confining phase for all values of the coupling constant $e^2$ \cite{polyakov-1977,polyakov-book}.

\subsection{Non-Linear Sigma Models and Antiferromagnetic Quantum Spin Chains}
\label{sec:quantum-spin-chains}

We will now discuss the one-dimensional case in which topology plays a crucial role. In one space dimension, the sites of the chain are labelled by an integer $j=1, \ldots, N$ (with $N$ even). After some simple manipulations, the continuum limit of the staggered Wess-Zumino terms of Eq.\eqref{eq:qafm-lattice-2} is
\begin{align}
\lim_{a_0 \to 0} S \sum_{j=1}^N (-1)^j \mathcal{S}_{WZ}[{\bm n}(j,t)] =&
\frac{S}{2} \int d^2x \; {\bm m}(x,t)  \cdot \partial_t {\bm m}(x,t) \times \partial_x {\bm m}(x,t) \nonumber \\
&+S \int d^2x \; {\bm l}(x,t)\cdot {\bm m}(x,t) \times \partial_t {\bm m}(x,t) 
\label{eq:cont-limit-1}
\end{align}
The continuum limit of the second term of Eq.\eqref{eq:qafm-lattice-2} is essentially the same as in Eq.\eqref{eq:cont-limit-2d-2}.

By putting together the results of Eq.\eqref{eq:cont-limit-1} and Eq.\eqref{eq:cont-limit-2d-2}, we find that  in the case of a 1D chain the continuum field theory has the Lagrangian density
\begin{equation}
\mathcal{L}[{\bm m},{\bm l}]=-2a_0JS^2{\bm l}^2+S{\bm l}\cdot {\bm m} \times \partial_t {\bm m}-\frac{a_0JS^2}{2} (\partial_x {\bm m})^2+\frac{S}{2} {\bm m} \cdot \partial_t {\bm m} \times \partial_x {\bm m}
\label{eq:cont-limit-2}
\end{equation}
Finally, we integrate-out the  ferromagnetic fluctuations represented by the massive field ${\bm l}(x,t)$ to find that the effective low-energy Lagrangian has the form 
\begin{equation}
\mathcal{L}=\frac{1}{2g} \left(\frac{1}{\varv_s} \left(\partial_t {\bm m}\right)^2-\varv_s \left(\partial_x {\bm m}\right)^2\right)+\frac{\theta}{4\pi}{\bm m} \cdot \partial_x {\bm m} \times \partial_t {\bm m}
\label{eq:haldane}
\end{equation}
which is  a non-linear sigma model with $N=3$ components (and hence has a global $O(3)$ symmetry) with a {\it topological term} whose coupling constant is $\theta$. This result
was first derived by Haldane \cite{haldane-1985}. 

In Eq.\eqref{eq:haldane} the effective  coupling constant $g$, the spin-wave velocity  $\varv_s$, and the $\theta$-angle are given by $g=\frac{2}{S}$, $\varv_s=2a_0JS$, and $\theta=2\pi $.
As in the 2D case, since $g=2/S$, large $S$ implies that the non-linear sigma model is in the weak coupling regime. 
 Provided that $\varv_s>0$, one can always rescale the time and space coordinates without affecting the form of the Lagrangian, including the coupling constant  or, as we will see, the value of the $\theta$ angle. 
 In what follows we will assume that we have done the rescaling in such a way that we set $\varv_s=1$ and, that time and space scale as lengths. Thus we will  use a relativistic notation and, after an analytic continuation to imaginary time, we  label the time coordinate by $x_2$ and the space direction by $x_1$. The partition function now takes the form
 \begin{equation}
 Z=\int \mathcal{D}{\bm m} \exp\Big(-\frac{1}{2g}\int_\Omega d^2x \;  \left(\partial_\mu {\bm n}\right)^2+i\theta\, \mathcal{Q}[\bm m]\Big)
 \label{eq:nlsm+theta}
 \end{equation}
 where $\Omega$ is the spacetime manifold.
 In this notation the first term of the exponent is called the Euclidean action of the non-linear sigma model. In Eq.\eqref{eq:nlsm+theta} we denoted by  $\mathcal{Q}[\bm m]$ the quantity
\begin{equation}
\mathcal{Q}[\bm m]=\frac{1}{8\pi} \int _\Omega d^2x\; \epsilon_{\mu \nu} {\bm m} \cdot \partial_\mu {\bm m} \times \partial_\nu {\bm m}
\label{eq:O(3)-top-inv}
\end{equation}

Here we will consider the case in which the spacetime manifold $\Omega$ is closed. In particular we will assume that it is a two-sphere $S_2$.
The quantity $\mathcal{Q}[\bm m]$  is the integral of a total derivative which counts the number of times the field configuration ${\bm m}(x_1,x_2)$ wraps around the sphere $S_2$. In other words, it yields a non-vanishing result only for ``large'' configurations which wind (or wrap) around the sphere $S_2$. Since $\mathcal{Q}[\bm n]$ is an integer, it has the same for all {\it smooth} the field configuration $\bm m$ that can be smoothly deformed into each other and are homotopically equivalent. If we demand that the field configurations $\bm m$ have finite Euclidean action, which requires that at infinity the configurations take the same (but arbitrary) value of $\bm m$, we have effectively compactified the $x_1-x_2$ plane into a two sphere $S_2$. On the other hand the field $\bm m$ is restricted by the constraint ${\bm m}^2=1$  to take values on a two-sphere $S_2$. Therefore, the  field configurations ${\bm m}(x_1,x_2)$ are smooth maps of the $S_2$ of the coordinate space to the $S_2$ of the target space of the field $\bm m$. Hence, the integer $\mathcal{Q}[\bm m]$ classifies the smooth field configurations  into a set of {\it equivalence classes} each labeled by the integer $\mathcal{Q}$. Under composition  homotopies form groups, and the equivalence classes themselves  also form a group which, in this case, is isomorphic to the group of integers, $\mathbb{Z}$. In Topology, these statements are summarized by the notation $\Pi_2(S_2) \simeq \mathbb{Z}$. The configurations with non-zero values of $\mathcal{Q}$ are called {\it instantons} which, in the quantum problem, represent tunneling processes of the non-linear sigma model.

Another consequence of  $\mathcal{Q}$ being an integer is that the contribution of its term to the weight of the path integral of Eq.\eqref{eq:O(3)-top-inv} is a periodic function of  $\theta$ angle. On the other hand, since the only allowed values of the $\theta$ are  $\theta=0$ (mod $2\pi$) for $S$ integer, or $\theta=\pi$ (mod $2\pi$) for $S$ a half-integer, the contribution of the topological invariant $\mathcal{Q}$ to the weight of the path integral is
\begin{equation}
\exp(i \theta \, \mathcal{Q}[\bm m])=(-1)^{2S \mathcal{Q}}
\label{eq:weight-Q}
\end{equation}
Therefore, for spin chains with $S$ integer, the weight is $1$ and the topological invariant does not contribute to the path integral. But, if $S$ is a half-integer, the weight is $(-1)^{\mathcal{Q}[\bm m]}$, and it does contribute. Moreover, its contribution is the same for all half-integer values of $S$.

These results have important consequences for the physics of spin chains which led Haldane to some startling conclusions \cite{haldane-1985}. In the weak coupling regime, $g \ll 1$ (equivalently, for large $S$), we can use the perturbative renormalization group and derive the beta function for all these $O(3)$ non-linear sigma models (with or without topological terms) and find that their beta functions are the same as in Eq.\eqref{eq:NLSM-RG} with $N=3$. Hence, for all $S$, the effective coupling flows to large values. We can make this inference since the topological term yields no contribution for all configurations which are related by smooth deformations. Thus, we infer that all spin chains with $S$ integer are in a massive phase with an exponentially small energy gap $\sim \exp(-2\pi S)$. This result is nowadays known as  the Haldane gap. 

On the other hand, these results also imply that all spin chain with half-integer spin $S$ are also the same and, in particular, th same as spin-$1/2$ chains. However, the Hamiltonian of the quantum spin chain with spin-1/2 degrees of freedom is an example of an integrable system and its spectrum is known to be gapless from its Bethe Ansatz solution \cite{bethe-1931,yang-1969}. However, the spin-1/2 chain is not only gapless but its low energy states are gapless solitons with a relativistic spectrum. The low energy description of a theory of this type must be described by a conformal field theory. Haldane concluded that the RG flow for spin-1/2 chains must have an IR stable fixed point at some finite (and large) value of the coupling constant. This conjecture was confirmed by Affleck and Haldane \cite{affleck-1987} who showed that the spin-1/2 Heisenberg chains are in the universality class of the SU(2)$_1$ Wess-Zumino-Witten model \cite{witten-1984} whose CFT was solved by Knizhnik and Zamolodchikov \cite{knizhnik-1984}.

\subsection{Topology and open integer-spin chains}
\label{sec:AKLT}

In the preceding section we showed that integer spin chains have a Haldane gap. We did that by showing that in that case the topological term is absent. However, the derivation is correct provided the spacetime manifold is closed, e.g. a sphere, a torus, etc. What happens if the system has a boundary? Let us denote  the boundary of $\Omega$ by $\Gamma=\partial \Omega$. For example we will take $\Gamma$ to be along the imaginary time direction and hence that it is a circle of circumference $1/T$, where $T$ is the temperature. The topological term has the same form as the Berry phase term of the path integral for spin, the ``Wess-Zumino'' term of Eq.\eqref{eq:WZ}, but its prefactor is 1/2 as big. Thus, for a system with an open boundary the topological term yields a net contribution equal to a Berry phase with a net prefactor of $S/2$. 

In other words, the boundary of the integer spin chain behaves as a localized degree of freedom whose spin is $1/2$ (mod an integer). Form the periodicity requirement we also see that if the spin chain is made of {\it odd-integer} degrees of freedom, there should be a spin 1/2 degree of freedom localized at the open boundary!. Conversely, if the chain is made of {\it even-integer} spins, there is no boundary degree of freedom! This line of argument implies that an antiferromagnetic chain with odd-integer spins must have a spin-1/2 degree of freedom at the boundary whereas a chain of even-integer spins should not. Notice that the ``bulk'' behavior is the same for both odd and even integer spin chains. The difference is whether or not they have a non-trivial ``zero-mode'' state at the boundary. In particular, the existence of this state is robust, i,.e. it cannot be removed by making smooth changes to the quantum Hamiltonian or, what is the same, the boundary state is {\it topologically protected}. 

A simple system that displays a protected spin-1/2 zero mode at the boundary is a generalized $S=1$ spin chain with Hamiltonian
\begin{equation}
H= \alpha \sum_{j=1}^N {\bm S}(j)\cdot {\bm S}(j+1)+\beta \sum_{j=1}^N \left({\bm S}(j) \cdot {\bm S}(j+1)\right)^2
\label{eq:alpha-beta}
\end{equation}
where ${\bm S}(j)=(S_x(j), S_y(j), S_z(j)$ are the spin 1 matrices at each lattice site $j$. This problem was examined in great detail by Affleck, Kennedy, Lieb and Tasaki \cite{affleck-1988c} who showed that at the special value of the parameters $\alpha=1/2$ and $\beta=1/6$ this Hamiltonian takes the form of a sum of projection operators
\begin{equation}
H=\sum_j P_2({\bm S}(j)+{\bm S}(j+1))
\label{eq:AKLT}
\end{equation}
where $P_2$ is an operator that projects out the spin 2 states. These authors constructed the exact ground state, known as the AKLT state, of this Hamiltonian by writing each spin 1 degree of freedom of two spin-1/2 degrees of freedom at each site. They showed that the ground state is a projected product state in which the ``constituent' spin-1/2 degrees of freedom (each labeled by $+$ and $-$ respectively) on nearby sites $j$ and $j+1$ are in a valence bond singlet state of the form $\frac{1}{\sqrt{2}} (|\uparrow_{j, +}, \;  \downarrow_{j+1,-}\rangle - \downarrow_{j,+},\; \uparrow_{j+1,-}\rangle$ (and symmetrizing at each site to project onto a spin 1 state). This state of the spin-1 chain  is translation invariant and gapped and hence agrees with Haldane's result. Moreover, it has a spin-1/2 degree of freedom at each open boundary \cite{hagiwara-1990}. 

The arguments discussed above show that the AKLT state is a gapped topological state. The gapless spin-1/2 boundary states are an example of  {\it spin fractionalization}. The spin-1/2 boundary degrees of freedom are an example of an {\it edge state} which are present in many, thought not all, topological phases of matter. One may ask what symmetry protects the gaplessness of these edge degrees of freedom. The only perturbation that would give a finite energy gap to the spin-1/2 edge states is an external magnetic field. However, this perturbation would break the global SU(2) symmetry of the Hamiltonian as well as time reversal invariance.

\section{Duality in Ising Models}
\label{sec:duality}

Duality plays a significant role of our understanding of statistical physics and of quantum field theory. Many seemingly unrelated correspondences between different theories have come to be called {\it dualities}.

\subsection{Duality in the 2D Ising Model}
\label{sec:2d-duality}

One of the earliest versions of duality transformations was used to relate the high-temperature expansion of the 2D classical Ising model and its low-temperature expansion \cite{kramers-1941}. In all dimensions, for concreteness we will think of a hypercubic lattice, the high temperature expansion is a representation of the partition function as a sum of contributions of closed loops on the lattice. At temperature $T$, a configuration of loops $\gamma$ contributes with a weight which in the Ising model has the form $C(\gamma) \; \tanh^{L(\gamma)}(\beta)$, where $\beta=1/T$ and $L(\gamma)$ is the length of the loop $\gamma$, i.e. the number of links on the loop, and $C(\gamma)$ is an entropic factor that counts the number of allowed loops with fixed perimeter $L(\gamma)$. However not all loop configurations are allowed as in the Ising model they satisfy constraints such as being non-overlapping, etc.

In section \ref{sec:TFIM} we noted that the partition function of the classical Ising model (in any dimension) can be interpreted as the path-integral of a quantum spin model in one dimension less on a lattice with a discretized imaginary time. In this picture, the loops $\gamma$ can be regarded as processes in which pairs of particles are created at some initial (imaginary) time, evolve and eventually are annihilated at a later (imaginary) time. In other words, the high temperature phase is a theory of a massive scalar field. The restrictions on the allowed loop configurations represents interactions among these particles. In the temperature range in which the expansion in loops is convergent the particles are massive as the loops are small. As the radius of convergence of the expansion is approached, longer and increasingly fractal-like loops begin to dominate the partition function, and concomitantly the mass of the particles decreases. This process is the signal of the approach to a continuous phase transition where the particles become massless. In fact, right at the critical point the particle interpretation is lost as the associated fields acquire anomalous dimensions.

Returning to the 2D classical Ising model, Kramers and Wannier also considered the low temperature expansion. This is an expansion around one of the broken symmetry state, e.g. the state with all spins up. In this low temperature regime the partition function is a sum of configurations of flipped spins. A typical configuration is a set of clusters of flipped spins. In the absence of a uniform field, a configuration of flipped spins has a energy cost only on links of the lattice with oppositely aligned spins (``broken bonds''). Thus a cluster of flipped spins costs an energy equal to $2$ (I assumed that I set $J=1$) for each broken bond at the {\it boundary} of the cluster. This boundary is {\it domain wall} which is a closed loop on the {\it dual lattice}. In 2D the dual of the square lattice is the square lattice of the dual sites (the centers of the elementary plaquettes of the 2D lattice). 
In 2D the links of the direct lattice pierce the links of the dual lattice. We can see that this analogous to the  the geometric duality of forms: sites (``0-forms") are dual to plaquettes (``2-forms'') and links (``1-forms'') are dual to links (also ``1-forms''). 

Thus, in 2D the low temperature expansion is an expansion in the loops $\gamma^*$ of the dual lattice that represent the domain walls. We can also regard the domain walls (the dual loops) as the histories of of pairs of particles on he dual lattice. However, the weight of each dual loop is $\exp(-2 \beta)$ per link of $\gamma^*$. Except for that, the counting and restrictions on the dual loops $\gamma^*$ are the same as those of $\gamma$. This means that there is a one-to-one correspondence between the two expansions with the replacement $\tanh \beta \leftrightarrow \exp(-2\beta^*)$.  This mapping means that the dual of the 2D classical Ising model (with global $\mathbb{Z}_2$ symmetry) at inverse temperature $\beta$  is a dual Ising model (also with global $\mathbb{Z}_2$ symmetry) on the dual lattice at inverse temperature $\beta^*$. This correspondence is in close analogy with what we discussed  in section \ref{sec:2d-vortices}.

In particular if one assumes that there is a transition at $\beta_c=1/T_c$ then there should also be a transition at $\beta^*_c=-1/2 \ln \tanh \beta_c$. Moreover, if one further assumes (as Kramers and Wannier did \cite{kramers-1941}) that there is a unique transition (correct in the Ising model but not in other cases), then the critical point must be such that $\exp(-2\beta_c)=\tanh \beta_c$, which yields the value $T_c=2/\ln(\sqrt{2}+1)$, which agrees with the Onsager result \cite{onsager-1944}.

In section \ref{sec:TFIM} we showed that the classical 2D Ising model is equivalent to the one-dimensional Ising model in a transverse field whose Hamiltonian is given in Eq.\eqref{eq:TFIM}. The 1D Ising model on a transverse field has spin degrees of freedom  defined on the sites of a one-dimensional chain labeled by an integer-valued variable $n$. The 1D Hamiltonian is expressed in terms of local operators, the Pauli matrices $\sigma_3(n)$ and $\sigma_1(n)$. The 1D chain has a dual lattice whose sites are the midpoints of the chain. Thus in 1D sites (``0-forms'') are dual to links (``1-forms'') and viceversa. 

We will now see that there is a Hamiltonian version of the Kramers-Wannier duality \cite{fradkin-1978}.  In Eq.\eqref{eq:kink} we introduced the kink creation operator which flips are $\sigma_3$ operators to the left and including site $j$. For clarity we will denote the kink creation operator operator as $\tau_3(\tilde n)$, where $\tilde n$ is the site of the dual lattice between the sites $n$ and $n+1$ of the original lattice. We will now define an operator $\tau_1(\tilde n)=\sigma_3(n) \sigma_3(n+1)$. The operators $\tau_1(\tilde n)$ and $\tau_3(\tilde n)$ satisfy the same algebra os the Pauli operators $\sigma_1(n)$ and $\sigma_3(n)$. Furthermore, we readily find the Hamiltonian of the dual theory is the same (up to boundary conditions) as the original Hamiltonian of Eq.\eqref{eq:TFIM} except that the dual coupling constant is $\lambda^*=1/\lambda$. So, once again, if we assume that there is a single (quantum) phase transition we require $\lambda_c^*=\lambda_c$ which is only satisfied by $\lambda_c=1$. In this language the kink creation operator plays the role of a disorder operator \cite{fradkin-1978}.

\subsection{The 3D duality: $\mathbb{Z}_2$ gauge theory}
\label{sec:3d-duality}

We will now discuss the role of duality in the 3D Ising model on a  cubic lattice. This case, and its generalizations to higher dimensions, was considered first by Franz Wegner \cite{wegner-1971}.

In section \ref{sec:2d-duality} we discussed the loop representation of the high temperature expansion and showed that it has the form form in all dimensions. Hence, in 3D the high temperature expansion is an expansion in closed loops $\gamma$ with weight of $\tanh \beta$ per unit link of loop. Hence, in 3D as well, the high temperature phase can be regarded as field theory of a massive scalar field. As in the 2D case, the weight of a loop configuration is $\tanh \beta$ for each link of the loop $\gamma$. 

However, the low temperature expansion has a radically different form and physical interpretation. Much as in 2D, the low temperature expansion is an expansion in clusters of flipped spins. Here too, in the absence of an external field, the only energy cost resides at the boundary of the clusters of overturned spins, the domain walls. But in 3D the clusters occupy volumes whose boundaries are closed {\it surfaces} $\Sigma^*$. In 3D a link (bond) is dual to a plaquette (expressing the fact that in 3D a 1-form is dual to a 2-form). Hence the closed domain walls of overturned spins is dual to a closed surface $\Sigma^*$ on the dual lattice.  The weight of each configuration of closed surfaces is $\exp (-2\beta)$ for each plaquette of the surface $\Sigma^*$. 

These facts mean that the dual of the 3D Ising model is a theory on the dual lattice with coupling constant 
$\beta^*$, with $\exp(-2\beta*)=\tanh \beta$, such that its expansion for small $\beta^*$ is a sum over configurations of closed surfaces $\sigma^*$, and for large $\beta^*$ is a sum of configurations of closed loops $\gamma^*$ on the dual lattice. The dual theory is the naturally defined on (dual) plaquettes, not on links. To this end, let us define a set of Ising-like degrees of freedom $\sigma_\mu(x)=\pm 1$ located on the links $(x,\mu)$ of the dual lattice. These degrees of freedom   are coupled on each plaquette of the lattice. Since each plaquette has four links, the coupling involves the degrees of freedom on all four links of each plaquette. The partition function of the dual theory is
\begin{equation}
Z=\sum_{\{ \sigma_\mu(x) \}} \exp\left(\beta^* \sum_{\rm plaquettes} \sigma_\mu(x) \sigma_\nu(x+e_\mu) \sigma_\mu(x+e_\mu+e_\nu)\sigma_\nu(x)\right)
\label{eq:Z2-gauge}
\end{equation}
where the sum in th exponent (the negative of the ``action'') runs on all the plaquettes of the dual lattice. 

The action of the theory of Eq.\eqref{eq:Z2-gauge} is invariant under the reversal of all six Ising degrees of freedom on links sharing a given site $x$. This is a {\it local symmetry}. Unlike the 3D Ising model, which has a global $\mathbb{Z}_2$ symmetry of flipping all spins simultaneously, this theory has {\it local} (or {\it gauge}) $\mathbb{Z}_2$ symmetry. This is the simplest example of a lattice gauge theory in which  the degrees of freedom are gauge fields that take values on the $\mathbb{Z}_2$ gauge group \cite{wegner-1971,wilson-1975,kogut-1979}.

We saw that in the spin model the high temperature expansion (i.e. the expansion in powers of $\tanh \beta$) is a sum over loop configurations which can be interpreted in terms of processes in which pairs of particles are created, evolve (in imaginary time) and then are destroyed (also in pairs). The analogous interpretation of the expansion of the $\mathbb{Z}_2$ gauge theory in powers of $\tanh \beta^*=\exp(-2\beta)$ as a sum over the configurations of closed surfaces is not in terms of the histories of {\it particles} but in terms of the histories of{ \it closed strings} which, as they evolve, sweep the closed surfaces. The physics is, however, more complex as the sum over surfaces runs over all surfaces of arbitrary topology with arbitrary
number of handles (or {\it genus}).  Thus, over (imaginary) time a closed strong is created, evolves, splits into two closed strings, etc. 

Therefore, the 3D Ising model, which has a  $\mathbb{Z}_2$ global symmetry is dual to a $\mathbb{Z}_2$ gauge theory which has a $\mathbb{Z}_2$ local symmetry. The duality maps the high temperature (disordered) phase of the Ising model to the strong coupling (small $\beta^*$)  phase of the gauge theory. In the gauge theory language this is the {\it confining} phase. This can be seen by computing the expectation value of the Wilson loop operator on the closed loop $\Gamma$, which here reads $\langle \prod_{(x,\mu) \in \Gamma}\sigma_\mu(x)\rangle$. In the small $\beta^*$ phase this expectation value decays exponentially with the size of the minimal surface bounded by the loop $\Gamma$. This behavior of the area law of the Wilson loop. Under duality the insertion of this operator is equivalent to an Ising model with a domain wall terminating on the loop $\Gamma$, and the area law of the Wilson loop is the consequence of the fact that if the Ising model has long range order, the free energy cost of the domain wall scales with its area. Moreover, in this phase of the gauge theory the closed strings are small meaning that the string tension (the energy per unit length of string) is finite. In the Ising model language, the string tension becomes surface tension of the domain wall.

In the preceding subsection we discussed a Hamiltonian version of the duality. We will briefly do the same in the case of the 2+1 dimensional Ising model. The hamiltonian of the Ising model in a transverse field on a square lattice is
\begin{equation}
H_{\rm 2D-TFIM}=-\sum_{\bm r} \sigma_1(\bm r) -\lambda \sum_{{\bm r}, j=1,2} \sigma_3(\bm r) \sigma_3({\bm r}+{\bm e}_j)
\label{eq:2D-TFIM}
\end{equation}
where $\bm r$ labels the sites of the square lattice and ${\bm e}_j$ (with $j=1,2$) are the two (orthonormal) primitive unit vectors of the lattice. Here too, at each site we have a two-level system (the spins), $\sigma_1$ and $\sigma_3$ are Pauli matrices acting on these states at each site and $\lambda$ is the coupling constant. Just as in the 1D case, this system has two phases: an ordered phase for $\lambda > \lambda_c$ and a disordered phase for $\lambda<\lambda_c$, where $\lambda_c$ is a critical coupling. This Hamiltonian is invariant under the  global $\mathbb{Z}_2$ symmetry generated by the global spin flip operator $Q=\prod_{\bm r} \sigma_1(\bm r)$ which commutes with the Hamiltonian, $[Q, H_{\rm 2D-TFIM}]=0$.

Let is consider now a $\mathbb{Z}_2$ gauge theory on the dual of the square lattice. We will now define a two-dimensional Hilbert space on each link, denoted by $(\tilde{\bm r}, j)$ (with $j=1,2$) of the dual lattice with sites labelled by $\tilde{\bm r}$. We will further define at each link $(\tilde{\bm r}, j)$  the Pauli operators $\sigma^1_j(\tilde{\bm r})$ and $\sigma^3_j(\tilde{\bm r})$. The Hamiltonian of the $\mathbb{Z}_2$ gauge theory is
\begin{equation}
H_{\mathbb{Z}_2 {\rm gauge}}=-\sum_{\tilde{\bm r}, j} \sigma^1_j(\tilde{\bm r})-
g \sum_{\tilde{\bm r}} \sigma^3_1(\tilde{\bm r})\sigma^3_2(\tilde{\bm r}+\tilde{\bm e}_1)\sigma^3_1(\tilde{\bm r}+\tilde{\bm e}_1+\tilde{\bm e}_2)\sigma^3_2(\tilde{\bm r})
\label{eq:H-Z2-gauge}
\end{equation}
where $\tilde{\bm e}_j=\epsilon_{jk} {\bm e}_k$ (with $j, k=1,2$).

Unlike the Hamiltonian of Eq.\eqref{eq:2D-TFIM}, the Hamiltonian of Eq.\eqref{eq:H-Z2-gauge} has a local symmetry. Let ${\tilde Q}(\tilde{\bm r})$ be the operator that flips the $\mathbb{Z}_2$ gauge degrees of freedom on  the four links that share the site $\tilde {\bm r}$,
\begin{equation}
 {\tilde Q}(\tilde{\bm r})=\prod_{j=1,2} \sigma^1_j(\tilde {\bm r}) \sigma^1_j({\tilde {\bm r}}-\tilde{\bm e}_j)
 \label{eq:local-Z2}
 \end{equation}
 These operators commute with each other, $[{\tilde Q}(\tilde{\bm r}), {\tilde Q}(\tilde{\bm r}')]=0$, and commute with the Hamiltonian, 
 $[H_{\mathbb{Z}_2 {\rm gauge}},{\tilde Q}(\tilde{\bm r})]=0$. The Hilbert space of gauge-invariant states are eigenstates of ${\tilde Q}(\bm r)$ with unit eigenvalue,  
 \begin{equation}
 {\tilde Q}(\bm r)|\rm Phys\rangle=|\rm Phys\rangle
 \label{eq:Z2-gauss}
 \end{equation}
  (for all $\bm r$). This constraint is the the $\mathbb{Z}_2$  Gauss Law.
  The Hamiltonian $H_{\mathbb{Z}_2 {\rm gauge}}$ has two phases:  a {\it confining} phase for $g<g_c$, and  a {\it deconfined } phase for $g>g_c$ (where $g_c$ is a critical coupling).
 
 The duality transformation is defined so that
 \begin{equation}
 \sigma_1(\bm r)= \prod_{{\rm plaquette}(\bm r)} \sigma^3_j(\tilde{\bm r})
 \label{eq:2D-duality1}
 \end{equation}
is the product of the $\sigma^3_j$ operators of the gauge theory on a dual plaquette centered at the site $\bm r$, and
\begin{equation}
\sigma^1_1(\tilde{\bm r})=\sigma_3(\bm r) \sigma_3({\bm r}+{\bm e}_2), \qquad \sigma^1_2(\tilde{\bm r})=\sigma_3(\bm r) \sigma_3({\bm r}+{\bm e}_1)
\label{eq:2D-duality2}
\end{equation}
These  identities imply that the gauge theory constraint of the $\mathbb{Z}_2$ Gauss Law, Eq.\eqref{eq:Z2-gauss}, is satisfied. This also means that duality is a mapping of the {\it gauge-invariant} sector of the $\mathbb{Z}_2$ gauge theory to the Hilbert space of the Ising model in a transverse field.

Eq.\eqref{eq:2D-duality1} and Eq.\eqref{eq:2D-duality2} imply that the Hamiltonians $H_{\rm 2D-TFIM}$ and $H_{\mathbb{Z}_2 {\rm gauge}}$ are equal to each other with the identification of the coupling constants $g=1/\lambda$. Hence, the {\it ordered phase} of the Ising model, $\lambda>\lambda_c$, maps onto the {\it confining phase} of the $\mathbb{Z}_2$ gauge theory and, conversely, the {\it disordered phase} of the Ising model maps onto the {\it deconfined phase} of the gauge theory.

 Eq.\eqref{eq:2D-duality2} also implies that the operator $\sigma_3(\bm r)$ of the Ising model can be identified with the operator
\begin{equation}
\sigma_3(\bm r)=\prod_{\gamma(\bm r)} \sigma_j^1(\tilde{\bm r})
\label{eq:dual-path}
\end{equation}
where the product is on the links of the dual lattice pierced by the path $\gamma(\bm r)$ on the direct lattice ending at the site $\bm r$. 

While $\sigma_3(\bm r)$ is of course just the order parameter of the Ising model, the dual operator defined by Eq.\eqref{eq:dual-path} anti-commutes with the plaquette term of the Hamiltonian of Eq.\eqref{eq:H-Z2-gauge} and creates an $\mathbb{Z}_2$ flux excitation at the plaquette. With some abuse of language, this operator can be regarded as creating 
a $\mathbb{Z}_2$ ``magnetic monopole''. Since in the confining phase this operator has an expectation value, we can regard this phase as a magnetic condensate.

We will now discuss the role of  boundary conditions which is different in both theories. Let us examine the behavior of the $\mathbb{Z}_2$ gauge theory with periodic boundary conditions. Periodic boundary conditions means that the 2D space is topologically a 2-torus. It is straightforward to show that the ground state in the confining phase is unique and insensitive to boundary conditions. The physics of the deconfined phase is more subtle. We will now see that on a 2-torus it has a four-fold degenerate ground state and that the degeneracy is not due to the spontaneous breaking of a global symmetry.

Let $\gamma_1$ and $\gamma_2$ be two non-contractible loops on the torus along the directions 1 and 2 respectively. Let us consider the (``electric'') Wilson loop operators along $\gamma_1$ and $\gamma_2$, $W[\gamma_1]=\prod_{({\bm r},j)\in \gamma_1} \sigma^3_j(\bm r)$  and similarly for $\gamma_2$. Similarly let us consider the (``magnetic'') 't Hooft operators $\tilde W[\tilde \gamma_1]$ and $\tilde W[\gamma_2]$ defined on the non-contractible closed paths of the dual lattice ${\tilde \gamma}_1$ and $\tilde\gamma_2$, such that $\tilde W[\tilde\gamma_1]=\prod_{\tilde \gamma_1} \sigma^1_2(\tilde{\bm r})$ and $\tilde W[\tilde\gamma_2]=\prod_{\tilde \gamma_2} \sigma^1_1(\tilde{\bm r})$. 
It is easy to see that the Wilson and't Hooft operators satisfy that $W[\gamma_i]^2=\tilde W[\tilde \gamma_j]^2=I$, and that
\begin{equation}
[W[\gamma_i],W[\gamma_j]=[\tilde W[\tilde \gamma_i],\tilde W[\tilde \gamma_j]]=0, \qquad\{W[\gamma_1],\tilde W[\tilde \gamma_2]\}=\{W[\gamma_2],\tilde W[\tilde \gamma_1]\}=0
\label{eq:dual-algebra}
\end{equation}

Let us consider the special limit of $g \to \infty$. In this limit the Wilson loop operators $W[\gamma_i]$ on the two non-contractible loops of the 2-torus commute with the Hamiltonian $H_{\mathbb{Z}_2 \textrm{gauge}}$ of Eq.\eqref{eq:H-Z2-gauge}. Therefore the eigenstates of the Hamiltonian can be chosen to be the eigenstates of these two Wilson loops. Since the  Wilson loop operators are hermitian and obey $W[\gamma_i]^2=I$, the spectrum of each loop is two-dimensional $|\pm 1\rangle_i$ ($i=1,2$) with eigenvalues $\pm1$, respectively. At $g \to \infty$ these states are also eigenstates of $H_{\mathbb{Z}_2 \textrm{gauge}}$. Thus,   the ground state of $H_{\mathbb{Z}_2 \textrm{gauge}}$  is four dimensional. The 't Hooft loops $\tilde W[\tilde \gamma_i]$ act as ladder operators in this restricted Hilbert space.

The conclusion is that, on a 2-torus, at least in the limit $g \to \infty$, the ground state of $H_{\mathbb{Z}_2 \textrm{gauge}}$ is degenerate. However, this degeneracy is not due to the spontaneous breaking of a global symmetry. Rather, it reflects the topological character of the theory. To see this, one can extend this analysis to a theory to be on a more general two-dimensional and instead of a 2-torus we can consider a surface with $g$ handles (not to be confused with the coupling constant!), e.g. $g=0$ for a sphere (or disk), $g=1$ for a 2-torus, $g=2$ for a pretzel, etc. For each of these two-dimensional surfaces the number different number of non-contractible loops is $g$, and the Wilson loops defined on them commute with each other (and with $H_{\mathbb{Z}_2 \textrm{gauge}}$). Hence, in the limit  $g \to \infty$ $H_{\mathbb{Z}_2 \textrm{gauge}}$ has a ground state degeneracy of $2^g$ which, clearly, depends only on the topology of the surface. We conclude that, at least as $g \to \infty$, the Hamiltonian $H_{\mathbb{Z}_2 \textrm{gauge}}$ is in a {\it topological phase}.

However, is  the exact degeneracy found in the limit $g \to \infty$ a property of the entire deconfined phase? For a finite system the expansion in powers of $1/g$ is  convergent. On the other hand, in the thermodynamic limit, $L \to \infty$,  the expansion has a finite radius of convergence with $g_c$ being an upper bound. Let us consider the ground state at $g=\infty$ with eigenvalue $+1$ for the Wilson loop $W[\gamma_1]$. The degenerate state with eigenvalue $-1$ is created by the 't Hooft loop $\tilde W[\tilde \gamma_2]$, i.e. $|-1\rangle_1=\tilde W[\tilde \gamma_2]\; |+1\rangle_1$. The 't Hooft operator is a product of $\sigma^1$ operators on links along the direction 1 crossed by the path $\tilde \gamma_2$ of the dual lattice, which involves (at least) $L$ links. Then, it takes $L$ orders in the expansion in powers of $1/g$ to mix the state $|+1\rangle_1$ with the state $|-1\rangle_1$, and this amplitude is of order $1/g^{L}$, which is exponentially small. The same argument applies to the mixing between all four states. For a finite but large system of linear size $L$, the degeneracy  is lifted but the energy splitting is exponentially small in the system size. Hence, the topological protection is a feature of the deconfined phase in the thermodynamic limit $L \to \infty$.

We conclude that the deconfined phase of the $\mathbb{Z}_2$ lattice gauge theory is a topological phase. This result extends to the case of a gauge theory with discrete gauge group $\mathbb{Z}$, the cyclic group of $k$ elements. In general spacetime dimension $D>2$  the $\mathbb{Z}_k$ gauge theory also has a confined and a deconfined phase and if $k\gtrsim 4$ it also has an intermediate ``Coulomb phase''\cite{elitzur-1979,ukawa-1980}. In section \ref{sec:BF} we will see that the low-energy (IR) regime of the deconfined phase is described by a topological quantum field theory known as the level $k$ BF theory.

\section{Bosonization}
\label{sec:1+1-bosonization}

The behavior of one-dimensional electronic systems is of great interest in Condensed Matter Physics for many reasons. One is that, even for infinitesimally weak interactions, these one-dimensional metals violate the basic principles of the Landau Theory of the Fermi Liquid \cite{baym-1991}. A central assumption of Femi Liquid theory is that at low energies the excitation energy of the fermion (electron) quasiparticle is always much larger than its width. Hence, at asymptotically low energies the quasiparticle excitations become increasingly sharp. 

A manifestation of this feature is that the quasiparticle propagator (the ``Green function'') has a pole on the real frequency axis with a finite residue $Z$. In one-dimensional metals this assumption always fails since the residue $Z$ vanishes, the Fermi field acquires a non-trivial anomalous dimension, and the pole is replaced by a branch cut. To this date this is the best example of what is called a "non-Fermi liquid". In one-dimensional metals this non-Fermi liquid is often called a ``Luttinger Liquid'' \cite{haldane-1981}. A detailed analysis can be found in chapter 6 of Ref. \cite{fradkin-2013}, and in other books.

Bosonization provides for a powerful tool to understand the physics of these non-trivial systems. Bosonization is a duality between a massless Dirac field in 1+1 dimensions and a (also massless) relativistic Bose (scalar) field \cite{mattis-1965,luther-1974,mandelstam-1975,coleman-1975}. In this context, bosonization is a set of operator identities relating observables between two different (dual) continuum field theories. These identities have a close resemblance to the Jordan-Wigner transformation which relates operators of a theory of spinless (Dirac) fermions on a  one-dimensional lattice to a theory of bosons with hard cores (i.e. spins) on the same lattice \cite{lieb-1961}. 

\subsection{Dirac fermions in one space dimensions}
\label{sec:Dirac-1D}

To understand how these operator identities come about and what these fields mean in the Condensed Matter context we will consider the simple problem of a system of non-interacting spinless fermions $c(n)$ on a one-dimensional chain of length $L$ (with $L \to \infty$), for simplicity with periodic boundary conditions. The Hamiltonian is 
\begin{equation}
H_0=-t \sum_{n=1}^L c(n)^\dagger c(n+1)+\textrm{h.c.}
\label{eq:H0-1D}
\end{equation}
In momentum space, and in the thermodynamic limit $L \to \infty$, the Hamiltonian becomes
\begin{equation}
H_0=\int_{-\pi}^\pi \frac{dp}{2\pi} (\varepsilon_0(p)-\mu) \; c^\dagger(p) c(p)
\label{eq:H0-1D-fourier}
\end{equation}
where $-\pi \leq p \leq \pi$, $\mu$ is the chemical potential (which fixes the fermion number) and $\varepsilon_0(p)$ is the (free) quasiparticle energy which, in this case, is
\begin{equation}
\varepsilon_0(p)=-2t\cos p
\label{eq:varepsilon0}
\end{equation}
This simple system has a global internal symmetry
\begin{equation}
c'(n)=e^{i \alpha} c(n), \qquad c'(n)^\dagger=e^{-i\alpha} c(c)
\label{eq:lattice-U1}
\end{equation}
where $\alpha$ is a constant phase (with period $2\pi$). This global symmetry reflects the conservation of fermion number
\begin{equation}
N_F=\sum_n c^\dagger(n) c(n)
\label{eq:lattice-fermion-number}
\end{equation}
The free fermion Hamiltonian of Eq.\eqref{eq:H0-1D} is also invariant under lattice translations. 

The ground state of this system is obtained by occupying all single particle states with energy below the chemical potential, $E\leq \mu \equiv EF$, which defines the Fermi energy $E_F$. In what follows we will redefine the zero of the energy at the Fermi energy. In the thermodynamic limit the one-particle states are labeled by momenta defined  in the  first Brillouin zone $[-\pi, \pi)$. The ground state $|G\rangle$ of this system is obtained by filling up all single-particle states with  momentum $p$ in the range $[-p_F, p_F)$, where $p_F$ is the Fermi momentum. Hence, the occupied states have energy $\varepsilon_0(p)\leq E_F$. The ground state is
\begin{equation}
|G\rangle=\prod_{|p|\leq p_F} c^\dagger(p)|0\rangle
\label{eq:filled-Fermi-sea}
\end{equation}
and is called the {\it filled Fermi sea}. We will further assume that the fermionic system is dense in the sense that the occupied states are a finite fraction of the available states, i.e. we assume that $|E_F|$ is a {\it finite} fraction of the bandwidth $W=4t$. 

The single-particle excitations have low energy if $|\varepsilon_0(p)-E_F|\ll |E_F|$. This range can only be accessed by quasiparticles with momenta $p$ close to $\pm p_F$. For states wit $p$ close to $p_F$ we can a linearized approximation   and write
\begin{equation}
\varepsilon_0(p) \simeq \varv_F (p-p_F)+\ldots
\label{eq:linearization}
\end{equation}
and similarly for the states near $-p_F$. Here $\varv_F=\frac{\partial \epsilon_0}{\partial p}\big|_{p_F}=2t \sin p_F$ is the Fermi velocity. Let $q$ being the momentum measured from $p_F$ (or $-p_F$ in the other case), This approximation is correct in a range of momenta $-\Lambda \leq q \leq \Lambda$, where $\frac{2\pi}{L} \ll \Lambda \ll \pi$. In this regime we can approximate the lattice fields $c(n)$ with two continuum right moving $\psi_R(x)$ and left moving $\psi_L(x)$ Fermi fields, such that (with $x=n a_0$)
\begin{equation}
c(n)\sim e^{i p_F x} \psi_R(x)+e^{-ip_F x} \psi_L(x)
\label{eq:decomposition-fermion}
\end{equation}
in terms of which the Hamiltonian takes the continuum form
\begin{align}
H_{\rm continuum} =&\int dx\; \left(\psi_R^\dagger(x) (-i\varv_F) \partial_x \psi_R(x)-\psi_L^\dagger(x) (-i\varv_F) \partial_x \psi_L(x)\right)\nonumber\\
			     =&\int \frac{dq}{2\pi} q \varv_F \left(\psi_R^\dagger(q)\psi_R(q)+\psi_R^\dagger(q)\psi_R(q)\right)
\label{eq:H-chiral-form}
\end{align}
In what follows I will rescale time and space in such a way as to set $\varv_F=1$.

Eq.\eqref{eq:H-chiral-form} is  the Hamiltonian of a massless Dirac field in 1+1 dimensions. It describes the effective low energy behavior of the excitations of the fermionic system defined on a lattice by the Hamiltonian of Eq.\eqref{eq:H0-1D}. Here low energy means asymptotically close to the Fermi energy (which we have set to zero) and momenta close to $\pm p_F$. We will see that this effective relativistic field theory gives a complete description of the universal low energy physics encoded in the microscopic model of Eq.\eqref{eq:H0-1D} up to some subtle issues associated with what are called quantum anomalies.

Let us denote by $\psi(x)$ the bi-spinor field $\psi(x)=(\psi_R(x), \psi_L(x))$ and the $2 \times 2$ Dirac gamma-matrices in terms of the three Pauli matrices are 
\begin{equation}
\gamma_0=\sigma_1, \qquad \gamma_1=-i\sigma_2, \qquad \gamma_5=\sigma_3
\label{eq:dirac-gamma-2D}
\end{equation}
which obey the Dirac algebra
\begin{equation}
\{ \gamma_\mu, \gamma_\nu \}=2g_{\mu \nu}
\label{eq:dirac-algebra}\end{equation}
where $g_{\mu \nu}$ is the metric tensor in 1+1-dimensional Minkowski spacetime
\begin{equation}
g_{\mu \nu}=
\begin{pmatrix}
1 & 0 \\
0 & -1
\end{pmatrix}
\label{eq:gmunu-2D}
\end{equation}

 Using the standard notation ${\bar \psi}(x)=\psi^\dagger(x) \gamma_0$ (where I left the spinor index $\alpha=1,2$ implicit) the Lagrangian density of the free massless Dirac fermion is
\begin{equation}
\mathcal{L}={\bar \psi} i \slashed{\partial} \psi
\label{eq:Dirac-Lagrangian-1+1}
\end{equation}
where we have used the Feynman slash notation which denotes the contraction of a vector field, say $A_\mu$ with the Dirac gamma matrices with a slash, $A_\mu \gamma^\mu \equiv \slashed{A}$. 

Formally, the Dirac lagrangian of Eq.\eqref{eq:Dirac-Lagrangian-1+1} (and the Dirac Hamiltonian of Eq.\eqref{eq:H-chiral-form}) are invariant under two separate global transformations. It has a global U(1) (gauge) symmetry transformation
\begin{equation}
 \psi'(x)=e^{i\theta} \psi(x)
\label{eq:global-U(1)}
\end{equation}
(where $\theta$ is constant) under which the two components of the bi-spinor $\psi$  transform in the same way
\begin{equation}
\psi_R'(x)=e^{i\theta} \psi_R(x), \qquad \psi_L'(x)=e^{i\theta} \psi_L(x)
\label{eq:global-U(1)-dirac}
\end{equation}

The massless Dirac theory is also invariant under a global gauge U(1) chiral transformation
\begin{equation}
\psi'(x)=e^{i \theta \gamma_5} \psi(x)
\label{eq:global-chiral-symmetry}
\end{equation}
which in components it reads
\begin{equation}
\psi'_R(x)=e^{i \theta} \psi_R(x), \qquad \psi'_L(x)=e^{-i \theta} \psi_L(x)
\label{eq:chiral-components}
\end{equation}

In general, if a system has a global continuous symmetry it should have a locally conserved current and a globally conserved charge. This is the content of Noether's Theorem. Since the massless Dirac theory has these two global symmetries one would expect that it  should have two separately conserved currents. 

The global U(1) symmetry has an associated current $j_\mu=(j_0, j_1)$ given by
\begin{equation}
j_\mu={\bar \psi} \gamma_\mu \psi
\label{eq:dirac-current}
\end{equation}
which is invariant under the global U(1) symmetry.
In terms of the right and left moving Dirac fields, $\psi_R$ and $\psi_L$, the components of the current are
\begin{equation}
j_0=\psi^\dagger_R \psi_R+\psi^\dagger_L\psi_L, \qquad j_1=\psi^\dagger_R \psi_R-\psi^\dagger_L\psi_L
\label{eq:dirac-current-components}
\end{equation}
This current  is locally conserved and satisfies the continuity equation
\begin{equation}
\partial_\mu j^\mu=0
\label{eq:conservation-dirac-current}
\end{equation}
It has an associated conserved global charge, which we will call fermion number
\begin{equation}
Q=\int_{-\infty}^\infty dx j_0(x)=\int_{-\infty}^\infty dx \left(\psi^\dagger_R \psi_R+\psi^\dagger_L\psi_L\right)
\label{eq:conserved-U(1)-charge}
\end{equation}
This is the continuum version of the conservation of fermion number $N_F$ of the free fermion lattice model discussed above.

Since this current is conserved it can be coupled to an electromagnetic field through a term in the Lagrangian
\begin{equation}
\mathcal{L}_{\rm int}=-e j_\mu A^\mu
\label{eq:gauge-coupling}
\end{equation}
where $A^\mu$ is the electromagnetic vector potential, which in 1+1 dimensions has only two components, $A^\mu=(A_0,A_1)$. Here $e$ is a coupling constant which is interpreted as the electric charge. This coupling amounts to making the U(1) symmetry a local gauge symmetry under which the fields transform as follows
\begin{equation}
 \psi'(x)=e^{i\theta(x)} \psi(x), \qquad A'_\mu(x)=A_\mu(x)+\frac{1}{e} \partial_\mu \theta(x)
\label{eq:local-U(1)}
\end{equation}
Now the conservation of fermion number becomes the conservation of the total electric charge
\begin{equation}
Q_e=-e Q
\label{eq:charge-identification}
\end{equation}

In a continuum non-relativistic model the Fermi field is $\psi(x)$ (ignoring spin), e.g. an electron gas in one dimension such as a quantum wire, the analog of decomposition shown in Eq.\eqref{eq:decomposition-fermion} of the electron field $\psi(x)$ becomes
\begin{equation}
\psi(x)=e^{i p_F x} \psi_R(x)+e^{-i p_F x} \psi_L(x)
\label{eq:decomposition-continuum}
\end{equation}
The electron field $\psi(x)$ transforms as  $\psi'(x)=e^{i \alpha} \psi(x)$ under the global U(1) gauge symmetry of Eq.\eqref{eq:global-U(1)-dirac}. In particular, the local  density operator $\rho(x)=\psi^\dagger(x) \psi(x)$ is invariant under this symmetry. Under both decompositions of Eqs. \eqref{eq:decomposition-fermion} and \eqref{eq:decomposition-continuum}, the local density operator becomes
\begin{equation}
\rho(x)=\psi^\dagger_R(x)\psi_R(x)+\psi^\dagger_L(x) \psi_L(x)+e^{i2p_F x} \psi^\dagger_R(x) \psi_L(x)+e^{-i2p_F x} \psi^\dagger_L(x) \psi_R(x)
\label{eq:density-decomposition}
\end{equation}
Clearly the non-relativistic density operator $\rho(x)$ is invariant under the global U(1) gauge symmetry. The same considerations applies to the lattice version of the density operator (the local occupation number).

Thus, the density operator $\rho(x)$ can be decomposed into a slowly varying part (the first two terms in Eq.\eqref{eq:density-decomposition}) and the last two terms which oscillate with wave vectors $Q=\pm 2p_F$. These observations imply that we can express $\rho(x)$ in the form
\begin{equation}
\rho(x)=\bar \rho+ j_0(x)+e^{iQx} \psi_Q(x)+e^{-i Q x} \psi_{-Q}(x)
\label{eq:fourier-decomposition}
\end{equation}
This decomposition can be interpreted as a Fourier expansion of the density operator in term of newly defined slowly-varying fields. In Eq.\eqref{eq:fourier-decomposition}  $Q=2p_F$ and $\bar \rho$ is the average density, where we  assumed that the Dirac density operator $j_0(x)$ has vanishing expectation value (i.e. it is normal-ordered). In Eq.\eqref{eq:fourier-decomposition} we defined the (bosonic) operators
\begin{equation}
\psi_Q(x)=\psi^\dagger_R(x) \psi_L(x), \qquad \psi_{-Q}(x)=\psi^\dagger_L(x) \psi_R(x)
\label{eq:chiral-order-parameters}
\end{equation}
which characterize the oscillatory component of the density. The operators $\psi_{\pm Q}(x)$ are the order parameters of a charge density wave state in one dimension and $Q$ is the ordering wavevector.

Repulsive interactions between the electrons can cause scattering process between the right and left moving components to become relevant (in the Renormalization Group sense) leading to the spontaneous breaking of translation invariance. The resulting state is known as a charge-density-wave (CDW). In this state the operators $\psi_{\pm Q}(x)$ (or a linear combination of them) acquire a non-vanishing expectation value, and the expectation value of the density operator $\rho(x)$ has a (static) modulated component, and the Dirac Hamiltonian  density  becomes
\begin{equation}
\mathcal{H} =\psi_R^\dagger(x) (-i) \partial_x \psi_R(x)-\psi_L^\dagger(x) (-i) \partial_x \psi_L(x)+ m \left(\psi^\dagger_R(x) \psi_L(x)+ \psi^\dagger_L(x) \psi_R(x) \right) 
\label{eq:massive-dirac}
\end{equation}
where $m$ is the Dirac mass.
The operator in the second term of Eq.\eqref{eq:massive-dirac} mixes the right and left moving components of the Dirac spinor. This hermitian operator is known as the Dirac mass term. In relativistic notation this operator is written 
\begin{equation}
{\bar \psi} \psi= \psi^\dagger_R(x) \psi_L(x)+\psi^\dagger_L(x) \psi_R(x)
\label{eq:dirac-mass}
\end{equation}
When $m\neq 0$, i.e. when $\langle {\bar \psi} \psi\rangle\neq 0$,  the fermionic spectrum has a mass gap, and the electronic states with momenta $\pm p_F$ have an energy gap. 
Alternatively we could have considered a state with the in which the (hermitian) operator that has an expectation value is 
\begin{equation}
i {\bar \psi} \gamma_5 \psi = i \left(\psi^\dagger_R(x) \psi_L(x)- \psi^\dagger_L(x) \psi_R(x) \right)
\label{eq:chiral-mass}
\end{equation}
which is known as the $\gamma_5$ mass term. In a more general CDW state both mass terms can be present.

\subsection{Chiral symmetry and chiral symmetry breaking}
\label{sec:global-chiral-symmetry}

The CDW states we  introduced have different symmetries.  To see this let us observe that the operators $\psi_{\pm Q}(x)$ transform non-trivially under a global U(1) chiral transformation
\begin{equation}
\psi_{\pm Q}'(x)=e^{\mp 2i \theta} \psi_{\pm Q}(x)
\label{eq:chiral-cdw-transf}
\end{equation}
Upon substituting this transformation in the expansion of the density $\rho(x)$ of Eq.\eqref{eq:fourier-decomposition}, we see that a chiral transformation is equivalent to a uniform displacement of the density operator  by $\theta/p_F$,
\begin{equation}
\rho(x) \to \rho\left(x+\theta/p_F\right)
\label{eq:displacement}
\end{equation}
Under a chiral transformation with arbitrary angle $\theta$, the Dirac and $\gamma_5$ mass term operators transform as an orthogonal transformation of a two-component vector. In particular for a chiral transformation with $\theta=\pi/4$,
\begin{equation}
\bar \psi \psi \mapsto i \bar \psi \gamma_5 \psi, \qquad  i \bar \psi \gamma_5 \psi \mapsto -\bar \psi \psi
\label{eq:chiral-transf-bilinears}
\end{equation}
which is equivalent to a displacement of the density by $1/4$ of the period of the CDW. In this sense these two states are equivalent, as is the state with a more general linear combination.

The microscopic lattice model (and the continuum non-relativistic model) are invariant under the spatial inversion symmetry $x \leftrightarrow -x$. This also implies a symmetry under the exchange of right and left moving components of the Dirac spinor, $\psi_R \leftrightarrow \psi_L$. In the massless Dirac theory this operation is equivalent to the multiplication of the spinor by the Pauli matrix $\sigma_1$. In the massless theory  the multiplication by $\sigma_2$ (followed by a chiral transformation with $\theta=\pi/2$) has the same effect. 

These symmetries have an important effect on the fermionic spectrum. To understand what they do let us consider the one-particle Dirac Hamiltonian with a Dirac mass $m$ (jn momentum space)
\begin{equation}
h=
\begin{pmatrix}
p & m\\
m & -p
\end{pmatrix}
\label{eq:h-dirac-1}
\end{equation}
This operator anti-commutes with the Pauli matrix $\sigma_2$, $\{ h, \sigma_2\}=0$. Let $|E\rangle$ be an eigenstate of the Hamiltonian $h$ with energy eigenvalue $E=\sqrt{p^2+m^2}$. Let us consider the state $\sigma_2 |E\rangle$. It is also an eigenstate of $h$  but with energy $-E$, i.e.
\begin{equation}
h \sigma_2 |E\rangle=-\sigma_2 h |E \rangle=- E |E \rangle, \quad \Rightarrow \sigma_2 |E\rangle=|-E\rangle
\label{eq:spectral-symmetry}
\end{equation}
Hence if $|E\rangle$ is an eigenstate of energy $E$, then $\sigma_2 |E\rangle$ is an eigenstate of energy $-E$. This means that the spectrum is invariant under charge conjugation symmetry. Notice that under this operation the spinor transforms as
\begin{equation}
\sigma_2 
\begin{pmatrix}
\psi_R\\
\psi_L
\end{pmatrix}=
\begin{pmatrix}
-i \psi_L\\
i \psi_R
\end{pmatrix}
=e^{-i \gamma_5 \pi/2}
\begin{pmatrix}
\psi_L\\
\psi_R
\end{pmatrix}
\label{eq:spinor-parity}
\end{equation}
In other words, this theory is invariant under charge conjugation $\mathcal{C}$ and parity $\mathcal{P}$ which, combined, it implies that it is invariant under time-reversal $\mathcal{T}$.

The same consideration applies in the case of a $\gamma_5$ mass term in which case the one-particle Dirac Hamiltonian now is
\begin{equation}
h=
\begin{pmatrix}
p& -im_5\\
i m_5&-p
\end{pmatrix}
\label{eq:h-dirac-2}
\end{equation}
This Hamiltonian now anti-commutes with the Pauli matrix $\sigma_1$ which also implies that the same charge conjugation symmetry $\mathcal{C}$, $|E\rangle \leftrightarrow |-E\rangle$,
 is present in the spectrum of the case of a $\gamma_5$ mass. We can also repeat the argument on parity invariance $\mathcal{P}$, which is now  multiplication by $\sigma_1$, Thus the theory is invariant under time-reversal $\mathcal{T}$.
 
 However, if the theory  has both a Dirac mass $m$ and a $\gamma_5$ mass $m_5$ these symmetries are broken. Indeed, the one-particle Dirac Hamiltonian now is
 \begin{equation}
h=
\begin{pmatrix}
p& m-im_5\\
m+i m_5&-p
\end{pmatrix}
=p \sigma_3+m \sigma_1+m_5 \sigma_2
\label{eq:h-dirac-3}
\end{equation}
which no longer has a spectral symmetry. In this case $\mathcal{C} \mathcal{P}$ is broken and, hence, so it  $\mathcal{T}$ since $\mathcal{C} \mathcal{P} \mathcal{T}$ remains unbroken (as it should).

In a lattice system this is a symmetry transformation only if $\theta=\pi n/p_F$ is a lattice displacement, which restricts the allowed values of the chiral angle to be discrete. Although this is true interactions play a significant role in the actual behavior. In fact, there are physical situations in which an effective continuous symmetry actually emerges in this he infrared (long-distance) limit. This is what happens when the CDW is incommensurate and, as we will see in the next subsection, it slides under the action of an electric field. 

In the case of a half-filled system (with only nearest-neighbor hopping matrix elements) the Fermi wave vectors are $\pm \pi/2$. In this case the allowed discrete chiral transformation has a chiral angle $\pi/2$ under which $\bar \psi \psi \mapsto - \bar \psi \psi$ (and similarly for $i \bar \gamma_5 \psi$) corresponding by a translation by one lattice spacing. The allowed four fermion operator is $(\bar \psi \psi)^2$. If the lattice fermions are spinless this operator reduces to
\begin{equation}
(\bar \psi \psi)^2=-2 j_R(x)  j_L(x) +\lim_{y \to x} \psi_R^\dagger(x)\psi_L^\dagger(x) \psi_L(y) \psi_R(y)+ {\rm h.c.}
\label{eq:psi_bar_psi-2}
\end{equation}
where introduced the  right and left moving (chiral) components of the current operator
\begin{equation}
j_R(x)=\frac{1}{2}(j_0(x)+j_1(x))=\psi^\dagger_R(x)\psi_R(x), \qquad j_L(x)=\frac{1}{2}(j_0(x)-j_1(x))= \psi^\dagger_L(x)\psi_L(x)
\label{eq:chiral-densities}
\end{equation}
The first term in Eq.\eqref{eq:psi_bar_psi-2} is known as the backscattering interaction and has scaling dimension 2. Hence, it is a marginal operator. The theory with only the first term is known in Condensed Matter Physics as the Luttinger model and in High-Energy Physics at the (massless) Thirring model. The second operator in Eq.\eqref{eq:psi_bar_psi-2}  formally violates momentum conservation as its total momentum is $ 4 p_F=2\pi$ which is a reciprocal lattice vector and, as such, it is equivalent to zero (mod $2\pi$). Such an operator is not formally allowed in a (naive) continuum theory. This operator is due to a lattice 
Umklapp process and breaks the formal continuous chiral symmetry to a discrete $\mathbb{Z}_2$ subgroup. Although the naive scaling dimension of this operator is 2 (and hence it is formally marginal). 
If the fermions are spinless, the  leading operator  actually vanishes and the leading non-vanishing operator actually has dimension 4, which is irrelevant. However, as shown above,  backscattering processes of the form $j_R(x) \, j_L(x)$ are part of this operator and are exactly marginal.

 If the interaction is strong enough the backscattering interaction  can make the Umklapp operator relevant. When this happens the fermionic system has a quantum phase transition to an insulating state with a period 2 (commensurate) CDW state.
On the other hand, for spin 1/2 fermions the operator $(\bar \psi \psi)^2$ is allowed and  is marginally relevant. If the interactions are repulsive the resulting state is an antiferromagnetic N\'eel state at quantum criticality, while for attractive interactions it is a period 2 CDW. This is what happens in the 1D Hubbard model (see, e.g. Ref. \cite{fradkin-2013} for a detailed analysis).

There are two theories in relativistic systems which are closely related to this problem. One if the Gross-Neveu model \cite{gross-1975} which is a theory of $N$ species of massless Dirac spinors with Lagrangian density 
\begin{equation}
\mathcal{L}_{\rm GN}={\bar \psi}_a i \slashed{\partial} \psi_a+g (\bar \psi_a \psi_a)^2
\label{eq:L-GN}
\end{equation} 
with $a=1, \ldots, N$ (summation of repeated induces is implies). The spin-1/2 Hubbard model corresponds to the case $N=2$. This Lagrangian is invariant only under the discrete chiral symmetry $\bar \psi_a \psi_a \mapsto - \bar \psi_a \psi_a$. This is a  discrete, $\mathbb{Z}_2$, symmetry  and as such it can be spontaneously broken  in 1+1 dimensions. For $N \geq 2$, the resulting state has a (dynamically) broken $\mathbb{Z}_2$ chiral symmetry and that there is a chiral condensate $\langle \bar \psi_a \psi_a \rangle \neq 0$ corresponding to a period 2 CDW.

The other theory is known as the chiral Gross-Neveu model whose Lagrangian density is
\begin{equation}
\mathcal{L}_{\rm cGN}={\bar \psi}_a i \slashed{\partial} \psi_a+g \left((\bar \psi_a \psi_a)^2-(\bar \psi_a \gamma_5 \psi_a)^2\right)
\label{eq:L-cGN}
\end{equation} 
which has the full continuous chiral symmetry. For $N=1$ this theory is equivalent to the Luttinger model (and to the Gross-Neveu model if we ignore the umklapp term). For $N\geq 2$ the chiral symmetry is formally broken. If this were true this theory would violate the Mermin-Wagner theorem. However a detailed study (most easily done using bosonization methods) shows that instead of long range order the correlator of both mass terms decay as a power law as a function of distance, consistent with the requirements by this theorem.

We close this subsection by noting that if the lattice model is not half filled but its density is either incommensurate or has a higher degree of commensurability, say $p/q$, the chiral symmetry is actually continuous (in the incommensurate case) or effectively continuous since the requisite umklapp terms are strongly irrelevant. However, if the lattice model is not at half filling charge conjugation symmetry $\mathcal{C}$ is broken at the lattice scale (in the UV), where it is equivalent to particle-hole symmetry, but it is recovered in the low-energy, IR, regime (up to irrelevant operators). In this sense, both the continuous chiral symmetry and charge conjugation symmetry can be regarded as emergent IR symmetries

\subsection{The chiral anomaly}
\label{sec:chiral-anomaly}

In section \ref{sec:global-chiral-symmetry} we showed  that the theory of massless Dirac fermions, in addition to a global U(1) gauge gauge symmetry, has a second conservation law which we called a global U(1) chiral symmetry, shown in Eq.\eqref{eq:global-chiral-symmetry}. This symmetry implies that there is a locally associated chiral current $j_\mu^5$, given by
\begin{equation}
j_\mu^5(x)={\bar \psi}(x) \gamma_\mu \gamma_5 \psi(x) 
\label{eq:chiral-current}
\end{equation}
which also satisfies a continuity equation
\begin{equation}
\partial^\mu j_\mu^5=0
\label{eq:continuity-chiral}
\end{equation}
and there is a globally conserved chiral charge $Q^5$
\begin{equation}
Q^5=\int_{-\infty}^\infty dx \; j_0^5(x)=\int_{-\infty}^\infty dx \left(\psi^\dagger_R \psi_R-\psi^\dagger_L\psi_L\right)
\label{eq:chiral-charge}
\end{equation}
It is easy to check that the Dirac current $j_\mu$ and the chiral current $j_\mu^5$ are related by
\begin{equation}
j_\mu^5=\epsilon_{\mu \nu} j^\mu
\label{eq:relation-currents}
\end{equation}
where $\epsilon_{\mu \nu}$ is the second rank Levi-Civita tensor.

The simultaneous conservation of both currents $j_\mu$ and $j_\mu^5$  in the massless Dirac theory implies that the right and left moving densities $j_R$ and $j_L$, defined in Eq.\eqref{eq:chiral-densities}, should be separately conserved. In fact, if the Dirac theory has a mass term
\begin{equation}
\mathcal{L}=\bar \psi i \slashed{\partial} \psi-m \bar \psi \psi
\label{eq:L-dirac-massive}
\end{equation}
 it is straightforward to show that
\begin{equation}
\partial^\mu j_\mu^5=2m i \bar \psi \gamma^5 \psi
\label{eq:non-conservation-axial-current}
\end{equation}
which means that in the massive theory the axial current is not conserved. This is easy to understand since the mass term mixes the  right and left moving components of the Dirac spinor and, hence, the right and left moving densities are not conserved.

What happens in the {\it massless} limit, $m \to 0$, is more subtle. This problem was investigated in the late 1960's in 3+1 dimensions by S. Adler \cite{adler-1969} and by J. S. Bell and R. Jackiw \cite{bell-1969} who were interested in the anomalous decay of a neutral pion into two photons, $\pi^0 \to 2\gamma$. This process appears at third order of perturbation theory and it involves the computation of a  triangle diagram (a fermion loop). In 3+1 dimensions this process has a UV divergence which needed to be regulated. These authors showed that it is not possible to find a regularization in which both the Dirac (gauge) current $j_\mu=\bar \psi \gamma_\mu \psi$ and the axial current $j_\mu^5=\bar \psi \gamma_\mu \gamma^5 \psi$ are conserved. In other words, if gauge invariance is preserved then the axial current is not and has an {\it anomaly} and results in a non-conservation of the axial current, $\partial^\mu j_\mu^5\neq 0$. On the other hand, at least in the case of the physical gauge-invariant regularization, the obtained expression for the anomaly in the axial current is {\it universal}, independent of the value of the UV regulator (the cutoff). Sometime later G. 't Hooft showed that in non-abelian gauge theories instant processes also lead to anomalies and, furthermore, the result was also universal \cite{thooft-1976,thooft-1976b}. Since the result is universal and, hence, independent of the UV scale, this led to the concept of anomaly matching conditions.

We will examine this problem in 1+1 dimensions (although it plays a key role in the theory of topological insulators in three space dimensions \cite{qi-2008}). As we noted above, the lattice model is gauge-invariant and has only one conserved current. The conserved axial current appeared only in the low-energy regime in which the lattice model is described by a theory of massless Dirac fermions. To understand this problem we will consider the theory of fermions in 1+1 dimensions coupled to a U(1) gauge field field. In the presence of a background (i.e. not quantized) electromagnetic field in the $A_0=0$ gauge the free fermion Hamiltonian of Eq.\eqref{eq:H0-1D} becomes
\begin{equation}
H[A]=-t \sum_{n=1}^L c^\dagger(n) e^{i eA(n,t)/\hbar} c(n)+\textrm{h.c.}
\label{eq:H[A]}
\end{equation}
An uniform and constant electric field is $E$ is represented by a vector potential $A=-c Et$ (with $c$ being the speed of light). The net effect of this gauge field is to shift of the momentum of the fermion quasiparticles $p \to p+ecEt/\hbar$ or, what is the same, to displace the fermion dispersion relation in momentum by the $ecEt/\hbar$. This means that the Fermi points are also 
shifted by that amount, $p_F \to p_F+ecEt/\hbar$ and $-p_F \to -p_F+ecEt/\hbar$. This means that the single particle states between $p_F$ and $p_F+ecEt/\hbar$ that were empty for $E=0$ are now occupied and the states between $-p_F$ and $-p_F+ecEt/\hbar$ that were occupied for $E=0$ are now empty. This means that number of right-moving fermions is increasing at a rate of $ecE/\hbar$ and that the number of left-moving fermions decreases at the same rate. This results in a net current. Throughout this process the total number of fermions is not changed, gauge invariance is satisfied, but the number of right and left moving fermions are not separately concerned. Notice that this is an effect that involved the entire Femi sea but the net effect is at low energies.

Let us  see now how this plays out in the effective Dirac theory. The massless Dirac Lagrangian density in the background of an unquantized electromagnetic field $A_\mu$ is
\begin{equation}
\mathcal{L}=\bar \psi (i \slashed{\partial}-e \slashed{A})  \psi
\label{eq:1D-Dirac-L-A}
\end{equation} 
Since there is no mass term the Dirac equation still decouples into two equations, for the right and left moving components of the Dirac spinor. In the $A_0=0$ gauge they are
\begin{equation}
i \partial_0 \psi_R=(-i\partial_1-A^1)\psi_R, \qquad i\partial_0 \psi_L=(i\partial_1-A^1) \psi_L
\label{eq:Dirac-1D-gauge}
\end{equation}
In the temporal gauge, $A_0=0$, a uniform electric field $E=\partial_0A^1$, and $A^1$ increases linearly with time. As $A^1$ increases, the Fermi momentum $p_F$ (which is equal to the Fermi energy $E_F$) also increases at the rate $eE$. The density of states of a system of length $L$ is $L/(2\pi)$. So, the rate of change of the number of right-moving fermions is
\begin{equation}
\frac{dN_R}{dx_0}=\frac{e}{2\pi} E
\label{eq:rate-right-movers}
\end{equation}
where we defined $N_R=\int_0^L dx \; j_R$ and similarly for $N_L$. If the UV regulator of the theory is compatible with gauge invariance, then  the total fermion number must be conserved and the total vacuum charge must remain equal to zero,
\begin{equation}
Q=\int_0^L dx\; j_0(x)=N_R+N_L=0
\label{eq:charge-conservation}
\end{equation}
Thus, if $N_R$ increases, then $N_L$ must decrease by the same amount. Or, equivalently, the electric field $E$ creates an equal number of particles $N_R$ and of antiparticles $\bar N_L=-N_L$.

On the other hand, the chiral charge $Q^5=N_R-N_L$ must increase at the rate
\begin{equation}
\frac{dQ^5}{dx_0}=\frac{dN_R}{dx_0}+\frac{d \bar N_L}{dx_0}=\frac{e}{\pi} E
\label{eq:axial-anomaly-1D}
\end{equation}
Again, the details of the UV regularization do not matter, only that it is gauge-invariant. We can also interpret Eq.\eqref{eq:axial-anomaly-1D} as the rate of particle-antiparticle pair creation by an electric field.

In a relativistic notation these results are expressed as
\begin{equation}
\partial^\mu j_\mu^5=\frac{e}{2\pi} \epsilon_{\mu \nu} F^{\mu \nu}
\label{eq:axial-anomaly}
\end{equation}
Hence, the formally conserved current $j_\mu^5$ has an anomaly and is not conserved due to quantum effects. Since it is not conserved, we cannot gauge the chiral symmetry.
In the next subsection we will see that the the anomaly is closely related with bosonization.

\subsection{Bosonization, anomalies and  duality}
\label{sec:bosonization-rules}

We will reexamine the problem at hand from the point of view of the fermionic currents of the Dirac theory $j_\mu$ as operators. Since the currents obey the continuity equation, $\partial_\mu j^\mu=0$ one expects that this would imply that it may be possible to write them as a curl of a scalar field, i.e. $j_\mu(x)=\epsilon_{\mu \nu} \partial^\nu \phi(x)$ where $\phi(x)$ should be a scalar field. Since this should be an operator identity we will need to understand how the currents act on the physical Hilbert space.

The Fermi-Bose mapping in 1D systems is closely related to the chiral anomaly we just discussed. Bosonization of a system of 1+1 dimensional massless Dirac fermions is a set of operator identities understood as matrix elements of the observables in the physical Hilbert space. These identities were first derived by Daniel Mattis and Elliott Lieb \cite{mattis-1965}, based on earlier work by Julian Schwinger \cite{schwinger-1959}. These identities were rediscovered (and their scope greatly expanded) by Alan Luther and Victor Emery \cite{luther-1974}, by Sidney Coleman \cite{coleman-1975}, by Stanley Mandelstam \cite{mandelstam-1975}, and by Edward Witten \cite{witten-1978}. A non-abelian version of bosonization was subsequently derived by Witten \cite{witten-1984}.

The physical Hilbert space is defined as follows. Let  $|{\rm FS}\rangle\equiv |0\rangle$ denote the {\it filled Fermi sea}. In what follows we will assume that the physical system is macroscopically large abd that local operators  create the physical states by acting {\it finitely} on the filled Fermi sea. Physical observables, such as the right and left moving densities $j_R(x)$ and $j_L(x)$, need to be {\it normal-ordered} with respect to the physical vacuum state, the filled Fermi (Dirac) sea $|0\rangle$. The normal ordered densities are $:j_R(x): \; \equiv j_R(x)-\langle 0|j_R(x)|0\rangle$ and $:j_L(x): \; \equiv j_L(x)-\langle 0|j_R(x)|0\rangle$. Since the densities are products of fermion operators they need to be defined as a limit in which the operators are separated by a short distance $\eta$.
Crucial to this construction is that the computation the expectation values be regularized in such a way that the charge (gauge) current $j_\mu(x)$ is locally conserved and satisfies the continuity equation $\partial_\mu j^\mu=0$ as an operator identity. In what follows all expectation values will refer to the filled Fermi sea state $|0\rangle$.

The propagators of the right and left moving Fermi fields are given by
\begin{equation}
\langle \psi_R^\dagger(x_0,x_1) \psi_R(0,0)\rangle=\frac{-i}{2\pi(x_0-x_1+i\epsilon)}, \qquad \langle \psi_L^\dagger(x_0,x_1) \psi_L(0,0)\rangle=\frac{i}{2\pi(x_0+x_1+i\epsilon)}
\label{eq:chiral-fermion-propagators}
\end{equation}
The expectation value of the currents at a space location $x_1$ are
\begin{equation}
\langle j_R(x_1)\rangle=\lim_{\eta \to 0} \langle \psi_R^\dagger(x_1+\eta) \psi_R(x_1-\eta)\rangle =\frac{i}{4\pi \eta}, \quad
\langle j_L(x_1)\rangle=\lim_{\eta \to 0} \langle \psi_L^\dagger(x_1+\eta) \psi_L(x_1-\eta)\rangle =\frac{-i}{4\pi \eta}
\label{eq:short-distance-chiral}
\end{equation}
which are divergent at short distances. 
It follows that the normal-ordered right and left moving current densities satisfy the equal-time commutation relations
\begin{equation}
[j_R(x_1),j_R(x'_1)]=-\frac{i}{2\pi} \partial_1 \delta(x_1-x'_1), \quad [j_L(x_1),j_L(x'_1)]=+\frac{i}{2\pi} \partial_1 \delta(x_1-x'_1)
\label{eq:schwinger-terms-chiral}
\end{equation}
These identities imply that the normal-ordered space-time components of the current $j_\mu=(j_0,j_1)$ satisfy the equal-time commutation relations
\begin{equation}
[j_0(x_1),j_1(x'_1)]=-\frac{i}{\pi} \partial_1 \delta(x_1-x'_1), \qquad [j_0(x_1),j_0(x'_1)]=[j_i(x_1),j_1(x'_1)]=0
\label{eq:U(1)-current-algebra}
\end{equation}
The non-vanishing right-hand sides of these commutators are known as Schwinger terms. These identities define the the U(1) (Kac-Moody) current algebra.

We should note that in a theory of {\it non-relativistic} Fermi fields $\psi({\bm x}, t)$ in all dimensions, the  charge density   $\rho({\bm x})$ and the  current  operators
${\bm j}(\bm x)=\frac{1}{2i}\left(\psi^\dagger(\bm x) {\bm \bigtriangledown}\psi(\bm x)-{\bm \bigtriangledown}\psi^\dagger(\bm x) {\bm \bigtriangledown}\psi(\bm x)\right)$ satisfy a similar expression (also at equal times)
\begin{equation}
[\rho({\bm x}),   j_k({\bm x}')]=-i \frac{e^2}{mc^2}\partial_k\left(\delta({\bm x}-{\bm x}') \rho({\bm x})\right), \quad [\rho(\bm x),\rho({\bm x}')]=0, \quad [j_k(\bm x), j_l({\bm x}')]=0
\label{eq:schwinger-nr}
\end{equation}
In one dimension, and in the regime in which the fermions have a macroscopic density so that $\rho(x) \simeq \langle \rho(x) \rangle$,  after a multiplicative rescaling of the operators, the non-relativistic identities of Eq.\eqref{eq:schwinger-nr} are equivalent to the U(1) current algebra of Eq.\eqref{eq:U(1)-current-algebra}.

The U(1) current algebra of Eq,\eqref{eq:U(1)-current-algebra} is reminiscent of the equal-time canonical commutation relations of a scalar field. Indeed, if $\phi(x)$ is a scalar field and $\Pi(x)=\partial_0 \phi(x)$ is its canonically conjugate momentum, then they obey the equal-time canonical commutation relations
\begin{equation}
[\phi(x_1), \Pi(x'_1)]= i \delta(x_1-x'_1)
\label{eq:scalar-ccr}
\end{equation}
We can then identify the charge density $j_0$ and the current density $j_1$ with the scalar field operators 
\begin{equation}
j_0(x)=\frac{1}{\sqrt{\pi}} \partial_1 \phi(x), \qquad j_1(x)=-\frac{1}{\sqrt{\pi}} \Pi(x)=-\frac{1}{\sqrt{\pi}} \partial_0 \phi(x)
\label{eq:op-id}
\end{equation}
which obey the U(1) current algebra of Eq.\eqref{eq:U(1)-current-algebra} as a consequence of the canonical commutation relations, Eq. \eqref{eq:scalar-ccr}. Furthermore, we can rewrite Eq.\eqref{eq:scalar-ccr} in the Lorentz covariant form
\begin{equation}
j_\mu(x)=\frac{1}{\sqrt{\pi}} \epsilon_{\mu \nu} \partial^\nu \phi(x)
\label{eq: bosonization-current}
\end{equation}
which is clearly consistent with the local conservation of the current $j_\mu$,
\begin{equation}
\partial_\mu j^\mu(x)=0
\label{eq:gauge-current-conservation}
\end{equation}

Let us examine now the question of the conservation of the chiral current $j_\mu^5$. In Eq.\eqref{eq:relation-currents} we showed that the gauge current and the chiral current are related by $j_\mu^5=\epsilon_{\mu \nu} j^\nu$. Therefore the divergence of the chiral current is
\begin{equation}
\partial^\mu j_\mu^5=\epsilon_{\mu \nu} \partial^\mu j^\nu=\frac{1}{\sqrt{\pi}} \epsilon_{\mu \nu} \epsilon^{\nu \lambda} \partial^\mu \partial_\lambda \phi=-\frac{1}{\sqrt{\pi}} \partial^2\phi
\label{eq:chiral-current-conservation-2}
\end{equation}
where we used the identification of the gauge current in terms of the scalar field $\phi$, Eq.\eqref{eq: bosonization-current}. Therefore  we conclude that
\begin{equation}
\partial^\mu j_\mu^5=0 \Leftrightarrow \partial^2 \phi=0
\label{eq:ju5-phi}
\end{equation}
This equation states that the  chiral current as an operator identity is conserved if and only if  the  field $\phi$ is a free massless scalar field, whose Lagrangian density  is
\begin{equation}
\mathcal{L}_B=\frac{1}{2} (\partial_\mu \phi)^2
\label{eq:L-free-scalar}
\end{equation}

In Eq.\eqref{eq:1D-Dirac-L-A} we considered the free massless Dirac Lagrangian coupled to a background (not quantized) gauge field $A_\mu$ through the usual minimal coupling which here is $\mathcal{L}_{\rm int}=-e j_\mu A^\mu$. using the bosonization identity for the gauge current, Eq.\eqref{eq: bosonization-current} we see that the Lagrangian density of the bosonized theory now becomes
\begin{align}
\mathcal{L}_B[A]=&\frac{1}{2} (\partial_\mu \phi)^2-\frac{e}{\sqrt{\pi}} \epsilon_{\mu \nu} \partial^\nu \phi (x) \; A^\mu(x)\nonumber \\
 \equiv & \frac{1}{2} (\partial_\mu \phi)^2+ J (x) \phi(x)
\label{eq:L_B-A}
\end{align}
where the source $J(x)$ is
\begin{equation}
J(x)=\frac{e}{\sqrt{\pi}} \epsilon_{\mu \nu} \partial^\nu A^\mu(x)=\frac{e}{\sqrt{4\pi}} F^*(x)
\label{eq:source-scalar}
\end{equation}
where $F^*(x)=\epsilon_{\mu \nu} F^{\nu \mu}(x)$ is the (Hodge) dual of the field strength $F_{\mu \nu}$. Hence, $F^*$ is (essentially) the source for the scalar field $\phi(x)$. This implies that the equation of motion of the scalar field must be
\begin{equation}
-\partial^2 \phi(x)=J(x)=\frac{e}{\sqrt{\pi}} \epsilon_{\mu \nu} \partial^\nu A^\mu
\label{eq:EOM-scalar+A}
\end{equation}
Retracing our steps we find that the chiral current $j_\mu^5$ obeys
\begin{equation}
\partial^\mu j_\mu^5=-\frac{1}{\sqrt{\pi}} \partial^2 \phi=\frac{e}{2\pi} \epsilon_{\mu \nu} F^{\mu \nu}
\label{eq:chiral-anomaly-bosonized}
\end{equation}
which reproduces the the chiral anomaly given in Eq.\eqref{eq:axial-anomaly-1D}.

These results suggest that the theory of a free massless Dirac spinor must be equivalent to the theory of the free massless scalar field. This statement is known as bosonization. However for this identification to be correct there must be an identification of the Hilbert spaces and of all the operators of each theory. We will not do this detailed analysis here but we will highlight the most significant statements.

Let us begin with the fermion number of the Dirac theory. Consider a system of fermions of total length $L$. Using the bosonization identity of Eq.\eqref{eq: bosonization-current} we find that the fermion number $N_F\equiv Q$ is given by
\begin{equation}
N_F=\int_0^L dx_1\; j_0(x_0,x_1)=\frac{1}{\sqrt{\pi}} \int_0^L dx_1 \; \partial_1 \phi(x_0,x_1)=\frac{1}{\sqrt{\pi}} (\phi(x_0,L)-\phi(x_0,0))
\label{eq:N_F-delta-phi}
\end{equation}
Thus, the vacuum sector of the Dirac theory, with $N_F=0$, corresponds to the theory of the scalar field with periodic boundary conditions, $\phi(x_1=0)=\phi(x_1=L)$. Furthermore, since the fermion number is quantized, $N_F \in \mathbb{Z}$, changing the fermion number is the same as twisting the boundary conditions of the scalar so that 
\begin{equation}
\phi(x_1+L)=\phi(x_1)+\sqrt{\pi} N_F
\label{eq:compactification-scalar}
\end{equation}
In String Theory \cite{polchinski-book} the scalar field is interpreted as the coordinate of a string. Compactifying the space where the string lives to be a circle of radius $R$ means that the string coordinate is defined modulo $2\pi R n$, where $n$ is an integer. We see that the condition imposed by Eq.\eqref{eq:compactification-scalar} is equivalent to say that the scalar field is {\it compactified} and that the compactification radius is $R=1/\sqrt{4\pi}$. This identification also imposes the restriction that  the allowed operators of the scalar field must obey the identification 
\begin{equation}
\phi(x)\sim \phi(x)+2\pi R n
\label{eq:compactification}
\end{equation}
as an equivalency condition.

The simplest bosonic operators that obey the compactification condition are the vertex operators $V_\alpha(x)$,
\begin{equation}
V_\alpha(x)=\exp(i \alpha \phi(x))
\label{eq:vertex-operators}
\end{equation}
The compactification condition then requires that the allowed vertex operators should have  $\alpha=n/R=\sqrt{4\pi} \; n$, where $n$ is an integer. Since the propagator of the scalar field in 1+1-dimensional (Euclidean) spacetime is
\begin{equation}
G(x-x')=-\frac{1}{2\pi} \ln \left(\frac{|x-x'|}{a}\right)
\label{eq:2D-propagator}
\end{equation}
where $a$ is a short-distance cutoff, we find that the scaling dimension of the vertex operator is $\Delta_\alpha=\alpha^2/(4\pi)=n^2$. We will see shortly that the vertex operator with $\alpha=\sqrt{4\pi}$ is essentially the Dirac mass operator (which has scaling dimension 1).

The free massless scalar field can be decomposed into right and left moving components, $\phi_R$ and $\phi_L$ respectively,
\begin{equation}
\phi=\phi_R+\phi_L, \qquad \vartheta=-\phi_R+\phi_L
\label{eq:scalar-chiral-decomposition}
\end{equation}
where 
\begin{equation}
\vartheta(x_0,x_1)=\int_{-\infty}^{x_1} dx'_1 \Pi(x_0,x'_1)
\label{eq:CR-dual}
\end{equation}
is the Cauchy-Riemann dual of the field $\phi(x)$ since they satisfy the Cauchy-Riemann equation
\begin{equation}
\partial_\mu \phi=\epsilon_{\mu \nu} \partial^\nu \vartheta
\label{eq:CR}
\end{equation}
The right and left moving component of the Dirac spinor are found to have the bosonized expression \cite{mandelstam-1975}
\begin{equation}
\psi_R(x)=\frac{1}{\sqrt{2\pi a} } :\exp(i \sqrt{4\pi} \phi_R(x)):, \qquad \psi_L(x)=\frac{1}{\sqrt{2\pi a} } :\exp(-i \sqrt{4\pi} \phi_L(x)):
\label{eq:bosonized-fermions}
\end{equation}
It is easy to check that the propagators of these operators agree with the expressions given in Eq.\eqref{eq:chiral-fermion-propagators}, and that they have scaling dimension $1/2$ and spin $1/2$.

How does a chiral transformation act on the scalar field? A chiral transformation by an angle $\theta$, c.f. Eq.\eqref{eq:chiral-components}, acts on the right and moving fermions as $\psi_R'=\exp(i\theta) \psi_R$ and $\psi_L'=\exp(-i \theta) \psi_L$.  From Eq.\eqref{eq:bosonized-fermions} we see that the right and left moving components of the scalar field transform as $\phi_R'=\phi_R+\theta/\sqrt{4\pi}$ and $\phi_L'=\phi_L+\theta/\sqrt{4\pi}$. This means that a chiral transformation by an angle $\theta$ of the Dirac fermion by an angle $\theta$ is equivalent to a translation (a shift) of the scalar field $\phi'=\phi+2\theta/\sqrt{4\pi}$.

We can use the Operator Product Expansion discussed in section \ref{sec:ope} to show that the fermion mass terms $\bar \psi \psi$ and $i \bar \psi \gamma^5 \psi$  are given by the following identifications
\begin{equation}
\bar \psi \psi=\frac{1}{2\pi a} :\cos(\sqrt{4\pi} \phi):, \qquad i \bar \psi \gamma^5 \psi=:\sin (\sqrt{4\pi}\phi):
\label{eq:bosonized-mass-terms}
\end{equation}
These operators have scaling dimension 1 and transform properly under chiral transformations. These identifications imply that a theory of free {\it massive} Dirac fermions 
\begin{equation}
\mathcal{L}_D=\bar \psi i \slashed{\partial}\; \psi-m \bar \psi \psi
\label{eq:massive-dirac-lagrangian}
\end{equation}
 is equivalent to the sine-Gordon field theory \cite{coleman-1975} whose Lagrangian is
 \begin{equation}
 \mathcal{L}_{{\rm SG}}=\frac{1}{2} (\partial_\mu \phi)^2-g :\cos(\sqrt{4\pi} \phi):
 \label{eq:sine-gordon}
 \end{equation}
 where $g=m/(2\pi a)$.

Given the central role played by the current algebra identities of Eqs. \eqref{eq:U(1)-current-algebra} one may wonder if a similar approach might apply in higher dimensions. Schwinger terms in current algebra play an important role in  relativistic field theories. However in higher dimensions  their structure is more complex and does not lead to identities of the type we have discussed. The reason at the root of this problem is largely kinematical. The Bose (scalar) field $\phi$ is qualitatively a bound state, a collective mode in the language of Condensed Matter Physics. In 1+1 dimensions this collective mode exhausts the spectrum at low energies due to the strong kinematical restriction on one spatial dimension.

The equivalency between the theory of free massive Dirac fermions and the sine-Gordon theory is an example of the power of bosonization. On the Dirac side the mapping the theory is free and its spectrum is well understood. But on the sine-Gordon side the theory is non-linear. In fact in the sine-Gordon theory the fermions are essentially solitons, domain walls of the scalar field. For these and many other reasons that we do not have space here bosonization plays a huge role in understanding the non-perturbative behavior of systems both in Condensed Matter Physics and in Quantum Field Theory in 1+1 dimensions. We will see in section \ref{sec:bosonization-dirac-2+1} that to an extent some of these ideas can and have been extended to relativistic systems and classical statistical mechanical systems in 2+1 dimensions.

\section{Fractional Charge}
\label{sec:fractional-charge}

\subsection{Solitons in one dimensions}
\label{sec:solitons}

We begin by returning to the equivalency between the theory of free massive Dirac fermions and sine-Gordon theory, in 1+1 dimensions. In Eq.\eqref{eq:N_F-delta-phi} we showed that the boundary conditions of the compactified scalar field $\phi(x)$ are determined by the fermion number $N_F$ of the dual Dirac theory, and that the vacuum sector of the Dirac theory maps onto the sine-Gordon theory with periodic boundary conditions. 
We will now examine the sector  with one fermion, $N_F=1$. This sector of the Dirac theory maps onto the sine-Gordon theory with {\it twisted} boundary conditions, 
\begin{equation}
\phi(L)-\phi(0)=\sqrt{\pi}
\label{eq:BC-1fermion}
\end{equation} 

The Hamiltonian of the sine-Gordon theory is
\begin{equation}
H_{\rm SG}=\int_{-\infty}^\infty dx\; \left[\frac{1}{2} \Pi^2(x)+\frac{1}{2} \left(\partial_x \phi(x)\right)+g \cos\left(\sqrt{4\pi} \phi(x)\right)\right]
\label{eq:H-SG}
\end{equation}
In the sector with periodic boundary conditions the classical  ground states are static and uniform configurations that minimize the potential energy. Since the potential energy is a periodic functional of $\phi(x)$ the classical minima are at $\phi_n(x)=(n+1/2) \sqrt{\pi}$, where $n$ is an arbitrary integer. The classical energy of these ground states is extensive and is given by $E_{\rm gnd}=-g L$ where $L$ is the linear size of the system. 

The classical ground state in the twisted sector is a {\it domain wall} (or {\it soliton}) which interpolates between the static and uniform ground states $\phi(x)=\pm \sqrt{\pi}/2$. The classical ground state in this sector is the static solution of the Euler-Lagrange equation 
\begin{equation}
\frac{d^2\phi}{dx^2}=-2g \sqrt{\pi} \sin\left(2\sqrt{\pi}\phi(x)\right)
\label{eq:E-L-soliton}
\end{equation}
such that asymptotically satisfies $\lim_{x \to \pm \infty}=\pm \sqrt{\pi}/2$. The solution is the classical soliton configuration
\begin{equation}
\phi(x)=\frac{2}{\sqrt{\pi}}\tan^{-1} \Big(\exp(2\sqrt{\pi g}(x-x_0))\Big)-\frac{\sqrt{\pi}}{2}
\label{eq:SG-soliton}
\end{equation}
The soliton solution represents a {\it domain wall} between two symmetry-related classical ground states with $\phi=\pm \sqrt{\pi}/2$.

The energy of the soliton (measured from the energy of the ground state in the trivial sector) is finite and is given by
\begin{equation}
E_{\rm soliton}=4 \sqrt{\frac{g}{\pi}}
\label{eq:E-soliton}
\end{equation}
where $x_0$ is a zero mode of the soliton solution and represents its coordinate. By coupling the bosonized theory to a weak electromagnetic field $A_\mu$, as given in Eq.\eqref{eq:L_B-A}, it is easy to check that it has electric charge $-e$ and, in this sense represents the electron. Then the identities of Eq.\eqref{eq:bosonized-fermions} can be used that as a quantum state it is indeed a fermion. 

\subsection{Polyacetylene}
\label{sec:polyacetylene}

In section \ref{sec:bosonization-rules} we saw that solitons of a scalar field can be regarded as being equivalent to electrons, fermions with charge $-e$. We will now see that in a theory of fermions coupled to a domain wall of a scalar field, the soliton carries fractional charge. This problem has been extensively studied in one-dimensional conductors such as polyacetylene, in particular by the work of Wu-Pei Su, J. Robert Schrieffer and Alan Heeger \cite{su-1979} and by Roman Jackiw and J. Robert Schrieffer \cite{jackiw-1981b}. In quantum field theory this problem was first discussed by Roman Jackiw and Claudio Rebbi \cite{jackiw-1976} and by Jeffrey Goldstone and Frank Wilczek \cite{goldstone-1981}.

In section \ref{sec:Dirac-1D} we showed that the physics of lattice fermions in one dimension at low energies is well described by a theory of massless Dirac fermions. In a one-dimensional conductor, such as polyacetylene, the fermions couple to the lattice vibrations (phonons). Su, Schrieffer and Heeger (SSH) \cite{su-1979} proposed a simple model in which the electrons couple to the lattice vibrations through a modulation of the hopping amplitude between two consecutive sites $n$ and $n+1$, instead of being a constant $t$, becomes $t_{n, n+1}=t-g (u_{n+1}-u_n)$, where $u_n$ is the displacement of the ion (a CH group in polyacetylene) at site $n$ from its classical equilibrium position and $g$ is the electron-phonon coupling constant. In polyacetylene there number of  electrons (which are spin-1/2 fermions) is equal to the number of sites of the lettuce and the electronic band is half-filled. At half filling this simple band structure is invariant under a particle-hole transformation. If the coupling to the lattice vibrations is included this symmetry remains respected provided the displacements  change sign $u_n \to -u_n$ for all lattice sites. In polyacetylene  the lattice dimerizes (a process known as a Peierls distortion) and the discrete translation symmetry by one lattice spacing is spontaneously broken: the system becomes a period 2 CDW on the bonds of the lattice. The broken symmetry state is still invariant under a particle-hole transformation. 

The effective field theory of this system is a theory of two Dirac spinors $\psi_{\alpha,\sigma}(x)$, where $\alpha=1, 2$ denotes right and left-moving fermions, and $\sigma=\uparrow,\downarrow$ are the two spin polarizations. The Lagrangian density of this system is
\begin{equation}
\mathcal{L}=\bar \psi_\sigma i \slashed{\partial} \psi_\sigma-g \phi(x) \bar \psi_\sigma(x) \psi_\sigma(x)-\frac{1}{2} \phi(x)^2
\label{eq:TLM}
\end{equation}
The real scalar field $\phi(x)$ represents the distortion field of the polyacetylene chain. Here we will assume that the chains has spontaneously distorted and we will regard the scalar field as static and classical. 
This is a good approximation since the masses of the CH complexes is much bigger than the electron mass. This continuum model is due to Takayama, Lin-Liu and Maki \cite{takayama-1980} and further developed by Campbell and Bishop \cite{campbell-1981,campbell-1982}. Many of the results fund in this (adiabatic) approximation  remain qualitatively correct upon taking into account the quantum dynamics of the chain, even in the  limit in which the ions are treated as being ``light'' (provided the spin of the fermions is taken into account) \cite{hirsch-1982,fradkin-1983}. 

In the field theory the discrete symmetry of displacements by one lattice spacing becomes the $\mathbb{Z}_2$ symmetry $\phi \to -\phi$. This is a symmetry of the electron phonon system once combined with the discrete chiral transformation $\psi \to \gamma_5 \psi$ under which $\bar \psi \psi \to -\bar \psi \psi$. The ground state is two fold degenerate $\pm \phi_0$ with 
\begin{equation}
\phi_0=\frac{2\Lambda \varv_F}{g} \exp  \left(-\frac{\pi \varv_F}{g^2}\right)
\label{eq:phi-0}
\end{equation}
and the Dirac fermion (the electron) has a exponentially small mass,  $m=g \phi_0$.

\subsection{Fractionally charged solitons}
\label{sec:soliton-fractional-charge}

Jackiw and Rebbi showed that the 1+1-dimensional classical $\phi^4$ theory has the following soliton  solution which interpolates between the two classically ordered  states at $\pm \phi_0$ \cite{jackiw-1976}
\begin{equation}
\phi(x)=\phi_0 \tanh \left(\frac{x-x_0}{\xi}\right)
\label{eq:phi4-soliton}
\end{equation}
where $\xi$ is the correlation length of $\phi^4$ theory, and $x_0$ is the (arbitrary) location of the soliton.
They further showed that, when coupled to a theory of massless relativistic fermions through a Yukawa coupling, as in Eq.\eqref{eq:TLM}, this soliton carries fractional charge. The argument goes as follows. The one-particle Dirac Hamiltonian for a Dirac fermion with a position-dependent mass $m(x)$ is
\begin{equation}
H=-i\sigma_1 \partial_x+m(x) \sigma_3=
\begin{pmatrix}
m(x) & -i\partial_x\\
-i \partial_x & -m(x)
\end{pmatrix}
\label{eq:1D-H-Dirac-soliton}
\end{equation}
where $m(x)=g \phi(x)$, with $\phi(x)$ being the soliton solution of Eq.\eqref{eq:phi4-soliton}. This Hamiltonian is hermitian and real. Furthermore, this Hamiltonian anti-commutes with the Pauli matrix $\sigma_2$. This implies that for every positive-energy state $|E\rangle$  with energy $+E$ there is a negative-energy eigenstate with energy $-E$ given by $\sigma_2|E\rangle$. Hence, the spectrum is particle-hole symmetric (or, what is the same, charge-conjugation invariant). In addition, and consistent with charge-conjugation symmetry, the Hamiltonian of Eq.\eqref{eq:1D-H-Dirac-soliton} has state with $E=0$, a zero-mode, with  a normalizable spinor wave function
\begin{equation}
\psi_0(x)=\frac{1}{\sqrt{2} }
\begin{pmatrix}
-i \\
1
\end{pmatrix}
\exp \left(-\textrm{sgn}(m) \int_0^x dx'\; m(x') \right)
\label{eq:zero-mode}
\end{equation}
which exists for an arbitrary function $m(x)$ which changes sign once at some location (which we took to be $x_0=0$). Jackiw and Rebbi further showed that the soliton (anti-soliton) carries fractional charge 
\begin{equation}
Q=\mp \frac{e}{2}
\label{eq:fractional-charge}
\end{equation}
This result follows from the spectral asymmetry identity of the density of states $\rho_S(E)$ in the presence of the soliton
\begin{equation}
Q=-\frac{e}{2} \int_0^\infty \left(\rho_S(E)-\rho_S(-E)\right)=-\frac{e}{2} \eta
\label{eq:spectral-asymmetry}
\end{equation}
where $\eta$ is the spectral asymmetry of the Dirac operator in the soliton background, and it is known as the APS $\eta$-invariant of Atiyah-Patodi-Singer \cite{atiyah-1975}.
Given the one-to-one correspondence that exists between positive and negative energy states in the spectrum, the spectral asymmetry follows from the condition that the zero mode be half-filled, which is required by normal-ordering or, what is the same, by charge neutrality. 
Another way to understand this result is that adding (or removing) a fermion of charge $-e$ results in the creation of a soliton-antisoliton pair, with each topological excitation carrying half of the charge of the electron. In other words, in the dimerized phase the electron is fractionalized. 
In this analysis we ignored the spin of the electron. If we take it into account the spin degree of freedom the soliton is instead a boson with charge $\mp e$.

There is an alternative, complementary, way to think about the charge of the soliton. Goldstone and Wilczek \cite{goldstone-1981} considered a theory in which the (massless) Dirac fermion is coupled to two real scalar fields, $\varphi_1$ and $\varphi_2$, with Lagrangian 
\begin{equation}
\mathcal{L}=\bar \psi i \slashed{\partial} \psi-g \varphi_1 \bar \psi \psi-i g\varphi_2 \bar \psi \gamma_5 \psi
		 \equiv \bar \psi i \slashed{\partial} \psi-g |\varphi | \bar \psi \exp(i \theta \gamma_5) \psi
\label{eq:L-GW}
\end{equation}
where $|\varphi |^2=\varphi_1^2+\varphi_2^2$ and $\theta=\tan^{-1}(\varphi_2/\varphi_1)$. They considered a soliton in which $g \varphi_1=m$ is the constant (in space) Dirac mass and $\varphi_2$ winds slowly between two values $\pm \varphi_0$ for $ x \to \pm \infty$. In this theory the one-particle Dirac Hamiltonian is 
\begin{equation}
H=-i\sigma_1 \partial_x+g \varphi_1 \sigma_3+g\varphi_2 \sigma_2
\label{eq:H-Dirac-CP}
\end{equation}
which is hermitian and complex and, hence, it violates CP invariance.

A perturbative calculation of the induced (gauge-invariant)  current $j_\mu$, which is given  by the triangle diagram of a fermion loop with two gauge field insertions and a coupling of the scalar fields, yields the result (with $a=1,2$)
\begin{equation}
\langle j_\mu(x)\rangle=\frac{1}{2\pi} \epsilon_{\mu \nu} \epsilon_{ab} \frac{\varphi_a \partial^\nu \varphi_b}{|\varphi |^2}=\frac{1}{2\pi} \epsilon_{\mu \nu} \partial^\nu \theta
\label{eq:GW-anomaly}
\end{equation}
which is locally conserved. Notice, however, that the induced axial current $j_\mu^5$ is not conserved, $\partial^\mu \langle j_\mu^5\rangle=-\frac{1}{2\pi} \partial^2 \theta\neq 0$ and, hence, this current is anomalous. 
We can now compute the total charge accumulated as the soliton is created adiabatically to be given by the Goldstone-Wilczek formula
\begin{equation}
Q=-e \frac{\Delta \theta}{2\pi}
\label{eq:GW}
\end{equation}
where $\Delta \theta=\theta( +\infty)-\theta(-\infty)$. Since $\lim_{x \to \pm \infty}\varphi_2(x)=\pm \varphi_0$, we obtain the result
\begin{equation}
Q=-\frac{e}{\pi} \tan^{-1}\left(\frac{g\varphi_0}{m}\right)
\label{eq:Q-general}
\end{equation}
In the limit $m=g\varphi \to 0$, where CP (or T) invariance is recovered, we get
\begin{equation}
\lim_{m \to 0} Q=-\frac{e}{2}
\label{eq:Q-CP}
\end{equation}
which is the Jackiw-Rebbi result for the fractional charge of a soliton of Eq.\eqref{eq:fractional-charge}. 

The results for the fractional charge of the soliton can also be derived using the bosonization identities of section \ref{sec:bosonization-rules}. Indeed, the bosonized expression for the Lagrangian of Eq.\eqref{eq:L-GW} is 
\begin{equation}
\mathcal{L}=\frac{1}{2} \left(\partial_\mu \phi\right)^2 -\frac{g|\varphi |}{2\pi a} \cos\left(\sqrt{4\pi} \phi-\theta\right)
\label{eq:L-GW-bosonized}
\end{equation}
Deep in the phase in which the Bose field $\phi$ is massive, the non-linear term in Eq.\eqref{eq:L-GW-bosonized} locks this field to the chiral angle $\theta$, i.e.
\begin{equation}
\phi=\frac{1}{\sqrt{4\pi}} \theta
\label{eq:phi-locked-theta}
\end{equation}
However, the bosonization identities also tell us that the gauge current $j_\mu$ is given by the curl of the scalar field  $\phi$. Therefore, in this state the current $j_\mu$ is
\begin{equation}
j_\mu=\frac{1}{\sqrt{\pi}}\epsilon_{\mu \nu} \partial^\nu \phi=\frac{1}{2\pi} \epsilon_{\mu \nu} \partial^\nu \theta
\label{eq:current-twist}
\end{equation}
which is the same as the Goldstone-Wilczek result of Eq.\eqref{eq:GW-anomaly}. That these two seemingly different approaches yield the same result is not accidental as they both follow from the axial anomaly.

\section{Fractional Statistics}
\label{sec:fractional-statistics}

A fundamental axiom of Quantum Mechanics is that identical particles are indistinguishable \cite{dirac-1930,landau-1957}. In non-relativistic Quantum Mechanics this leads to the requirement that the quantum states of a system of identical particles must be eigenstates of the pairwise particle exchange operator. Since two exchanges are equivalent to the identity operation this implies that the states must be even or odd under pairwise exchanges. This result, in turn, implies that particles can be classified as either being bosons (whose states are invariant under pairwise exchanges) or fermions (whose states change sign under pairwise particle exchanges). A consequence is that bosons obey the the Bose-Einstein (and can condense into a single particle state) whereas fermions obey the Fermi-Dirac distribution and must obey the Pauli exclusion principle. It is an implicit assumption of this line of reasoning that all relevant states of a system of identical particles can be efficiently represented by a (suitably symmetrized or antisymmetrized) product state.
	
	This classification is present at an even deeper level in (relativistic) Quantum Field Theory where locality, unitarity and Lorentz invariance require that the fields be classified as representations of the Lorentz group and obey the Spin-Statistics Theorem \cite{weinberg-book}. The Spin-Statistics Theorem is actually an {\it axiom} of local relativistic Quantum Field theory which requires that fields that transform with an integer spin representation of the Poincar\'e group (i.e. scalars, gauge fields, gravitational fields, etc) must be bosons while fields that transform with a half-integer spin representation (i.e. Dirac spinors, etc) must be fermions. This spin-statistics connection is intrinsic to the construction of String Theory \cite{polchinski-book}.
	
	Given these considerations there was a general consensus that fermions and bosons were the only possible types of statistics. Nevertheless several exceptions to this rule were known to exist. One is the construction of the magnetic monopole in 3+1 dimensional gauge theory  by Tai-Tsun Wu and Chen-Ning Yang who showed that a scalar coupled to a Dirac magnetic monopole behaves as a Dirac spinor \cite{wu-1976}. This was an early example of statistical transmutation by coupling a matter field to a non-trivial configuration of a gauge field. As we will note below, the construction of anyons (particles with fractional statistics) in 2+1 dimensions has a close parentage to the Wu-Yang example. Examples of statistical transmutation were known to exist in 1+1 dimensional theories where a system of hard-core bosons was shown to be equivalent to a theory of free fermions using the Jordan-Wigner transformation \cite{lieb-1961} which represents a fermion as a composite operator of a hard-core boson and an operator that creates a kink (or soliton). This construction also underlies the fermion-boson mapping in 1+1 dimensional field theories \cite{mattis-1965,luther-1974,coleman-1975,mandelstam-1975} that we discussed in section \ref{sec:bosonization-rules}. Finally, in the late 1970s it was found that 1+1 dimensional $\mathbb{Z}_N$ spin systems harbor operators known as parafermions which obey the same algebra shortly afterwards found to be obeyed by anyons in 2+1 dimensions \cite{fradkin-1980}. 
	
\subsection{Basics of fractional statistics}
\label{sec:basics}
	
Jon Magne Leinaas and Jan Myrheim wrote an insightful paper in 1977 in which they examined the structure of the configuration space of the histories of a system of N identical particles \cite{leinaas-1977}. Using the Feynamn path-integral approach, they showed that if the worldlines of the identical particles are not allowed to cross, then the configuration space is topologically non-trivial. Through a detailed analysis they showed that the three and higher dimensions under a pairwise particle exchange the states must be either even or odd and hence the particles are either bosons or fermions. Leinaas and Myrheim also showed that in one and two space dimensions the wave functions can change by a {\it phase}, nowadays known as the statistical angle. In retrospect this result could have been anticipated (but was not) in an earlier paper by Michael G. G. Laidlaw and C\'ecile Morette DeWitt \cite{laidlaw-1971} who did a similar analysis of the configuration space of identical particles in the Feynman path integral.

Frank Wilczek generalized the Aharonov-Bohm effect \cite{aharonov-1959}  to describe the quantum mechanics of composite objects made of electric charge and magnetic flux in two space dimensions \cite{wilczek-1982b}. Wilczek showed that composite objects made of a non-relativistic  particle of charge $q$ bound to  a magnetic flux of (magnetic) charge $\Phi$  behaves as an object with fractional angular momentum $q\Phi/2\pi$. Here $\Phi$ is measured in units of the flux quantum $2\pi$ (in units in which $\hbar=c=e=1$). Furthermore, in a subsequent paper Wilczek \cite{wilczek-1982a}  showed that, upon an adiabatic process in which the two composites exchange positions without their worldlines coinciding, the wave function of two identical flux-charge composites changes by a phase factor $\exp(\pm i q\Phi)$. For instance if $\Phi=\pi$ (i.e. a half-flux quantum) the acquired phase is equal to $\exp(i\pi)=-1$. thus if the particle was a boson, the composite becomes a fermion and viceversa. Wilczek coined the term {\it anyon} to describe the behavior of an arbitrary charge-flux composite. Furthermore, this construction also implies that not only {\it fractional statistics} but also {\it fractional spin}, consistent with a generalization of the Spin-Statistics Theorem. In other terms, ``flux-attachment'' implies fractional statistics (and fractional spin). Clearly Wilczek's construction gave an explicit physical grounding to the general arguments of the 1977 paper by Leinaas and Myrheim. It is worth note the close analogy between this construction in 2+1 dimensions and the Wu-Yang construction in 3+1 dimensions (whose consistency with the Spin-Statistics Theorem had been shown earlier on by Alfred Goldhaber \cite{goldhaber-1976}). 

In this description the statistics of the composites (the anyons) enters in the form of complex weights (phases) given in terms of the {\it linking numbers} of the worldlines. Hence, the concept of fractional statistics is intimately related to the theory of knots and of the representations of the braid group. These concepts were originally introduced in physics to describe statistical transmutation in the theory of solitons \cite{finkelstein-1968} in the context of the Skyrme model \cite{skyrme-1962}. Yong-Shi Wu \cite{wu-1984} developed an explicit connection between  the  work of Leinaas and Myrheim and Wilczek's work on anyons  in terms of operations acting on the worldlines of the anyons and described by the Braid Group. Wu's work and the somewhat early paper by Wilczek and Zee on the statistics of solitons \cite{wilczek-1983} marked the definite entry of the theory of knots (and of the braid group) in physics in general and in condensed matter physics in particular. The classification of anyons in terms of representations of the braid group labeled by the fractional statistics (determined by linking numbers) as well as of fractional (or topological) spin (determined by the writhing number of the worldlines) leads to a rich set of physical consequences. As it will turn out, Wilczek's anyons are described one-dimensional representations of the braid group. As we will see below, additional and intriguing (non-abelian) representations will also play a role.

\subsection{What is a topological field theory}
\label{sec:TQFT}

We will now consider a special class of gauge theories known as topological field theories. These theories often (but not always) arise as the low energy limit of more complex gauge theories. In general, one expects that at low energies the phase of a gauge theory be either confining or deconfined. While confining phases have (from really good reasons!) attracted much attention, deconfined phases are often regarded as trivial, in the sense that the general expectation is that their vacuum states be unique and the spectrum of low lying states is either massive or massless. 

Let us consider a gauge theory whose action on a manifold $\mathcal{M}$ with metric tensor tensor $g_{\mu \nu}(x)$ is
\begin{equation}
S=\int_{\mathcal{M}} d^Dx\; \sqrt{g} \; \mathcal{L}(g, A_\mu)
\end{equation}
At the classical level, the the energy-momentum tensor $T^{\mu \nu}(x)$ is the linear response of the action to an infinitesimal change of the local metric, 
\begin{equation}
T^{\mu \nu}(x)\equiv \frac{\delta S}{\delta g_{\mu \nu}(x)} 
\label{eq:Tmunu-sym2}
\end{equation}
That a  theory is topological means that depends only on  the topology of the space in which is defined and, consequently,  
it is independent of the local properties that depend on the metric, e.g. distances, angles, etc. 
Therefore, at least at the classical level, the energy-momentum tensor of a topological field theory must vanish identically,
\begin{equation}
T^{\mu \nu}=0
\end{equation}
In particular, if the theory is topological, the {\it  energy} (or Hamiltonian) is also zero. 
Furthermore, if the theory is independent of the metric, it is invariant under {\it  arbitrary} coordinate transformations. 
Thus, if the theory is  a gauge theory, the expectation values of Wilson loops will be independent of the size and shape of the loops. 
Whether or not a theory of this type can be consistently defined at the quantum level is a subtle problem which we will briefly touch on below.

It turns out that, due to the non-local nature of the observables of a gauge theory, the low energy regime of a theory in its deconfined phase 
can have non-trivial global properties. In what follows, we will say that a gauge theory is topological if all local excitations are massive 
(and in fact we will send their mass gaps to infinity). The remaining Hilbert space of states is determined by global properties of the theory, 
including the topology of the manifold of their space-time. In several cases, the effective action of a topological  field theory does not depend 
on the metric of the space-time, at least at the classical level. In all cases, the observables are non-local objects, Wilson loops and their 
generalization.

\subsection{Chern-Simons Gauge Theory}
\label{sec:CS}
	
Gauge theories play a key role in physics. In 2+1 dimensions it is possible to define a special gauge theory which is odd under time reversal invariance and parity: Chern-Simons gauge theory. Originally introduced in Quantum Field Theory in1982 by Stanley Deser, Roman Jackiw and  Stephen Templeton \cite{deser-1982a,deser-1982b}, Chern-Simons gauge theory can be defined for any compact Lie group, as well as an extension of Einstein's gravity in 2+1 dimensions. In 1989 Edward Witten \cite{witten-1989} showed that Chern-Simons gauge theory computes the expectation values of configurations of Wilson loops, regarded as the worldlines of  heavy particles, in terms of a set of topological invariants known as the Jones polynomial that classify knots in three dimensions.

 We will consider the simplest case, the U(1) Chern-Simons gauge theory. The Chern-Simons action for a U(1) gauge field $A_\mu$  in 2+1 dimension is
\begin{equation}
S[A]=\frac{k}{4\pi} \int_\Omega d^3x \, \epsilon_{\mu \nu \lambda} A^\mu \partial^\nu A^\lambda+\int_\Omega d^3x\, J_\mu A^\mu
\label{eq:CS-U1}
\end{equation}
where $J_\mu$ is a set of conserved currents (representing the worldlines of a set of heavy particles). On a closed 3-manifold $\Omega$ (e.g. a sphere, a torus, etc) the Chern-Simons action is gauge invariant provided the parameter $k$ (known as the level) is an integer. If the manifold $\Omega$ has a boundary, the action is not gauge invariant at the boundary. Gauge invariance is restored by additional boundary degrees of freedom. This structure is general, and not just a feature of the U(1) theory. 

Since the Chern-Simons action is first order in derivatives it is not invariant under time reversal and under parity (which in 2+1 dimensions is a reflection). These  symmetries make this theory relevant to the description of the fractional quantum Hall effect. In the absence of external sources, $J_\mu=0$, at the classical level the Chern-Simons action is invariant under arbitrary changes of coordinates. This means that the theory is, at least classically, a topological field theory.

For a general non-abelian gauge group $G$ the Chern-Simons action becomes
\begin{equation}
S=\frac{k}{4\pi} \int_{\mathcal{M}} d^3x \; \textrm{tr} \Big(AdA+\frac{2}{3} A\wedge A \wedge A \Big)
\label{eq:CS-non-abelian-2+1}
\end{equation}
Here, the cubic term is shorthand for 
\begin{equation}
\textrm{tr}\Big(A \wedge A \wedge A\Big)\equiv \textrm{tr}\Big(\epsilon_{\mu \nu\lambda} A^\mu A^\nu A^\lambda \Big)
\end{equation}
for a gauge field $A^\mu$ that takes values on the algebra of the gauge group $G$. 

\subsection{BF gauge theory}
\label{sec:BF}

A closely related (abelian) gauge theory is the so-called $BF$ theory \cite{horowitz-1989} which, in a general spacetime dimension $D$ (even or odd), 
is a theory of a vector field  $A^\mu$ (a one-form) and an antisymmetric tensor field $B$ with $D-2$ Lorentz indices (a $D2$ form), 
known as a Kalb-Ramond field. In 2+1 dimensions the action of the BF theory  is 
\begin{equation}
S=\frac{k}{2\pi} \int_{\mathcal{M}} d^3x\; \epsilon_{\mu \nu \lambda} B^{\lambda} \partial^\mu A^\nu
\label{eq:BF}
\end{equation}
where, once again, $k$ is an integer.  

The BF gauge theory has the same content as the topological sector of a discrete $\mathbb{Z}_k$ gauge theory. To see this we will consider the theory of compact electrodynamics which is a U(1) gauge theory minimally coupled to charged scalar field $\phi$. The will assume that the gauge theory is defined for a compact U(1) gauge group (meaning that the gauge flux is quantized) and that the complex scalar field has charge integer $k\in\mathbb{Z}$.  As usual minimal coupling is implemented by introducing the covariant derivative $D_\mu-\partial_\mu+i k A_\mu$, where $A_\mu$ is the U(1) gauge field. Deep in the phase in which the global U(1) symmetry is spontaneously broken, usually called the Higgs regime, the amplitude of the scalar field is frozen at its (real) vacuum expectation value $\phi_0$ but its phase $\omega$, representing the Goldstone mode,  is unconstrained. In this limit the Lagrangian of this theory becomes
\begin{equation}
\mathcal{L}=|\phi_0|^2 \left(\partial_\mu \omega-k A_\mu\right)^2-\frac{1}{4e^2} F_{\mu\nu}^2
\label{eq:compact-QED}
\end{equation}
where $e$ is the coupling constant of the gauge field (the electric charge) and $F_{\mu \nu}$ is the field strength of the gauge field $A_\mu$. In general spacetime dimension $D$ the charge $e$ has units  of length$^{(D-4)/2}$. In this limit the gauge field becomes massive (this is the Higgs mechanism). This theory has fluxes  quantized in units of $2\pi/k$ and has only $k$ distinct fluxes. This is the $\mathbb{Z}_k$ gauge theory.

Here we will consider this theory in 2+1 dimensions. We will use a gaussian (Hubbard-Stratonovich) decoupling of the first term of Eq.\eqref{eq:compact-QED} in terms of a gauge field $C_\mu$ to write the Lagrangian in the equivalent form
\begin{equation}
\mathcal{L}=
-\frac{1}{4 |\phi_0|^2} C_\mu^2 + C^\mu  (\partial_\mu \omega-k A_\mu)-\frac{1}{4e^2} F_{\mu\nu}^2
\label{eq:compact-QED-HS}
\end{equation}
Up to an integration by parts, we see that  the phase field $\omega$ plays the role of a Lagrange multiplier field which forces the vector field $B_\mu$  to obey the constraint $\partial_\mu C^\mu=0$. This constraint is solved by writing $C_\mu$  as
\begin{equation}
C_\mu=\frac{1}{2\pi} \epsilon_{\mu \nu \lambda} \partial^\nu B^{\lambda}
\label{eq:CB}
\end{equation} 
of a 1-form gauge field $B_{\mu}$. Upon solving the constraint the Lagrangian of Eq.\eqref{eq:compact-QED-HS} becomes
\begin{equation}
\mathcal{L}=\frac{k}{2\pi} \epsilon_{\mu \nu \lambda} A^\mu \partial^\nu B^\lambda-\frac{1}{4e^2} F_{\mu \nu}(A)^2-\frac{1}{32\pi^2|\phi_0|^2} F_{\mu \nu}(B)^2
\label{eq:BF2}
\end{equation}
For spacetime dimensions $D<4$ the IR the Maxwell terms for the fields $A^\mu$ and $B_\mu$ are irrelevant in the IR and, in this limit,  this theory reduces to the BF theory at level $k$ of Eq.\eqref{eq:BF}. Therefore, $\mathbb{Z}_k$ gauge theory is equivalent to the BF theory at level $k$. In fact, this result is essentially valid in all dimensions with the main difference being that the field $B_\mu$ is, in general, a rank $D-2$ Kalb-Ramond antisymmetric field.

\subsection{Quantization of Abelian Chern-Simons Gauge Theory}
\label{sec:CS-quantization}

By expanding the action of Eq.\eqref{eq:CS-U1}  the Lagrangian density becomes
\begin{equation}
\mathcal{L}=\frac{k}{4\pi} \epsilon_{ij} A_i \partial_0 A_j+A_0 \left(\frac{k}{2\pi} B-J_0\right) -J_i A_i
\label{eq:CS-L}
\end{equation}
where $B=\epsilon_{ij} \partial_i A_j$ is the local flux, $J_0$ is a local classical density and $J_i$ a local classical current.
Then, the first term of Eq.\eqref{eq:CS-L} implies that the  spatial components of the gauge field obey equal-time canonical commutation relations
\begin{equation}
[A_1(x), A_2(x')]=i \frac{2\pi}{k} \delta(x-x')
\label{eq:CS-CCR}
\end{equation}

The second term of the Lagrangian enforces the Gauss Law  which for this theory simply implies that the states in the physical Hilbert space obey the constraint
\begin{equation}
B(x)=\frac{2\pi}{k} J_0(x)
\label{eq:gauss-U1}
\end{equation}
Thus the Gauss Law requires that a charge density necessarily has a magnetic flux attached to it. In other terms, the physical states are charge-flux composites as postulated in Wilczek's theory \cite{wilczek-1982a}. This is the theoretical basis to the concept of flux attachment which, as we will see in section \ref{sec:CS-FQHE},  is widely used in the theory of the fractional quantum Hall effect.

The third term in Eq.\eqref{eq:CS-L} simply states that the Hamiltonian density is just
\begin{equation}
\mathcal{H}=J_i A_i
\label{eq:H-CS}
\end{equation}
Hence, in the absence of sources the Hamiltonian vanishes, $\mathcal{H}=0$.

The Chern-Simons action is locally gauge-invariant, up to boundary terms. To see this let us perform a gauge transformation, 
$A_\mu \to A_\mu +\partial_\mu \Phi$, where $\Phi(x)$ is a smooth, twice differentiable function. Then,
\begin{align}
S[A^\mu+\partial^\mu \Phi]=&\int_{\mathcal{M}} (A^\mu+\partial^\mu \Phi) \epsilon_{\mu \nu \lambda} \partial^\nu (A^\lambda+\partial^\lambda \Phi)\nonumber\\
=&\int_{\mathcal{M}} d^3x\; \epsilon_{\mu \nu \lambda} A^\mu \partial^\nu A^\lambda+\int_{\mathcal{M}} d^3x\; \epsilon_{\mu \nu \lambda} \partial^\mu \Phi \partial^\nu A^\lambda
\end{align}
Therefore, the change is 
\begin{equation}
S[A^\mu+\partial^\mu \Phi]-S[A^\mu]=\int_{\mathcal{M}} d^3x\; \partial^\mu \Phi F^*_\mu
= \int_{\mathcal{M}} d^3x\; \partial^\mu(\Phi F_\mu^*)-\int_{\mathcal{M}} d^3x\; \Phi \partial^\mu F^*_\mu
\end{equation}
where $F_\mu^*=\epsilon^{\mu \nu \lambda} \partial_\nu A_\lambda$, is the dual field strength. 
However, in the absence of magnetic monopoles, this field satisfies the Bianchi identity, 
$\partial^\mu F^*\mu=\partial^\mu (\epsilon_{\mu \nu \lambda} \partial^\nu A^\lambda)=0$.
Therefore, using the Gauss Theorem, we find that the change of the action is a total derivative and integrates to the boundary
\begin{equation}
\delta S=\int_{\mathcal{M}} d^3x \; \partial^\mu(\Phi F_\mu^*)=\int_{\Sigma} dS^\mu \Phi F^*_\mu
\end{equation}
where $\Sigma=\partial \mathcal{M}$ is the boundary of $\mathcal{M}$. In particular, if $\Phi$ is a non-zero constant function on 
$\mathcal{M}$, then the change of the action under such a gauge transformation is
\begin{equation}
\delta S=\Phi \times \textrm{flux}(\Sigma)
\end{equation}
Hence, the action is not invariant if the manifold has a boundary, and the theory must be supplied with additional degrees of 
freedom at the boundary.

 Indeed, the flat connections, i.e. the solution of the equations of motion, $F_{\mu \nu}=0$, are pure gauge transformations, 
 $A_\mu =\partial_\mu \phi$, and have an action that integrates to the boundary. 
 Let the $\mathcal{M}=D \times \mathbb{R}$ where $D$ is a disk in space and $\mathbb{R}$ is time. 
 The boundary manifold is $\Sigma=S^1\times \mathbb{R}$, where $S^1$ is a circle. 
 Thus, in this case, the boundary manifold $\Sigma$ is isomorphic to a cylinder. The action of the flat configurations reduces to
\begin{equation}
S=\int_{S^1 \times \mathbb{R}} d^2x \frac{k}{4\pi}  \partial_0\phi \partial_1 \phi
\end{equation}
This implies that the dynamics on the boundary is that of a scalar field on a circle $S^1$, and obeys periodic boundary conditions. 

Although classically the theory does not  depend on the metric, it is invariant under arbitrary transformations of the coordinates. 
However, any gauge fixing condition will automatically break this large symmetry. 
For instance, we can specify a gauge condition at the boundary in the form of a boundary term of the form 
$ \mathcal{L}_{{\rm gauge\; fixing}} =A_1^2$. In this case, the boundary action of the field $\varphi$ becomes
\begin{equation}
S[\varphi]=\int_{S^1\times \mathbb{R}}  d^2x\; \frac{k}{4\pi}\; \Big[ \partial_0 \phi \partial_1\phi- (\partial_1 \phi^2)\Big]
\label{eq:chiral-U(1)-level-k-CFT}
\end{equation}
The solutions of the equations of motion of this compactified scalar field have the form $\phi(x_1\mp  x_0)$, 
and are right (left) moving chiral fields depending of the sign of $k$. This boundary theory is not topological but is is conformally invariant.

A similar result is found in non-abelian Chern-Simons gauge theory. In the case of the $SU(N)_k$ Chern-Simons theory  on a manifold $D \times \mathbb{R}$, where $D$ is a disk whose boundary is $\Gamma$, and $\mathbb{R}$ is time, the action is
\begin{equation}
S_{\rm CS}[A]=\int_{D \times \mathbb{R}} d^3x \; \Bigg[\frac{k}{8\pi} \textrm{tr}\Bigg(\epsilon_{\mu \nu \lambda} A^\mu \partial^\nu A^\lambda+\frac{2}{3} \epsilon^{\mu \nu \lambda} A_\mu A_\nu A_\lambda\Bigg)\Bigg]
\label{eq:non-abelian-CS2}
\end{equation}
This theory integrates to the boundary, $\Gamma \times \mathbb{R}$ where it becomes the chiral (right-moving) $SU(N)_k$ Wess-Zumino-Witten model (at level $k$) at its IR fixed point, $\lambda_c^2=4\pi/k$
\begin{equation}
S_{WZW}[g]=\frac{1}{4\lambda_c^2} \int_{\Gamma \times \mathbb{R}}  d^2x \; \textrm{tr} \left(\partial_\mu g \partial^\mu g^{-1}\right)+ 
\frac{k}{12\pi} \int_B \epsilon^{\mu \nu \lambda} \textrm{tr} \left( g^{-1} \partial_\mu g \;  g^{-1} \partial_\nu g\;  g^{-1} \partial_\lambda g \right)
\label{eq:WZW2}
\end{equation}
Here, $g \in SU(N)$ parametrizes the flat configurations of the Chern-Simons gauge theory. The boundary theory is a non-trivial CFT, the chiral Wess-Zumino-Witten CFT \cite{witten-1989}.

\subsection{Vacuum degeneracy a torus}
\label{sec:degeneracy-torus}

We will now construct the quantum version of the U(1) Chern-Simons gauge theory on a manifold $\mathcal{M}=T^2\times \mathbb{R}$, 
where $T^2$ is a spatial torus, of linear size $L_1$ and $L_2$.
Since this manifold does not have boundaries, the flat connections, 
$\epsilon_{ij} \partial_i A_j=0$ do not reduce to local gauge transformations of the form $A_i=\partial_i \Phi$. 
Indeed, the holonomies of the torus $T^2$, i.e. the Wilson loops on the two non-contractible cycles of the torus 
$\Gamma_1$ and $\Gamma_2$ are gauge-invariant observables:
\begin{equation}
\int_0^{L_1} dx_1 A_1\equiv {\bar a}_1, \qquad \int_0^{L_2} dx_1 A_2\equiv{\bar a}_2
\end{equation}
where ${\bar a}_1$ and ${\bar a}_2$ are time-dependent. Thus, the flat connections now are
\begin{equation}
A_1=\partial_1 \Phi + \frac{{\bar a}_1}{L_1}, \qquad A_2=\partial_2 \Phi +\frac{{\bar a}_2}{L_2}
\end{equation}
whose action is
\begin{equation}
S=\frac{k}{4\pi} \int dx_0 \epsilon_{ij} {\bar a}_i \partial_0 {\bar a}_j
\end{equation}
Therefore, the global degrees of freedom ${\bar a}_1$ and ${\bar a}_2$ at the quantum level become operators 
that satisfy the commutation relations
\begin{equation}
\left[{\bar a}_1, {\bar a}_2\right]=i \frac{2\pi}{k}
\end{equation}

We find that the flat connections are described by the quantum mechanics of ${\bar a}_1$ and ${\bar a}_2$. A representation of this algebra is 
\begin{equation}
{\bar a}_2 \equiv -i \frac{2\pi}{k} \frac{\partial}{\partial {\bar a}_1}
\end{equation}
Furthermore, the Wilson loops on the two cycles become
\begin{equation}
W[\Gamma_1]=\exp\left(i \int_0^{L_1} A_1\right)\equiv e^{i{\bar a}_1},\quad W[\Gamma_2]=\exp\left(i \int_0^{L_2} A_2\right)\equiv e^{i{\bar a}_2}
\end{equation}
and satisfy the algebra
\begin{equation}
W[\Gamma_1] W[\Gamma_2]=\exp(-i 2\pi/k) W[\Gamma_2]W[\Gamma_1]
\end{equation}

Under large gauge transformations 
\begin{equation}
{\bar a}_1 \to{\bar a}_1+2\pi, \qquad {\bar a}_2 \to {\bar a}_2+2\pi
\end{equation}
Therefore, invariance under large gauge transformations on the torus implies that ${\bar a}_1$ and ${\bar a}_2$ define a two-torus target space. 

Let us define the unitary operators
\begin{equation}
U_1=\exp(i k {\bar a}_2), \quad U_2=\exp(-i k {\bar a}_1)
\end{equation}
which satisfy the algebra
\begin{equation}
U_1 U_2=\exp(i 2\pi k) U_2U_1
\end{equation}
The unitary transformations $U_1$ and $U_2$ act as shift operators on ${\bar a}_1$ and ${\bar a}_2$ by $2\pi$, 
and  hence generate the large gauge transformations. Moreover, the unitary operators $U_1$ and $U_2$ leave the Wilson loop operators 
on  non-contractible cycles invariant,
\begin{equation}
U_1^{-1} W[\Gamma_1]U_1=W[\Gamma_1], \quad U_2^{-1} W[\Gamma_2]U_2=W[\Gamma_2]
\end{equation}
Let $|0\rangle$ be the eigenstate of $W[\Gamma_1]$ with eigenvalue $1$, i.e. $W[\Gamma_1]\|0\rangle=|0\rangle$. 
The state $W[\Gamma_2]|0\rangle$ is also an eigenstate of $W[\Gamma_1]$ with eigenvalue $\exp(-i 2\pi/k)$, since
\begin{equation}
W[\Gamma_1]W[\Gamma_2]|0\rangle=e^{i 2\pi/k} W[\Gamma_2]W[\Gamma_1]|0\rangle=e^{-i 2\pi/k} W[\gamma_2 |0\rangle
\end{equation}
More generally, since
\begin{equation}
W[\Gamma_1] W^p[\Gamma_2]\|0\rangle=e^{-i 2\pi p/k} W^p[\Gamma_2] |0\rangle
\end{equation}
we find that, provided $k \in \mathbb{Z}$, there are $k$ linearly independent vacuum states $|p\rangle=W^p[\Gamma_ 2]|0\rangle$, for the U(1) 
Chern-Simons gauge theory at level $k$. It is denoted as the U(1)$_k$ Chern-Simons theory. 
Therefore the finite-dimensional topological space on a two-torus is $k$-dimensional. It is trivial to show that, on a surface of genus $g$,
 the degeneracy is $k^g$.

We see that in the abelian U(1)$_k$ Chern-Simons theory the Wilson loops must carry $k$ possible values of the unit charge. 
This property generalizes to the non-abelian theories, which are technically more subtle. We will only state some important results. 
For example, if the gauge group is SU(2) we expect that the Wilson loops will carry the representation labels of the group SU(2),
 i.e. they will be labelled by $(j, m)$, where $j=0, \frac{1}{2}, 1, \ldots$ and the $2j+1$ values of $m$ satisfy $|m|\leq j$. 
 However, it turns out SU(2)$_k$ Chern-Simons theory has fewer states,  and that the values of $j$ are restricted to the range 
 $j=0,\frac{1}{2}, \ldots, \frac{k}{2}$.

\subsection{Fractional Statistics and Braids}
\label{sec:braids}

Another aspect of the topological nature of Chern-Simons theory is the behavior of expectation values of products of Wilson loop operators.
Let us compute the expectation value of a product of two Wilson loop operators on two positively oriented closed contours $\gamma_1$ and $\gamma_2$. We will do this computation in the abelian Chern-Simons theory U(1)$_k$ in 2+1-dimensional Euclidean space. Note that the Euclidean Chern-Simons action is pure imaginary since the action is first-order in derivatives.
The expectation value to be computed is
\begin{equation}
W[\gamma_1 \cup \gamma_2]=\Big< \exp \Big(i \oint_{\gamma_1 \cup \gamma_2} dx_\mu A_\mu\Big) \Big>_{\rm CS}
\end{equation}
The result changes depending on whether the loops $\gamma_1$ and $\gamma_2$ are linked or unlinked. In this section we will compute the contribution to  this expression for a pair 
of contours $\gamma_1$ and $\gamma_2$. Here we  will not include the contribution to this expectation value for each contours. 
We will return to this problem in section \ref{sec:bosonization-dirac-2+1} where we discuss the problem of fractional spin.

The expectation value of a Wilson loop on the union of two contours, as in the present case, $\gamma$ can be written as
\begin{equation}
\Big< \exp \Big(i \oint_{\gamma} dx_\mu A_\mu\Big) \Big>_{\rm CS}=\Big< \exp \Big(i \int d^3x  J_\mu A_\mu\Big) \Big>_{\rm CS}
\label{eq:Wilson-CS}
\end{equation}
where the current $J_\mu$ is
\begin{equation}
J_\mu(x)=\delta(x_\mu-z_\mu(t)) \frac{dz_\mu}{dt}
\end{equation}
Here $z_\mu(t)$ is a parametrization of the contour $\gamma$. Therefore, the expectation value of the Wilson loop is \cite{witten-1989}
\begin{equation}
\Big< \exp \Big(i \oint_{\gamma} dx_\mu A_\mu\Big) \Big>_{\rm CS}\equiv \exp(i I[\gamma]_{\rm CS})
=\exp\Big(-\frac{i}{2} \int d^3x\; \int d^3y\; J_\mu(x) G_{\mu \nu}(x-y) J_\nu(y)\Big)
\end{equation}
where $G_{\mu \nu}(x-y)=\langle A_\mu(x) A_\nu(y)\rangle_{\rm CS}$ is the propagator of the Chern-Simons gauge field. 
Since the loops are closed, the current $J_\mu$ is conserved, i.e. $\partial_\mu J_\mu=0$, 
and the effective action $I[\gamma]_{\rm CS}$ of the loop $\gamma$ is gauge-invariant. 

The Euclidean propagator of Chern-Simons gauge theory (in the Feynman gauge) is
\begin{equation}
G_{\mu \nu}(x-y)=\frac{2\pi}{k} G_0(x-y)\epsilon_{\mu \nu \lambda} \partial_\lambda \delta(x-y)
\label{eq:CS-propagator}
\end{equation}
where $G_0(x-y)$ is the propagator of the massless Euclidean scalar field, which satisfies
\begin{equation}
-\partial^2 G_0(x-y)=\delta^3(x-y)
\end{equation}
Using these results, we find the following expression for the effective action
\begin{align}
I[\gamma]_{\rm CS}=&\frac{\pi}{k} \int d^3x \; \int d^3y\; J_\mu(x) J_\nu(y) G_0(x-y) \epsilon_{\mu \nu \lambda} \partial_\lambda \delta(x-y)\nonumber\\
=&\frac{\pi}{k} \oint_\gamma dx_\mu \oint_\gamma dy_\nu \epsilon_{\mu \nu \lambda} \partial_\lambda G_0(x-y)
\end{align}
Since the current $J_\mu$ is conserved, it can be written as the curl of a vector field, $B_\mu$, as
\begin{equation}
J_\mu=\epsilon_{\mu \nu \lambda} \partial_\nu B_\lambda
\end{equation}
In the Lorentz gauge, $\partial_\mu B_\mu=0$, we can write
\begin{equation}
B_\mu=\epsilon_{\mu \nu \lambda} \partial_\nu \phi_\lambda
\end{equation}
Hence,
\begin{equation}
J_\mu=-\partial^2 \phi_\mu
\end{equation}
where
\begin{equation}
\phi_\mu(x)=\int d^3y \; G_0(x-y) J_\mu(y)
\end{equation}
Upon substituting this result into the expression for $B_\mu$, we find
\begin{equation}
B_\mu=\int d^3y \epsilon_{\mu \nu \lambda} \partial_\nu G_0(x-y) J_\lambda(y) 
=\oint_\gamma \epsilon_{\mu \nu \lambda} \partial_\nu G_0(x-y) dy_\lambda
\end{equation}
Therefore, the effective action $I[\gamma]_{\rm CS}$ becomes
\begin{equation}
I[\gamma]_{\rm CS}= \frac{\pi}{k} \oint_\gamma dx_\mu \oint_\gamma dy_\nu \epsilon_{\mu \nu \lambda} \partial_\lambda G_0(x-y)
=\frac{\pi}{k} \oint_\gamma dx_\mu B_\mu(x)
\label{eq:Igamma1}
\end{equation}
Let $\Sigma$ be an oriented open surface of the Euclidean three-dimensional space whose boundary is the oriented loop (or union of loops) $\gamma$, i.e. $\partial \Sigma=\gamma$. Then, using Stokes Theorem we write in the last line of Eq.\eqref{eq:Igamma1} as 
\begin{equation}
I[\gamma]_{\rm CS}=\frac{\pi}{k} \int_\Sigma dS_\mu \epsilon_{\mu \nu \lambda} \partial_\nu B_\lambda=\frac{\pi}{k} \int_\Sigma dS_\mu J_\mu
\end{equation}
The integral in the last line of this equation is the flux of the current $J_\mu$ through the surface $\Sigma$. 
Therefore, this integral  counts the number of times $n_\gamma$  the Wilson loop on $\gamma$ pierces the surface $\Sigma$ (whose boundary is $\gamma$), 
and therefore it is an integer, $n_\gamma \in \mathbb{Z}$. We will call this integer the {\it linking number} (or Gauss invariant) of the configuration of  loops. In other words,   
the expectation value of the Wilson loop operator is
\begin{equation}
W[\gamma]_{\rm CS}=e^{i \pi n_\gamma /k}
\label{eq:linking-number}
\end{equation}
The linking number is a {\it topological invariant} since, being an integer, its value cannot be changed by smooth deformations of the loops, provided they 
are not allowed to cross.

We will now see that this property of Wilson loops in Chern-Simons gauge theory leads to the concept of fractional statistics. 
Let us consider a scalar matter field that is massive and charged under the Chern-Simons gauge field. 
The excitations of this matter field are particles that couple minimally to the gauge field. Here we will be interested in the case in which these particles are very heavy.  
In that limit, we can focus on states that have a few of this particles which will be in their non-relativistic regime.

Consider, for example, a state with two particles which in the remote past, at time $t=-T \to -\infty$, are located at two points $A$ and $B$. 
This initial state will evolve to a final state at time $t=T \to \infty$, in which the particles either go back to their initial locations (the direct process), 
or to another one in which they exchange places,  $A \leftrightarrow B$. At intermediate times, the particles follow smooth worldlines. 
These two processes, direct and exchange. 
There we see that the direct process is equivalent to a history with two unlinked loops (the worldlines of the particles), whereas in the exchange process the two loops form a link.
It follows from the preceding discussion that the two amplitudes differ by the result of the computation of the Wilson loop expectation value for the loops 
$\gamma_1$ and $\gamma_2$. Let us call the first amplitude $W_{\rm direct}$ and the second $W_{\rm exchange}$. 
The result is
\begin{equation}
W_{\rm exchange}=W_{\rm direct} e^{\pm i \pi /k}
\end{equation}
where the sign depends on how the two worldlines wind around each other.

An equivalent interpretation of this result is that if $\Psi[A,B]$ is the wave function with the two particles at locations $A$ and $B$, 
the wavefunction where their locations are exchanged is
\begin{equation}
\Psi[B,A]=e^{\pm i \pi/k} \Psi[A,B]
\end{equation}
where the sign depends on whether the exchange is done counterclockwise or clockwise. Clearly, for $k=1$ the wave function is antisymmetric and the particles are fermions, while for $k \to \infty$ they are bosons. 
At other values of $k$ the particles obey {\it fractional statistics}  and are called {\it anyons} \cite{wilczek-1982b,leinaas-1977}. The phase factor $\phi=\pm \pi/k$ is called the statistical phase.

Notice that, while for fermions and bosons the statistical phase $\varphi=0, \pi$ is uniquely defined (mod $2\pi$), for other values of $k$ the statistical angle is specified up to a 
sign that specifies how the worldlines wind around each other. 
Indeed, mathematically the exchange process  is known as a {\it braid}.
Processes in which the worldlines wind clock and counterclockwise are braids that are inverse of each other. 
Braids can also be stack sequentially yielding multiples of the phase $\varphi$. 
In addition to  stacking braids, Wilson loops can be fused: seen from some distance, a pair of particles will behave as a new particle with a well defined behavior under braiding. 
This process of fusion is closely related to the concept of fusion of primary fields in Conformal Field Theory.

Furthermore, up to regularization subtleties \cite{grundberg-1990}, the self-linking terms (those with $a=b$) yield a topological spin $1/2k$, consistent with the spin-statistics connection \cite{polyakov-1988}. For $k=1$ this means that the flux-charge composites have spin 1/2. 

What we have just described is a mathematical structure called the {\it Braid} group. 
The example that we worked out using abelian Chern-Simons theory yields one-dimensional representations of the Braid group with the phase $\varphi$ 
being the label of the representations. For U(1)$_k$ there are $k$ types of particles (anyons). 
That these representations are abelian means that, in the general case of U(1)$_k$,  acting on a one-dimensional representation $p$ (defined mod $k$) 
with a one-dimensional representation $q$ (also defined mod $k$) yields the representation one-dimensional $p+q$ (mod $k$). 
We will denote the operation of {\it fusing} these representations (particles!) as $[q]_{{\rm mod}\; k} \times [p]_{{\rm mod} \; k}=[q+p]_{{\rm mod}\; k}$.
These representations are in one-to-one correspondence with the inequivalent charges of the Wilson loops, and with the vacuum degeneracy of the U(1)$_k$
 Chern-Simons theory on a torus.

A richer structure arises in the case of the non-abelian  Chern-Simons theory at level $k$ \cite{witten-1989}, such as SU(2)$_k$. 
For example, for SU(2)$_1$ the theory has only two representations, both are one-dimensional, and have statistical angles $\varphi=0, \pi/2$.  

However, for SU(2)$_k$, the content is more complex. 
In the case of SU(2)$_2$  the theory has a) a trivial representation $[0]$ (the identity, $(j, m)=(0,0)$), b) a  (spinor) representation $[1/2]$ ($(j, m)=(1/2, \pm 1/2)$), 
and c) a the representation $[1]$ ($(j, m)= (1, m)$, with $m=0, \pm 1$). 
These states will fuse obeying the following rules: $[0] \times [0]=[0]$, $[0] \times [1/2]=[1/2]$, $[0]\times [1]=[1]$, $[1/2] \times [1/2]=[0]+[1]$, $[1/2]\times [1]=[1/2]$, 
and $[1]\times[1]=[0]$ (note the truncation of the fusion process!).

Of particular interest is the case $[1/2]\times[1/2]=[0]+[1]$. 
In this case we have two fusion channels, labeled by $[0]$ and $[1]$. 
The braiding operations now will act on a two-dimensional Hilbert space and are represented by $2 \times 2$ matrices. 
This is an example of a non-abelian representation of the braid group.
These rather abstract concepts have found a physical manifestation in the physics of the fractional quantum Hall fluids, whose excitations are vortices that carry 
fractional charge and anyon (braid) fractional statistics.

Why this is interesting can be seen by considering a Chern-Simons gauge theory with four quasi-static Wilson loops. For instance in the case of the SU(2)$_2$ Chern-Simons theory the Wilson loops  carry the spinor representation, $[1/2]$. If we call the four particles $A$, $B$, $C$ and $D$, we would expect that their quantum state would be completely determined by the coordinates of the particles. This, however, is not the case since, if we fuse $A$ with $B$, the result is either a state $[0]$ or a state $[1]$. Thus, if the particles were prepared originally in some state, braiding (and fusion) will lead to a linear superposition of the two states. This braiding process defines a unitary matrix, a representation of the Braid Group. The same is true with the other particles. However, it turns out that for four particles there are only two linearly independent states. This two-fold degenerate Hilbert space of topological origin is called a topological qubit. 

Moreover, if we consider a system with $N$ (even) number of such particles, the dimension of the topologically protected Hilbert space is $2^{\frac{N}{2}-1}$. Hence, for large $N$, the entropy per particle grows as $\frac{1}{2} \ln 2=\ln \sqrt{2}$. Therefore the qubit is not an ``internal'' degree of freedom of the particles but a collective state of topological origin. Interestingly, there are physical systems, known as non-abelian fractional quantum Hall fluids that embody this physics and  are accessible to experiments! For these reasons, the non-abelian case has been proposed as a realization of a topological qubit \cite{kitaev-2003,dassarma-2007}.

\section{Topological Phases of Matter}
\label{sec:gauge}


\subsection{Topological Insulators}
\label{sec:TI}

We will now give a brief discussion of the physics of Topological Insulators from a field-theoretic perspective. Topological insulators are systems whose electronic states (band structures) have special topological properties which manifest in the existence of symmetry-protected edge states. For this reason these systems are known as symmetry-protected topological states (or SPTs). The simplest example is found in one space dimension where it is related to the fascinating problem of fractionally charged solitons and electron fractionalization. In several ways many of the concepts involved in these 1+1-dimensional systems can and have been extended to higher dimensions.

\subsubsection{Dirac fermions in 2+1 dimensions}
 \label{sec:dirac-2+1}

We will now see that, in spite of the formal similarities withe the 1+1 dimensional case, this theory has different symmetries, particularly concerning parity and time reversal invariance. 
In addition, in 2+1 dimensions it is not possible to define a $\gamma^5$ Dirac matrix which implies that there is no chiral symmetry and no chiral anomaly. 
 We will consider first a theory of a Dirac field which in 2+1 dimensions is a bi-spinor (as is in 1+1 dimensions). The Lagrangian of the Dirac theory coupled to a background gauge field $A_\mu$ is
\begin{equation}
\mathcal{L}=\bar \psi i \gamma^\mu \partial_\mu \psi-m \bar \psi \psi-eA_\mu \bar \psi \gamma_\mu \psi\equiv \bar \psi i \gamma^\mu D_\mu(A) \psi-m \bar \psi \psi
\label{eq:Dirac-2+1}
\end{equation}
where $\bar \psi =\psi^\dagger \gamma_0$, $\bar \psi \gamma_\mu \psi \equiv j_\mu$ is the gauge-invariant (and conserved) Dirac current, and $D_\mu(A)=\partial_\mu+i e A_\mu$ is the covariant derivative.

In Eq.\eqref{eq:Dirac-2+1},  $\gamma_\mu$ are the three $2 \times 2$ Dirac matrices which obey the   algebra, $\{ \gamma_\mu, \gamma_\nu \}=2 g_{\mu \nu}$, 
where $g_{\mu \nu}=\textrm{diag}(1, -1, -1)$ is the metric of 2+1 dimensional Minkowski spacetime.  The Dirac matrices $\gamma_\mu$ can be written in terms of the three $2 \times 2$ Pauli matrices. 
For instance we can choose $\gamma_0=\sigma_3$, $\gamma_1=i \sigma_2$ and $\gamma_2=-i\sigma_1$ which satisfy the Dirac algebra. Since the three gamma matrices involve  all tree Pauli matrices, it is not possible to define a  $\gamma_5$ matrix which would anticommute with the  gamma matrices. In this sense, it is not possible to define a chirality for bi-spinors in 2+1 dimensions 

We could have also chosen a different set of gamma matrices. For instance, we could have chosen $\gamma_0=\sigma_3$, 
$\gamma_1=i \sigma_2$ and $\gamma_2=+i\sigma_1$ which also satisfy the Dirac algebra. These two choices are equivalent to the 2D  parity transformation $x_1 \to x_1$ and $x_2 \to -x_2$. 
In other words, we can choose the three gamma matrices to be defined in terms of a right handed or a left handed frame  (or triad). Thus, the choice of handedness of the frame  used to define the gamma matrices can be regarded as a chirality. 

The parity transformation is also equivalent to a unitary transformation 
(a change of basis) $U=\gamma_1=i\sigma_1$ which flips the sign of both $\gamma_2$ and $\gamma_0$. Thus, a parity transformation is equivalent to a change of the sign of $\gamma_0$ 
or, what is the same, as changing the {\it sign} of the Dirac mass $m \to -m$.
Since the Dirac theory is charge conjugation invariant, the parity transformation is equivalent to time reversal. It is easy to see that in 2+1 dimensions the massive Dirac theory is not invariant under 
time reversal since this the single particle Dirac Hamiltonian $H(\bm p)$ for momentum $\bm p$ involves all three Pauli matrices,
\begin{equation}
H(\bm p)={\bm \alpha} \cdot {\bm p}+\beta m
\label{eq:single-particle-dirac}
\end{equation}
where, as usual, ${\bm \alpha}=\gamma_0 \bm \gamma$ and $\beta=\gamma_0$.
 In fact, time reversal is equivalent to  the change of the sign of the Dirac mass. 
These considerations also apply to the massless Dirac theory coupled to a background gauge field.

\subsubsection{Dirac Fermions and Topological Insulators}
\label{sec:Dirac-TI}

In addition systems in one space dimension, discussed in section \ref{sec:Dirac-1D}, in which Dirac fermions naturally describe the low energy electronic degrees of freedom,  Dirac fermions also play a role in many other system in Condensed Matter Physics. Such systems range from the nodal Bogoliubov fermionic excitations of d-wave superconductors in 2D and p-wave superfluids in 3D, to Chern and $\mathbb{Z}_2$ topological insulators in 2D and $\mathbb{Z}_2$ topological insulators and Weyl semimetals in 3D. The also play a role in spin liquid phases of frustrated spin-1/2 quantum  antiferromagnets, such as Dirac and chiral spin liquids. Dirac fermions play also a significant role in our understanding of the compressible limit of 2D electron gases (2DEG)  in large magnetic fields (see section \ref{sec:CS-FQHE}). We will not cover here all of these examples. Instead we will focus of the 2D Chern topological insulators (which exhibit the anomalous quantum Hell effect) and on the 3D $\mathbb{Z}_2$ topological insulators.

In Condensed matter Physics the important low energy electronic degrees of freedom belong to the electronic states close to the Fermi energy. In such systems the lattice periodic potential determines the properties of their band structures. The only significant exception to this general rule is the case of the 2DEGs in GaAs-AlAs heterostructures (the most commonly used platform for the study of quantum hall effects) whose  electronic densities are low enough so that the lattice periodic potential can for practical purposes almost always  be ignored.

 In  general,  Dirac (and Weyl) fermions arise when the conduction and valence bands cross at isolated points of the Brillouin zone. In such situations, the near the crossing points the low energy electronic states locally (in momentum space) look like cones and, ignoring possible anisotropies, look like the states of a Dirac-Weyl theory.  For these reasons, Dirac fermions play a central role in the theory of graphene type materials \cite{castro_neto-2009}  and, more significantly, in the theory of topological insulators \cite{qi-2011}.

Dirac fermions play a central role in Quantum Electrodynamics and in the Standard Model of Particle Physics. The problem of quark confinement requires an understanding of these theories in a  regime inaccessible to Feynman-diagrammatic perturbation theory and an intrinsically non-perturbative formulation, known as Lattice Gauge Theory \cite{wilson-1974b,kogut-1975,susskind-1977}, needed to be developed. For this reason in the 1970s fermionic Hamiltonians with crossings at points became of great interest in High Energy Physics as a way to describing the short-distance dynamics of quarks  in Lattice Gauge Theory.  

This program  run into difficulties when extended to the theory of Weak Interactions which have an odd number of species of chiral (Weyl) Dirac fermions. All local discretizations of the Dirac equation yield an even number of species. This fact came to be known as the fermion doubling problem. These results are special cases of a general theorem, due to Holger Nielsen and Masao Ninomiya \cite{nielsen-1981a, nielsen-1981b,nielsen-1983} (and extended by Daniel Friedan \cite{friedan-1982}), which proves that for systems whose kinetic energy is {\it local} in space (as it must be) there must always be an {\it even} number of crossings. Therefore,  it is impossible to write a local theory  with an odd number of chiral fermionic species.

In the context of Condensed Matter Physics, the simplest example of Dirac fermions is found in the electronic structure of graphene (a single layer of graphite)  discussed by A. H. Castro Neto and coworkers \cite{castro_neto-2009}. Graphene is an allotrope of crystalline carbon in which the carbon atoms are arranged in a 2D hexagonal lattice, which is a lattice with two inequivalent sites in its unit cell, labeled by $A$ and $B$. Let $\bm r_A$ and $\bm r_B$ be the sites of the two sublattices. Each site $\bm r_A$ has three nearest neighbors on the $B$ sublattice located at $\bm r_B^i=\bm r_A +\bm d_i$, where $i=1, 2, 3$ and 
\begin{equation}
\bm d_1=\left(\frac{1}{2\sqrt{3}}, \frac{1}{2}\right), \qquad \bm d_2=\left(\frac{1}{2\sqrt{3}}, -\frac{1}{2}\right), \qquad \bm d_3=\left(-\frac{1}{\sqrt{3}}, 0\right)
\label{eq:hexagonal-lattice}
\end{equation}
 (in units in which the lattice spacing is $a=1$). The nearest neighbor $A$ sites are related by the vectors 
 \begin{equation}
 \bm a_1=\left(\frac{\sqrt{3}}{2},-\frac{1}{2}\right), \qquad \bm a_2=\left(0,1\right), \qquad \bm a_3=\left(-\frac{\sqrt{3}}{2}, -\frac{1}{2}\right)
 \label{eq:primitive}
 \end{equation}
 where $\bm a_1$ and $\bm a_2$ are the primitive lattice vector of the hexagonal lattice. The nearest neighbor sites of the $B$ sublattice are also related by these same three vectors.
 
 The low energy electronic degrees of freedom of graphene can be described by a single fermionic state (ignoring spin) at each carbon atom . Let $c^\dagger (\bm r_A)$ and $c^\dagger (\bm r_B)$ be the fermion operators that creates an electron at the site $\bm r_A$ and at the site nearest neighbor sites $\bm r_B$, respectively. The Hamiltonian  is 
\begin{equation}
H=t_1 \sum_{\bm r_A, i=1, 2, 3} c^\dagger(\bm r_A) c(\bm r_A+\bm d_i) +\textrm{h.c.}
\label{eq:H0-graphene}
\end{equation}
where $t_1$ is the hopping matrix element between nearest neighbor $A$ and $B$ sites. 

 In Fourier space 
 \begin{equation}
 c(\bm r_A)=\int_{\rm BZ} \frac{d^2k}{(2\pi)^2} \psi_A(\bm k) \exp(i {\bm k} \cdot \bm r_A), \quad  c(\bm r_B)=\int_{\rm BZ} \frac{d^2k}{(2\pi)^2} \psi_B(\bm k) \exp(i {\bm k} \cdot \bm r_B)
 \label{eq:FT-fermions-graphene}
 \end{equation}
 where BZ is the first Brillouin zone of the hexagonal lattice, a hexagonal region of reciprocal space with vertices at $\pm 2\pi/\sqrt{3}(1,1/\sqrt{3})$ (which are denoted by $K$ and $K'$, respectively)  and their two images under $2\pi/3$ rotations. 
 In momentum space the Hamiltonian is \cite{semenoff-1984}
 \begin{equation}
 H=t_1 \int_{\rm BZ} \frac{d^2k}{(2\pi)^2}  
 \begin{pmatrix}
 \psi_A(\bm k) & \psi_B(\bm k)
 \end{pmatrix}
 \begin{pmatrix}
 0 & \sum_{j=1,2,3} \exp(i \bm k \cdot \bm d_j)\\
 \sum_{j=1,2,3} \exp(-i \bm k \cdot \bm d_j)& 0
 \end{pmatrix}
 \begin{pmatrix}
 \psi_A(\bm k)\\
 \psi_B(\bm k)
 \end{pmatrix}
 \label{eq:H0-graphene-FT}
 \end{equation}
 The one-particle states of this Hamiltonian have eigenvalues $E(\bm k)=\pm \Big|\sum_{j=1,2,3} \exp(i {\bm k} \cdot \bm d_j\Big|$. Clearly $E(\bm k)$ vanishes at the $K$ and $K'$ points, the corners of the Brillouin zone. Hence, there is a crossing of the two bands at the $K$ and $K'$ points near which the energy eigenvalues are a two cones at $K$ and $K'$, respectively.
 Thus, the low energy states of graphene consists of two massless (gapless) Dirac bi-spinors, with opposite chirality/parity.
 
 Other examples with similar fermion content are a theory of fermions on a 2D square lattice with a $\pi$ magnetic flux (1/2 of the flux quantum)  per plaquette which arises, for instance,  in the theory of the integer quantum Hall effect on lattices by David Thouless, Mahito Kohmoto, Peter Nightingale and Marcel den Nijs \cite{thouless-1982} (albeit for more general flux per plaquette), and in the theory of flux phases of 2D spin liquids  of Ian Affleck and J. Brad Marston \cite{affleck-1988b}. It also arises in the theory of the chiral spin liquid of Xiao-Gang Wen, Frank Wilczek and Anthony Zee \cite{wen-1989}. We will discuss these theories in  section \ref{sec:gauge}.

 \subsubsection{Chern invariants}
 \label{sec:chern}
 
 The single-particle quantum states of free fermions (electrons) in the  periodic potential of a crystal are Bloch wave functions labeled by the lattice momentum $\bm k$ and band indices $m$. The Bloch states are periodic functions of the lattice momenta whose periods are the Brillouin zones. Topologically each Brillouin zone is a torus: in 1D is a 1-torus, in 2D a 2-torus, etc. The symmetries (and shapes) of the Brillouin zone are dictated by the symmetries of the host crystal.
 
 In this section we will consider the quantum states of systems of fermions on a 2D lattice with $M$ bands with eigenvalues $\{ E_m(\bm k)\} $ with $m=1,\ldots, M$. The eigenstates are Bloch states $\{ |u_m(\bm k)\rangle \}$ such that the wave functions ar $\psi_m(\bm x)=u_m(\bm k) \exp(i \bm k \cdot \bm x)$, where $\bm k$ is a (quasi) momentum in the first Brillouin zone \cite{bloch-1929}. We will assume that the band spectrum is such that all the bands are separated from each other by a finite gap for all momenta in the first Brillouin zone. Hence, the eigenvalues obey the inequality $|E_m(\bm k)-E_n(\bm k)|>0$. We will assume that the system of interest has $N<M$ filled bands and, consequently, that there is a gap separating the top-most occupied band $N$ (the ``valence band'') and the lowest unoccupied band $N+1$ (the ``conduction band'') that does not close everywhere in the first Brillouin zone.

 Let us consider specifically a 2D system and a Bloch state $|u_m(\bm k)\rangle$ at a momentum $\bm k$, and let $\partial_{\bm k} |u_m(\bm k)\rangle$ be an infinitesimally close Bloch state with momentum $\bm k'=\bm k + d\bm k$. The inner product of these two infinitesimally close Bloch states is the vector field (defined on the first Brillouin zone for each band $m$)
 \begin{equation}
 \mathcal{A}^{(m)}_j(\bm k)=i \langle u_m(\bm k)| \partial_j |u_m(\bm k)\rangle
 \label{eq:Berry-connection}
 \end{equation}
 where $j=1,2$ are the orthogonal direction in momentum space and we have used the notation $\partial_i=\partial/\partial k_i$. This vector field  is known as the {\it Berry Connection} of the electronic states in the $m$th band.

The theory of the quantum states of non-interacting electrons in periodic potentials as developed by Felix Bloch \cite{bloch-1929} is the foundation of much of what we know about the physics of many semiconductors and simple metals. This theory is the core subject of many textbooks \cite{kittel-1953,ashcroft-1976}. This theory is based on the classification of the electronic states as representations of the group of lattice translations and point (and spatial) group symmetry transformations. A key unstated assumption in much of this body of work is that the Bloch states are globally well-defined functions on the Brillouin zone of a given crystal. It is rather remarkable that this assumption was not stated explicitly since the original work by Bloch in 1929 until the work by Thouless, Kohmoto, Nightingale and den Nijs on the quantum states of electrons on 2D lattices in the presence of an uniform magnetic field \cite{thouless-1982}. The realization that there is a topological obstruction to define the electronic states globally in the Brillouin zone is the key to the development of the the theory of topological insulators and, more generally, of the modern theory of band structures \cite{bradlyn-2017,cano-2021}.

In quantum mechanics states are defined up to a {\it phase}. This means that Bloch states that differ by a phase describe the same quantum state. In other words each quantum state $|u_m(\bm k)\rangle$ is a member of a {\it ray} (or {\it fiber}) of states each labeled by a phase, 
 \begin{equation}
 |u_m)(\bm k)\rangle \mapsto \exp(i f_m(\bm k))\; |u_m(\bm k)\rangle
 \label{eq:ray}
 \end{equation}
  Changing the states by phase factors defines a U(1)$^M$ unitary transformation of physically equivalent states. Since we have a fibre at each point $\bm k$, this amount at defining a {\it fibre bundle}.
  However, under this transformation the Berry connection $\mathcal{A}_j^{(m)}(\bm k)$ changes by a gradient of the phases
 \begin{equation}
 \mathcal{A}_j^{(m)}(\bm k) \mapsto \mathcal{A}_j^{(m)}(\bm k) +\partial_{j} f_m(\bm k)
 \label{eq:gauge-transf-fiber}
 \end{equation}
 where we assumed that the phases $f_m(\bm k)$ are continuous and differentiable functions on the entire first Brillouin zone. Since the physical states cannot change by redefinitions of the phases of the basis states we are led to the condition that only the data that is invariant under the gauge transformations defined by Eqs. \eqref{eq:ray} and \eqref{eq:gauge-transf-fiber} is physically meaningful, which is encoded in the  gauge-invariant pseudo-scalar quantity $\mathcal{F}^{(m)}(\bm k)$
 \begin{equation}
 \mathcal{F}^{(m)}(\bm k)=\epsilon_{ij} \partial_{i} \mathcal{A}_j^{(m)}(\bm k)
 \label{eq:Berry-curvature}
 \end{equation}
 which is known as the {\it Berry curvature}. 
 
 In what follows we will be interested in the quantity 
 \begin{equation}
 C_1^{(m)}=\frac{1}{2\pi} \oint_\Gamma dk_j \mathcal{A}_j^{(m)}(\bm k)=\frac{1}{2\pi} \int_{\rm BZ} d^2k\; \mathcal{F}^{(m)}(\bm k)
 \label{eq:C1-m}
 \end{equation}
 where $\Gamma$ is the boundary of the 1st Brillouin zone (which we denote by BZ). 
 We will now show that the quantity $C_1^{(m)}$, which measure the {\it flux} of the Berry connection $\mathcal{A}_j^{(m)}(\bm k)$ through the first Brillouin zone (in units of $2\pi$),  is an integer independent of the particular connection $\mathcal{A}_j^{(m)}(\bm k)$ that we have chosen. This integer-valued quantity is a topological invariant known as the first Chern number.
 
 However, since the first Brillouin zone is a 2-torus ,which is a closed manifold, a Berry connection with a net flux cannot obey periodic boundary conditions. Instead, we must allow for generalized (large) gauge transformations that wrap around the 2-torus of the first Brillouin zone. This problem is similar (and closely related to) the problem of the wave functions of a charged particle moving in the presence of a magnetic monopole \cite{wu-1976}. We will consider a 2D system whose first Brillouin zone is spanned by two reciprocal lattice vectors $\bm b_1$ and $\bm b_2$, related by the two primitive lattice vectors $\bm a_1$ and $\bm a_2$ by the relation $\bm b_i \cdot \bm a_j=2\pi \delta_{ij}$ (with $i,j=1,2$) The generalized gauge transformations now are 
 \begin{align}
 |u_m(\bm k+\bm b_1)\rangle=&\exp(i f_m^{(1)}(\bm k)) \; |u_m(\bm k)\rangle, \qquad
  |u_m(\bm k+\bm b_2)\rangle=\exp(i f_m^{(2)}(\bm k))\;  |u_m(\bm k)\rangle\nonumber\\
  \mathcal{A}_j^{(m)}(\bm k+\bm b_1)=&\mathcal{A}_j^{(m)}(\bm k)+\partial_j f_m^{(1)}(\bm k),
  \qquad
   \mathcal{A}_j^{(m)}(\bm k+\bm b_2)=\mathcal{A}_j^{(m)}(\bm k)+\partial_j f_m^{(2)}(\bm k)
   \label{eq:Berry-generalized-BCs}
   \end{align}
   where $f_m^{(1)}(\bm k)$ and $f_m^{(2)}(\bm k)$ are smooth functions of $\bm k$.
   
   We will now show that the Bloch states cannot be defined globally over the Brillouin zone if the flux of the Berry connection does not vanish. More specifically let us assume that at some point $\bm k_0 \in T$ we  know the Bloch state $|u_m(\bm k_0)\rangle$ which satisfies the generalized periodic boundary conditions of Eq.\eqref{eq:Berry-generalized-BCs}. The question is, can we  determine the Bloch state at some other also {\it arbitrary} point $\bm k_0'$? We will now that there is a topological obstruction that does not allow it if the flux of the Berry phase is not zero. The reason is that the {\it phase} of the Bloch state cannot be defined since, in general, the Bloch state will vanish at some point of the Brillouin zone where the phase is undefined.
   
   We will prove that this si true by following an elegant construction due to Mahito Kohmoto which goes as follows \cite{kohmoto-1985}. The Brillouin zone is a torus, that we will denote by $T$, can be regarded as the tensor product of two circles, $T\equiv S_1 \times S_1$. Let us split the torus $T$ into two disjoint subsets (or patches) $H_I$ and $H_{II}$ such that $T=H_I \bigcup H_{II}$. We will assume that in region $H_I$. Let us assume that the Bloch state vanishes at some point $\bm k_0 \in T_I$ and that the Bloch state does not vanish for all points $\bm k \in H_{II}$. This means that we can choose the Bloch state $|u_m(\bm k)\rangle$ to be real for all $\bm k \in H_{II}$. On the other hand we can always assign some arbitrary phase to the Bloch state at $\bm k_0 \in T_I$. Once we have done that we can extend the definition of the phase to a neighborhood of $\bm k_0$ wholly included in $H_I$. If we the assume that there is only one zero, then the phase can be defined over all of $H_I$. The result is that we now have two different definitions of the phase of the Bloch state on $H_I$ and on $H_{II}$.  Let $|u_m(\bm k)\rangle_I$ and $|u_m(\bm k)\rangle_{II}$ be the two resulting definitions of the Bloch state. Let the closed curve $\gamma$ be the common boundary of regions $H_I$ and $H_{II}$,  $\gamma=\partial H_I=\partial H_{II}$. However on the common boundary $\gamma$ these the two definitions of the Bloch state must be  a gauge transformation,
   \begin{equation}
   |u_m(\bm k)\rangle_I=\exp(i f_m(\bm k)) \; |u_m(\bm k)\rangle_{II}
   \label{eq:transition-function}
   \end{equation}
   on all points $\bm k \in \gamma$. The gauge transformation $f_m(\bm k)$, known as the transition function,  is a smooth periodic function  of the points $\bm k$ on the closed curve $\gamma$. 
    Likewise, the Berry connection $\mathcal{A}_j^{(m)}(\bm k)$ has two definitions on the regions $H_I$ and $H_{II}$ which differ by a gauge transformation on all the points $\bm k$ of the common boundary $\gamma$,
   \begin{equation}
   \mathcal{A}_j^{(m)}(\bm k)\Big|_I-\mathcal{A}_j^{(m)}(\bm k)\Big|_{II}=\partial_j f_m(\bm k)
   \label{eq:transition-function-2}
   \end{equation}
   The transition function defines a mapping of the closed curve $\gamma$, which is topologically equivalent to a circle $S_1$, to the phase of the Bloch state which is defined mod $2\pi$ and hence is also a circle $S_1$. Hence the transition functions $f_m(\bm k)$ are homotopies which can be classified by the homotopy group $\Pi_1(S_1)\simeq \mathbb{Z}$. We recognize that the classification of the the transition functions is the same that we used for the vortices of section \ref{sec:2d-vortices}. Hence, the change of the transition function on a full revolution of the closed curve $\gamma$ must be an integer multiple of $2\pi$.
       
 We will now compute the quantity $C_1^{(m)}$, given in Eq.\eqref{eq:C1-m}, which measures the  flux of the Berry connection through the 2-torus $T$ that defines the first Brillouin zone. Using the partition of the torus $T=T_I \bigcup T_{II}$ we can write
 \begin{equation}
 C_1^{(m)}=\frac{1}{2\pi}\int_T d^2k \; \mathcal{F}^{(m)}(\bm k)=\frac{1}{2\pi} \left(\int_{H_I} d^2k \;\mathcal{F}^{(m)}(\bm k)+\int_{H_{II}} d^2k \;\mathcal{F}^{(m)}(\bm k)\right)
 \label{eq:C_1-1}
 \end{equation}
 Using Stokes Theorem on the two regions $H_I$ and $H_{II}$ we get
\begin{equation}
 C_1^{(m)}=\frac{1}{2\pi} \oint_\gamma dk_j \left(\mathcal{A}_j(\bm k) \Big|_I-\mathcal{A}_j(\bm k) \Big|_{II}\right)
 \label{eq:C_1-2}
 \end{equation}
 Since the two definitions of the Berry connection differ by a gauge transformation, Eq.\eqref{eq:transition-function-2}, we can express $C_1^{(m)}$ in terms of the transition function $f_m(\bm k)$ on the curve $\gamma$
 \begin{equation}
 C_1^{(m)}=\frac{1}{2\pi} \oint_\gamma d\bm k \cdot \bm \partial f_m(\bm k)=\frac{1}{2\pi} \left(\Delta f_m(\bm k)\right)_\gamma=n
 \label{eq:C_1-3}
 \end{equation}
 where we used that the transition functions are classified by the integer-valued quantity $n \in \mathbb{Z}$.

We conclude that the flux of the Berry curvature  $C_1^{(m)}$ takes only integer values.  It is known of the first Chern number $C_1^{(m)}$ which is a topological invariant since it cannot change by a smooth redefinition of the curve $\gamma$. Since $C_1^{(m)} \neq 0$ implies that the Bloch states must vanish at least at some point (or points) $\bm k \in H_I$, then the integer-valued Chern number can only change if a redefinition of the curve $\gamma$ crosses at least one point $\bm k$ at which the Bloch state vanishes.

The conclusion of this analysis is that whenever the flux of the Berry curvature through the Brillouin zone does not vanish the Bloch states cannot be defined globally on the Brillouin zone.  A direct consequence is that in this case the band is characterized by a topological invariant called the first Chern number. This number is a property of a given band and, in general, it is different for each band. This is a particular case of a more general topological classification of the states of free fermions on a lattice which depends on on the dimensionality of the system \cite{schnyder-2008,kitaev-2009,moore-2007}.

\subsubsection{The quantum Hall effect on a lattice}
\label{sec:TKNN}

 We will now show that 2D free fermion systems with Chern bands, i.e. bands characterized by a  non-vanishing Chern number, are insulators that have a quantized Hall conductivity. We will do this for the theory of the integer quantum Hall effect on a 2D lattice of Thouless, Kohmoto, Nightingale and den Nijs (TKNN) \cite{thouless-1982, kohmoto-1985} (see, also, Ref. \cite{fradkin-1987}). Although this problem played a key role in the development of the theory of topological phases of matter, for many years it was viewed as an academic problem since it would require gigantic magnetic fields to be in the regime of interest for typical solids. The situation changed with the development of twisted bilayer graphene and similar systems which have unit cells large enough for this physics to be observable. These new materials has allowed to study this problem  experimentally.

TKNN  considered a system of free fermions on a planar (actually square) lattice in a uniform magnetic field $B$ with flux  $2\pi p/q$ per plaquette (in units in which the flux quantum $\phi_0=e\hbar/c=1$) with $p$ and $q$ two co-prime integers. The tight-binding Hamiltonian for this simple problem is
 \begin{equation}
 H=\sum_{\bm r, j=1,2} t_j \; c^\dagger(\bm r)\; e^{i A_j(\bm r)} \; c(\bm r + \bm e_j)+\textrm{h.c.}
 \label{eq:2d-lattice-B}
 \end{equation}
 where $c(\bm r)$ and $c^\dagger(\bm r)$ are fermion creation and annihilation operators on the lattices sites $\{ \bm r =(m, n)\}$, $\bm e_j$ are lattice unit vectors, and $t_j$, with $j=x,y$,  is a hopping matrix element between nearest neighbor sites of the square lattice along the $x$ and $y$ directions. On each link of the lattice we defined a vector potential $A_j(\bm r)$, where the oriented sum of the  lattice vector potentials on each plaquette $2\pi \phi$, where $\phi=\frac{p}{q}$,  is the flux on each plaquette of the square lattice. In the axial (Landau) gauge $A_1(m, n)=0$, $A_2(m, n)=2\pi m \phi $  is a periodic function of $m$ with period $q$. In this gauge the Hamiltonian becomes
 \begin{align}
 H=&t \sum_{m,n} \left(c^\dagger(m+1,n)c(m,n)+c^\dagger(m,n)c(m+1,n)\right)\nonumber \\
   +&t\sum_{m,n} \left(c^\dagger (m,n+1) \; e^{\displaystyle{i 2\pi  \phi m  }}\; c(m,n)+ c^\dagger(m,n) \; e^{\displaystyle{-i2\pi \phi m}}  \; c(m,n+1)\right)
   \label{eq:Hofstadter}
   \end{align}
 In this gauge the problem reduces to a system with a theory  with a $q \times 1$ unit cell with $q$ inequivalent sites. In momentum space, the ({\it magnetic})  Brillouin zone of this system is $-\frac{\pi}{q}\leq k_1 \leq \frac{\pi}{q}$ and $-\pi \leq k_2 \leq  \pi$. The Fourier transform of the operators $c^\dagger(\bm r)$ and $c(\bm r)$ are, respectively, $c^\dagger(\bm k)$ and $c(\bm k)$ which obey the standard anticommutation relations, $\{ c(\bm k), c(\bm q) \}= \{ c^\dagger (\bm k), c^\dagger (\bm q) \}=0$, and $\{ c^\dagger (\bm k), c(\bm q) \}= \delta(\bm k -\bm q)$.   In momentum space the Hamiltonian becomes
 \begin{equation}
 H=q \int_{-\pi/q}^{\pi/q} \frac{dk_x}{2\pi} \int_{-\pi}^\pi \frac{dk_y}{2\pi} \; \widetilde{H}(k_x, k_y)
 \label{eq:H-FT}
 \end{equation}
 where
 \begin{align}
 \widetilde{H}(k_x, k_y)=&\frac{1}{q}\sum_{n=0}^{q-1}\; 
 				\Big\{ 2 t_x \; \cos(k_x+2\pi \phi n)\; c^\dagger(k_x+2\pi \phi n, k_y)\; c(k_x+2\pi \phi n, k_y)\nonumber\\
                  & \!\!\!\!\!  \!\!\!\!\!   \!\!\!\!\!   \!\!\!\!\!  \!\!\!\!\!  \!\!\!\!\! + t_y \Big[ 
                   e^{\displaystyle{-ik_y}} \; c^\dagger (k_x+2\pi (n+1) \phi, k_y)\;c(k_x+2\pi n \phi, k_y) + e^{\displaystyle{ik_y}}\; c^\dagger (k_x+2\pi (n-1) \phi, k_y) \; c(k_x+2\pi n \phi, k_y)
		    \Big]
		    \Big\}
\label{eq:H-tilde}
\end{align}
The spectrum of this system was first investigated by Hofstadter \cite{hofstadter-1976} consists of $q$ bands which for $q$ an odd integer are separated by finite energy gaps. 

 For fixed values of $k_x$ and $k_y$ (in the magnetic Brillouin zone), the Hamiltonian $\widetilde H(k_x, k_y)$ is the tight-binding model of a one-dimensional chain on the 1D lattice of $q$ sites located at $k_x+2\pi n\phi$  with nearest neighbor hopping. The single-particle (Bloch) states $u_n (\bm k)$ of this 1D model obeys the Schr\"odinger Equation
 \begin{equation}
2  t_x\; \cos(k_x+2\pi n \phi) u_n(\bm k)+t_y\left(e^{\displaystyle{-ik_y}}\; u_{n-1}(\bm k)+e^{\displaystyle{ik_y}}\; u_{n+1}(\bm k) \right)=E_n u_n
\label{eq:harper}
\end{equation}
 which is known as Harper's equation. In general this equation does not admit an analytic solution. However, the nature qualitative features of the spectrum can be obtained by an expansion either on $t_x/t_y$ or on $t_y/t_x$. TKNN used degenerate perturbation theory to show that the spectrum has $q$ bands and that the $r$-th band is characterized by two integers $s_r$ and $\ell_r$ which are the solution of the Diophantine equation
\begin{equation}
r=q \; s_r+ p \; \ell_r
\label{eq:diophantine}
\end{equation}
with $\ell_0=s_0=0$. It isl also obvious that for $r=q$ $s_q=1$ and $\ell_q=0$ (for all $p$).

Furthermore, TKNN showed that there was an, until then unsuspected, relation between the Hall conductivity $\sigma_{xy}$ of a gapped system of electrons in periodic potentials in a uniform magnetic field  and the Berry connection of the filled bands. This relation implies the quantization of the Hall conductivity and its computation in terms of a topological invariant, the Chern number of the occupied bands. 
They showed that in this context each band has a non-trivial Berry connection $\mathcal{A}_j^{(r)}=i \langle u_r(\bm k)| \partial_j | u_r(\bm k\rangle$ (with $\partial_j=\partial_{k_j}$) of the form of Eq.\eqref{eq:Berry-connection} with a non-vanishing (first) Chern number $C_1^{(r)}$, i.e. the flux through the magnetic Brillouin zone. 

The conductivity tensor characterizes the electrical properties of a physical system. Linear Response Theory provides a framework for computing the conductivity tensor by perturbing the system with a weak external electromagnetic field $A_\mu$  and computing the currents that they induce \cite{martin-1967, fradkin-2021}. The expectation value of the gauge-invariant current operator $J_\mu(x)$  is 
\begin{equation}
\langle J_\mu(x)\rangle=-\frac{i}{\hbar} \frac{\delta}{\delta A_\mu(x)} \ln Z[A_\mu]
\label{eq"induced-current}
\end{equation}
where $Z[A_\mu]$ is the partition function in the presence of a background (i.e. classical) electromagnetic field $A_\mu(x)$. In general spacetime dimension $D=d+1$, the  lowest order in the vector potential $A_\mu$, the induced current is given in terms of the polarization tensor $\Pi_{\mu \nu}(x, y)$,
\begin{equation}
\ln Z[A_\mu]=\frac{i}{2} \int d^Dx \int d^Dy \; A_\mu(x) \; \Pi^{\mu \nu}(x, y)\; A_\nu(y)+O(A^3)
\label{eq:Pimunu-general}
\end{equation}
Therefore, the induced current is related to the external field by
\begin{equation}
\langle J_\mu(x)\rangle \equiv J_\mu(x)=\int d^Dy\;  \Pi_{\mu \nu}(x, y) \; A^\nu(y)
\label{eq:kubo}
\end{equation}
Expressions of this type are know as a {\it Kubo formula}. The polarization tensor can be regarded as a {\it generalized susceptibility}.

Although  we are using a continuum relativistic notation, these expressions are generally valid, even in lattice systems.
Gauge invariance requires that the polarization tensor be conserved,
\begin{equation}
\partial_\mu \Pi^{\mu \nu}(x, y)=0
\label{eq:conservation-Pimunu}
\end{equation}
In general, the (retarded) polarization tensor $\Pi^{\mu \nu}(x, y)$ is related to the the (retarded) current-current correlation function $\mathcal{D}^R_{\mu \nu}(x, y)$,
\begin{equation}
\mathcal{D}^{\mu \nu}(x, y)=-\frac{i}{\hbar} \Theta(x_0-y_0) \langle [J_\mu(x), J_\nu(y)] \rangle
\label{eq:ret-Dmunu}
\end{equation}
by the identity
\begin{equation}
\Pi_{\mu \nu}^R(x, y)=\mathcal{D}^R_{\mu \nu}(x, y)-i \hbar \Big< \frac{\delta J_\mu(x)}{\delta A_\nu(y)}\Big>
\label{eq:Pi-D}
\end{equation}
The last term in Eq.\eqref{eq:Pi-D}, usually called a ``contact term''. This term  vanishes only for a theory of relativistic fermions (in the continuum). In all other cases, lattice or continuum, relativistic or not, the contact term does not vanish and its form depends on the specific theory. In non-relativistic systems, e.g. in a Fermi liquid, this term is the origin of the $f$-sum rule \cite{martin-1967,nozieres-1966}.

When the external field $A_\mu$  represents is a (locally) uniform electric field $\bm E$,  the induced current  is $J_i=\sigma_{ij} E_j$, where $\sigma_{ij}$ is the conductivity tensor, which can be obtained from the polarization tensor as the limit
\begin{equation}
\sigma_{ij}=\lim_{\omega \to 0} \frac{1}{i \omega} \lim_{{\bm q} \to 0} \Pi_{ij}(\omega, \bm q)
\label{eq:conductivity-tensor}
\end{equation}
In a metal, which has a Fermi surface, the order of limits in which $\omega$ and $\bm q$ vanish matters and only the order of limits specified above is the correct one to take. Ina Dirac systems, that we will discuss below, the order does not matter due to relativistic invariance. The other case in which the order does not matter is the Chern insulator that we are interested in. In general, in an isotropic system,  the conductivity tensor has a symmetric part and an antisymmetric part. The symmetric part of the conductivity tensor  yields the longitudinal conductivity which has all the effects of dissipation. The antisymmetric part does not vanish in the presence of a magnetic field or, more generally, if time reversal symmetry is broken, and yields the Hall conductivity. 

A Chern insulator is an insulator and as such is a state with an energy gap. In such a state the longitudinal conductivity vanishes since there is no dissipation in a gapped state. But in a Chern insulator time reversal invariance is broken. In the system that we are discussing is broken by the magnetic field. The Hall conductivity can be calculated from the antisymmetric part of the polarization tensor as the limit
\begin{equation}
\sigma_{xy}=\lim_{\omega \to 0} \frac{i}{\omega} \Pi_{xy}(\omega,0)
\label{eq:sigma-xy}
\end{equation}
In addition, the contact term does not contribute to the Hall conductivity. As a result the hall conductivity can be computed in terms of the antisymmetric component of the current-correlation function. 

let us now return to the theory of free fermions on a lattice in a (commensurate) magnetic field takes. As we saw the electronic states are split into $q$ bands with single particle states $|\psi_n(\bm k)\rangle$ where $\bm k$ takes values on the first magnetic Brillouin zone. We will compute the Hall conductivity for this system assuming that the Fermi energy $E_F$ lies in the gap between the $n$-th and the $n-1$-th bands. Let us label by $\alpha$ the occupied bands and by $\beta$ the unoccupied bands. Hence $E_\alpha (\bm k) < E_F < E_\beta(\bm k)$. In this case, the Kubo formula for the Hall conductivity becomes
\begin{equation}
\sigma_{xy}=i \frac{e^2}{\hbar} \sum_{E_\alpha < E_F <E_\beta} \int_{BZ}\frac{d^2k}{(2\pi)^2}\; 
\frac{\langle u_\alpha(\bm k)|J_y|u_\beta(\bm k)\rangle \langle u_\beta(\bm k)|J_x|u_\alpha(\bm k)\rangle  - \langle u_\alpha(\bm k)|J_x|u_\beta(\bm k)\rangle \langle u_\beta(\bm k)|J_y|u_\alpha(\bm k)\rangle }{(E_\alpha(\bm k)-E_\beta(\bm k) )^2}
\label{eq:kubo1}
\end{equation}
In the lattice model the current operator is simply
\begin{equation}
{\bm J}=\frac{e}{\hbar} \partial_{\bm k} \widetilde{H}(\bm k)
\label{eq:current-Htilde}
\end{equation}
Using this form of the current and the Schr\"odinger equation of Eq.\eqref{eq:harper} we can rewrite the Kubo formula for the Hall conductivity, Eq.\eqref{eq:kubo1}, as a sum over only the occupied states labeled by $\alpha$. The expression for the Hall conductivity now takes  the more compact form
\begin{equation}
\sigma_{xy}=\frac{e^2}{\hbar} \sum_{\alpha}\int_{BZ} \frac{d^2k}{(2\pi)^2} \epsilon_{j l} \partial_j \langle u_\alpha(\bm k)| \partial_k |u_\alpha(\bm k)\rangle 
		  =\frac{e^2}{h} \sum_\alpha \frac{1}{2\pi} \int_{BZ} \mathcal{F}^{(\alpha)}(\bm k)
	          =\frac{e^2}{h} \sum_\alpha C_1^{(\alpha)}
	          \label{eq:kubo2}
	          \end{equation}
Therefore, we conclude that each occupied band contributes to the Hall conductivity an amount equal to its first Chern number (multiplied by $e^2/h$). This is the main result of TKNN \cite{thouless-1982}. 

In the specific case of the model that we are discussing here the Chern number is given explicitly in terms of the solution of the Diophantine equation, Eq.\eqref{eq:diophantine}
\begin{equation}
 C_1^{(r)}=\frac{1}{2\pi} \oint_\Gamma dk_j \mathcal{A}_j^{(r)}(\bm k)=\ell_r-\ell_{r-1}
 \label{eq:Chern-TKNN}
 \end{equation}
 where the integers $\ell_r$ and $\ell_{r-1}$ are the solutions of the Diophantine equation for the bands $r$ and $r-1$ (A detailed derivation can be found in Ref.\cite{fradkin-2013}, chapter 12). The TKNN integers $\ell_r$ are topological invariants of the bands. 
 
 We conclude, with TKNN,  that for a system with $r$ filled bands the Hall conductivity $\sigma_{xy}^{(r)}$ is given by the expression 
 \begin{equation}
 \sigma_{xy}^{(r)}=\frac{e^2}{h} \sum_{n=1}^r (\ell_r-\ell_{r-1})=\frac{e^2}{h}\; \ell_r
 \label{eq:hall-chern}
 \end{equation}
 thus showing that this system exhibits and integer quantum Hall effect and that the Hall conductivity is (up to the ratio of universal constants $e^2/h$) given by a topological invariant, the first Chern number of the band.

 We conclude with two comments. The first is that when {\it all} the bands are occupied the Hall conductivity vanishes. This obvious physical constraint is automatically satisfied by the Diophantine equation. The other comment concerns the case of what happens when $q$ is an even integer. In this case it is possible to show that there are band crossings. This is what happens for $q=2$ which is the $\pi$ flux lattice. in this case one has a theory of two species of Dirac bi-spinors. As we will see in section \ref{sec:aqhe} what happens depends on how the gapless state is reached.

 \subsubsection{The Anomalous Quantum Hall Effect}
\label{sec:aqhe}  
 
 We will now discuss a simple and insightful model of a Chern insulator that displays the quantum anomalous Hall effect, that is a Hall effect in a  system with broken time reversal symmetry in the absence of a uniform magnetic field. In order to do that we will reconsider the graphene model with additional terms in the Hamiltonian of Eq.\eqref{eq:H0-graphene-FT} that will open up energy gaps in its spectrum.
  There are several ways to do this. One way requires breaking the symmetry between the two sublattices $A$ and $B$, by assigning them a site energy of $\pm M$, which breaks the mirror reflection symmetry of the hexagon. This is what happens with graphene on a boron nitrade substrate \cite{semenoff-1984} (there are other ways that involve breaking the point group symmetry \cite{ryu-2009}). Another way to open a gap is to break time reversal symmetry which is accomplished by complex hopping amplitudes $t_2 \exp(i\phi)$ between nearest neighbor $A$ sites (and between nearest neighbor $B$ sites as well). The phase $\phi$ is oriented from $\bm r_A$ to $\bm r_A+\bm a_i$, where $\bm a_1=\bm d_2-\bm d_3$, etc. (and similarly for the $B$ sites). Physically, the oriented sum of the phases $\phi$ on the triangles generated by the hopping terms $t_1$ and $t_2$ represent the magnetic fluxes through each triangle such that the total flux on the elementary hexagon is zero. This model was considered by Haldane \cite{haldane-1988b}. With these changes the one-particle Hamiltonian becomes $2\times 2 $ Hermitian matrix 
 \begin{equation}
 H(\bm k)=h_0(\bm k)\; \mathbf{1}+{\bm h} (\bm k) \cdot \bm \sigma
 \label{eq:Hdek}
 \end{equation}
 where $\mathbf{I}$ is the $2 \times 2$ identity matrix and $\bm \sigma=(\sigma_1, \sigma_2, \sigma_3)$ is a vector made of the  three Pauli matrices. The functions $h_0(\bm k)$ and 
 $\bm h(\bm k)=(h_1(\bm k), h_2(\bm k), h_3(\bm k)$ are, respectively, given by
 \begin{align}
h_0(\bm k)=&2t_2 \cos \phi \sum_{i=1,2,3} \cos (\bm k \cdot \bm a_i), 
\qquad
h_1(\bm k)=t_1 \sum_{i=1,2,3}  \cos (\bm k \cdot \bm d_i), \nonumber\\
h_2(\bm k)=&t_1 \sum_{i=1,2,3} \sin (\bm k \cdot \bm d_i), 
\qquad
h_3(\bm k)=M-2t_2 \sin \phi \sum_{i=1,2,3} \sin (\bm k \cdot \bm a_i)
 \label{eq:Hs}
 \end{align}
 
  Provided $|t_2/t_1|<1/3$ the spectrum of the Hamiltonian of Eq.\eqref{eq:H0-graphene-FT} has two bands that do not overlap and have a finite energy gap unless they cross at the corners $K$ and $K'$ the Brillouin zone where the gap is smallest. The two bands cross at $K$ and/or  $K'$ if $M=\pm 3\sqrt{3} t_2 \sin \phi$ (for $K$ and $K'$, respectively). When either one of these conditions are not met the spectrum is gapped at either $K$ or $K'$ or, more generally,  at both. For values of the parameters for which the band gaps are small, the low energy states of the Hamiltonian of Eq.\eqref{eq:H0-graphene-FT} is that of a theory of two Dirac fields, that we will label as $\psi_1(x)$ and $\psi_2(x)$, whose one-particle Hamiltonians have the form of Eq.\eqref{eq:single-particle-dirac} but with generally different masses $m_1 \propto M-3 \sqrt{3} t_2\sin \phi$ and $m_2 \propto M+3\sqrt{3} t_2 \sin \phi$. So, in general, both fields will have non-vanishing masses which can have the same sign or opposite sign. 
  
We conclude that the effective low-energy theory consist of two species of massive Dirac bi-spinors, albeit generally with different masses. To examine the electromagnetic response we will couple the lattice model and the effective low energy theory to a slowly varying background electromagnetic field $A_\mu$. The  Lagrangian of the low energy theory is
\begin{equation}
\mathcal{L}=\bar \psi_1 (i \slashed{\partial}-m_1) \psi_1+ \bar \psi_2 (i \slashed{\partial}-m_2) \psi_2-e A_\mu (\bar \psi_1 \gamma^\mu \psi_1+\bar \psi_1 \gamma^\mu \psi_1)
\label{eq:eff-low-energy-aqh}
\end{equation}
In general the two masses will be different in magnitude and sign. Equivalently the Dirac fermions may have the same or opposite chirality. We will examine the electromagnetic response for both cases in section \ref{sec:parity-anomaly} where we will see that in the phase in which the masses of both Dirac fermions have the same sign this model describes a topological insulator that exhibits the anomalous quantum Hall effect, whereas in the phase where the signs are opposite it is a conventional insulator.

However, by varying the parameters we can tune to special values of these parameters where one of the two masses  vanishes.  In section \ref{sec:parity-anomaly} we will see that these special values of the parameters  correspond to quantum phase transitions at which the topological properties of the ground state change.  We will see that this is a quantum critical point with special properties. In particular, time reversal invariance is explicitly broken at this quantum critical point. In this regime the heavy Dirac fermion plays the role of the heavy regulating field as in the Pauli-Villars regularization of the theory of a single Dirac fermion. In the lattice model this is  a manifestation of the Nielsen-Ninomiya theorem which requires that there should be an even number of Dirac fermions \cite{nielsen-1981a,nielsen-1981b,friedan-1982}. In this context, the heavy fermion is the ``doubler''.

\subsubsection{The Parity Anomaly}
\label{sec:parity-anomaly}

We saw in section \ref{sec:chiral-anomaly} that Dirac fermions in 1+1 dimensions have a chiral anomaly. That statement mean that although the theory of free massless Dirac fermions has a global chiral symmetry, the associated current, which we called $j_\mu^5$, is not conserved when the system is coupled to an external gauge field. This effect happened because the right and left moving (Weyl) components of the massless Dirac fermion cannot be separately conserved in the presence of a gauge field. 

Analogs of the 1+1-dimensional chiral  anomaly exist on 3+1 dimensions (and, more generally, in all even spacetime dimensions). The cancelation of these anomalies is a key condition for the consistency of gauge theories \cite{thooft-1976}. We will see in section \ref{sec:gauge} that these anomalies play a role on the theory of 3D topological insulators. More generally, anomalies play a central role in the theory of symmetry-protected topological (SPT) phases of matter \cite{chen-2012,senthil-2015,witten-2016}.

Dirac fermions in 2+1 dimensions do not have an analog of the chiral anomaly for the simple reason that the chiral symmetry does ot exist. There is, however, an anomaly involving parity or, equivalently, time reversal invariance, known as the parity anomaly. The parity anomaly was discovered in the computation of the effective action of the electromagnetic field in a theory of massive Dirac fermions in 2+1 dimensions by Redlich \cite{redlich-1984a,redlich-1984b}. 

Consider theory on a single massive Dirac fermion (a bi-spinor) $\psi(x)$ in 2+1 dimensions coupled to a background U(1) gauge field $A_\mu$. The Lagrangian is
\begin{equation}
\mathcal{L}[\psi,\bar \psi, A_\mu]=\bar \psi i \slashed{D} \psi-m \bar \psi \psi
\label{eq:Dirac-2+1-gauge}
\end{equation}
where the covariant derivative is $D_\mu=\partial_\mu + i A_\mu$. The effective action $S_{\rm eff}[A_\mu]$ of the gauge field $A_\mu$ is formally obtained by computing the path integral
\begin{equation}
Z[A_\mu]=\exp(i S_{\rm eff}[A_\mu])=\int \mathcal{D}\psi \mathcal{D}\bar \psi \exp(i \int d^3x\; \mathcal{L}[\psi, \bar \psi, A_\mu])=\textrm{Det} (i \slashed{D}[A_\mu]-m)
\label{eq:ZA-Dirac-2+1}
\end{equation}
where we used that the theory is free to write the partition function as a functional determinant. In 1+1 dimensions, in the case of a massless theory, this determinant can be computed exactly using the heat kernel  (or zeta-function) technique  \cite{fujikawa-1979,gamboa_saravi-1984}. Aside from important technicalities, the reason for why this works is that as a consequence of the chiral anomaly, the effective action computed at one loop order in the fermions is exact in 1+1 dimensions. In higher dimensions this is no longer true.

We will sketch the computation of the effective action to one loop order, which means quadratic in the gauge field $A_\mu$, which has the form
\begin{equation}
S_{\rm eff}[A_\mu]= \frac{1}{2} \int d^3x \int d^3y\; A_\mu(x)\, \Pi^{\mu \nu}(x-y)\,  A_\nu(y)+\ldots
\label{eq:one-loop-effective-action}
\end{equation}
where the ellipsis contains  the higher order contributions. The polarization operator $\Pi^{\mu \nu}(x-y)$ in momentum space is (here $p_\mu$ is the momentum transfer) by the one-loop result
\begin{equation}
\Pi^{\mu \nu}(p)=\int \frac{d^3q}{(2\pi)^3} \textrm{tr} \left[S(q+p/2) \gamma^\mu S(q-p/2) \gamma^\nu\right]
\label{eq:one-loop-Pimunu-2+1}
\end{equation}
The symbol $\textrm{tr}$ signifies the trace over the spinor labels and  $S(p)$ is the (massive) Feynman propagator
\begin{equation}
S(p)=\frac{1}{\slashed{p}-m+i\epsilon}
\label{eq:S(p)}
\end{equation}
The polarization tensor $\Pi^{\mu \nu}(p)$ has the explicit decomposition into a parity even and a parity odd contributions
\begin{equation}
\Pi^{\mu \nu}(p)=(p^2 g^{\mu \nu}-p^\mu p^\nu)\; \Pi_0(p^2)-i \epsilon^{\mu \nu\lambda} p_\lambda \; \Pi_A(p^2)
\label{eq:Pimunu-transverse}
\end{equation}
where $p^2=p_0^2-{\bm p}^2$. The polarization tensor  is manifestly transverse, $p_\mu \Pi^{\mu \nu}(p)=0$ and, hence, the one-loop effective action of Eq.\eqref{eq:one-loop-effective-action} is gauge-invariant (as it should be). The parity even $\Pi_0(p^2)$ and parity odd $\Pi_A(p^2)$ kernels are given by
\begin{align}
\Pi_0(p^2)=&-\frac{|m|}{4\pi p^2}+\frac{1}{8\pi \sqrt{p^2}} \left(\frac{4m^2}{p^2}+1\right) \sinh^{-1}\left(\frac{\sqrt{p^2}}{\sqrt{4m^2-p^2}}\right) \label{eq:Pi0-even}\\
\Pi_A(p^2)=&-\frac{m}{2\pi\sqrt{p^2}} \sinh^{-1}\left(\frac{\sqrt{p^2}}{\sqrt{4m^2-p^2}}\right)
\label{eq:Pi0-odd}
\end{align}
These two results are obtained in the limit in which the regulator scales are sent to infinity leaving behind these finite results, which hold provided the Dirac mass $m$ does not vanish.

In the low energy regime, $|p^2|\ll m^2$, the kernels take the asymptotic values
\begin{equation}
\Pi_0(p^2)\simeq \frac{1}{16\pi |m|}  \qquad \Pi_A(p^2)\simeq -\frac{1}{4\pi} \textrm{sgn}(m)
\label{eq:Pi0PiA-low-energy}
\end{equation}
This result is the leading contribution to the parity even term (the first term) up to corrections of the order of $p^2/m^2$. However, the second term, which is parity odd, is the exact contribution in the limit $p \to 0$ for all non-vanishing values of the Dirac mass.  In this regime, the effective Lagrangian of the gauge field $A_\mu$ takes the local form of the sum of a Maxwell term and a Chern-Simons term
\begin{equation}
\mathcal{L}_{\rm eff}[A_\mu]=-\frac{1}{4\pi |m|} F_{\mu \nu} F^{\mu \nu} \pm  \frac{1}{2}\; \frac{1}{4\pi} \; \epsilon_{\mu \nu \lambda} \; A^\mu \partial^\nu A^\lambda
\label{eq:LA-eff_action}
\end{equation}
where $F_{\mu \nu}=\partial_\nu A_\nu-\partial_\nu A_\mu$ is the field strength of the gauge field.
Clearly, since the Maxwell term has two derivatives while the Chern-Simons term has one derivative, in the long distance regime the Maxwell term is irrelevant (subdominant). Also, while the prefactor of the Chern-Simons term is a pure number, the prefactor of the Maxwell term is proportional to $\frac{1}{|m|}$ (as required by dimensional analysis). 

The second term in the Lagrangian of  Eq.\eqref{eq:LA-eff_action} is a Chern-Simons term for the background gauge field $A_\mu$ \cite{deser-1982a, deser-1982b}. This term is gauge-invariant (up to boundary terms) but breaks parity and time reversal invariance. This phenomenon is  called the {\it parity anomaly} \cite{redlich-1984a}, although Witten \cite{witten-2016} has argued that it is better to call this phenomenon a {\it time-reversal anomaly}. The {\it sign} of the prefactor of the Chern-Simons term depends of the {\it sign} of the Dirac mass $m$: positive for $m>0$ and negative for $m<0$ or, alternatively for positive and  negative chirality of the Dirac (bi-spinor) field which fixes the sign of the breaking of time reversal invariance. 

 In addition to the sign, the prefactor of the Chern-Simons term is equal to $\frac{1}{2}$, and  it is not an integer. In section \ref{sec:degeneracy-torus} we showed that for a 
  {\it dynamical} Chern-Simons gauge theory to be  defined consistently on a closed surface this coefficient must be quantized and should take {\it integer} values. The fact that the coefficient is  half-quantized means that the classical global symmetry of a theory of a single Dirac bi-spinor cannot be gauged, which means that the gauge theory cannot be quantized. This is an example of an obstruction to the quantization of a gauge theory due to an anomaly \cite{thooft-1976}.

We will now use these results to compute the effective low-energy action for the theories of lattice Dirac fermions of section \ref{sec:Dirac-TI}. In those systems the low energy theory is that of {\it two} Dirac (bi-spinors) with different masses $m_1$ and $m_2$. Since both Dirac fields are coupled (minimally) to the same background gauge field $A_\mu$, the effective Lagrangian is just the sum of the contribution for each Dirac fermion
\begin{equation}
\mathcal{L}_{\rm eff}[A_\mu]=-\frac{1}{4\pi |m|_{\rm eff}} F_{\mu \nu}F^{\mu \nu}+\frac{1}{2} \left(\textrm{sgn}(m_1)+\textrm{sgn}(m_2)\right) \frac{1}{4\pi} \epsilon_{\mu \nu \lambda} A^\mu \partial^\nu A^\lambda
\label{eq:eff-action-doubler}
\end{equation}
where $|m|_{\rm eff}$ is 
\begin{equation}
\frac{1}{|m|_{\rm eff}}=\frac{1}{|m_1|}+\frac{1}{|m_2|}
\label{eq:m_eff}
\end{equation}
This result implies that, as anticipated in section \ref{sec:Dirac-TI}, this theory has two phases: a parity even phase and a parity-odd phase. In the regime in which the signs of the two mass terms are equal and opposite, $\textrm{sgn}(m_1)=-\textrm{sgn}(m_2)$, the coefficient of the Chern-Simons term cancels and the low-energy effective Lagrangian is a Maxwell term (generally with an effective speed of light much smaller than that in vacuum): this phase is a conventional insulator. 

Conversely, in the phase in which the two masses have the same sign, $\textrm{sgn}(m_1)=\textrm{sgn}(m_2)$, the coefficient of the Chern-Simons terms does not cancel and it is given by $\pm \frac{1}{4\pi}$, where the sign is the sign of both mass terms. This phase is also an insulator but one in which time-reversal invariance is broken. Moreover, in this phase the induced current $j_\mu$ by the background field $A_\mu$ in the long-distance limit is controlled by the Chern-Simons term and it is given by
\begin{equation}
j_\mu=\frac{\delta \mathcal{L}_{\rm eff}}{\delta A_\mu(x)}=\pm \frac{e^2}{2\pi \hbar} \epsilon_{\mu \nu\lambda} \partial^\nu A^\lambda
\label{eq: induced-CS-current}
\end{equation}
From this result we see that in the parity-broken anomalous quantum Hall phase the system has a correctly quantized and non-vanishing Hall  conductivity
\begin{equation}
\sigma_{xy}=\pm \frac{e^2}{h}
\label{eq:sigma_xy-QAH}
\end{equation}
Therefore, the phase with $\textrm{sgn}(m_1)=\textrm{sgn}(m_2)$ displays the {\it quantum anomalous Hall} effect with a Hall conductivity whose sign equals the signs of both masses. Notice that this is true regardless the magnitudes of the masses $m_1$ and $m_2$, and that only their signs matter. In section \ref{sec:TI} we will identify this phase with a topological phase of matter known as the {\it Chern Insulator}.

We showed that the Hall conductivity of the anomalous quantum Hall phase is, as expected, equal to $e^2/h$ using  the effective low energy Dirac theory. One may wonder if this approximation may be missing some contributions to the Hall conductivity. Fortunately there is an alternative quite elegant way of computing the Hall conductivity as a property of the entire occupied band. This approach, originally introduced by Gregory Volovik \cite{volovik-1987} in the context of superfluid $^3$He-A, and adapted to the theory of the quantum anomalous Hall effect by Viktor Yakovenko \cite{yakovenko-1990}, by Maarten Golterman, Karl Jansen and David Kaplan for a theory of Wilson fermions in odd  dimensional hypercubic lattices \cite{golterman-1993}, and by Xiao-Liang Qi, Yong-Shi Wu and Shou-Cheng Zhang \cite{qi-2006},  involves the derivation of a topological invariant for a two-band model.

As we saw above the Hall conductivity is obtained from the $xy$ component of the polarization operator as the limit
\begin{equation}
\sigma_{xy}=\lim_{\omega \to 0} \frac{i}{\omega} \lim_{\bm q \to 0} \Pi_{xy}(\omega, \bm q)
\label{eq:sigma-xy-Pi}
\end{equation}
For  a  free fermion system $\Pi_{xy}(\omega,0)$ is given by the current correlator
\begin{equation}
\Pi_{xy}(\omega, 0)=\int_{\rm BZ} \frac{d^2k}{(2\pi)^2} \int_{-\infty}^\infty \frac{d\Omega}{2\pi}\; \textrm{tr}\left[J_x(\bm k) G(\bm k, \omega+\Omega) J_y(\bm k)G(\bm k, \Omega)\right]
\label{eq:Pixy-2band}
\end{equation}
where $G(\bm k, \Omega)$ is the propagator for a two-band free fermion system. A generic two-band free fermion system has a one-particle Hamiltonian of the form given in Eq.\eqref{eq:Hdek}. The (one-body) current operator for such a system is obtained as
\begin{equation}
 J_l(\bm k)=\frac{\partial h_0(\bm k)}{\partial k_l}\; \mathbf{1}+\frac{\partial h_a (\bm k)}{\partial k_l} \; \sigma_a
 \label{eq:2band-currents}
 \end{equation}
The propagator $G(\bm k, \omega)$ is the $2\times 2$ matrix (in band indices)
\begin{equation}
G(\bm k, \omega)=\frac{1}{\omega \mathbf{1}-{\bm h}(\bm k) \cdot \bm \sigma+i\epsilon}=\frac{P_+(\bm k)}{\omega-E_+(\bm k)+i\epsilon}+\frac{P_-(\bm k)}{\omega-E_-(\bm k)+i\epsilon}
\label{eq:G-2band}
\end{equation}
where $P_\pm(\bm k)$ are the  operators that project onto the (empty) conduction band and the (filled) valence band whose energies are, respectively $E_\pm(\bm k)$,
\begin{equation}
P_\pm(\bm k)=\frac{1}{2} \left(\mathbf{1}\pm \hat {\bm h}(\bm k) \cdot \bm \sigma\right)
\label{eq:Ppm}
\end{equation}
where $\hat {\bm h}(\bm k)$ is the unit vector defined for every  momentum $\bm k$ of the first Brillouin zone
\begin{equation}
\hat {\bm h}(\bm k)=\frac{\bm h (\bm k)}{||{\bm h}(\bm k)||}
\label{eq:H-hat}
\end{equation}
Upon performing the frequency integration and the band traces in the expression for $\Pi_{xy}(\omega,0)$ of Eq.\eqref{eq:Pixy-2band}, we find that the Hall conductivity takes the form
\begin{equation}
\sigma_{xy}=\frac{e^2}{2\hbar} \int_{\rm BZ}\frac{d^2k}{(2\pi)^2}\; \epsilon_{abc} \frac{\partial \hat h_a(\bm k)}{\partial k_x} \frac{\partial \hat h_b(\bm k)}{\partial k_y}\hat h_c(\bm k) \left(n_+(\bm k))-n_-(\bm k)\right)
\label{eq:sigma-xy-integral}
\end{equation}
where $n_\pm(\bm k)$ are the Fermi functions (at zero temperature) for the two bands. Since 
\begin{equation}
E_+(\bm k)-E_-(\bm k)=2||{\bm h}(\bm k)||=2\sqrt{{\bm h}^2(\bm k)}>0
\label{eq:2band-gap}
\end{equation}
there is a non-vanishing gap on the entire Brillouin zone between the occupied valence band, with $n_-(\bm k)=1$, and the unoccupied conduction band, with $n_+(\bm k)=0$. For the insulating state the Fermi energy lies inside this gap and the expression for the Hall conductivity reduces to the following
\begin{equation}
\sigma_{xy}=-\frac{e^2}{2\hbar} \; \int_{\rm BZ}\frac{d^2k}{(2\pi)^2}\; \epsilon_{abc} \frac{\partial \hat h_a(\bm k)}{\partial k_x} \frac{\partial \hat h_b(\bm k)}{\partial k_y}\hat h_c(\bm k) 
\label{eq:sigma-xy-invariant}
\end{equation}
 In our discussion of quantum antiferromagnets in one space dimensions in section \ref{sec:quantum-spin-chains} we found that their effective low-energy action contained a crucial topological term proportional to the integer-valued topological invariant  $\mathcal{Q}$ (the winding number) of Eq.\eqref{eq:O(3)-top-inv} which classifies the smooths maps (homotopies) of the 2D surface (say a sphere $S^2$) onto the target space of a three-component unit vector field which also a sphere $S^2$. These equivalence classes are represented by the notation $\Pi_2(S^2) \simeq \mathbb{Z}$.
  In the case at hand the unit vector $\hat {\bm h}(\bm k)$ are points on a 2-sphere. Hence, $\hat {\bm h}(\bm k)$ is a map of the first Brillouin zone (which is a 2-torus) to the sphere $S^2$. Such maps are also classified by the same integer-valued topological invariant defined in Eq.\eqref{eq:O(3)-top-inv}. These results imply that  the Hall conductivity of the two-band system is 
  \begin{equation}
  \sigma_{xy}=\frac{e^2}{2\pi \hbar} \mathcal{Q}[\hat {\bm h}]
  \label{eq:sigma-xy-2band-topo}
  \end{equation}
  In other words we have shown that the Hall conductivity is given in terms of a topological invariant of the occupied band (in units of $e^2/h$), the winding number $\mathcal{Q}$.  In the two-band model the topological invariant $\mathcal{Q}$ plays the same role as the Chern number does in the work of TKNN \cite{thouless-1982,kohmoto-1985}
  When $\mathcal{Q} \neq 0$, the two-band system exhibits the quantized anomalous quantum Hall effect. This si a property of the entire band of occupied states, and not just a consequence of the low energy approximation. This result implies that the low energy approximation captures all of the topology of the band. It also implies that in these lattice models the Berry curvature is highly concentrated near the points in momentum space where the two bands are close in energy.
  
  .

We will now consider the problem of the quantum phase transition between the trivial and the Chern insulator.
To address this problem  we will tune the parameters of the lattice model discussed in section \ref{sec:Dirac-TI}, the phase $\phi$ and the ratio of hopping amplitudes $t_2/t_1$, to the point at which the mass of one of the two species of Dirac fermions, say the fermion $\psi_1$, is zero, $m_1=0$, while keeping the fermion $\psi_2$ massive, $m_2\neq 0$. In the low energy regime we have a massless fermion and a massive fermion. This point in parameter space is a quantum phase transition between a trivial insulator and a Chern insulator. In the low energy regime the fermion $\psi_1$ is massless. A massless fermion is a scale-invariant system in the sense that the correlators of all its observables exhibit power law behavior (free field in this case). 

We will now discuss briefly the electromagnetic response of this system at the quantum phase transition where one fermion becomes massless.  In particular, it is natural to ask if the coefficient of the parity-odd (Chern-Simons) term non-vanishing {\it at the quantum phase transition}.  To answer this question we will look at the behavior of the parity-even kernel $\Pi_0(p^2)$ and the parity-odd kernel $\Pi_A(p^2)$ in the massless limit for the light fermion, $m_1 \to 0$. We find that the {\it total} contribution of both the light fermion and  of the heavy fermion to the polarization kernels is (assuming $m_2/m_1 \to \infty$)
\begin{equation}
\lim_{m_1 \to 0} \Pi_0(p^2)=\frac{i}{16 \sqrt{p^2}}, \qquad \lim_{|m_2| \to \infty} \Pi_A(p^2)=-\frac{1}{4\pi} \textrm{sgn}(m_2)
\label{eq:massless-Pi}
\end{equation}
The important conclusion is that at this quantum critical point the parity-even kernel $\Pi_0(p^2)$ is {\it non-local}  and that the heavy regulator fermion (the ``doubler'') yields the leading finite non-vanishing and local contribution to the parity-odd kernel $\Pi_A(p^2)$.

We can now use these results to compute the conductivity {\it tensor} $\sigma_{ij}$ at the quantum critical point where $m_1 \to 0$. Since the system is spatially isotropic the conductivity tensor has the form
\begin{equation}
\sigma_{ij}=
\begin{pmatrix}
\sigma_{xx} & \sigma_{xy}\\
-\sigma_{xy} & \sigma_{xx}
\end{pmatrix}
\label{eq:cinductivity-tensor}
\end{equation}
where we used that $\sigma_{yy}=\sigma_{xx}$ since the system is isotropic. The {\it longitudinal} conductivity $\sigma_{xx}$ and the Hall conductivity $\sigma_{xy}$ are
\begin{equation}
\sigma_{xx}=\frac{\pi}{8} \frac{e^2}{h}, \qquad \sigma_{xy}=\pm \frac{1}{2} \frac{e^2}{h}
\label{eq:conductivities-qcp}
\end{equation}
in other words, {\it at} the quantum critical point the system has a finite (and universal) longitudinal conductivity. This result may seem surprising as there is no disorder in this model. A finite universal longitudinal conductivity is a standard occurrence in 2D systems at a quantum critical point. For example, at the superconductor-insulator transition the conductivity is (conjectured) to be) $\sigma_{xx}=e^2/2h$ (with $\sigma_{xy}=0$).   In addition, it also has finite and also universal Hall conductivity. The Hall conductivity at the quantum critical point is due to the heavy fermionic  ``doubler'' and is equal to 1/2 (in units of $e^2/h$). So, the quantum critical point is not time-reversal invariant since this symmetry is broken at the UV (lattice).

We can examine the massless theory using  a more formal approach \cite{witten-2016,seiberg-2016}.  In a gauge-invariant regularization  of the massless theory the partition function of a Dirac fermion coupled to a background (unquantized) U(1) gauge field, the partition function is not time-reversal invariant. It is given by 
  \begin{equation}
  Z[A_\mu]=\int \mathcal{D} \bar \psi \mathcal{D} \psi \; \exp\left(i \int d^3x \; \bar \psi i\slashed{D}[A_\mu]\psi\right)=\textrm{det} (i\slashed{D}[A_\mu])
  =\Big| \textrm{det}(i\slashed{D}[A_\mu])\Big| \; \exp\left(\pm i \frac{\pi}{2} \eta[A_\mu]\right)
  \label{eq:T-anomaly}
  \end{equation}
  where the sign depends on how time-reversal invariance is broken by the choice of regularization. 
The quantity $\eta[A_\mu]$ that appears in the phase factor is  the Atiyah-Patodi-Singer $\eta$-invariant \cite{atiyah-1975} which we already encountered in the discussion of the fractionally charged solitons in section \ref{sec:soliton-fractional-charge}. With some caveats \cite{witten-2016,seiberg-2016}, the phase factor of the partition function of Eq.\eqref{eq:T-anomaly} is commonly written in the form of a 1/2-quantized Chern-Simons term
\begin{equation}
\frac{\pi}{2} \eta[A_\mu] \equiv \pm \frac{1}{2} \int d^3x\;  \frac{1}{4\pi} \epsilon_{\mu \nu \lambda} A^\mu \partial^\nu A^\lambda
\label{eq:eta-1/2}
\end{equation}
often denoted as a U(1)$_{1/2}$ Chern-Simons term. This term plays the same role as the contribution of the heavy fermion doubler in the lattice theory.

  However, we can also wonder if there is a way to have a time-reversal invariant theory of a {\it single}  massless Dirac fermion, $m \to 0$. This case cannot be realized in a y 2D lattice model but, as will see, it can be realized on the surface states of a 3D time-reversal invariant $\mathbb{Z}_2$ topological insulator, which we will discuss in section \ref{sec:TI}.

\subsection{Three-dimensional $\mathbb{Z}_2$ topological insulators}
\label{sec:3dTI}

The classification of topological insulators in terms of a Chern number is only possible in even space dimensions in systems with broken time reversal symmetry: in $d=2$ the Berry connection is abelian and the topological invariant is the first Chern number while in in $d=4$  the Berry connection in a non-abelian SU(2) gauge field and the topological invariant is the second Chern number \cite{qi-2008}, etc. 
We will now discuss the time-reversal invariant topological insulators \cite{fu-2007,moore-2007,qi-2008} and, in particular, those that are invariant under inversion symmetry. Such states exist in both two and space dimensional insulator with strong spin-orbit coupling.

\subsubsection{$\mathbb{Z}_2$ Topological Invariants}
\label{sec:Z2-topo}

Let $\{ |u_n(\bm k)\rangle \}$ be the Bloch states. We will represent time reversal by the anti-unitary operator $\Theta$  that acts on the single particle (Bloch) states  by  complex conjugating the state and reverses the spin. For spin-1/2 fermions $\Theta=\exp(i \pi \sigma_2) \mathcal{K}$, where $\mathcal{K}$ is the complex conjugation operator. In this case $\Theta^2=-\mathbf{1}$. Let us assume that we have two occupied Bloch bands for each point $\bm k$ f the Brillouin zone. In this case the states form a rank-2 vector bundle over the torus of the Brillouin zone. In time reversal invariant systems the anti-unitary time reversal transformation $\mathcal{T}$ induces an involution in the Brillouin zone that identifies the points $\bm k$ and $-\bm k$. Time reversal the acts on the one-particle (Bloch) Hamiltonian as $\Theta H(\bm k) \Theta^{-1}=H(-\bm k)$. The states  $|u_n(\pm \bm k)\rangle$ are related by time reversal as $|u_n(-\bm k)\rangle = \Theta |u_n(\bm k)\rangle$ which implies that the bundle is real. The condition $\Theta^2=-\mathbf{1}$ implies that the bundle is real. In algebraic topology these bundles are classified by an integer (here the number of occupied bands) and a $\mathbb{Z}_2$ index that will allow us to classify these states. 

In a periodic lattice there exists a set of points $\bm Q_i$ of the Brillouin zone with the property that they differ by their images under the action of time reversal by a reciprocal lattice vector, $-\bm Q_i=\bm Q_i+\bm G$. In $d=2$ there four such points and $d=3$ there are eight points, and are given by $\bm Q_i=\frac{1}{2} \sum_j n_j \bm b_j$, where $n_j=0, 1$, $j=1,2$ in $d=2$ and $j=1,2,3$ in $d=3$. Here $\bm b_j$ are the primitive lattice vectors.  Kane and Mele \cite{kane-2005} defined the $2N \times 2N$ antisymmetric matrix $\varw_{m, n}(\bm k)=\langle u_m(-k)|\Theta |u_n(\bm k)\rangle$. They showed that at each time reversal invariant point $\bm Q_i$ one can define an index $\delta_i$
\begin{equation}
\delta_i=\frac{\sqrt{\textrm{det} [\varw(\bm Q_i)]}}{\textrm{Pf}[\varw(\bm Q_i)]}=\pm 1
\label{eq:deltas}
\end{equation}
where $\textrm{det} [\varw]$ and $\textrm{Pf}[\varw]$ are, respectively,  the determinant and the Pfaffian of the matrix $\varw$, and $\textrm{det} [\varw]=\textrm{Pf}[\varw]^2$. The sign of the quantities $\delta_i$ can be made unambiguous by requiring that the Bloch states be continuous. In addition, the quantities $\delta_i$ are gauge-dependent. However the  products 
\begin{equation}
(-1)^\nu =\prod_{i=1}^4 \delta_i
\label{eq:Z2-index-2d}
\end{equation}
in $d=2$, and
\begin{equation}
(-1)^{\nu_0}=\prod_{i=1}^8 \delta_i, \qquad (-1)^{\nu_k}= \prod_{n_k=1, n_{j \neq k}=0, 1} \delta_i(n_1,n_2,n_3)
\label{eq:Z2-indices-3d}
\end{equation}
are gauge and are also topological invariant. The $\mathbb{Z}_2$-valued indices $\nu$ and $\nu_0$ are robust to disorder and are called strong topological indices. 
Furthermore, Fu and Kane showed that if $\xi_n(\bm Q_i)=\pm 1$ are the parity eigenvalues of the occupied parity eigenstates, the quantities $\delta_i$ are given by 
$\delta_i=\prod_{m=1}^N \xi_{2m}(\bm Q_i)$. In $d=3$ the index $\nu_0$ does not rely on the existence of inversion symmetry.

In the case of a two-band model in $d=2$ and in $d=3$, the states are four-component spinors reflecting the two bands and the two spin components. In the context of these systems with strong spin-orbit coupling spin is actually the $z$-component of the atomic total angular momentum $\bm J$ of the electrons with energies close to the Fermi energy. In these systems the one-particle Hamiltonian $H(\bm k)$  is a $4 \times 4$ Hermitian matrix which can be expanded  as a linear combination of Dirac matrices. 
A simple and very useful model of systems of this type is the Wilson fermion model (with continuous time) \cite{wilson-1974b,bernevig-2006,qi-2008} of a square (cubic) lattice with $4$ states per site (parity and spin) whose Hamiltonian three dimensions is
\begin{equation}
H(\bm k)=\sin \bm k \; \cdot \bm \alpha+M(\bm k)\; \beta
\label{eq:wilson-fermion}
\end{equation}
where $\bm \alpha=\gamma_0 \bm \gamma$ and $\beta=\gamma_0$ are the conventional $4\times 4$  Dirac matrices. The $\gamma$ matrices satisfy the Clifford algebra $\{ \gamma_\mu, \gamma_\nu \}=2 g_{\mu \nu} \mathbf{1}$, where $g_{\mu \nu}=\textrm{diag}(1, -1, -1, -1)$ is the metric tensor of four-dimensional Minkowski space time. An additional $\gamma$ matrix of interest is $\gamma_5=i\gamma_0\gamma_1\gamma_2\gamma_3$. 

The (Wilson) mass term $M(\bm k)$ in two dimensions is $M(\bm k)=M+\cos k_1 + \cos p_2 - 2$, and in three dimensions is $M(\bm k)=M+\cos k_1+\cos k_2 + \cos k_3- 3$. In Eq.\eqref{eq:wilson-fermion} we see that, consistent with the requirements of the Nielsen-Ninomiya theorem \cite{nielsen-1981a,nielsen-1981b}, in addition to a possible low energy Dirac fermion (if $M$ is small) there are three more massive Dirac fermions (in $d=2$) and seven other Dirac fermions in $d=3$, so that the total number of Dirac fermions is always even, $8$ in this case.  With Wilson's mass term the additional Dirac fermions (the ``doublers'') are always heavy even if the Dirac fermion near the $\Gamma$ point $\bm Q=0$ is light.

 In the Dirac basis time reversal is the operation 
$\Theta=(i \sigma_2 \otimes I) \mathcal{K}$, where $\mathcal{K}$ is complex conjugation, and parity is $P=\beta$. The matrices $\bm \alpha$ and $\beta$ commute with $P \Theta$. 
At the time-reversal and parity invariant points of the Brillouin zone $\{ \bm Q_i\}$ the Hamiltonian  only depends of the the matrix $\beta$,
$H(\bm Q_i)=M(\bm Q_i) \beta$.
Since the parities of the spinors are the eigenvalues of the matrix $\beta$, we conclude that in the two-band models the quantities $\delta_i$ are simply equal to the sign of the mass of the fermions defined at the time-reversal-invariant points $\bm Q_i$ of the BZ \cite{fu-2007,qi-2008},
$\delta_i=\delta(\bm Q_i)=-\textrm{sgn} \; M(\bm Q_i)$. Using this  result it follows that in two dimensions the system is a $\mathbb{Z}_2$ topological insulator with index $\nu=1$ (mod 2) if $0<M<4$ while it is trivial for other values of $M$, i.e. $\nu=0$ mod 2. Similarly, in three dimensions a $\mathbb{Z}_2$ (strong) topological insulator exists only if $0<M<2$ (in the other regimes this system is in a weak topological insulator state or in a trivial one).

We conclude that both in two and three dimensions the low energy theory of the $\mathbb{Z}_2$ topological insulators consists of a single Dirac fermion whose mass is small compared to that of the fermion doublers and has the opposite sign. In both cases there is a quantum phase transition between a trivial insulator and the time-reversal invariant topological insulator at the point where he mass of the light fermion vanishes. 

\subsubsection{The Axial Anomaly and the Effective Action}
\label{sec:axial-anomaly}

We will now look at the electromagnetic response of a $\mathbb{Z}_2$ topological insulator in 3+1 dimensions.
 This problem can be addressed more easily using  the continuum field theory description which is valid in the regime where the mass $M$ is weak. 
 In this regime one species of Dirac fermions is light (i.e. its mass is small) while the Dirac doublers remain heavy. Much as we did in our discussion of the Chern insulators 
 and the {\it parity anomaly} in section \ref{sec:parity-anomaly} we will keep in mind that the fermion doublers play the role of heavy regulators, such as in the Pauli-Villars scheme,  in Quantum Field Theory. 
 Here too, we will see that a field-theoretic anomaly, known as the {\it axial anomaly} plays a central role in the physics. The analysis is very similar to what we did with the {\it chiral anomaly}
  in 1+1 dimensions in section \ref{sec:chiral-anomaly}.

Let us begin with a theory of a single massive Dirac fermion in 3+1 dimensions.  The Lagrangian of the  free massive Dirac theory is
\begin{equation}
\mathcal{L}=\bar \psi \left(\; i \slashed{\partial}-m \right) \psi
\label{eq:dirac-lagrangian}
\end{equation}
The equation of motion of the spinor field operator $\psi(x)$ (I omit the spinor indices here) is the Dirac equation
\begin{equation}
\left(i \slashed{\partial}-m\right)\psi=0
\label{eq:dirac-3=1}
\end{equation}
The Dirac Lagrangian has a global gauge symmetry $\psi \to e^{i \theta} \psi$ which requires that 
 the  Dirac current $j_\mu=\bar \psi \gamma_\mu \psi$ is locally conserved, $\partial_\mu j^\mu=0$. The {\it massless} Dirac theory can be decomposed into a theory of two Weyl bi-spinor fields which 
 obey separate Dirac Lagrangians. Furthermore, the massless theory has the additional global symmetry under the transformation $\psi \to e^{i \theta \gamma_5} \psi$ and  
 the additional formally locally conserved {\it axial} current $j_\mu^5=i \bar \psi \gamma_\mu \gamma^5 \psi$. However the conservation law of the axial current is violated in the {\it massive} Dirac theory
\begin{equation}
\partial^\mu j_\mu^5=-2m i \bar \psi \gamma^5 \psi
\label{eq:axial-non-conservation}
\end{equation}
since the two Weyl bi-spinors transmute into each other in the presence of a mass term. This is the origin of the phenomenon of neutrino oscillations. In the condensed matter physics context there is a similar phenomenon in systems of Weyl semimetals which have  crossings between the valence and conducting bands at two locations $\pm \bm Q$ of the BZ, with each crossing associated with each  Weyl bi-spinor. A  charge density wave with ordering wavevector $2\bm Q$ mixes the two Weyl fermions which become gapped, becoming effectively a single massive Dirac fermion \cite{gooth-2019}. This state is often called an ``axionic''-charge-density-wave. 

We will now show that the axial symmetry has an anomaly and cannot be gauged. Thus, we will consider the problem of a Dirac theory coupled to a background U(1) gauge field $A_\mu$ and reexamine the putative conservation of the axial current $j_\mu^5$. This question can be addressed in different ways. Quite early on Steven Adler \cite{adler-1969}, and John Bell and Roman Jackiw \cite{bell-1969} examined this problem by computing a Dirac fermion triangle diagram for the process of a neutral pion decaying into two photons, $\pi^0 \to 2\gamma$. In particle physics the pion is the Goldstone boson of the spontaneously broken chiral symmetry. The analog  of this problem in condensed matter physics is the phase mode of an incommensurate charge density wave. The relativistic Lagrangian for this problem is a theory of Dirac fermions coupled to a complex scalar field $\phi=\phi_1+i \phi_2$ through two Yukawa couplings
\begin{equation}
\mathcal{L}=\bar \psi i \slashed{D} \psi+g \phi_1 \bar \psi \psi+i g \phi_2 \bar \psi \gamma^5 \psi-V(\phi_1, \phi_2)=\bar \psi i \slashed{D}  \psi +g |\phi | \bar \psi e^{i \gamma_5 \theta} \psi-V(|\phi |^2)
\label{eq:chiral-Yukawa}
\end{equation}
where $|\phi |^2=\phi_1^2+\phi_2^2$, $\tan \theta=\phi_2/\phi_1$, and $D_\mu=\partial_\mu+i e A_\mu$ is the covariant derivative. Here we are regarding the gauge field $A_\mu$ as a background probe field.

The triangle Feynman diagram computes the polarization tensor for the electromagnetic field $A_\mu$ with an insertion of the coupling to the complex scalar field in an otherwise massless theory. Assuming a gauge-invariant regularization of the diagram, this computation finds that the axial current $j_\mu^5$ is anomalous and is not conserved even in the massless theory \cite{adler-1969,bell-1969}
\begin{equation}
\partial^\mu j_\mu^5=-\frac{e^2}{16\pi^2} F^{\mu \nu} F^*_{\mu \nu}
\label{eq:axial-anomaly-3+1}
\end{equation}
where $F^*_{\mu \nu}=\frac{1}{2} \epsilon_{\mu \nu \lambda \rho}F^{\lambda \rho}$ is the dual of the electromagnetic field tensor. This is the {\it axial anomaly}.

To see how the axial anomaly arises we will follow the physically transparent approach of Nielsen and Ninomiya \cite{nielsen-1983}, that we also employed in section \ref{sec:chiral-anomaly} in 1+1 dimensions. We will consider a theory of free massless dirac fermions coupled to a background electromagnetic field. Since the theory is massless, the Dirac equation decouples into an equation to the right handed Weyl fermion $\psi_R$ (with positive chirality $\gamma_5 \psi_R=+ \psi_R$) and a left handed Weyl fermion $\psi_L$ (with negative chirality, $\gamma_5 \psi_L=-\psi_L$). In the gauge $A_0=0$, the Dirac equations become
\begin{equation}
[i \partial_0-(-i\bm \partial - e \bm A)\cdot \bm \sigma]\psi_R=0, \qquad [i \partial_0-(i \bm \partial - e \bm A)\cdot \bm \sigma]\psi_L=0
\label{eq:diracs-3+1}
\end{equation}
Let us now consider look at the solutions of the  Weyl equation for right-handed fermions $\psi_R$, Eq.\eqref{eq:diracs-3+1}. The left-handed fermions $\psi_L$ are analyzed similarly. Wd will consider a gauge field $A_1=0$ and $A_2=B x_1$ representing a uniform static magnetic field of strength $B$ pointing along the $x_3$ direction. In this this gauge the eigenstates are plane waves along the directions $x_2$ and $x_3$ and harmonic oscillator states along the direction $x_1$. The eigenvalue spectrum consists of Landau levels with energies 
\begin{equation}
E(n, p_3, \sigma_3)=\pm \left[2eB \left(n+\frac{1}{2}\right)+p_3^2+eB\sigma_3\right]^{1/2}
\label{eq:relativistic-landau-3+1}
\end{equation}
for $n=0, 1, 2, \ldots$, except for the {\it zero mode} with $n=0$ and $\sigma_3=-1$, for which
\begin{equation}
E(n=0,p_3,\sigma_3=-1)=\pm p_3
\label{eq:zero-mode-LL-3+1}
\end{equation}
where the $+$ sign holds for $\psi_R$ and the $-$ sign for $\psi_L$. Just as in the case of non-relativistic fermions the relativistic Landau levels are degenerate.
We will consider a system of Dirac fermions at charge neutrality and, hence, $E_F=0$. The positive and negative energy states are charge conjugate of each other and the negative energy states are filled. For right-handed fermions, the zero mode states with $p_3<0$ are filled while for right handed states the zero modes with $p_3>0$ are filled. The density of states of the zero  modes is $LeB/(4\pi^2)$ (where $L$ is the linear size of the system).

Let us consider now turning on an external electric field $E$ {\it parallel} to the magnetic field $B$. Just as we saw in 1+1 dimensions in section \ref{sec:chiral-anomaly}, the electric field leads to pair creation by shifting the Fermi momentum to $p_F$ for the zero modes. There is no particle creation for the states with $n \neq  0$ and they do not contribute to the anomaly. The rate of creation of right-handed fermions $N_R$,
\begin{equation}
\frac{dN_R}{dx_0}=\frac{1}{L}\frac{LeB}{4\pi^2} \frac{p_F}{dx_0}=\frac{e^2}{4\pi^2} EB
\label{eq:anomaly-right-handed}
\end{equation}
The annihilation rate of left handed particles is
\begin{equation}
\frac{dN_L}{dx_0}=-\frac{1}{L}\frac{LeB}{4\pi^2} \frac{p_F}{dx_0}=-\frac{e^2}{4\pi^2} EB
\label{eq:anomaly-left-handed1}
\end{equation}
and the creation rate of left handed anti-particles is
\begin{equation}
\frac{d\bar N_L}{dx_0}=\frac{1}{L}\frac{LeB}{4\pi^2} \frac{p_F}{dx_0}=\frac{e^2}{4\pi^2} EB
\label{eq:anomaly-right-handed2}
\end{equation}
the axial anomaly is the total rate of creation of right handed particles and of left-handed antiparticles:
\begin{equation}
\frac{dQ_5}{dx_0}=\frac{dN_R}{dx_0}+\frac{d\bar N_L}{dx_0}=\frac{e^2}{2\pi^2} EB
\label{eq:axial-anomaly-total}
\end{equation}
which agrees with  the expression of Eq.\eqref{eq:axial-anomaly-3+1}.

We now turn to the  $\mathbb{Z}_2$  topological insulator. As we saw this is a system with two species of ($4$ component) Dirac spinors. In the topological phase the sign of the Dirac mass term one of the Dirac fermions (the one near the $\Gamma$ point in the lattice model) is opposite (negative) to the sign of the mass term of the of the other Dirac fermion (the fermion doubler). Explicit calculations \cite{qi-2008,essin-2009,hosur-2010} on the lattice model obtain the result that the effective low-energy action for the electromagnetic gauge field $A_\mu$ in the topological phase is
\begin{equation}
S_{\rm eff}[A_\mu]=\int d^4x \left[-\frac{1}{4e^2} F_{\mu \nu}F^{\mu \nu}+\frac{\theta}{32 \pi^2} e^2 \epsilon_{\mu \nu \lambda \rho} F^{\mu \nu} F^{\lambda \rho}\right]+\ldots
\label{eq:Seff-A-theta}
\end{equation}
In a time-reversal invariant system the  allowed values of the $\theta$ angle of Eq.\eqref{eq:Seff-A-theta} are restricted to be $\theta=n \pi$, with $n \in \mathbb{Z}$. The case $\theta=0$ (mod $2\pi$) represents a trivial insulator whereas $\theta=\pi$ (mod $2\pi$) holds for a $\mathbb{Z}_2$ time-reversal invariant topological insulator.

The second term in the effective action of the electromagnetic gauge field of Eq.\eqref{eq:Seff-A-theta},  known as the $\theta$ term, has been extensively discussed in the high-energy physics literature \cite{fujikawa-1979,wilczek-1987,peskin-1995}. The derivation of this term is subtle. As it stands, unless $\theta$ is varying in space-time (in which case this is known a the axion field)  this term is a total derivative. In non-abelian Yang-Mills gauge theories this term is proportional to a topological invariant known as the Pontryagin index which counts the instanton number of the gauge field configurations \cite{callan-1976,coleman-1985}. In the context of the Lagrangian of Eq.\eqref{eq:chiral-Yukawa} this term is induced by the coupling of the Dirac fermion to the Yukawa coupling of the complex scalar field $\phi$ to the Dirac and $\gamma_5$ mass terms. In this case, the lowest order contribution is given by the triangle diagram. The fact that this term is exact at lowest order reflects the fact that the axial anomaly is in fact a non-perturbative effect which is the same in both the weak coupling and the strong coupling regimes \cite{thooft-1976}. 

In the phase where the potential $V(|\phi |^2)$ in the Lagrangian of Eq.\eqref{eq:axial-anomaly} has a minimum at $|\phi_0| \exp(i \theta)$ the chiral symmetry is spontaneously broken and the phase field $\theta(x)$ is the Goldstone boson of the spontaneously broken chiral symmetry. In this phase the Dirac fermions become massive through the Yukawa couplings to the complex scalar field, and the phase of the field $\phi$ enters in the effective action as an axion field which couples to the gauge field $A_\mu$ through the $\theta$ term of the effective action. We should note that in a recently studied axionic CDW state of a Weyl semimetal \cite{gooth-2019} the phase of the CDW plays the role of the axion field.

\subsubsection{Theta terms, and Domain walls: Anomaly  and the Callan-Harvey Effect}
\label{sec:CH-effect}

We will now discuss some remarkable behaviors of three-dimensional $\mathbb{Z}_2$ topological insulators. We will begin with the electromagnetic response encoded in the effective action of Eq.\eqref{eq:Seff-A-theta}. We will assume that the $\theta$ angle is effectively a slowly varying Goldstone mode of the spontaneously broken $U(1)$ chiral symmetry (i.e. the axion field) present in the Lagrangian of Eq.\eqref{eq:chiral-Yukawa}.  If the field $\theta$ is constant, the $\theta$ term is a total derivative and it does not contribute to the local equations of motion. We will see below that this term [lays a key role in the physics of a domain wall, which we will regard as the interface of a $\mathbb{Z}_2$ topological insulator and a trivial insulator. In the general case in which $\theta$ varies slowly (as Goldstone modes do) its presence leads to interesting modification of Maxwell's equations,  known as axion electrodynamics \cite{wilczek-1987}:
\begin{align}
\bm \bigtriangledown \cdot \bm E=&\tilde \rho -e^2\bm \bigtriangledown \theta \cdot \bm B, & \bm \bigtriangledown \times \bm E=&-\partial_t \bm B \nonumber\\
 \bm \bigtriangledown \times \bm B=&\partial_t \bm E+\tilde {\bm j}+e^2\left(\partial_t \theta+\bm \bigtriangledown \theta \times \bm E\right), & \bm \bigtriangledown \cdot \bm B=&0
\label{eq:axion-qed}
\end{align}
where $\tilde {\bm \rho}$ and $\tilde {\bm j}$ are external probe electric charge and current densities. The equations of axion electrodynamics have many remarkable properties. Here we will focus on effects: the topological magnetoelectric effect \cite{qi-2008} and the Witten effect \cite{witten-1979}. 

Let us consider a $\mathbb{Z}_2$ topological insulator with a flat open boundary (the $x_1-x_2$ plane) perpendicular to the direction $x_3$. We will assume that the topological insulator lies at $x_3 <0$. This means that for $x_3<0$, and far from the  surface, $ \theta(x_3) \to \pi$, while in the trivial vacuum, and also  far from the surface, $\theta(x_3) \to 0$. We will assume that the change of $\theta$ from $0$ to 
$\pi$ occurs on a short distance $\xi$. We will call this configuration an axion domain wall. In the region where $\theta(x_3)$ is changing, $|x_3|\lesssim \xi$, an applied uniform magnnetic field $\bm B$ induces a uniform electric field $\bm E$ parallel to $\bm B$ whose magnitude is proportional to the change in $\theta$. This is the topological magnetoelectric effect \cite{qi-2008}.

A similar striking effect is obtained by considering the case of a magnetic monopole of magnetic charge $2\pi/e$ (as required by Dirac quantization)  inside a sphere of the trivial region of radius $R$, with $\theta=0$, surrounded by a region with $\theta\neq 0$. The two regions are separated by a thin (axion) wall in which $\theta$ changes for $0$ to $\pi$ in a narrow shell of thickness $\xi\ll R$. The equations of axion electrodynamics imply that the magnetic monopole induces an electric charge $Q_e$ on the surface of the sphere 
\begin{equation}
Q_e=e \frac{\Delta \theta}{2\pi}
\label{eq:witten-effect}
\end{equation}
Thus, a magnetic monopole acquires an electric charge and becomes a {\it dyon}. This is the Witten effect \cite{witten-1979}. In the particular case of a $\mathbb{Z}_2$ topological insulator, time reversal invariance requires that $\Delta \theta=\pi$, which implies that a monopole with unit magnetic charge has an electric charge $e/2$. A similar argument implies that an external magnetic field perpendicular to the open surface of the $\mathbb{Z}_2$ topological insulator (which is essentially an axion domain wall) induces an {\it electric charge polarization} on the surface proportional to the total magnetic flux, which is another manifestation of the topological magnetoelectric effect.

The reader should readily recognize that the result of Eq.\eqref{eq:witten-effect} for the  electric charge induced my the magnetic monopole is the same as the Goldstone-Wilczek equation for the fractional charge for a one dimensional soliton of Eq.\eqref{eq:GW}. We will now see that this is not just an analogy. We will follow here the general approach of Curtis Callan and Jeffrey Harvey \cite{callan-1985} who extended the earlier work of Goldstone and Wilczek \cite{goldstone-1981}. Let us consider the bulk of a $\mathbb{Z}_2$ topological insulator and assume that $\theta(x)$ is slowly varying. We can use the triangle diagram calculation to find that an electromagnetic gauge field $A_\mu$ induces a current  $\langle J_\mu(x)$ given by 
\begin{equation}
\langle J_\mu(x)\rangle=-i \frac{e}{16 \pi^2} \epsilon_{\mu \nu \lambda \rho} \frac{\phi^*(x) \partial^\nu \phi(x)-\phi(x) \partial^\nu \phi^*(x)}{|\phi(x)|^2} F^{\lambda \rho}(x)=\frac{e}{8\pi^2} \epsilon_{\mu \nu \lambda \rho} \partial^\nu \theta(x) F^{\lambda \rho}(x)
\label{eq:callan-harvey}
\end{equation}
This result implies that, in the case of a domain wall in the $x_1-x_2$ plane, a magnetic field perpendicular to the wall induces a current towards the wall and, hence, a charge accumulation on the wall. Where does this charge come from?  To understand this problem we will consider a $\mathbb{Z}_2$ topological insulator occupying a slab of macroscopic size $L$  between a wall at $x_3=0$ and a far way ``anti-wall'' at $x_3=L$. In this configuration a magnetic field normal to the wall(s) induces a transfer of charge from one wall to the other. Similarly, an electric field {\it parallel} to the wall induces a current also parallel to the wall and {\it perpendicular} to the electric field, i.e. a Hall current. 

In 1+1 dimensions we saw that soliton configurations acquire a fractional charge associated to states bound to the soliton (which are zero modes when $\Delta \theta=\pi$). We will see that the surfaces of the 3D $\mathbb{Z}_2$ topological insulators also have zero modes which ar Weyl fermions propagating on the wall. To see how this works we will consider a 3+1 dimensional Dirac fermion with a Dirac mass that changes sign at $x_3=0$. The Lagrangian now ill be
\begin{equation}
\mathcal{L}=\bar \psi i \slashed{\partial} \psi+g \phi(x) \bar \psi \psi
\label{eq:varying-mass}
\end{equation}
where $\phi(x)$ is now a {\it real} scalar field that has the asymptotic behaviors $\phi(x_3)=\phi_0 \; f(x_3)$ such that $\lim_{x_3 \to \pm \infty}f(x_3)=\pm 1$. We will assume that $f(x_3)$ is a monotonous function of $x_3$, but its actual dependence on $x_3$ is immaterial aside from the requirement that it should change sign at some point which we will take to be $x_3=0$. This is a special case of Eq.\eqref{eq:chiral-Yukawa} with $\phi_1=\phi$ and $\phi_2=0$. We will now recognize that this just a Dirac fermion with a position-dependent  Dirac mass $m(x_3)=g \phi(x_3)$. 
The one-particle Dirac Hamiltonian for this system is
\begin{equation}
H=-i \bm \alpha \cdot \bm \bigtriangledown +m(x_3) \beta
\label{eq:3D-dirac-variable-mass}
\end{equation}
where $\bm \alpha$ and $\beta$ are the four $4 \times 4$ Dirac matrices. By symmetry, this Hamiltonian can be split into two Hamiltonians, $H=H_{\rm wall}+H_\perp$, 
\begin{equation}
H_{\rm wall}=-i \alpha_1 \partial_1-i \alpha_2 \partial_2, \qquad H_\perp=-i \alpha_3 \partial_3+m(x_3) \beta
\label{eq:dirac-split}
\end{equation}
Let the spinor $\psi_\pm$ be and eigenstate of the anti-hermitian Dirac matrix $\gamma_3=\beta \alpha_3$ with eigenvalues $\pm i$, respectively
\begin{equation}
\gamma_3 \psi_\pm=\pm i \psi_\pm
\label{eq:psi-pm}
\end{equation}
We seek a spinor solution $\psi_\pm$ of the Dirac equation Eq.\eqref{eq:3D-dirac-variable-mass} such that $H_\perp \psi_\pm=0$ 
\begin{equation}
\pm \partial_3 \psi_\pm+m(x_3) \psi_\pm=0
\label{eq:zero-mode-3D}
\end{equation}
which is also a solution of 
\begin{equation}
(i \gamma_0-i \gamma_1 \partial_1-i\gamma_2 \partial_2)\psi_\pm=0
\label{eq:weyl}
\end{equation}
In other words, it is a solution of the massless Dirac equation in 2+1 dimensions. The requirement that $\psi_\pm$ be an eigenstate of $\gamma_3$  reduces the number of 
spinor components from four to two. The full solution has the form
\begin{equation}
\psi_\pm=\eta_\pm (x_0, x_1, x_2) F_\pm (x_3), \qquad \pm \partial_3 F_\pm (x_3)=-m(x_3) F_\pm(x_3)
\label{eq: solution1}
\end{equation}
For a domain wall with $\lim_{x_3 \to \infty}m(x_3)=+m$, the normalizable solution is  $F_+(x_3)$ is
\begin{equation}
F_+(x_3)=F(0)\; \exp\left(-\int_0^\infty dx'_3\; m(x'_3)\right)
\label{eq:Fpm}
\end{equation}
where $F(0)$ is a constant. For the anti-domain wall, for which $\lim_{x_3 \to \infty} m(x_3)=-m$, the normalizable solution is $F_-(x_3)$.

We conclude that there is a 2+1-dimensional massless Dirac theory that describes the quantum states  bound to the wall which propagate along the wall. 
These states are the generalization of the fractionally charged mid-gap states of solitons in one dimension discussed in section \ref{sec:soliton-fractional-charge}. 
The energy of these states is $E(\bm p)=\pm |\bm p|$, where $\bm p=(p_1, p_2)$ (where I set the Fermi velocity to unity). Experimental evidence for 2+1 dimensional massless Dirac fermions 
on the surface of the 3D $\mathbb{Z}_2$ topological insulator Bi$_2$Te$_3$ has been found in spin-polarized angle-resolved photoemission studies of the surface states which showed that 
they have a linear energy-momentum relation (expected of massless Dirac fermions) as well as the  spin-momentum locking  characteristic of these spinor states \cite{hsieh-2009}.

Having succeeded in showing that the surface of the $\mathbb{Z}_2$ time-reversal invariant 3D topological insulator has a two-component massless Dirac spinor we now want to determine its electromagnetic response. In fact we have already discussed this problem in our discussion of the parity anomaly in section \ref{sec:parity-anomaly} where we showed that the effective action for the electromagnetic field $A_\mu$ of Eq.\eqref{eq:eta-1/2} of a single massless Dirac bi-spinor is a Chern-Simons term with a prefactor which is $1/2$ of the allowed value. In that purely 2+1-dimensional context we saw that time reversal invariance is actually broken. However, the 3D problem is time-reversal invariant so there must be a contribution that cancels this time-reversal anomaly. The answer is that the requisite cancellation is supplied by the bulk. 

To see how this can happen we return to the bulk effective action of the 3D topological insulator of Eq.\eqref{eq:Seff-A-theta} where we observed that the $\theta$-term is a total derivative. Suppose that the system has a boundary at $x_3=0$ and that the topological insulator exists for $x_3>0$ (where the fermion mass  is negative, $m<0$) and trivial for $x_3<0$ (where $m>0$). Only the region with $m<0$ contributes to the $\theta$-term.  The region of four-dimensional space time occupied by the topological insulator is $\mathcal{M}$ and in this region the $\theta$-term becomes
\begin{equation}
S_{\theta}[A]=\frac{\theta}{32\pi^2} \int_\mathcal{M} d^4x\; \epsilon_{\mu \nu\lambda\rho}F^{\mu \nu} F^{\rho\lambda}=\frac{\theta}{8\pi^2} \int_\mathcal{M} d^4x \; \epsilon_{\mu \nu\lambda\rho} \partial^\mu A^\nu \partial^\lambda A^\lambda=\frac{\theta}{8\pi^2} \int_{\partial \mathcal{M}} d^3x\;  \epsilon_{\mu\nu\lambda} A^\mu \partial^\nu A^\lambda
\label{eq:Sboundary}
\end{equation}
where $\partial \mathcal{M}\equiv \Sigma \times \mathbb{R}$ is the boundary of the region $\mathcal{M}$, $\Sigma$ is the surface of the 3D topological insulator and $\mathbb{R}$ is time. 

Thus we see that the $\theta$-term of the effective action $S_\theta[A]$  integrates to the boundary where it has the form of a 2+1-dimensional Chern-Simons term. 
Since for a time-reversal-invariant 3D topological insulator $\theta=\pi$, we see that in this case the boundary Chern-Simons term becomes
\begin{equation}
S_{\theta=\pi}[A]=\frac{1}{8\pi} \int_{\partial \mathcal{M}} d^3x \;  \epsilon_{\mu\nu\lambda} A^\mu \partial^\nu A^\lambda
\label{eq:boundary-CS}
\end{equation}
with a coefficient which is 1/2 of the allowed value. 
This bulk contribution either cancels the parity anomaly of the boundary state, rendering the full system time-reversal invariant. In other words, in this system time-reversal symmetry is realized by cancellation of the anomaly between the bulk of the topological insulator and the boundary or, equivalently, by an {\it inflow} of the parity anomaly between the boundary and the bulk of the system. Another way to phrase this result is the statement that a single Dirac fermion cannot exist on its own in 2+1 dimensions but it can as the boundary state of a 3+1-dimensional system whose anomaly cancels the anomaly of the boundary.

\subsection{Chern-Simons Gauge Theory and The Fractional Quantum Hall Effect}
\label{sec:CS-FQHE}

We will now turn to the problem of the quantum Hall effects. This is a problem that revealed the existence of profound and far reaching connections between condensed matter physics, quantum field theory, conformal field theory and topology. In particular, the {\it fractional} quantum hall effect is the best studied and best understood topological phase of matter. As such, it has become the conceptual springboard for its manifold generalizations.

The integer (IQHE) and fractional (FQHE) quantum Hall effects are fascinating phenomena observed in fluids of electrons in two dimensions in strong perpendicular magnetic fields. The integer quantum Hall effect was discovered by Klaus von Klitzing in 1980 \cite{klitzing-1980} in transport measurements of the longitudinal and Hall resistivity of the surface states of metal oxide field-effect transistors (MOSFET) in magnetic field of up to 15 Tesla. The effect that von Klitzing discovered (for which he was awarded the 1985 Nobel Prize in Physics) was that in the high field regime the measured Hall conductivity showed a series of sharply defined plateaus at which took the values $\sigma_{xy}=n e^2/h$, where $n$ is an integer, and the longitudinal conductivity appeared to vanish $\sigma_{xx} \to 0$ as the the temperature was lowered down to $T \simeq 1.5$ K. Remarkably the measured value of the Hall conductivity was  obtained with a precision of $\sim 10^{-9}$. To this date the measurement of the Hall conductivity in the IQHE yields the most precise definition of the fine structure constant.

Subsequent transport experiments in ultra-high-purity GaAs-AlAs heterostructures by Dan Tsui, Horst St\"ormer and Art Gossard found that, in addition to the IQHE, two-dimensional electron fluids in high magnetic fields exhibit the {\it fractional} quantum Hall effect meaning that there are equally sharply-defined plateaus of the Hall conductivity at the values $\sigma_{xy}=\frac{p}{q}\frac{e^2}{h}$, where $p$ and $q$ are co-prime integers \cite{tsui-1982}. Much as in the IQHE case, in the FQHE the longitudinal conductivity vanishes at low temperatures. It is important to note that both in the IQHE and in the FQHE the observed temperature dependence in the highest purity samples of the longitudinal conductivity is activated, $\sigma_{xx}\sim \exp(-W/T)$. The observed value of the energy scale $W$ is the experimental estimate of an energy gap in the electron fluid in the quantum Hall states.

In addition to MOSFETS and GaAs-AlAs heterostructures both the IQHE and the FQHE have been seen in several other experimental platforms, particularly in graphene and other 2D materials \cite{du-2009,zibrov-2017}. The explanation of these effects and of a panoply of startling consequences that were uncovered in the course of understanding this phenomenon is the focus of this section.

\subsubsection{Landau levels and the Integer Hall effect}
\label{sec:LL}

At some level the integer quantum Hall effect can be explained by the Landau quantization of the energy levels of a  free charged moving in two dimensions in a perpendicular magnetic field \cite{landau-1930}. The  Hamiltonian for a  non-relativistic particle of charge $-e$ and mass $M$ in a perpendicular magnetic field $B$ is
\begin{equation}
H=\frac{1}{2M} \left(-i \hbar \bm \bigtriangledown +i\frac{e}{ c} \bm A(\bm x)\right)^2
\label{eq:H-LL}
\end{equation}
for a uniform perpendicular magnetic field $\bm B=B\;  \hat {\bm e_z}=\bm \bigtriangledown \times \bm A(\bm x)$. 

In the circular gauge the vector potential is $A_i=-\frac{1}{2} B \epsilon_{ij} x_j$. We will assume that the 2D plane has linear size $L$. The total magnetic flux is $\Phi=BL^2$ and we will assume that there is an integer number $N_\phi$ of magnetic flux quanta piercing the plane, $\phi=N_\phi \phi_0$, where $\phi_0=hc/e$ is the flux quantum. In units such that $\hbar=e=c=1$ the flux quantum is $\phi_0=2\pi$ and $\Phi=2\pi N_\phi$. 

In the presence of a magnetic field the components of the canonical momentum operator $\bm p=-i \hbar \bm \bigtriangledown-\frac{e}{c} \bm A$ do not commute with each other,
\begin{equation}
[p_i, p_j]-i \frac{e\hbar}{c} B \epsilon_{ij}
\label{eq:pi-pj}
\end{equation}
This means that translations in two directions do not commute with each other. However, the components of the operator $\bm k=\bm p(-B)$ commute with $\bm p$ (and hence with the one-particle Hamiltonian) but do not commute with each other: the commutator is the same as in Eq. \eqref{eq:pi-pj}. Since $[\bm k, H]=0$ they act as symmetry generators of the group of {\it magnetic translations}. For arbitrary displacements $\bm a$ and $\bm b$ the translation operators $t(\bm a)=\exp(i \bm a \cdot \bm k/\hbar)$ (and similarly with $\bm b$) satsify
\begin{equation}
t(\bm a) t(\bm b)=\exp(i \bm a \times \bm b \; \cdot \bm e_z/\ell_0^2) t(\bm b) t(\bm a)
\label{eq:magnetic-translations}
\end{equation}
Magnetic translations only commute with each other is the area subtended by $\bm a$ and $\bm b$ contains an integer number of magnetic flux quanta.

Given the rotational symmetry of the circular gauge it is natural to work in complex coordinates $z=x_1+ix_2$. We will also use the notation $\partial_z=(\partial_1-i \partial_2)/2$ and $\partial_{\bar z}=(\partial_1+i \partial_2)/2$. In this gauge, up to a normalization, the eigenstate wave functions have the form
\begin{equation}
\psi(z, \bar z)=f(z, \bar z) \; \exp\left(-\frac{|z|^2}{4\ell_0^2}\right)
\label{eq:landau-wf}
\end{equation}
where $\ell_0=\sqrt{\frac{\hbar c}{e|B|}}$ is the magnetic length. In this gauge (and in complex coordinates) the angular momentum operator $L_z=-i\hbar (x_1 \partial_2-x_2\partial_1)=\hbar(z\partial_{z}-\bar z \partial_{\bar z})$. 

Any analytic function $f(z)$ is an eigenstate with energy $E_0=\frac{1}{2} \hbar \omega_c$. A complete basis of analytic functions are the monomials $f_n(z)=z^n$ and 
have energy $E_0$ and angular momentum $L_z=n \hbar$. This is the lowest Landau level whose wave functions are $\psi_n(z)=z^n \exp(-|z|^2/4\ell_0^2)$.  
On the other hand, an anti-analytic function $f_N=\bar z^N$ is an eigenstate of energy $E_N=\hbar \omega_c \left(N+\frac{1}{2}\right)$, where $\omega_c=\frac{e|B|}{Mc}$ 
is the cyclotron frequency, and angular momentum $L_z=-N \hbar$. States with angular momentum $n\hbar$  have the same energy and the degeneracy is equal to the number 
of flux quanta $N_\phi$. For the most part we will be interested in the states in the lowest Landau level. 

In the absence of disorder the Landau levels have an extensive degeneracy equal to the number of flux quantum $N_\phi$. If we consider a system of $N$ electrons 
in a Landau level the natural measure of density is not the areal density $\rho=\frac{N_e}{L^2}$ but the fraction $\nu=\frac{N_e}{N_\phi}$ the states in the Landau level which are occupied by 
electrons. The many-body state in which all $N_\phi$ states of the lowest Landau level has filling fraction $\nu=1$. The wave function for this state is the Slater determinant of the 
Landau states in the $m=0$ level. After some simple algebra the wave function of this state is found to be 
\begin{equation}
\Psi_{\nu=1}(z_1, \ldots, z_{N_e})= \prod_{i<j} (z_i-z_j) \; \exp \left(-\sum_{i=1}^{N_e} \frac{|z_i|^2}{4\ell_0^2}\right)
\label{eq:vandermonde}
\end{equation}
This is the ground state of the non-interacting system and it is non-degenerate. It is easy to see that this state has a finite energy gap. Indeed, in the Hilbert space with a 
fixed number of electrons, the lowest energy excitation is a particle-hole pair in which the particle is in an unoccupied state of the first excited Landau level, with $m=1$ 
and energy $E_1=\frac{3}{2} \hbar \omega_c$ and a hole a single-particle state in the lowest Landau level, with energy $E_0=\hbar \omega_c$. 
The excitation energy of the electron-hole pair is just $\hbar \omega_c$ which is finite. However, in the free particle system this excited states has a huge degeneracy as the 
particle and the hole can be in any single particle state of the Landau level.

At a very naive level one can estimate the Hall conductivity for a translationally invariant system filling up $n$ Landau level by the following simple argument. If $n$ Landau levels are  
filled, the number of electrons $N=nN_\phi$ and the total charge is then $Q=eN=e\; n \; N_\phi=e \; n \; BL^2/\phi_0=n e^2 BL^2/hc$. If a weak and uniform in-plane electric field 
$\bm E$ is applied  in the presence of a perpendicular magnetic field $\bm B=B\; \bm e_z$, then  the entire system (its center of mass) moves at the drift velocity $\bm \varv$ such that 
$\bm \varv \times \bm B=-\bm E c$. Thus, there is a Hall current $\bm J=Q \bm \varv$. The current density is $\bm j=\bm J/L^2=Q\bm \varv/L^2$. Hence,
$j_i=\frac{Qc}{BL^2} \; \epsilon_{ij}\; E_j$, which implies that the Hall conductivity is $\sigma_{xy}=Qc/BL^2=n e^2/h$.

While the above argument is formally correct and yields the correct value of the Hall conductivity in this highly idealized setting, at a very basic level it is faulty. 
To begin with, it assumes exact translational invariance and, hence, Galilean symmetry, which are not obeyed in any realistic system. 
The second objection is that, even in this idealized system, as the number of electrons is varied, the chemical potential (and hence the Fermi energy) 
jumps discontinuously from one Landau level to the next resulting in a linearly increasing Hall conductivity (as predicted classically) without any of the observed plateaus. 
Furthermore, this argument does not explain which the Hall conductivity is so precisely quantized whereas, a priori, one would expect that being a transport coefficient the 
Hall conductivity would depend on lots of complicated materials details. But, it it turns out that it does not!

Let us first discuss the observed {\it universality} of the Hall conductivity, namely its robustness and independence of microscopic details. In a remarkable paper 
Qian Niu, David Thouless and Yong-Shi Wu showed that, provided the ground state is separated by a finite energy gap from the excited states, 
the Hall conductivity is actually a topological invariant \cite{niu-1985}. The argument has a similarity with what we discussed in the case of the Hall effect on a 2D lattice 
(in section \ref{sec:TKNN}) but in this more general system one considers the full many-body wave function, rather than the one-particle states. 
To show that this true they considered a system of $N$ electrons with periodic boundary conditions, i.e. on a two-dimensional torus,  $T^2=S_1 \times S_1$, with $N_\phi$ 
flux quanta going through the torus. The torus is a rectangle of dimensions $L_1$ and $L_2$ with opposite ends identified. This is necessary to allow for a Hall current to be 
globally allowed. The electromagnetic gauge field for a system ona  torus with uniform non-vanishing flux cannot obey periodic boundary conditions but rather accross the torus 
the gauge fields must differ by (large) gauge transformations
\begin{equation}
A_1(x_1, x_2+L_2)=A_1(x_1, x_2)+\partial_1 \beta_2(x_1, x_2), \qquad A_2(x_1+L_1, x_2)=A_2(x_1, x_2)+\partial_2 \beta_1(x_1, x_2)
\label{eq:twisted-torus-A}
\end{equation}
while the wave functions themselves obey twisted boundary conditions
\begin{align}
\Psi([\bm x^{(j)}+L_1 \bm e_1])=&\exp\left(-i\frac{e}{\hbar c} \sum_{j=1}^{N_e} \beta_1([\bm x^{(j)}])+i \theta_1\right) \times \Psi([\bm x^{(j)}])
\nonumber\\
\Psi([\bm x^{(j)}+L_2 \bm e_2])=&\exp\left(-i\frac{e}{\hbar c} \sum_{j=1}^{N_e} \beta_2([\bm x^{(j)}])+i \theta_2\right) \times \Psi([\bm x^{(j)}])
\label{eq:twisted-torus-Psi}
\end{align}
where $\bm e_j$ (with $j=1,2,$) are two unit vectors on the torus, and  $\theta_1$ and $\theta_2$ are two  angles that {\it twist} the boundary conditions of the wave functions. 

 In order to have a current on the torus they assumed that, in addition to the uniform flux going through the torus,  there  a weak  uniform electric field $\bm E$ on the torus 
 represented by the constant in space vector potential $\delta \bm A=\bm E \; t=\bm \bigtriangledown [U(\bm x) t]$ whose circulation on the two non-contractible circles 
 $\Gamma_1$ and $\Gamma_2$, which wrap around the directions $x_1$ and $x_2$ of the torus $T^2$, are 
\begin{equation}
I_j=\oint_{\Gamma_j} \delta \bm A\cdot d\bm x=t \oint_{\Gamma_j}\bm E \cdot d\bm x=i t E_j L_j
\label{eq:twists}
\end{equation}
Line integrals of a gauge field on non-contractible loops in space (or space-time) are called {\it holonomies}.  By inspection we see that the angles $\theta_1$ and $\theta_2$ 
are given by
\begin{equation}
\theta_j=\frac{e}{\hbar c} I_j
\label{eq:twist-angles}
\end{equation}
Alternatively, the angles $\bm \theta=(\theta_1, \theta_2)$ can be interpreted as magnetic fluxes (in units of the flux quantum $\phi_0$) through the two non-contractible circles of the torus $T^2$.

The angles $\bm \theta$ are defined mod $2\pi$ since they are phase factors that twist the phase of the wave functions, and define a 2-torus of boundary conditions. By comparing with what we did in section \ref{sec:TKNN} in the case of the single-particle wave functions in the lattice model, we see that the {\it many-body wave function} $\Psi_{\bm \theta}$ and the twist angles $\bm \theta$ is the same as the relation between the phase of the Hofstadter wave functions with the momentum $\bm k$ of the magnetic BZ (which is also a 2-torus). Indeed, as it is always the case, while the phase of (in this case) the many-body wave function $\Psi_{\bm \theta}$ is arbitrary, the changes of this phase as the the twist angle $\bm \theta$ is changed are not. To quantify this dependence, for a given many-body state $\Psi_{\bm \theta}^{(\alpha)}$, where $\alpha$ labels the state, we define a Berry connection $\bm {\mathcal{A}}^{(\alpha)}(\bm \theta)$ on the torus of boundary conditions  $\bm \theta$ 
\begin{equation}
\bm {\mathcal{A}}^{(\alpha)}(\bm \theta)=i \Big< \Psi^{(\alpha)}_{\bm \theta}\Big| \partial_{\bm \theta}\Big|\Psi^{(\alpha)}_{\bm \theta} \Big>
\label{eq:QTW-Berry}
\end{equation}
Under a redefinition of the phase of the state 
\begin{equation}
\Psi_{\bm \theta}^{(\alpha)}\to \exp(i f(\bm \theta))\; \Psi_{\bm \theta}^{(\alpha)}
\label{eq:phase-transformation}
\end{equation}
the Berry connection $\bm {\mathcal{A}}^{(\alpha)}(\bm \theta)$  changes by a gauge transformation
\begin{equation}
\bm {\mathcal{A}}^{(\alpha)}(\theta) \to \bm {\mathcal{A}}^{(\alpha)} (\theta) -\partial_{\bm \theta} f(\bm \theta)
\label{eq:Berry-gauge-transf}
\end{equation}

Niu, Thouless and Wu \cite{niu-1985} showed that the expression of the Kubo formula for the Hall conductivity of the state $\Psi_{\bm \theta}^{(\alpha)}$, averaged over the boundary conditions $\bm \theta$, is given in terms of the flux of the Berry connection $\bm {\mathcal{A}}^{(\alpha)}(\bm \theta)$ through the 2-torus of boundary conditions, by the gauge-invariant quantity
\begin{equation}
\langle (\sigma_{xy})_\alpha \rangle_{\bm \theta}=\frac{e^2}{\hbar} \int_0^{2\pi }\frac{d\theta_1}{2\pi} \int_0^{2\pi}\frac{d\theta_2}{2\pi} 
\left(\partial_1 \mathcal{A}^{(\alpha)}_2-\partial_2 \mathcal{A}^{(\alpha)}_1\right)
=\frac{e^2}{h} \frac{1}{2\pi} \oint_{\gamma} \bm {\mathcal A}^{(\alpha)}(\bm \theta) \cdot d\bm \theta
\label{eq:NTW}
\end{equation}
where $\gamma$ is the square contour with corners at $(0,0)$, $(2\pi,0)$, $(0,2\pi)$ and $(2\pi, 2\pi)$, that defines the 2-torus of boundary conditions $\bm \theta$. 
This result is known as the Niu-Thouless-Wu formula. 

Eq.\eqref{eq:NTW} implies that for the Hall conductivity to be non-vanishing the Berry connection $\bm {\mathcal{A}}^{(\alpha)}(\bm \theta)$ must have a non-vanishing flux through the torus of boundary conditions. For this to be true, just as we did in section \ref{sec:chern}, we must consider large gauge transformations of the form
\begin{align}
\mathcal{A}_1(\bm \theta+2\pi \bm e_2)=&\mathcal{A}_1(\bm \theta)+\partial_1 f_2(\bm \theta)\nonumber\\
\mathcal{A}_2(\bm \theta+2\pi \bm e_1)=&\mathcal{A}_2(\bm \theta)+\partial_2  f_1(\bm \theta)\nonumber\\
\Psi^{(\alpha)}(\{ \bm x^{(j)} \}; \bm \theta+2\pi \; \bm e_1)=&\exp(i f_1(\bm \theta))\; \Psi^{(\alpha)}(\{ \bm x^{(j)} \}; \bm \theta)\nonumber\\
\Psi^{(\alpha)}(\{ \bm x^{(j)} \}; \bm \theta+2\pi \; \bm e_2)=&\exp(i f_2(\bm \theta))\; \Psi^{(\alpha)}(\{ \bm x^{(j)} \}; \bm \theta)
\label{eq:doubly-twisted}
\end{align}
where $\bm e_1$ and $\bm e_2$ are two orthogonal unit vectors on the torus of boundary conditions. We can now repeat the analysis we did in section \ref{sec:chern} which, in this context, implies that the wave function $\Psi^{(\alpha)}_{\bm \theta}$ cannot be globally well defined on the torus of boundary conditions, where it must have zeros, and where it should be defined in patches. The end result is that the wave functions labeled by $\alpha$ are classified by a {\it topological invariant}, the first-valued  Chern number $C_1^{(\alpha)}$, which here it is given by
\begin{equation}
C_1^{(\alpha)}=\frac{1}{2\pi} \oint_\gamma \mathcal{A}^{(\alpha)}(\bm \theta) \cdot d \bm \theta
\label{eq:Chern-number}
\end{equation}
and, hence, that the Hall conductivity in he state $\alpha$ (averaged over the twisted boundary conditions) exhibits the integer quantum Hall effect,
\begin{equation}
\langle (\sigma_{xy})^{(\alpha)}\rangle_{\bm \theta}=C_1 \frac{e^2}{h}
\label{eq:topo-invariance-QH}
\end{equation}
Expressing the Hal conductivity in terms of the first Chern number, which is a topological invariant, proves that the value of the conductivity cannot be changed by local physics effects, such as disorder, etc. The topological nature of the Hall conductivity is the reason for the robustness and high precision of the measured value of the Hall conductivity. 
In subsequent work Niu and Thouless showed that, provided the state has a finite energy gap to all excitations, in the thermodynamic limit the Hall conductivity averaged over boundary conditions and for a given boundary condition are equal. 

The result of Eq.\eqref{eq:topo-invariance-QH} seemingly implies that there can only be an {\it integer} quantum Hall effect which, as we see shortly, it is not the case: 
there is a fractional quantum Hall effect in which the Hall conductivity is a {\it fraction}, $\sigma_{xy}=\frac{p}{q} \frac{e^2}{h}$, where $p$ and $q$ are co-prime integers. 
This value  of the Hall conductivity is also found experimentally to be obeyed with the same precision as in the integer quantum Hall effect. 
In other words, the Hall conductivity in the fractional case must also have a topological character. To understand why this can be the case we need an implicit assumption that we made in our derivation. 
In fact, in deriving the result of Eq.\eqref{eq:topo-invariance-QH} we assumed (implicitly) that for each value of $\bm \theta$ there is only one many body state $\Psi_{\bm \theta}^{(\alpha)}$ 
(up to gauge transformations). As we will see below, in the {\it fractional} quantum Hall effect on a torus (and, in fact on any closed surface except the sphere)  the ground state 
is degenerate in the thermodynamic limit. We will also see that traversing the torus of boundary conditions once maps one degenerate state to another one.  
In general, we will find that  on the 2-torus in the thermodynamic limit there are $m \in \mathbb{Z}$ exactly degenerate states. 
In this case, one returns to the original state after sweeping $m$ times the torus of boundary conditions. 
We will also see below that the degenerate states are labeled by quantum numbers related to the {\it anyons} that they support.

While the topological argument proves the robustness (and universality) of the value of the Hall conductivity, it does not provide an insight for why there are plateaus in its magnetic field dependence. 
If for some value of the filling fraction $\nu=N/N_\phi$ the electron fluid is exactly at a ground state $\Psi^{(\alpha)}$ upon changing either the number of particles $N$ or the magnetic field $B$ (but not both) 
the fluid will now will be at the state $\Psi^{(\alpha)}$ plus or minus some number of electrons. 
The existence of the plateau in $\sigma_{xy}$ can be understood if the extra electrons (or holes) do not contribute to the Hall conductivity. 
This happens in the presence of disorder as these extra particles will be localized by the disorder and localized states, whose wave functions decay exponentially are insensitive to boundary conditions and, 
hence, these states do not contribute to the conductivity. 
This argument was first put forth by Laughlin \cite{laughlin-1981} who build on the result that in a disordered system in two dimensions all states are localized \cite{abrahams-1979} 
but made the key assumption that the state at the center of the Landau level (broadened by disorder) is extended and contributes to the conductivity if this state is filled. 
This picture implies that this is actually a quantum phase transition from an insulator (dubbed a Hall insulator) to a state with a quantized Hall conductivity. 

This intuitive and appealing proposal has been the focus of research for many years and it is largely an unsolved problem. 
There is in fact a proposed field theory for this quantum phase transition based on a a non-linear sigma model (in replica space) with a topological $\theta$-term \cite{levine-1983,pruisken-1984}, 
analogous to the theory we discussed in section \ref{sec:quantum-spin-chains} for quantum antiferromagnets in 1+1 dimensions. in this proposal the (Boltzmann) longitudinal conductivity 
$\sigma_{xx}$ plays the role of the inverse of the coupling constant of the non-linear sigma model while $\sigma_{xy}$ plays the role of the coefficient of the $\theta$-term. 
Although the RG flows suggested by this approach qualitatively agree with currently existing experimental data, and actual  derivation is still wanting. 

However, aside from the fact that is is still an unsolved problem, there is the problem that this explanation ignores the fact that in a Landau level interactions are dominant and are the key to the understanding 
of the fractional case even in the same samples! A symptom of this problem is that, at least in the cleaner samples, as we noted above the longitudinal conductivity vanishes exponentially with temperature. 
This dependence means that there is a clean energy gap (and not just a mobility gap) which suggests that the state with additional excitations may have some sort of at least local order which would provide 
for a local energy gap. An actual theory based on this picture has yet to be developed.

\subsubsection{The Laughlin Wave Function}
\label{sec:laughlin}

We will turn now to the {\it fractional} quantum Hall effect (FQHE).  The FQHE is observed in fractionally filled Landau levels. These states have fractional filling fractions, namely that 
$\nu=\frac{N_e}{N_\phi}$ is a fraction. Most of the observed states are in the lowest Landau level, with $N=0$, but a few  states with very intriguing properties are seen in the 
first excited Landau Level with $N=1$. 

Let us consider for now the states in the lowest Landau level, $N=0$. We will assume that there is no disorder and, for now, we will ignore the effects of boundaries of the sample. 
Physically, the important interaction here is the Coulomb interaction although, in some samples with screening layers, short-range interactions may be important as well. 
Since there is a large magnetic field, the Zeeman interaction is typically a large energy scale and the electros are expected to be fully polarized. 
However, there are situations under which the gyromagnetic factor can be made small (and even zero) and spin unpolarized states have to be considered. 
In some platforms, such as graphene, additional quantum numbers are present, such as ``flavor'' associated with the different Dirac states. 
In our discussion we will not cover these richer possibilities. 

Under these circumstances we have a problem in which all $N_\phi$ available one-particle states are exactly degenerate (the Landau level is ``flat'') but only a fraction of them  are filled.  
In this limit, this system as as strongly interaction as it gets. Not surprisingly, in a system of this type time-honored standard approximations fail, such as Hartree-Fock.  
Given the macroscopic degeneracy and the nature of the states in a magnetic field, a system of this type may harbor many different types of actual ground states depending 
of the types of interactions that are considered.

Tsui, St\"ormer and Gossard \cite{tsui-1982} did transport experiments in the two-dimensional electron gas (2DEG) in high-purity GaAs-AlAs heterostructures and reported the 
discovery of a state with a highly precise value of the Hall conductivity of $\sigma_{xy}=\frac{1}{3} \frac{e^2}{h}$. Such a state cannot be understood in any obvious way
 in terms of the integer quantum Hall effect. The observation of a clean gap in the low temperature longitudinal resistivity $\sigma_{xx} \to 0$ as $T \to 0$ implied that the 
 2DEG is incompressible and quite likely uniform. 
 
 The first breakthrough was Laughlin's unique insight which led to his explanation of the experiments in terms of a novel 
 quantum liquid of fully spin polarized electrons at filling fractions $\nu=\frac{1}{m}$ (with $m$ and odd integer) whose wave function he proposed to be 
\begin{equation}
\Psi_m(z_1,\ldots,z_N)=\prod_{i<j} (z_i-z_j)^m \; \times \; \exp\left(-\sum_{i=1}^{N_e} \frac{|z_i|^2}{4 \ell_0^2}\right)
\label{eq:laughlin}
\end{equation}
where $N=N_e$ is the number of electrons, $\{ z_i \}$ are their (complex) coordinates and $m$ is an odd integer (which makes the wave function antisymmetric,
as required by Fermi statistics).

Laughlin extracted the physics encoded in this wave function by considering the probability $\Big| \Psi_m(z_1,\ldots,z_N)\Big|^2$ of finding the $N$ electrons in the locations $\{ z_i \}$ (with $i=1,\ldots,N$). 
The norm of the Laughlin wave function is
\begin{equation}
\Big|\Big|\Psi_m\Big|\Big|^2=\int d^2z_1 \ldots d^2z_N \Big|\Psi_m(z_1,\ldots,z_N)\Big|^2
\label{eq:norm-laughlin}
\end{equation}
The Laughlin wave function $\Psi_m(z_1,\ldots,z_N)$  is an eigenstate of angular momentum with eigenvalue $L_m=\frac{1}{2} mN(N-1)$. This remarkable wave function has a large overlap 
($\sim 98\% $)  with the exact wave function with $V(R)=\frac{e^2}{\varepsilon R}$ (Coulomb) interactions in systems of up to eight particles ($\varepsilon$ is the dielectric constant).

The norm of the state can ten be interpreted as the classical partition function of a system of a system of $N$ particles  at the locations $\{ z_i \}$ at inverse temperature $\beta=m$  with the effective interaction 
\begin{equation}
U(z_1,\ldots,z_N)=-2 \sum_{1\leq i <j\leq N} \ln|z_1-z_j|+\frac{1}{2m\ell_0^2} \sum_{j=1}^N |z_j|^2
\label{eq:one-component-plasma}
\end{equation}
This classical partition function is a one-component plasma, of a gas of particles with unit (negative) charge  interacting with each other with the repulsive  
2D classical (logarithmic!) Coulomb interaction, $V_{\rm CG}=-\ln |z_i-z_j|$ (not $1/R$!). The last term represents the contribution of a neutralizing 
background positive charge (which here it originates in the gaussian factor of the wave function). In other words, this is a one-component plasma. We can check that this is correct since 
$\bigtriangledown^2\left(\frac{|z|^2}{2m\ell_0^2}\right)=\frac{2}{m\ell_0^2}$ which corresponds to a uniform (areal) charge density 
$\rho_0=\frac{1}{2\pi m \ell_0^2}$.
Classical Monte Carlo simulations of this probability distribution showed that this wave function describes an incompressible fluid  state if $m \lesssim 7$, 
while for larger values it represents a crystalline state. This approach is known as the plasma analogy.

\subsubsection{Quasiholes have fractional charge}
\label{sec:quasihole-charge}

Laughlin considered a vortex-like excitation in the fluid created by the adiabatic insertion of an infinitesimally thin solenoid at some coordinate $z_0$ carrying a magnetic flux $\Phi$, a {\it fluxoid}. 
In the presence of this solenoid the single particle state $z^n \exp(|z|^2/4\ell_0^2)$ acquires a branch cut and becomes the state $z^{n+\alpha}  \exp(|z|^2/4\ell_0^2)$ where $\alpha=\Phi/\phi_0$, 
with $\phi_0=hc/e$ being the flux quantum. This analytic structure is the Aharonov-Bohm effect \cite{aharonov-1959}.

However, if the solenoid carries just one  flux quantum $\alpha=1$ and the net effect is that the state becomes $z^{n+1} \exp(-|z|^2/4\ell_0^2)$ which has one more unit of angular 
momentum. For these reasons, in the presence of a solenoid with one flux quantum inserted at $z_0$, the Laughlin wave function for the 2DEG in a large magnetic field now becomes
\begin{equation}
\Psi_m^{qh}(z_0;z_1,\ldots,z_N)=\prod_{i=1}^N (z_i-z_0) \; \times \; \Psi_m(z_1,\ldots,z_N)
\label{eq:laughlin-qh}
\end{equation}
A calculation using the plasma analogy reveals that the net effect of the solenoid is to expel a charge equal to $-e/m$ from the vicinity of the location of the solenoid or, 
equivalently, to create a {\it fractionally charged quasihole} at $z_0$ with charge $Q_{\rm qh}=+e/m$. The charge distribution is uniform at long distances but is depleted (hence a quasihole) on a length scale 
$\xi$. One can check that this state costs a finite energy $\varepsilon_0$ which depends on the particular interaction.

On the other hand, if we consider a system in which the solenoid is inserted adiabatically so that after a long time $t$ there is a quasihole at $z_0$, an amount of charge equal to $-e/m$ 
will flow radially outwards to the outer edge of the 2DEG. The adiabatic insertion of the solenoid generates a azimuthal electric field (an e.m.f.) and  a radial Hall current, 
and a Hall conductivity $\sigma_{xy}=\frac{1}{m} \frac{e^2}{h}$. 
Hence, the Laughlin wave function exhibits the fractional quantum Hall effect and its excitations are vortices with fractional charge $e/m$. 

It is important to note the non-local nature of the Laughlin quasihole. The non-locality is inherited from the fluxoid (the solenoid) inserted at $z_0$: even though the magnetic field of the fluxoid 
vanishes away from it, $B(\bm x)=\Phi \; \delta^2(\bm x-\bm x_0)$, its vector potential $\bm A(\bm x)$ does not since its circulation on any non-contractible loop $\gamma$ that contains the location 
of the fluxoid inside must be equal to the flux $\Phi$ carried by the fluxoid. Although in principle one can choose a singular gauge in which $\bm A(\bm x)$ vanishes locally, this cannot be true 
everywhere as it must leave behind  a Dirac-like string on a curve $\Gamma$ ranging form the location $z_0$ of the fluxoid to the boundary of the system with $\bm A$ taking a singular value on 
$\Gamma$. This feature has the same form as the 2D vortex that we discussed in section \ref{sec:2d-vortices}. Kivelson and Ro\u{c}ek \cite{kivelson-1985} showed that the existence of this 
Dirac string causes the {\it phase} of a quantum state that carries charge $e^*$ to have a branch cut on the curve $\Gamma$ and to change by $\exp(i 2\pi (e^*/e) \Phi/\phi_0)$, 
as required by the Aharonov-Bohm effect \cite{aharonov-1959}.  This effect is unobservable for integer-charged states, i.e. electrons. 

\subsubsection{The Jain States}
\label{eq:jain}

The construction of the Laughlin quasihole motivated Jainendra Jain \cite{jain-1989,jain-1989b} to rewrite the Laughlin wave function in the form
\begin{equation}
\Psi_m(z_1,\ldots,z_N)=\prod_{i<j} (z_i-z_j)^{m-1} \times \; \Psi_{\nu=1}(z_1,\ldots,z_N)
\label{eq:jain1}
\end{equation}
Here the prefactor represents particles each attached with an even number, $m-1$, of flux quanta. The second factor,$ \Psi_{\nu=1}(z_1,\ldots,z_N)$, 
is the wave function of fermions filling up a lowest Landau level, i.e. the Vandermonde determinant of Eq.\eqref{eq:vandermonde}. In this form, the inverse of the filling factor, $1/\nu$, 
which is the number of flux quanta per particle, is written as 
$\frac{1}{\nu}=(m-1)+1=m$. The interpretation now is that the $(m-1)\phi_0$ magnetic flux carried by each  particle, which partially screens the uniform field down to an effective field 
$B_{\rm eff}=B-(m-1)\phi_0$ corresponding to one effective flux quantum per particle, i.e. a filled lowest Landau level of the effective field $B_{\rm eff}$. 
In other words, the wave function for the $\nu=1/m$ 
FQH state is reinterpreted as the $\nu_{\rm eff}=1$ {\it integer} QH state of $N$ {\it composite fermions}, each being an electron carrying attached an even integer $m-1$ flux quanta. 

This procedure is known as flux attachment. Physically it means that the strong interactions between the electrons for other electrons to be further away which, 
in virtue of being in a strong magnetic field, is equivalent to an increase of their relative angular momentum, as if there was a local change in the magnetic field. 
There is no reason to believe that the composite fermions are weakly coupled, even though this is frequently stated in the literature without any real supporting evidence.

Jain then generalized this reinterpretation of the Laughlin state to a whole sequence of FQH states obtained by filling $p$ levels of these partially screened magnetic fluxes
\begin{equation}
\Psi_{m, p}(z_1,\ldots,z_N)=P_{LLL} \prod_{i<j} (z_i-z_j)^{m-1}  \Psi_{\nu=p}(z_1,\ldots,z_N)
\label{eq:jain-seq}
\end{equation}
where $\Psi_{\nu=p}(z_1,\ldots,z_N)$ is the wave function of $p$ filled Landau levels (of composite fermions), and $P_{LL}$ is an operator that projects onto the lowest Landau level. By counting fluxes we see that the filling fractions for the Jain states satisfy
\begin{equation}
\frac{1}{\nu(m, p)}=m-1\pm \frac{1}{p}
\label{eq:jain2}
\end{equation}
where we allowed for the possibility that the external fluxes may be over-screened by the composite fermions. Alternatively we can write
\begin{equation}
\nu(m, p)=\frac{p}{p(m-1)+\pm1}
\label{eq:jain-fractions}
\end{equation}
The Jain states with $p=1$ are the laughlin states and each is called the primary state of a Jain sequence.
Almost all the observed FQH states belong to one of these sequences (and its generalizations). In particular, the most prominent states (i.e. those with larger plateaus) belong to the sequence
 $\frac{1}{3}, \frac{2}{5}, \frac{3}{7}, \frac{4}{9}, \frac{5}{11}, \ldots$ and to the reversed sequence $1, \frac{2}{3}, \frac{3}{5}, \frac{4}{7}, \ldots$. For  $p \to \infty$, the Jain sequences converge to the values 
of the filling fraction $\lim_{p \to \infty}\nu(m, p)=\frac{1}{m-1}$. In this limit, the $m-1$ fluxes attached to each particle exactly cancels th external flux leading us to a (possibly) 
Fermi liquid of composite fermions \cite{halperin-1993}.

\subsubsection{Quasiholes have fractional statistics}
\label{sec:fractional-statistics-qh}

The non-locality and topological nature of the Laughlin quasihole implies that this state must be regarded as a soliton (or vortex) of the 
charged quantum fluid. The vortex-like state with wave function $\Psi_m^{qh}(z_0;z_1,\ldots,z_N)$ is called the Laughlin quasihole. This state represents a composite object of a flux quantum and a 
fractional charge. This structure of this quantum state is strongly reminiscent of the flux-charge composite objects considered by Frank Wilczek 
who showed that such objects should be {\it anyons}, states that exhibit {\it fractional statistics} \cite{wilczek-1982b}. Although at a conceptual level anyons were proposed 
(almost) prior to the discovery of the FQHE, it is in this setting that fractional statistics entered actual physics. The actual experimental observation took work by many people, 
and was only confirmed in 2020 \cite{nakamura-2020}.

Fractional statistics dictates the analytic form of the wave functions of two or more quasiholes \cite{halperin-1984}. Let us consider a laughlin-type wave function for two quasiholes located at (complex) coordinates $u$ and $\varw$. Naively we expect the wave function for two quasiholes in the $\nu=1/m$ Laughlin state to have the form
\begin{equation}
\Psi(u, \varw; z_1, \ldots,z_N)= N(u, \varw)\; \prod_{j=1}^N (z_j-u)(z_j-\varw) \; \Psi_m(z_1,\ldots,z_N)
\label{eq:2qh-wf}
\end{equation}
The prefactor $N(u, \varw)$ must be chosen to account for the fact that theres is an additional change of angular momentum due to the additional fluxoid at $\varw$. 
We also expect that as the quasihole with charge $e/m$ is adiabatically carried around the the other quasihole (also with charge $e/m$) there will be ac accrued phase in the wave 
function due to the Aharonov-Bohm effect of the charge of one quasihole circling around the flux of the other (or, equivalently, crossing the branch cut) \cite{kivelson-1985}. The requirement of translation invariance and analyticity are met by the choice \cite{halperin-1984} (ignoring a multiplicative normalization constant)
\begin{equation}
N(u, \varw)= (u-\varw)^{1/m} \exp\left(-\frac{1}{4\ell_0^2 m} (|u|^2+|\varw|^2)\right)
\label{eq:halperin84}
\end{equation}
which has a branch cut stretching from $u$ to $\varw$. A calculation using the plasma analogy represents this state as a set of $N$ charge $-1$ particles interacting with two additional particles of charge 
$-1/m$ at $u$ and $\varw$ (and a neutralizing background). The branch cut implies that dragging a quasihole around the other during a $\pi$ rotation followed by  a translation (i.e. an exchange), 
 induces a monodromy   
in the wave with a jump in its phase of $\exp(\pm i \pi/m)$ (with the sign determined by the orientation of the monodromy) in such a way the that wave function changes by a phase factor
\begin{equation}
\Psi(u, \varw; z_1, \ldots,z_N) \mapsto e^{\pm i \frac{\pi}{m}}\; \Psi(u, \varw; z_1, \ldots,z_N)
\label{eq:frac-stat-2qh}
\end{equation}
In other words, the quasihole is an {\it anyon} with fractional statistics $\pi/m$.
Daniel Arovas, J. Robert Schrieffer and Frank Wilczek  \cite{arovas-1984} further refined this argument by showing that under such a slow adiabatic change the wave function of two quasiholes acquires a Berry phase which accounts for the fractional statistics. An explicit path-integral derivation of this effect, which uses the fact that the quasihole states are coherent states can be found in Ref. \cite{fradkin-2013}, chapter 13.

\subsubsection{Hydrodynamic Effective Field Theory}
\label{sec:QFT-FQH-hydro}

We will now turn to the approaches to the FQHE using the methods of quantum field theory. As we will see the concept of flux attachment plays a key role. In section \ref{sec:CS-quantization} we showed that Chern-Simons gauge theory is in fact a theory of flux attachment. Thus we expect that Chern-Simons gauge theory should play a key role as well. 

It is useful to ask first why should Chern-Simons gauge theory play a central role at least in the description of the low energy physics. 
There is a simple, yet powerful, phenomenological argument due to J\"org Fr\"ohlich and Anthony Zee that shows why this should be the case \cite{frohlich-1991b}. 
This in essence a hydrodynamic argument. The fractional quantum Hall effect occurs in a 2DEG in the presence of a large magnetic field which breaks explicitly time reversal invariance. 
In the regime of the FQHE the 2DEG is a uniform charged incompressible fluid in which charge is conserved. This means that the charge 3-current $j_\mu=(j_0, \bm j)$ 
(where $j_0$ is the charge density and $\bm j$ is the charge current density) must be locally conserved, 
\begin{equation}
\partial_\mu j^\mu=0
\label{eq:conservation-charge-current}
\end{equation}
which is the continuity equation. The solution of this equation is that the conserved current can be written in terms of a {\it dual} (in the Hodge sense)  vector field $\mathscr{B}_\mu$such that
\begin{equation}
j_\mu=\frac{1}{2\pi} \epsilon_{\mu \nu \lambda} \partial^\nu \mathscr{B}^\lambda
\label{eq:current-gauge-duality}
\end{equation}
The factor of $\frac{1}{2\pi}$ is introduced for later convenience. 
The current field $j_\mu$ is not changed by a smooth redefinition of the vector field $\mathscr{B}_\mu \to \mathscr{B}_\mu +\partial_\mu \Phi$, where $\Phi(x)$ is a non-singular field. 
This means that the vector field $\mathscr{B}_\mu$ is a gauge field.

The effective action of this theory is a functional of the current distribution $\j_\mu$ and, hence, of the gauge field $\mathscr{B}_\mu$. 
It should be gauge-invariant, and odd under parity and time-reversal. 
Since the fluid is incompressible and uniform, at long distances and low energies the action must be a local functional of the gauge field which should be at least Galilean invariant.
 In addition, the coupling of the fluid to an external electromagnetic probe field $A_\mu$ must be of the usual form $-e  j^\mu A_\mu$. 
 To lowest orders in derivatives, there the unique local and gauge-invariant Lagrangian density which is odd under time reversal (and parity) is the Chern-Simons theory
\begin{equation}
\mathcal{L}_{\rm eff}[\mathscr{B}_\mu]=\frac{m}{4\pi} \epsilon_{\mu\nu\lambda} \mathscr{B}^\mu \partial^\nu \mathscr{B}^\lambda-\frac{1}{4g^2} \mathscr{F}_{\mu \nu}^2 
-\frac{e}{2\pi} A^\mu \epsilon_{\mu \nu \lambda} \partial^\nu \mathscr{B}^\lambda+\ldots
\label{eq:eff-action-Ascr}
\end{equation}
The first term is the Chern-Simons term of the Lagrangian. The coefficient $m$ is a dimensionless integer which, as we saw in section \ref{sec:CS} is required for the theory to be invariant on a closed manifold.
The second is a Maxwell term. Here $\mathscr{F}_{\mu \nu}=\partial_\mu \mathscr{B}_\nu-\partial_\nu \mathscr{A}_\mu$ is the field strength tensor of the gauge field $\mathscr{B}_\mu$. 
By power counting we see that the coupling constant $g^2$ has units of length$^{-1}$ and, hence this term is irrelevant at long distances and will be neglected.
The third term is the coupling of the current $j_\mu$ to the electromagnetic field $A_\mu$. Upon an integration by parts this term can be written as $\mathscr{B}_\mu \mathscr{J}^\mu$ where 
\begin{equation}
\mathscr{J}_\mu=-\frac{e}{2\pi} \epsilon_{\mu \nu\lambda} \partial^\nu A^\lambda
\label{eq:Jscr}
\end{equation}
is a current minimally coupled to $\mathscr{B}_\mu$.

The equation of motion of the (dynamical) gauge field $\mathscr{B}_\mu$ is
\begin{equation}
\frac{\delta \mathcal{L}}{\delta \mathscr{B}}=0
\label{eq:EOM}
\end{equation}
which yields the relation
\begin{equation}
\frac{m}{2\pi} \epsilon_{\mu\nu\lambda} \mathscr{B}^\mu \partial^\nu \mathscr{B}^\lambda=\mathscr{J}_\mu
\label{eq:EOM-CS-Ascr}
\end{equation}
From the definition of the current $\mathscr{J}_\mu$, Eq.\eqref{eq:Jscr}, we see that the solution of the equation of motion of Eq.\eqref{eq:EOM-CS-Ascr}, up to a gauge transformation, is
\begin{equation}
m \mathscr{B}_\mu=-e A_\mu
\label{eq:sol-EOM}
\end{equation}
By plugging this relation back into the Lagrangian for the gauge field $\mathscr{B}_\mu$, Eq \eqref{eq:eff-action-Ascr} (which si equivalent to integrate out the gauge field $\mathscr{B}_\mu$) we find that the effective Lagrangian of the electromagnetic field $A_\mu$ is just a Chern-Simons term
\begin{equation}
\mathcal{L}_{\rm eff}[A_\mu]=\frac{e^2}{4\pi m} \epsilon_{\mu\nu\lambda} A^\mu \partial^\nu A^\lambda+\ldots
\label{eq:eff-L-A}
\end{equation}
This result allows us to read-off the Hall conductivity of the fluid
\begin{equation}
\sigma_{xy}=\frac{1}{m} \frac{e^2}{2\pi \hbar}
\label{eq:fqh-hydro}
\end{equation}
where we restored units such that $\hbar$ is not unity. In other words, this fluid exhibits the fractional quantum Hall effect for a fluid at filling fraction $\nu=1/m$. 

We should note that this heuristic argument applies equally to systems of fermions, for which $m$ is odd,  as well as to systems of bosons, for which $m$ is even.

\subsubsection{Composite Boson Field Theory }
\label{sec:CB-QFT-FQH}

We will now discuss tow field-theoretic approaches to the FQHE. 
These approaches use the concept of flux attachment as a mapping of fermions to bosons and as a mapping of fermions to fermions. 
These approaches are a form of duality transformation which engineers a form of statistical transmutation. 
We will first show that Chern-Simons theory can be used to effect such a mapping. 
That this is possible should not be surprising since, as we saw in section \ref{sec:braids}, Chen-Simons theory is a theory of fractional statistics and, hence, of anyons.

We should make some important comments on both approaches before we get into a detailed description.  While the mapping of fermions to composite bosons and fermions to composite bosons is correct, their approximate mean field descriptions violate symmetries of the 2DEG in a magnetic field. At the root level is the fact that the flux attachment is local in space time and approximate descriptions of these theories bring about a large amount of Landau level mixing, even in the large field limit. For instance, at the mean field theory level the composite boson approach is equivalent to a Bose-Einstein condensate which is invariant under translations and under  time reversal. In this picture the breaking of time reversal is present in the coupling to a Chern-Simons gauge field. Likewise, the mean field theory of the composite fermion theory is a problem of composite fermions in a partially screened. While a partially screened magnetic field still breaks time reversal, it does not behave properly under magnetic translations. As we will see, these problems will be solved, at least in the low energy regime, by quantum fluctuations which restore the symmetries. In both cases, in the fractional states one recovers an effective topological field theory which captures the universal physics of these states. Needless to say it, both approaches do poorly in the computation of non-universal dimensionful quantities such as energy gaps, etc. These difficulties become very severe in the compressible states whose low energy behavior is not topological.

Unlike the wave function approaches which project these states into a specific Landau level (usually the lowest), the field theoretic approach does not effect such projection. Several papers have been written attempting to do a field theory projected into a Landau level. These approaches transformed the problem into quantum field theory on a non-commutative plane, which is inherent the nature of the states in a magnetic field. Although some significant progress has been made in this direction, these theories remain  poorly understood \cite{pasquier-1998,read-1998,susskind-2001,polychronakos-2001,fradkin-2002,dong-2020,goldman-2022}

In this section we will focus on the composite boson theory.
Let us consider a theory in 2+1 dimensions with two dynamical abelian gauge fields $\mathscr{A}_\mu$ and $\mathscr{B}_\mu$, whose Lagrangian is a sum of a Chern-Simons term at level $k$ and a $BF$ term:
\begin{equation}
\mathcal{L}=\frac{k}{4\pi} \epsilon_{\mu\nu\lambda} \mathscr{A}^\mu \partial^\nu \mathscr{A}^\lambda+\frac{1}{2\pi} \epsilon_{\mu\nu\lambda} \mathscr{A}^\mu \partial^\nu \mathscr{B}^\lambda
\label{eq:L-A+B}
\end{equation}
The equation of motion for the field $\mathscr{A}_\mu$ is
\begin{equation}
\frac{k}{2\pi} \epsilon_{\mu\nu\lambda}  \partial^\nu \mathscr{A}^\lambda+ \frac{1}{2\pi} \epsilon_{\mu\nu\lambda}  \partial^\nu \mathscr{B}^\lambda=0
\label{eq:eom-A+B}
\end{equation}
whose solution (up to a gauge transformation) is 
$k \mathscr{A}_\mu=-\mathscr{B}_\mu$.
Plugging this relation into the Lagrangian of Eq.\eqref{eq:L-A+B} we find that the effective Lagrangian for the field $\mathscr{B}_\mu$ is
\begin{equation}
\mathcal{L}[\mathscr{B}_\mu]=-\frac{1}{4\pi k} \epsilon_{\mu\nu\lambda} \mathscr{B}^\mu \partial^\nu \mathscr{B}^\lambda
\label{eq:A-B-duality}
\end{equation}
which states that exchanging $\mathscr{A}_\mu \leftrightarrow \mathscr{B}_\mu$ is equivalent to the ``duality'' $k \leftrightarrow -1/k$.

In section \ref{sec:braids} we showed that particles coupled to a Chern simons gauge field at level $k$ have fractional statistics $\exp(\pm i \pi /k)$. 
Hence, we see that particles coupled to the field $\mathscr{B}_\mu$ have fractional statistics $\exp(\pm i \pi k)$. In other words, for $k$ odd this coupling amounts to exchanging a theory of fermions 
to a theory of bosons coupled to the gauge field $\mathscr{B}_\mu$. This is a form of bosonization. 
Alternatively, for $k$ even it maps a theory of fermions to another theory of fermions coupled to a gauge field $\mathscr{B}_\mu$. 
These observations have led to distinct, but ultimately  equivalent description of the quantum Hall effect. 

We will begin the approach of mapping fermions to bosons and use it in the problem of the FQHE. This is the Landau-Ginzburg theory (or {\it composite boson theory}) of Shoucheng Zhang, Hans Hansson and Steven Kivelson \cite{zhang-1989} (ZHK). In this approach the problem of fermions coupled to a magnetic field interacting with each other through a two-body potential (that we will assume is ultra  with coupling constant $\lambda$, for simplicity) becomes the same problem but for a theory of bosons which are also coupled to a Chern-Simons gauge field, which we will denote by $\mathscr{A}_\mu$. The Lagrangian density for this equivalent system of {\it composite bosons} is 
\begin{equation}
\mathcal{L}_{\rm CB}= \phi^*(x) [i D_0 +\mu] \phi(x) +\frac{1}{ 2 M}|{\bm D}\phi(x)|^2 - \lambda (|\phi(z)|^4+\frac{1 }{ 4\pi m}
 \epsilon_{\mu \nu \lambda}\mathscr{A}^{\mu} \partial^\nu  \mathscr{A}^{\lambda} 
\label{eq:action-second-quantized}
\end{equation}
where $x=(x_0,\bm x)$ are the space-time coordinates, $\mu$ is the chemical potential, $M$ is the mass of the fermions, and $m$ is an arbitrary odd integer. The covariant derivative 
\begin{equation}
D_\mu=\partial_\mu+i \frac{e}{\hbar c} A_\mu+i \mathscr{A}_\mu
\label{eq:covariant-derivative-bosons}
\end{equation}
which effects the minimal coupling of the complex scalar field $\phi(x)$ to the background electromagnetic field $A_\mu$  and to the Chern-Simons gauge field $\mathscr{A}_\mu$, known in this context os known as the statistical gauge field. 

ZHK showed that the FQH state can be thought as being closely related to to a phase in which the complex scalar field condenses, much as in the case of a superfluid even though  the FQH i{\it is not} a superfluid. To see how this works we will write
\begin{equation}
\phi(x)=\sqrt{\rho(x)} \; \exp(i \omega(x))
\label{eq:BEC}
\end{equation}
The classical equations of motion of the theory of composite bosons, Eq.\eqref{eq:action-second-quantized}, are
\begin{align}
\frac{\delta \mathcal{L}_{\rm CB}}{\delta \phi^*(x)}=&0 \qquad &\Rightarrow \qquad & (i D_0+\mu)\phi(x)-\frac{1}{2M} \bm D^2\phi(x)-2\lambda |\phi(x)|^2 \phi(x)=0
\label{eq:CEM-CB1}\\
\frac{\mathcal{L}_{\rm CB}}{\delta \mathscr{A}_0(x)}=&0 \qquad &\Rightarrow \qquad & \frac{1}{2\pi m} \epsilon_{ij} \partial_i \mathscr{A}_j+|\phi(x)|^2=0 
\label{eq:CEM-CB2}\\
\frac{\mathcal{L}_{\rm CB}}{\delta \mathscr{A}_i(x)}=&0 \qquad &\Rightarrow \qquad & \frac{1}{2\pi m} \epsilon_{i\alpha \beta} \partial^\alpha \mathscr{A}^\beta+\frac{i}{2M}\left[\phi^*(x) D_i \phi(x)-(D_i \phi(x))^*\phi(x)\right]=0 
\label{eq:CEM-CB3}\\
\frac{\mathcal{L}_{\rm CB}}{\delta \mu}=&\rho_0 L^2T \qquad &\Rightarrow \qquad & \int d^3x \; |\phi(x)|^2=\rho_0 L^2T
\label{eq:CEM-CB4}
\end{align}
where $\rho_0$ is the areal density of electrons.

A uniform solution of Eqs. \eqref{eq:CEM-CB1}-\eqref{eq:CEM-CB4} with constant amplitude of the composite bosons, i.e. a uniform and static condensate $\bar \phi$, with uniform statistical gauge field strength $\bar {\mathscr{B}}$, 
requires that the there should be no current in the ground state and that each term of Eq.\eqref{eq:CEM-CB3} should be zero. This condition implies that the external field is canceled by the average statistical field, $\frac{eB}{\hbar c} +\bar {\mathscr{B}}=0$. This condition together with Eqs.\eqref{eq:CEM-CB2} and \eqref{eq:CEM-CB4} imply that 
\begin{equation}
\rho_0=\frac{1}{m} \frac{eB}{\hbar c}=\frac{1/m}{2\pi \ell_0^2} 
\label{eq:filling-fraction-CB}
\end{equation}
and we conclude that the filling fraction is
$\nu=\frac{1}{m}$,
which are the Laughlin states.
In addition we get that the chemical potential is $\mu=2\lambda \rho_0$ and that $|\bar \phi |^2=\rho_0$, with $\rho_0$ given in Eq.\eqref{eq:filling-fraction-CB}.

However, these mean field results seeming  imply that this system is a Bose-Einstein condensate.  To understand why this is incorrect, and to determine what it actually is, we need to go beyond the mean field theor that we just described and evaluate the effects of quantum fluctuations and write
\begin{equation}
\phi(x)=\left(\rho_0+\delta \rho(x)\right)^{1/2} \; \exp(i\omega(x)), \qquad \frac{eA_\mu}{\hbar c} +\mathscr{A}_\mu=\delta \mathscr{A}_\mu
\label{eq:fluctuations-CB}
\end{equation}
The partition function of this theory is a path integral over the fluctuating density fields $\delta \rho$, the Goldstone field $\omega$ and the fluctuation of the Chern-Simons gauge field 
$\delta \mathscr{A}_\mu$. Upon integrating out the massive density fluctuations to lowest (quadratic) order we find that the effective Lagrangian for $\omega$ and $\delta \mathscr{A}_\mu$ is
\begin{equation}
\mathcal{L}_{\rm eff}[\omega,\delta \mathscr{A}_\mu]= \frac{\kappa}{2} \left(\partial_0 \omega- \delta \mathscr{A}_0\right)^2
-\frac{\rho_s}{2} \left(\partial_i \omega - \delta \mathscr{A}_i \right)^2
+\frac{1 }{ 4\pi m} \epsilon_{\mu \nu \lambda}\delta \mathscr{A}^{\mu} \partial^\nu  \delta \mathscr{A}^{\lambda} +\ldots
\label{eq:eff-L-CB}
\end{equation}
The first two terms of Eq.\eqref{eq:eff-L-CB} is the effective Lagrangian of the Goldstone mode $\omega$ in the Bogoliubov theory of superfluidity a compressibility $\kappa$ and a superfluid stiffness $\rho_s$ given by
\begin{equation}
\kappa=\frac{1}{2\lambda}, \qquad 
\rho_s=\frac{\rho_0}{M}=\frac{\nu}{2\pi} \hbar \omega_c
\label{eq:L-CB-coeffs}
\end{equation}
However, as we see, this is not a superfluid since the phase field is ``eaten'' by the Chern-Simons gauge field $\delta \mathscr{A}_\mu$ which now acquires a mass term. In this sort of Higgs mechanism the fluctuating gauge field has a massive longitudinal mode and a massive transverse mode. Thus, this state doe not have any massless modes.

To see that this theory describes the fractional quantum Hall effect we need to compute the linear response to a weak electromagnetic perturbation $\delta A_\mu(x)$. This can be accomplished by considering the new Lagrangian
\begin{equation}
\mathcal{L}_{\rm eff}[\omega,\delta \mathscr{A}_\mu, \delta A_\mu]= \frac{\kappa}{2} \left(\partial_0 \omega- \delta \mathscr{A}_0-\frac{e}{\hbar c} \delta A_0\right)^2
-\frac{\rho_s}{2} \left(\partial_i \omega - \delta \mathscr{A}_i -\frac{e}{\hbar c} \delta A_i\right)^2
+\frac{1 }{ 4\pi m} \epsilon_{\mu \nu \lambda}\delta \mathscr{A}^{\mu} \partial^\nu  \delta \mathscr{A}^{\lambda} +\ldots
\label{eq:eff-L-CB-deltaA}
\end{equation}
and integrate out the phase field $\omega$ (which can be set to zero in the London/unitary gauge) and the fluctuation of the statistical field $\delta \mathscr{A}_\mu$ to find that effective electromagnetic action is
\begin{equation}
S_{\rm eff}[\delta A_\mu]=\frac{1}{4\pi m}\frac{e^2}{\hbar}  \int d^3x \; \epsilon_{\mu \nu \lambda}\delta {A}^{\mu} \partial^\nu  \delta {A}^{\lambda} +\ldots
\label{eq:eff-action-em-CB}
\end{equation}
from which we conclude that the Hall conductivity predicted by the composite boson theory is 
\begin{equation}
\sigma_{xy}=\frac{1}{m} \frac{e^2}{h}
\label{eq:FQHE-CB}
\end{equation}
as it should be for a FQHE at filling fraction $\nu=1/m$.

We conclude by looking at vortex states in the composite boson theory. A vortex state is a time-independent solution of the equations of motion Eqs.\eqref{eq:CEM-CB1}-\eqref{eq:CEM-CB4} with the asymptotic behavior (in the temporal gauge $\delta \mathscr{A}_0=0$)
\begin{align}
\lim_{|\bm x|\to \infty} \phi(\bm x)=&\sqrt{\rho_0}\; e^{i \varphi(\bm x)}\\
\lim_{|\bm x|\to \infty} \delta \mathscr{A}_i(\bm x)=&\pm \partial_i \varphi(\bm x)=\pm \epsilon_{ij} \frac{x_j}{|\bm x|^2}
\label{eq:vortex-CB}
\end{align}
where $\varphi(\bm x)$ is the azimuthal angle on the plane
\begin{equation}
\varphi(\bm x)=\tan^{-1}\left(\frac{x_2}{x_1}\right)
\label{eq:azimuthal}
\end{equation}
The energy of a neutral vortex (i.e. not coupled to a gauge field) is logarithmically divergent, $E_{\rm vortex}\simeq \frac{\rho_s}{2} \ln (R/a_0)$ where $R$ is the linear size of the system and $a_0$ is a short-distance cutoff (see section \ref{sec:2d-vortices}). However, the situation here is different since the complex scalar field $\phi(x)$ is coupled to a dynamical Chern-Simons gauge field $\mathscr{A}_\mu$. Except for the Chern-Simons nature of this gauge field, the problem we are dealing with is similar to that of a superconductor coupled to a dynamical gauge field or to the Abelian-Higgs model in quantum field theory. In the case of interest here we have finite energy vortex solutions which satisfy the asymptotic condition
\begin{equation}
\lim_{|\bm x|\to \infty} \Big| \Big(i \partial_j -\delta \mathscr{A}_j\Big)\phi(\bm x)\Big|^2=0
\label{eq:asymptotic-condition}
\end{equation}
which is obeyed by configurations that obey Eq.\eqref{eq:vortex-CB}.  Thus, at long distances, the circulation of the Chern-Simons gauge field on a large closed contour $\Gamma$ that contains the vortex satisfies
\begin{equation}
\oint_\Gamma \delta \mathscr{A}_j dx_j=\pm 2\pi
\label{eq:winding}
\end{equation}
This vortex carries charge. To see this we compute the local charge density $j_0(\bm x)$
\begin{equation}
j_0(\bm x)=- \frac{\delta S_{\rm eff}}{\delta A_0(x)}=-e \frac{\delta S_{\rm eff}}{\delta \delta \mathscr{A}_0(x)}=+e \frac{\delta S_{\rm CS}}{\delta \delta \mathscr{A}_0(x)}=\frac{e}{2\pi m} \epsilon_{ij} \partial_i \delta \mathscr{A}_j(x)
\label{eq:vortex-charge-density}
\end{equation}
and compute the total charge of the vortex on a larger region $\Sigma$ whose boundary is $\Gamma$ and obtain
\begin{equation}
Q=e\int d^2x j_0(\bm x)=\frac{e}{2\pi m} \int_\Sigma d^2x\; \epsilon_{ij} \partial_i \delta \mathscr{A}_j(\bm x)=\frac{e}{m} \frac{1}{2\pi} \oint_\Gamma dx_j \delta \mathscr{A}_j(x)=\pm \frac{e}{m}
\label{eq:vortex-charge}
\end{equation}
Therefore, the vortex of the composite boson theory has the same charge as the Laughlin quasihole.

To determine the statistics of the vortex we go back to the effective Lagrangian written in the form of Eq.\eqref{eq:L-A+B}, (from now on we set $\delta \mathscr{A}^\mu \equiv \mathscr{A}^\mu$)
\begin{equation}
\mathcal{L}_{\rm eff}=\frac{\kappa}{2} (\partial_0 \omega -\mathscr{A}_0)^2-\frac{\rho_s}{2} (\partial_j \omega-\mathscr{A}_j)^2 
+\frac{1}{2\pi} \epsilon_{\mu \nu \lambda} \mathscr{A}^\mu \partial^\nu  \mathscr{B}^\lambda 
-\frac{e}{2\pi}  \epsilon_{\mu \nu \lambda} {A}^\mu \partial^\nu  \mathscr{B}^\lambda 
-\frac{m}{4\pi}  \epsilon_{\mu \nu \lambda} \mathscr{B}^\mu \partial^\nu  \mathscr{B}^\lambda
\label{eq:vortex-action-CB1}
\end{equation}
where $A_\mu$ is the external electromagnetic probe field, and where we used units with $\hbar=c=1$. The first two terms make the statistical gauge field $\mathscr{A}_\mu$ massive. In the low energy limit the statistical gauge field is frozen to the vortex configurations of the complex scalar field $\phi$ or, equivalently, to the vortex singularities of its phase field $\omega$.
The vorticity current $\Omega_\mu$ is
\begin{equation}
\Omega_\mu=\epsilon_{\mu \nu \lambda} \partial^\nu \mathscr{A}^\lambda
\label{eq:vortex-current}
\end{equation}
The effective Lagrangian for the field $\mathscr{B}_\mu$ is
\begin{equation}
\mathcal{L}_{\rm eff} [\mathscr{B}_\mu] = -\frac{m}{4\pi}  \epsilon_{\mu \nu \lambda} \mathscr{B}^\mu \partial^\nu   \mathscr{B}^\lambda  
-\frac{e}{2\pi} \; A^\mu \epsilon_{\mu \nu \lambda} \partial^\nu  \mathscr{B}^\lambda + \Omega_\mu \mathscr{B}^\mu
\label{eq:FQH-TQFT}
\end{equation}
which, except for the coupling to the electromagnetic field, has the same form as in Eq.\eqref{eq:CS-U1}, and of the effective action introduced in section \ref{sec:QFT-FQH-hydro} on phenomenological grounds.

The effective topological field theory of Eq.\eqref{eq:FQH-TQFT} encodes al the {\it universal} data of fractional quantum Hall fluids. 
Being topological this effective field theory has no energy scales and describes the physics at energies low compared to any excitation energy gap.  
In addition to yielding the correct Hall conductivity $\sigma_{xy}=\nu e^2/h$ (with $\nu=1/m$) and the fractional vortex charge $Q=e/m$, 
this theory will allow us to draw  important additional results about the low energy physics. 

By  using the results of section \ref{sec:braids}, we  conclude 
that the vortices of the composite boson are anyons with fractional statistics $\exp(\pm i \pi /m)$, consistent with the conclusions of section \ref{sec:fractional-statistics-qh}. 
In addition, on a 2-torus the ground state of this system is $m$-fold degenerate. 
This feature, characteristic of systems with {\it topological order}, is that the ground state degeneracy depends on the topology of the surface on which the 2DEG resides. 
The ground state  on a torus degeneracy was shown by Duncan Haldane and Edward Rezayi \cite{haldane-1985b} by an explicit construction of a model wave function for the Laughlin states on a torus. 
On a surface with $g$ handles (i.e. genus $g$) the degeneracy is $m^g$. 
Xiao-Gang Wen and Qian Niu \cite{wen-1990} showed that these results hold for any FQH state which are, quite generally, topological quantum fluids.  

Finally, we may ask how many {\it distinct} types of vortices does this theory have. The fundamental (Laughlin) vortex has charge $e/m$ and statistics $\pi/m$. 
In principle we could have a vortex with any  integer topological charge (winding number) $n \in \mathbb{Z}$. Such a vortex has charge $ne/m$ and statistics $\pi n^2/m$. 
However, a vortex with topological charge $n=m$ has charge $e$ and statistics $m \pi$. This vortex is indistinguishable from a hole (a missing electron) since it has the same charge and statistics. 
We conclude that a Laughlin state has $m$ distinct vortices with $n=1,2, \ldots, m-1$. We notice that this number is the as the number of degenerate states on a torus. 
In section \ref{sec:FQH+CFT} we will see  that this fact is closely related to the number of primary fields of a conformal field theory associated with the FQH states.

\subsubsection{Composite Fermion Field Theory}
\label{sec:CF-QFT-FQH}

At the beginning of section \ref{sec:CB-QFT-FQH} we noted that flux attachment can be used to define equivalent theories: a) by attaching an {\it odd} number, $m$, fluxes to each electron thereby 
becoming a composite boson, and b)  by attaching  an {\it even} number, $m-1$, fluxes by which the electrons turn into {\it composite fermions}, as in Jain's construction \cite{jain-1989}. 
In this section we  discuss the most salient features of the field theory approach to composite fermions, introduced by Ana L\'opez and me in 1991\cite{lopez-1991,lopez-1993} 
as a theory of all the Jain states. This approach was extended in 1993  by Bertrand Halperin, Patrick Lee and Nicholas Read \cite{halperin-1993} to the case of the compressible states.

By following the same approach that we used in section \ref{sec:CB-QFT-FQH}, but adapted to the case in which we map to a theory of a composite Fermi field $\psi(x)$, 
we find their dynamics is described by the effective action \cite{lopez-1991}
\begin{align}
\mathcal{S}_{\rm CF} = &\int  d^3 x \; \left\{ \psi^*(x) [i D_0 +\mu] \psi(x) +\frac{1}{2 M} |\bm D \psi(x)|^2 \right\}
+ \frac{1}{4\pi n}  \int d^3x \epsilon_{\mu \nu \lambda} \mathscr{A}^{\mu} \partial^{\nu} \mathscr{A}^\lambda 
\nonumber\\ 
&- \frac{1}{2} \int  d^3x \int d^3x' \;( |\psi(x)|^2-\rho_0) V(x-x') (|\psi(x')|^2-\rho_0)
\label{eq:eff-action+fermions+CS1}
\end{align}
where $n=m-1$ is an even integer. Here $V(x-x')=\delta(x_0-x'_0) V(|\bm x - \bm x'|)$ is the instantaneous electron-electron repulsive interaction, 
and $\rho_0$ is the average areal neutralizing  charge density.
As before, the covariant derivative is $D_\mu=\partial_\mu+i e A_\mu+\mathscr{A}_\mu$, where $A_\mu$ is the electromagnetic field (including the uniform magnetic field $B$), 
and $\mathscr{A}_\mu$ is the statistical gauge field. We are using units such that $\hbar=c=1$.

We can simplify somewhat the form of the actions for the composite fermions by using the fact that the Gauss law of  Chern-Simons gauge theory states that the particle density 
and the gauge flux are rigidly tied together, $|\psi(x)|^2=\frac{1}{2\pi n} \mathscr{B}\equiv \epsilon_{ij} \partial_i \mathscr{A}_j$ (this is the flux attachment) as a operator identity. 
Thus, we can write the action as
\begin{align}
\mathcal{S}_{\rm CF}=&\int  d^3 x \;
 \left\{
\psi^*(x) [i D_0 +\mu] \psi(x) +\frac{1}{2M} |\bm D \psi(x)|^2 \right\}
+ \frac{1}{4\pi n}  \int d^3x \; \epsilon_{\mu \nu \lambda} \mathscr{A}^{\mu} \partial^{\nu } \mathscr{A}^\lambda 
\nonumber\\ 
&- \frac{1}{2} \int  d^3x \int d^3x' \; \left(\frac{\mathscr{B}(x)}{2\pi n}-\rho_0)\right) V(x-x') \left(\frac{\mathscr{B}(x')}{2\pi n}-\rho_0 \right)
\label{eq:eff-action+fermions+CS2}
\end{align}
Since this action is a quadratic form in the Fermi (Grassmann) fields $\psi$, we can integrate them out to obtain the following  effective action for the statistical gauge field $\mathscr{A}_\mu$
\begin{equation}
S_{\rm eff}[\mathscr{A}_\mu]=-i \textrm{tr} \left( iD_0+\mu+\frac{\bm D^2}{2M}\right)+\frac{1}{4\pi n} \int d^3x\, \epsilon_{\mu \nu \lambda} \;  (\mathscr{A}^{\mu}-eA^\mu)  
\partial^{\nu} \; (\mathscr{A}^\lambda -eA^\lambda)+ S_{\rm int}[\mathscr{A}_\mu-eA_\mu]
\label{eq:eff-action-Ascr-CF}
\end{equation}
where $A_\mu$ is a probe external electromagnetic field (with vanishing average) and does not include the uniform magnetic field. The interaction term is
\begin{equation}
S_{\rm int}[\mathscr{A}_\mu-eA_\mu]=- \frac{1}{2} \int  d^3x \int d^3x' \; \left(\frac{(\mathscr{B}(x)-eB(x))}{2\pi n}-\rho_0)\right)\; V(x-x')\; \left(\frac{(\mathscr{B}(x')-eB(x'))}{2\pi n}-\rho_0 \right)
\label{eq:Sint-CF}
\end{equation}
Here too $B(x)$ is an external probe with vanishing average.

We will investigate the properties of the path integral for the statistical gauge field 
\begin{equation}
Z[A_\mu]=\int \mathcal{D} \mathscr{A}_\mu \; \exp(i S_{\rm eff}[\mathscr{A}_\mu, A_\mu])
\end{equation}
using a saddle point expansion. The saddle-point condition of the effective action of Eq.\eqref{eq:eff-action-Ascr-CF}
\begin{equation}
\frac{\delta S_{\rm eff}}{\delta \mathscr{A}_\mu(x)}=0
\label{eq:spe}
\end{equation}
leads to the equation of motion for  the field $\mathscr{A}_\mu$
\begin{equation}
\langle j_\mu^F(x)\rangle+\frac{1}{4\pi n}\epsilon_{\mu \nu \lambda}\left[\mathscr{F}^{\nu \lambda}(x)-e F^{\nu \lambda}\right]=0
\label{eq:EON-Ascr-CF}
\end{equation}
where $\langle j_\mu^F\rangle$ is the expectation value of the fermionic current. In addition we need to impose the condition for the particle density to be uniform and equal 
to the neutralizing background charge $\langle j_0^F(x)\rangle=\rho_0$. 

We will assume that the electromagnetic field $A_\mu$ describes just a static uniform magnetic field of strength $B$. 
Under these conditions, the ground state should also be static and uniform. This means that the field strength of the statistical gauge field should 
be constant  and uniform value $\langle \mathscr{B} \rangle$, and that the current $\langle {\bm j}^F\rangle$ should vanish in the ground state. 
On the other hand, the Gauss Law of the Chern-Simons gauge field implies that 
$\langle \mathscr{B} \rangle=-2\pi n \; \rho_0$,
while the zero current condition requires the the (statistical) electric field vanishes, $\langle \mathscr{E}\rangle=0$. Since the composite fermion couples in the same way to the 
electromagnetic field $A_\mu$ and to the statistical field $\mathscr{A}_\mu$ we conclude that they experience an effective magnetic field 
\begin{equation}
B_{\rm eff}=B+\frac{1}{e} \langle \mathscr{B}\rangle=B- 2\pi n \; \frac{\rho_0}{e}
\label{eq:effective-field-CF}
\end{equation}
If the total number of electrons is $N$, then $\rho_0/e=N/L^2$ where $L$ is the linear size of the system. Let us denote by $N_\phi^{\rm eff}$
 the total number of flux quanta of the effective magnetic field  (in units of $\hbar=c=1$ in which the flux quantum is $\phi_0=2\pi$), and $2\pi N_\phi=BL^2$ is the total flux. Then, the total effective flux is
\begin{equation}
2\pi N_\phi^{\rm eff}=2\pi N_\phi- 2\pi n \; N
\label{eq:effective-flux-CF}
\end{equation}
Let $\nu=N/N_\phi$ be the filling fraction and $\nu^{\rm eff}=N/N_\phi^{\rm eff}$ the effective filling fraction. Eq.\eqref{eq:effective-flux-CF} implies that these filling fractions are related by
\begin{equation}
\frac{1}{\nu^{\rm eff}}=\frac{1}{\nu}-n
\label{eq:nu-eff-nu}
\end{equation}
Recall that $n$ is an {\it even} integer. However, the system will be incompressible only if $\nu^{\rm eff}=p \in \mathbb{Z}$. 
In other words, the composite fermions fill an integer number $p$ of the partially screened Landau levels, 
which is Jain's condition. For this condition to be satisfied the filling fraction $\nu(p, n)$ is
\begin{equation}
\nu_\pm(n, p)=\frac{p}{np\pm 1}
\label{eq:Jian-fractions}
\end{equation}
which are the Jain fractions. In the special case $p=1$ the Jain fractions are the 
Laughlin fractions, with $n=m-1$. 
Similarly we find that $B_{\rm eff}$ is
\begin{equation}
B_{\rm eff}=\pm \frac{B}{np\pm 1}
\label{eq:Jian-eff-B}
\end{equation}
So we see that the external field $B$ is partially screened to the smaller value $B_{\rm eff}$ which  can  be parallel (screened) to $B$ or anti-parallel (overscreened) to $B$. 
For the same reason, at this mean field level the cyclotron frequency $\omega_c$ is reduced by the same amount to $\omega_c^{\rm eff}=\omega_c/(np\pm 1)$.

We will discuss now the effects of  quantum fluctuations for the action of Eq.\eqref{eq:eff-action-Ascr-CF} about the saddle-point solutions. 
We will work to the lowest order (quadratic) in the fluctuations, which can be regarded as a semi-classical (or ``RPA'') treatment of this theory. 
The quadratic effective action for the fluctuations of statistical gauge field, which we will still denote by $\mathscr{A}_\mu$ is
\begin{align}
S^{(2)}_{\rm eff}[\mathscr{A}_\mu]=&\frac{1}{2} \int d^3x \int d^3y \; \mathscr{A}^\mu(x) \Pi^{CF}_{\mu \nu}(x, y) \; \mathscr{A}^\nu(y)\nonumber\\
&+\frac{1}{4\pi n} \int d^3x \; \epsilon_{\mu \nu \lambda} (\mathscr{A}_\mu(x)-eA_\mu(x)) \partial^\nu (\mathscr{A}_\nu(x)-eA_\nu(x))+S_{\rm int}(\mathscr{A}_\mu-eA_\mu)
\label{eq:quadratic-fluctuations}
\end{align}
Here, $\Pi_F^{\mu \nu}(x, y)$ is the polarization tensor of the composite fermions for $p$ filled effective Landau levels. The interaction term of the action is given in Eq.\eqref{eq:Sint-CF}.

As required by gauge invariance, the composite fermion polarization tensor is transverse, $\partial^\mu \Pi^{CF}_{\mu \nu}=0$, which fixes the tensorial structure. 
In momentum and frequency space the composite fermion polarization tensor $\Pi^{CF}_{\mu \nu}(\bm q, \omega)$ depends on three kernels 
$\Pi^{CF}_0(\bm q, \omega)$, $\Pi_1^{CF}(\bm q, \omega)$ and $\Pi_2^{CF}(\bm q, \omega)$, 
where $\Pi^{CF}_0$ and $\Pi^{CF}_2$ contain the parity-even response of the composite fermions and $\Pi_1^{CF}$  is their parity-odd response. 
Each kernel  is given as a series terms representing particle-hole processes with simple poles at the excitations energies 
$\omega_{rs}=(r-s)\omega_c^{\rm eff}$, with $r>p$ (particles) and $s \leq p$ (holes). Each term has a residue which is an integer power of $\bm q^2$ times  a 
Laguerre polynomial in $\bm q^2$. 
Details of these kernels can be found in Ref.\cite{lopez-1991} and in chapter 13 of 
Ref.\cite{fradkin-2013}. However, what will be important here is that, in the gapped states these kernels have a low energy and low momentum limit which is local. 
This will enable us to find a local topological effective action only for the gapped states.

After integrating out the statistical gauge field $\mathscr{A}_\mu$ we find the effective action for the external electromagnetic probe field $A_\mu$, which has the standard form
\begin{equation}
S_{\rm eff}[A_\mu]=\frac{1}{2} \int d^3x \;\int d^3y\;  A_\mu(x)\; \Pi^{\mu \nu}(x-y)\; A_\nu(y)
\label{eq:full-em-response}
\end{equation}
which encodes the full electromagnetic response of the system, not just of the composite fermions. Once again, gauge invariance requires that the full polarization tensor be transverse, 
$\partial^\mu \Pi_{\mu \nu}=0$. 

In section \ref{sec:TKNN} we showed that there is a relation between the polarization tensor $\Pi_{\mu \nu}$ and the current-current correlation function $\mathcal{D}_{\mu \nu}$. 
There we mentioned the these correlators are required (by gauge invariance) to satisfy a Ward identity known as the f-sum rule:
\begin{equation}
\int_{-\infty}^\infty \frac{d\omega}{2\pi}\; i \omega \; \mathcal{D}^R_{00}(\omega, \bm q)=\frac{\rho_0}{M} \bm q^2
\label{eq:sum-rule}
\end{equation}
where the (retarded) density-density correlation function is related to the (retarded) polarization tensor, $\mathcal{D}^R_{00}(\omega, \bm q)=\Pi_{00}^R(\omega, \bm q)$.
In the limit of low momentum $\bm q$, at fixed frequency $\omega$, The expression for $\Pi_{00}^R(\omega, \bm q)$ is
\begin{equation}
\Pi_{00}^R(\omega, \bm q)\simeq \bm q^2 \Pi_0^R(\omega, \bm q)=-\frac{\rho_0}{M}\; \frac{\bm q^2}{(\omega+i\epsilon)^2-\omega_c^2}
\label{eq:Pi00-low-q}
\end{equation}
where $\omega_c=eB/(Mc)$ is the cyclotron frequency of the {\it electrons}. This expression saturates the sum rule. 
This means that whichever corrections Eq.\eqref{eq:Pi00-low-q} may have must vanish in the low $\bm q$ limit. 
In other words, in this limit the approximation that we made is actually exact. This is well known result in the theory of the Fermi liquid \cite{nozieres-1966}.

 In addition, the result of Eq.\eqref{eq:Pi00-low-q} implies that there is a particle-hole (density) collective mode which at low momentum has energy $\hbar \omega_c$, without corrections. 
 This result is known as Kohn's theorem \cite{kohn-1961} which states that in a Galilean invariant system the 2DEG must have an exact eigenstate at the cyclotron frequency. 
 More intuitively, the center of mass of the 2DEG executes a cyclotron motion regardless of whether the particles are free, interacting or not, fermions, bosons, or anyons. 
 This is a general result. Notice that the composite fermions do not satisfy Kohn's theorem as they cyclotron mode would be at the effective cyclotron frequency $\omega_c^{\rm eff}$, 
 and that the leading quantum fluctuations have restored the magnetic symmetry.

In the limit of low energy $\omega \to 0$ and low momentum  $\bm q \to 0$  of the kernels $\Pi_0^{CF}$, $\Pi_1^{CF}$ and $\Pi_2^{CF}$ become
\begin{equation}
\Pi_0^{CF}(0,0)=\frac{p}{2\pi} \frac{M}{B_{\rm eff}} \equiv \varepsilon^{CF}, \qquad 
\Pi_1^{CF}(0,0)=\pm \frac{p}{2\pi}\equiv \sigma_{xy}^{CF}, \qquad
\Pi_2^{CF}(0,0)= -\frac{1}{2\pi} \frac{p^2}{M}\equiv -\chi^{CF}
\label{eq:response-CF}
\end{equation}
where $\varepsilon^{CF}$ is the ``dielectric constant'' of the composite fermions, $\chi^{CF}$ is their effective ``permeability'', 
and $\sigma_{xy}^{CF}$ is the (integer) composite fermion Hall conductivity. We then find that  low-energy effective action of the statistical gauge field $\mathscr{A}_\mu$ is
\begin{align}
S_{\rm eff}[\mathscr{A}_\mu]=&\frac{1}{2} \left(\sigma_{xy}^{CF}+\frac{1}{2\pi n}\right) \int d^3x  \; \epsilon_{\mu \nu \lambda} \mathscr{A}^\mu(x)\; \partial^\nu \; \mathscr{A}^\lambda(x)\nonumber\\
-&\frac{e}{2\pi n} \int d^3x  \; \epsilon_{\mu \nu \lambda} \mathscr{A}^\mu(x)\; \partial^\nu \; A^\lambda(x)+\frac{e^2}{4\pi n}  \int d^3x  \; \epsilon_{\mu \nu \lambda} {A}^\mu(x)\; \partial^\nu \; {A}^\lambda(x)\nonumber\\
+&\int d^3x \frac{1}{2}\left(\varepsilon^{CF}\;{\mathscr{E}_i}^2(x)-\chi^{CF}\; \mathscr{B}^2(x)\right)\nonumber\\
-&\frac{1}{8\pi^2n^2} \int d^3x \int d^3y \left(\mathscr{B}(x)-eB(x))\; V(x-y)\; (\mathscr{B}(y)-eB(y)\right)
\label{eq:quadratic-effective-action-CF}
\end{align}
Notice that, except possibly for the last term, the effective action is a local. This is possible because the mean field theory state is gapped.

In what follows we will focus only on the leading terms of the effective action and neglect the subleading terms, the least two terms of the effective action which have the form of a 
Maxwell term and an additional (possibly non-local) term. The remaining leading terms have a Chern-Simons form and a BF form whose effective Lagrangian is
\begin{equation}
\mathcal{L}_{\rm eff}[\mathscr{A}_\mu]=\frac{1}{4\pi} \left(\pm p+\frac{1}{n}\right)  \; \epsilon_{\mu \nu \lambda} \mathscr{A}^\mu(x)\; \partial^\nu \; \mathscr{A}^\lambda(x)
-\frac{e}{2\pi n}  \; \epsilon_{\mu \nu \lambda} \mathscr{A}^\mu(x)\; \partial^\nu \; A^\lambda(x)+\frac{e^2}{4\pi n}    \; \epsilon_{\mu \nu \lambda} {A}^\mu(x)\; \partial^\nu \; {A}^\lambda(x)
\label{eq:eff-L-jain}
\end{equation}
If now now integrate ot the statistical gauge field $\mathscr{A}_\mu$ we find the the electromagnetic field $A_\mu$ has the effective Lagrangian
\begin{equation}
\mathcal{L}_{\rm eff} [A_\mu]=\frac{\sigma_{xy}}{2}\; \epsilon_{\mu \nu \lambda} {A}_\mu(x)\; \partial^\nu \; {A}_\nu(x)
\label{eq:eff-L-em-FQH}
\end{equation}
from where we  find that the  Hall conductivity $\sigma_{xy}$ {\it of the 2DEG} is
\begin{equation}
\sigma_{xy}=\frac{e^2}{2\pi \hbar} \frac{p}{pn\pm 1}
\label{eq:hall-jain}
\end{equation}
which is the Hall conductivity of the Jain states (in standard units). Here, as before, $n$ is an even integer.

We will now return to the effective Lagrangian of Eq.\eqref{eq:eff-L-jain}. As it stands, since the Chern-Simons term has fractional level, it is not invariant under large gauge transformations and, as such, cannot be defined on a closed manifold. To remedy this problem  we can now write this Lagrangian in terms of a theory with two dynamical gauge  fields $\mathscr{A}_\mu$ and $\mathscr{B}_\mu$ as follows:
\begin{align}
\mathcal{L}_{\rm eff}[\mathscr{A}_\mu,\mathscr{B}_\mu]=&\frac{p}{4\pi} \epsilon_{\mu \nu \lambda} \mathscr{A}^\mu(x)\; \partial^\nu \; \mathscr{A}^\lambda(x)-\frac{n}{4\pi} \epsilon_{\mu \nu \lambda} \mathscr{B}^\mu(x)\; \partial^\nu \; \mathscr{B}^\lambda(x)\nonumber\\
+&\frac{1}{2\pi} \epsilon_{\mu \nu \lambda} \mathscr{A}^\mu(x)\; \partial^\nu \; \mathscr{B}^\lambda(x)
-\frac{e}{2\pi} \epsilon_{\mu \nu \lambda} \mathscr{B}^\mu(x)\; \partial^\nu A^\lambda
\label{eq:jain-tqft}
\end{align}
This is a topological quantum field theory for all the Jain states \cite{lopez-1999}. This theory can be defined on any manifold no matter its topology. 
It is easy to see that upon integrating out the field $\mathscr{B}_\mu$ we recover the effective field theory for $\mathscr{A}_\mu$ of Eq.\eqref{eq:eff-L-jain}. 
Also, in the special case of the Laughlin states, for which $p=1$, we can integrate out the field $\mathscr{A}_\mu$, 
and arrive to the effective topological field theory for the field $\mathscr{B}_\mu$ of Eq.\eqref{eq:FQH-TQFT} that we derived in the composite boson theory in 
section \ref{sec:CB-QFT-FQH}.

Let us rewrite the Lagrangian of Eq.\eqref{eq:jain-tqft} in a more compact yet general form. 
To this end we define a multicomponent gauge field $\mathscr{A}_\mu^I$ with $I=1, \ldots, L$, an $L$-component  {\it charge} vector $\bm t$, and an $L\times L$ second rank symmetric matrix $K_{IJ}$. 
Wen and Zee have given a general classification of  a large class of fractional quantum Hall states, said to be {\it abelian}, whose topological field theory is defined  by the Lagrangian \cite{wen-1992,wen-1995}
\begin{equation}
\mathcal{L}=\frac{1}{4\pi} K_{IJ} \epsilon^{\mu \nu \lambda} \mathscr{A}^I_\mu(x)\; \partial^\nu \; \mathscr{A}^J_\lambda(x)-\frac{e}{2\pi} t_I \epsilon^{\mu \nu \lambda} {A}_\mu(x)\; \partial^\nu \; \mathscr{A}_I^\lambda(x)
\label{eq:K-matrix}
\end{equation}
In the case of the theory of Eq.\eqref{eq:jain-tqft} which has two components, $L=2$, with $\mathscr{A}_\mu^1=\mathscr{A}_\mu$ and $\mathscr{A}_\mu^2=\mathscr{B}_\mu$. The two-component vector is $\bm t=(0, 1)$ and the matrix  $2\times 2$ matrix is
\begin{equation}
K=
\begin{pmatrix}
-p & 1\\
1 & n
\end{pmatrix}
\label{eq:K-matrix-jain}
\end{equation}
Wen and Zee showed that, quite generally, a theory of the $K$-matrix form has a vacuum degeneracy on a torus of $|\textrm{det} K|$ and, on a  surface of genus $g$, the degeneracy is $|\textrm{det} K|^g$. They also showed that the Hall conductivity is $\sigma_{xy}=\nu\; e^2/h$ where the filling fraction is 
\begin{equation}
\nu=\sum_{I, J=1}^L t_I \; K^{-1}_{IJ} \; t_J
\label{eq:K-matrix-nu}
\end{equation}

The properties  of the quasiparticles can be determined by computing the appropriate correlator. In the composite fermion theory, whose Lagrangian is given in Eq.\eqref{eq:eff-action+fermions+CS2}, the {\it gauge-invariant} composite fermion propagator is
\begin{equation}
G_{CF}(x-y;\gamma(x, y))=\Big< \psi^\dagger(x) \exp(i \int_{\gamma(x, y)} dz_\mu (eA^\mu(z)+\mathscr{A}^\mu)) \psi(y)\Big>
\label{eq:Gcf-gauge-invariant}
\end{equation}
which, as in any gauge theory, it is gauge-invariant but path-dependent. 
Here $\gamma$ is an oriented open path with endpoints that the space-time locations $x$ and $y$. The composite fermion propagator has information about many of the properties of the quasiparticles, including their electric charge. To find what is the statistics of the quasiparticles we need to compute a two-particle correlator (a four-point function) which is defined in terms of two open oriented paths, say $\gamma_1(x, y)$ and $\gamma_2(u, w)$, with endpoints at the locations of the two particles, respectively. 

The computation of these correlators is done using a feynam sum over trajectories with different weights which depend on the gauge field configurations. To do these calculations is, in general, a complicated problem. However, in a gapped state, the quasiparticles are massive and in the low energy regime these expressions are dominated by a classical trajectory on an oriented path $\tilde \gamma(x, y)$ with the same endpoints $x$ and $y$. The end result reduces to the computation of the expectation value of a Wilson loop operator 
\begin{equation}
W[\Gamma]=\Big<\exp(i \oint_\Gamma \; dz_\mu(eA_\mu(z)+\mathscr{A}^\mu(z))\Big>_{\mathscr{A}^\mu}
\label{eq:wilson-loop-CF}
\end{equation}
on the {\it closed} oriented path $\Gamma=\gamma \bigcup \tilde \gamma^-$ (where $\tilde \gamma^-(y, x)$ has the opposite orientation as $\tilde \gamma(x, y)$ and runs from $y$ to $x$. The explicit dependence on the external electromagnetic field yields the value of the charge of the quasiparticle through the Aharanov-Bohm effect. Likewise, the two-particle correlator is reduced, in the asymptotic low energy regime, to the computation of an expectation value of two Wilson loops in the effective Chern-Simons theory, as was done in section \ref{sec:braids}, which yields the fractional statistics of the quasiparticles. The interested reader can find details in chapter 13 of Ref.\cite{fradkin-2013}.

A system described by a theory of the form of Eq.\eqref{eq:K-matrix} has different types of quasiparticles. In general we can assign an integer-valued quantum number $\ell_I\in \mathbb{Z}$ as the charge of the quasiparticle with respect to the gauge field $\mathscr{A}^\mu_I$. The general coupling of a quasiparticle current has the form $\mathcal{L}_{qp}=j_\mu \ell_I \mathscr{A}_I^\mu$. Then, the charge $Q[\bm \ell]$ and the statistics $\delta[\bm \ell]$ of a general quasiparticle defined by the integer-valued $L$-component vector $\bm \ell$ are
\begin{equation}
Q[\bm \ell]=-e t_I K^{-1}_{IJ} \ell_J, \qquad \frac{\delta[\bm \ell]}{\pi}= \ell_I K^{-1}_{IJ} \ell_J+1
\label{eq:qp-charge-statistics}
\end{equation}
where the $+1$ is required to refer the statistics to fermions (not bosons) \cite{lopez-1999}. These results are general and hold for any abelian FQH state \cite{wen-1992}.

In the case of the Jain states, the charge vector is $\bm t=(0,1)$ and the quasiparticle is labeled by the vector $\bm \ell=(1,0)$ which yield the results for the charge $Q$ and the statistics $\delta$
\begin{equation}
Q=\frac{-e}{np+1}, \qquad \delta=\pi \left(\frac{-n}{np+1}+1\right)
\label{eq:qp-qn-jain}
\end{equation}
which are the quantum numbers of the quasiparticles of the Jain states. For the Laughlin state at $\nu=1/3$ we obtains $Q=-e/3$ and $\delta=\pi/3$ (as we should), 
which for the first Jain state at $\nu=2/5$ the charge $Q$ and the statistics $\delta$ are  $Q=-e/5$ and $\delta=3\pi/5$. With some caveats, the predictions for the quasiparticle charge in the 
FQH states with $\nu=1/3$ and $\nu=2/5$ have been confirmed in noise measurements at a quantum point contact by R. de Picciotto and coworkers \cite{de-picciotto-1997} 
and by  L. Samindayar and coworkers \cite{samindayar-1997}. Fractional statistics has been measured experimentally in both FQH states at $\nu=1/3$ and $\nu=2/5$ in (Fabry-Perot) 
interferometers by J. Nakamaura, S. Liang, G. Gardner, and M. Manfra \cite{nakamura-2020}.

\subsubsection{The Compressible States}
\label{sec:compressible}

At large values $p \to \infty$ the Jain sequences converge to the even-denominator fractions $\nu_\infty(n)=1/n$. In this limit, $B_{\rm eff} \to 0$ and $\omega_c^{\rm eff} \to 0$. 
In other words, at the filling fractions $\nu_\infty(n)$ the average effective field experienced by the composite fermions is zero. 
This mean field theory then would predict that this state is essentially a Fermi liquid of composite fermions which is a compressible state. 
In addition, one can compute the effective charge of the composite fermion and find that it is $e/(np\pm 1)$, which also vanishes in the compressible limit. 
It is straightforward to see that the Fermi sea of the composite fermions is a disk with Fermi momentum  $p_F=(2\nu_{\infty}(n))^{1/2} \hbar/\ell_0$ 
and that the Fermi energy is $E_F=\nu_\infty(n)\hbar/(M\ell_0^2)$. Halperin, Lee, and Read \cite{halperin-1993} 
did an in-depth analysis of the consequences of this compressible state which are largely in agreement with experiment. 

In our discussion of the incompressible FQ states, both in the composite boson  and in the composite fermion approaches in sections \ref{sec:CB-QFT-FQH} and 
\ref{sec:CF-QFT-FQH} we saw that the mean field approximation violated symmetries that were restored by quantum fluctuations. This fact allowed us the derive effective filed theories which are 
asymptotically exact in the  low energy IR regime. What allowed us to obtain these exact  results is the existence of a gap. However, the compressible states are gapless and in some sense should 
be regarded theories of a quantum critical system. 

In fact, if we reexamine the field theory of composite fermions of section \ref{sec:CF-QFT-FQH} we see that, in the compressible limit $p \to \infty$, the action of Eq.\eqref{eq:eff-action+fermions+CS2} 
describes a system of fermions at finite density coupled to the {\it fluctuations} $\mathscr{A}_\mu$ of the statistical gauge field with a Chern-Simons term, but without an external uniform field 
(which has been canceled by the flux of the average statistical gauge field). This theory has two salient features. one is that the only trace left of the explicitly broken time-reversal symmetry
 is in the presence of the Chern-Simons term. This means that while the mean field theory looks like a Fermi liquid, the consequences of the broken time reversal invariance only enters in the 
 quantum fluctuations. In addition, in this regime this theory also breaks the particle-hole symmetry of a half-filled Landau level (in the high magnetic field regime).

However, these contributions, already at one-loop order, typically contain IR divergencies. These IR divergencies are due to singular forward scattering processes of the composite fermions 
by the gauge fields. They are most severe in the computation of the fermion self energy and depend on the form of the electron-electron interactions. For instance if the interaction potential $V(R)$ 
is short range, the one-loop contribution to the fermion self-energy has an imaginary part withe the behavior $\Sigma''(\omega)\sim \omega^{2/3}$ which is {\it much larger} than the real part 
$\Sigma'(\omega)$ as $\omega \to 0$ and the the Fermi energy $E_F$ is approached \cite{halperin-1993}. 
This behavior, which is found in many metallic systems at quantum criticality \cite{hertz-1976,millis-1993,sachdev-1999} implies that the width of the quasiparticle is asymptotically 
much larger than the energy as the Femi surface is approached. This one-loop result means that perturbation theory fails in the IR and that actual behavior may be non-perturbative. 
This problem is present even in the case of unscreened $\frac{e^2}{R}$ (Coulomb) interactions where the singular behavior is milder with $\Sigma''(\omega) \sim \omega \ln \omega$ 
(known as Marginal Fermi Liquid scaling \cite{varma-1989}). 

At any rate, these IR singular behaviors imply that the quasiparticle picture is inadequate and the that in these regimes there may be no quasiparticles. 
These systems are generally known as ``Strange Metals''. Many strongly interacting physical systems of interest, ranging form high $T_c$ superconductors to (largely conjectured) spin liquids with 
spinon Fermi surfaces to even quark-gluon plasmas share these types of singular IR behaviors \cite{varma-1989,polchinski-1994}. 
In spite of a large body of theoretical work, it has remained largely unsolved problem.  Perturbative renormalization group methods used to justify the 
Landau theory of the Fermi liquid \cite{shankar-1994,polchinski-tasi} generally fail in these regimes, 
although in some cases long range interactions have allowed some degree of control \cite{nayak-1994,nayak-1994b}. 
Remarkably,  the AdS/CFT Gauge/Gravity Duality \cite{maldacena-1998} has brought new insights into strange metal behaviors \cite{faulkner-2010}.
We will see below that relativistic dualities have given additions insights into the physics of compressible states.

\subsubsection{Fractional Quantum Hall Wave Functions and Conformal Field Theory}
\label{sec:FQH+CFT}

The Laughlin wave function and  its generalizations share the remarkable feature of being essentially universal. Aside form the dependence on the magnetic length $\ell_0$  in the 
gaussian factors which ensure that the function is integrable at long distances (a feature inherited from the single-particle states in the Landau level), the FQH ``ideal'' wave functions 
do not have any length scales. In addition, in almost all cases, the ideal wave functions are the exact ground state wave functions of some suitable ultra-local Hamiltonian projected onto the lowest Landau level. 
Another feature, closely related to their universality and analyticity, is that these wave functions look similar to correlation functions in two-dimensional critical (chiral) critical systems. 
We will now see that these features are not accidental and point to a deep relation between quantum Hall states and Conformal Field Theory (CFT). 

Although the relation between the simpler case of the Laughlin state was known to several people, such as Sergio Fubini \cite{fubini-1991}, this deep connection with {\it rational} CFT was 
articulated in considerable generality in a somewhat earlier paper by Gregory Moore and Nicholas Read \cite{moore-1991}. This approach showed its full power in the formulation of the 
non-abelian quantum Hall states.

Let us begin with the simpler case of the Laughlin states with $\nu=1/m$, where  $m$ is odd for fermions and even for bosons. 
We will consider a U(1) {\it Euclidean} CFT of a compact boson (a scalar field) $\phi(x)$ closely related to our  discussion of bosonization in 1+1 dimensions  in section \ref{sec:bosonization-rules}. This theory has been extensively studied in the CFT literature \cite{ginsparg-1988,difrancesco-1997,polchinski-book}
The Euclidean action for this theory is 
\begin{equation}
S=\frac{1}{8\pi} \int d^2x\; \left(\partial_\mu \phi\right)^2
\label{eq:U1-CFT}
\end{equation}
This theory is compactified by the condition that the only allowed observables are invariant under the discrete symmetry
\begin{equation}
\phi(x) \mapsto \phi(x)+2\pi R n
\label{eq:compactification2}
\end{equation}
where $R$ is the compactification radius and $n$ is an arbitrary integer. The observables that satisfy this condition are vertex operators of the form
\begin{equation}
V_p(x)=\exp\left(i \frac{p}{R} \phi(x)\right)
\label{eq:vertex-operators2}
\end{equation}
where $p$ is an integer.

The correlators of this theory are products of holomorphic (right-moving in Minkowski spacetime) and antiholomorphic (left-moving in Minkowski spacetime) factors. In the context of the FQH states we will be interested in a {it chiral} theory, whose correlators are holomorphic (analytic) functions in complex coordinates $z=x_1+ix_2$. Formally, the field $\phi(z, \bar z)$ is decomposed into a sum of a holomorphic field $\phi(z)$ and and antiholomorphic field $\phi(\bar z)$, so that the propagators is
\begin{equation}
\langle \phi(z, \bar z) \; \phi(\varw, \bar {\varw})\rangle=-\ln(z-\varw)-\ln (\bar z -\bar {\varw})
\label{eq:propagator-compactified-boson}
\end{equation}

In what follows we will focus on the chiral(holomorphic) field $\phi(z)$ whose propagator is
\begin{equation}
\langle \phi(z) \phi(\varw)=-\ln (z-\varw)
\label{eq:chiral-propagator}
\end{equation}
The correlator of the chiral current operator 
\begin{equation}
j(z)=\frac{i}{R} \partial_z \phi(z)
\label{eq:chiral-current-op}
\end{equation}
is given by
\begin{equation}
\langle i \partial_z \phi(z) \; i \partial_{\varw} \phi(\varw)\rangle\sim \frac{1}{(z-\varw)^2}
\label{eq:current}
\end{equation}
which tells us that this operator has (chiral) scaling dimension $\Delta=1$. Similarly the correlator of two {\it chiral} vertex operators 
\begin{equation}
V_p(z)=\exp\left( i \frac{p}{R} \phi(z)\right)
\label{eq:chiral-vertex-operator}
\end{equation}
is
\begin{equation}
\langle V_p(z) \; V_{-p}(\varw)\rangle=\frac{1}{(z-\varw)^{p^2/R^2}}
\label{eq:correlator-chiral-vertex-operators}
\end{equation}
whose (chiral) scaling dimension is $\Delta_p=p^2/(2R^2)$.  
These expressions are very similar to the operators that we used in section \ref{sec:2d-vortices} to describe vortices in 2D superfluids (and planar magnets). 
The main difference is that here the fields and their correlators are holomorphic where is section \ref{sec:2d-vortices} the correlators are the product of a holomorphic and antiholormophic factors.

For general $p$ and $R$ the correlator of Eq.\eqref{eq:correlator-chiral-vertex-operators} has a branch cut from $z$ to $\varw$. This means that the {\it monodromies} of the operators,
 i.e. dragging an operator around the other, induces a change in the {\it phase} of the correlator equal to $2\pi p^2/R^2$. 
 This behavior is reminiscent of the analytic structure the the quasiparticle wave function in the FQH states and of the {\it braiding} operation between anyons.

Moore and Read \cite{moore-1991} showed that the following correlator in a U(1) compactified boson CFT (with compactification radius $R=\sqrt{m}$) 
is equal to the Laughlin wave function at filling fraction $\nu=1/m$,
\begin{align}
\Psi_m(z_1,\ldots. z_N)=&\Big<\left(\prod_{i=1}^N \exp\left(i \sqrt{m} \phi(z_i)\right)\right)\; \exp\left(-i \int d^2z' \sqrt{m} \;\rho_0\; \phi(z')\right)\Big>\\
=& \prod_{i<j} (z_i-z_j)^m\; \exp\left(-\frac{1}{4}\sum_{i=1}^N |z_i|^2 \right)
\label{eq:laughlin-U(1)-CFT}
\end{align}
where we have used units in which the magnetic length is $\ell_0=1$. 

In section \ref{sec:2d-vortices} we saw that correlators of vertex operators are non-vanishing only if their total charge (vorticity) is zero. 
The reason for this condition is that the (Euclidean) action is invariant under arbitrary uniform shifts of the field $\phi$. Now, in the correlator on the right hand side of Eq.\eqref{eq:laughlin-U(1)-CFT} 
we have a product of $N$ vertex operators, each with the same charge $\sqrt{m}$. 
Then, under an arbitrary shift  by $\alpha$ of the field $\phi(z)$, the product of vertex operators change by a phase equal to $N \sqrt{m} \alpha$. 
On the other hand the other operator in the correlator, which describes a continuum background charge, changes by a phase $-\sqrt{m} \rho_0L^2 \alpha$, where $L^2$ is the area. 
Since $\rho_0$ is the areal particle density, $N/L^2$, the two changes of the phase cancel each other exactly. 
Thus, the operator in the expectation value of Eq.\eqref{eq:laughlin-U(1)-CFT} is charge neutral and has a non-vanishing expectation value. 

The other important feature of the correlator of Eq.\eqref{eq:laughlin-U(1)-CFT} is the choice of compactification radius, $R=\sqrt{m}$, and of the vertex operator $V_p(z)$ with $p=m$: 
\begin{equation}
V_m(z)=\exp(i \sqrt{m} \phi(z))
\label{eq:chiral-electron-operator}
\end{equation}
whose correlation function is
\begin{equation}
\langle V_m(z) \; V_{-m}(\varw)\rangle=\frac{1}{(z-\varw)^m}
\label{eq:chiral-electron-correlator}
\end{equation}
Which is invariant under a $2\pi$ rotation (since $m \in \mathbb{Z}$). However under a rotation by $\pi$ (i.e. an {\it exchange}) it changes by $(-1)^m$. 
Hence, for $m$ odd it changes sign, and the vertex operator $V_m$ behaves as a {\it fermion}, while for $m$ even the vertex operator behaves as a {\it boson}. 
We will call the vertex operator $V_m\equiv V_e$ the electron operator.

In other words, the correlator of Eq.\eqref{eq:laughlin-U(1)-CFT} is the expectation value of $N$ vertex operators each describing an {\it electron} (for $m$ odd) and a neutralizing background charge. 
Then, an elementary calculation yields the expression of the Laughlin wave function. 

In this formulation the ground state wave function for the integer QH state at $\nu=1$, which has the for of a Vandermonde determinant shown in Eq.\eqref{eq:vandermonde}, is interpreted as the correlator of $N$ electron operators in the U(1) CFT with compactification radius $R=1$.
In this simple case, the electron operator is the vertex operator $V_1(z)=\exp(i \phi(z))$, and there are no other operators (aside from the current $j(z)=i \partial_z \phi$). The propagator of this vertex operator is
\begin{equation}
\langle V_1(z) V_{-1}(\varw)\rangle=\frac{1}{z-\varw}
\label{eq:fermion-nu1}
\end{equation}
which indeed represents a fermion with electric charge $e$.

We will now see that the vertex operator of Eq.\eqref{eq:chiral-vertex-operator} with $p=1$ (and $R=\sqrt{m}$)
\begin{equation}
V_1(z)=\exp\left(\frac{i}{\sqrt{m}} \phi(z) \right)
\label{eq:qp-chiral-vertex-op}
\end{equation}
represents the Laughlin quasihole. The correlator of this vertex operator is
\begin{equation}
\langle V_1(z) \; V_{-1}(\varw)\rangle=\frac{1}{(z-\varw)^{1/m}}
\label{eq:qp-vertex-correlator}
\end{equation}
which has a branch cut stretching from $z$ to $\varw$ and, as a result, under a rotation by $\pm \pi$ its phase changes by $\pm \pi/m$, just as the Laughlin anyons do. 
To see that this operator indeed is related to the Laughlin quasihole we will compute the effect of inserting the vertex operator $V_1(z)$ in the correlator of Eq.\eqref{eq:laughlin-U(1)-CFT} and find
\begin{align}
&\Big<\exp\left(\frac{i}{\sqrt{m}} \phi(\varw)\right)\left(\prod_{i=1}^N \exp\left(i \sqrt{m} \phi(z_i)\right)\right)\; \exp\left(-i \int d^2z' \sqrt{m} \;\rho_0\; \phi(z')\right)\Big>=\nonumber \\
&= \prod_{i=1}^N(z_i-\varw) \prod_{i<j} (z_i-z_j)^m\; \exp\left(-\frac{1}{4}\sum_{i=1}^N |z_i|^2-\frac{1}{4m} |\varw|^2 \right)\nonumber \\
&=\Psi_m(\varw;z_1,\ldots,z_N)
\label{eq:qh-wf-CFT}
\end{align}
which is, indeed, the wave function for the Laughlin quasihole of Eq.\eqref{eq:laughlin-qh}. 
The insertion  of an additional vertex operator  $V_1(u)$ at $u$ inside the expectation value of Eq.\eqref{eq:qh-wf-CFT}  yields the two-quasihole wave function of Eq.\eqref{eq:2qh-wf}, proposed by 
Halperin \cite{halperin-1984} and discussed in section \ref{sec:fractional-statistics-qh}, with the same branch cut shown in Eq.\eqref{eq:halperin84} and, therefore carry fractional statistics $\pi/m$.

The operator product expansion discussed in section \ref{sec:ope}  of the vertex operators of this CFT yields new insight on the quasiholes. Indeed  the OPE of two vertex operators of charges $p$ and $q$ is
\begin{equation}
\lim_{\varw \to u} V_p(\varw) V_q(u)= \lim_{\varw \to u} \frac{1}{(\varw -u)^{\Delta_p+\Delta_q-\Delta_{[p+q]_m}}}V_{[p+q]_m}(u)
\label{eq:OPE-vertex-op}
\end{equation}
where $[p]_m$ is the integer $p$ modulo a multiple of $m$, i.e. if $p=mr+s$ (with $r$ a non-negative integer and $0\leq s<m$, then $[p]_m=s$. 
Eq.\eqref{eq:OPE-vertex-op} is understood in the sense that the contribution of all additional operators to the right hand side vanish as $\varw \to u$. 
In other words, the vertex operators are primary fields of this CFT and there are only $m$ primary fields. The physical process described by the OPE is called {\it fusion}. 

A CFT with a finite number of primaries is called a rational CFT \cite{ginsparg-1988,difrancesco-1997}. 
This result is, of course, the same statement  that we made is section \ref{sec:CB-QFT-FQH} that the FQH state has $m$ distinct anyons (vortices). 
This result also means that a vertex operator of charge $m$, i.e. an electron, is indistinguishable from the {\it identity operator}, $V_0$. 
In this sense, the allowed primaries are fields which are {\it local} respect to the electron operator $V_m$. 
From the point of view of the FQH state, an operator that creates or destroys an electron (at fixed filling fraction) has no effect, as the FQH state is a fluid made of electrons 
This also means that all physical operators must braid trivially with the electron operator.
In a CFT an operator of this type is said to generate and extended symmetry algebra \cite{moore-1989a,moore-1989b}. 

Furthermore, the OPE of the chiral current $j(z)$ with the vertex operator $V_1(z)$ is
\begin{equation}
\lim_{\varw \to u} j(\varw) V_1(u)=\frac{1/m}{z-w} V_1(u)
\label{eq:current-vertex-ope}
\end{equation}
which implies that the vertex operator represents a state with charge $1/m$ (in units of the electric charge $e$).

The CFT approach to construct FQH states has become a very powerful tool.
We will see in section \ref{sec:edge} that FQH states on an open geometry (e.g. a disk) have edge states which can be understood a chiral CFTs.  
In addition to providing a deeper understanding of more general abelian (muliti-component) FQH states with a $K$-matrix structure, 
new classes of of FQH states with non-abelian (multi-dimensional) representations of the Braid Group has been proposed. 
The anyons of these states are of particular interest since they have been proposed originally by Kitaev in 1997 \cite{kitaev-2003} that anyons can be used as physical qubits for 
topologically-protected quantum computation \cite{kitaev-2003,freedman-2000,freedman-2002,preskill-2004,dassarma-2007}.

So far we discussed the case of abelian anyons. Abelian anyons have the property that the fusion of two anyons is another anyon. 
Similarly,  a braiding operation between two anyons is equivalent (up to a phase factor) to another anyon. Mathematically this means that abelian anyons are one-dimensional representations of the Braid group. 
The fractional statistics of an anyon is then a label of the representation of the Braid group. As we saw, the same holds under fusion. 

However, the Braid group generally admits multi-dimensional representations. In this more general case the fusion (or braiding) of two anyons is represented 
by a linear combination (superposition) of anyon states. Linear combinations of anyon states are represented by unitary matrices of rank greater than 1. 
Since matrices generally do not commute braiding and/or fusion processes correspond to a multiplication of these matrices. Such unitary processes are then regarded as quantum gates. 
This concept is the physical basis of topological quantum computation. It is currently the cutting edge of research in the field.

We will now discuss the simplest non-abelian FQH states, the Moore-Read (MR) states \cite{moore-1991}. The Moore-Read wavefunctions are
\begin{equation}
\Psi_{MR}(z_1,\ldots,z_N)=\textrm{Pf}\left(\frac{1}{z_i-z_j}\right) \prod_{i<j} (z_i-z_j)^n \; \exp(-\frac{1}{4\ell_0^2}\sum_{i=1}^N|z_i|^2)
\label{eq:MR-wf}
\end{equation}
which describes a FQH of electrons at filling fraction $\nu=1/n$, with $n$ an {\it even} integer. Here $\textrm{Pf}\left(1/(z_i-z_j\right)$ is the Pfaffian of the matrix $1/(z_i-z_j)$.

This state was motivated by the discovery of an unexpected FQH state in the $N=1$ Landau level with an {\it even denominator} filling fraction $\nu=5/2$ by Willett and coworkers in 1987 
\cite{willett-1987,eisenstein-1988,pan-2008}. Until then (and since then) this is the only FQH state with an even denominator filling fraction. 
The Pfaffian factor has a simple pole when two coordinates approach each other. However, provided $n \geq 1$ the ``Laughlin factor'' in the MR wavefunction cancels the singularity.
We will shortly discuss the CFT construction of the Moore-Read state.
The poles in the MR wavefunction suggest that in this state the electrons can get closer to each other than in a Laughlin state. 
This observation suggests that there may be some form of attraction (or suppression of repulsion) between the electrons in the MR state and 
 motivated the notion that the observed even-denominator plateau may be physically related to some sort of pairing interaction. 
 Shortly after the Moore-Read proposal was made, a paired wave function at $\nu=1/2$ with a Pfaffian factor was also proposed  by Greiter, Wen and Wilczek \cite{greiter-1991,greiter-1992,wen-1991} 
 who suggested that it reflects electrons pairing in the $\ell=1$ angular momentum channel in time-reversal breaking condensate with symmetry $p_x+ip_y$.

Although, following the logic behind the Kohn-Luttinger mechanism for paring in angular momentum state $\ell geq 1$  with repulsive interactions \cite{kohn-1965,chubukov-1993,raghu-2011} 
it may possible to get a paired state even in the lowest Landau level (although there is no evidence for it, so far), the $N=1$ Landau level may be more hospitable to such a state. 
Indeed, in the $N=1$ Landau level  the single-particle states i have angular momentum greater or equal than 1, and  their probability distributions are suppressed both at short and long distances; 
these states look like smoke-rings.  The matrix elements of the Coulomb potential in the $N=1$ Landau level are parametrically suppressed at short distances 
(compared with the states in the lowest, $N=0$, Landau level). In this scenario, the $\nu=5/2$ plateau is interpreted as a $\nu=1/2$ fully polarize state of the $N=1$ landau level, with an $N=0$ 
Landau level filled with electrons with both spin polarizations in a state with $\nu=2$. 
There is numerical evidence that an instability into a p-wave paired state such as the MR state is favored if the short-distance repulsion between the {\it composite fermions} is softened \cite{park-1998}.

In fact, in 1983 Halperin \cite{halperin-1983} considered generalizations of the Laughlin state in which due to very strong attractive interactions electrons would form clusters which then condense into a
Laughlin-type state. The simplest example was Laughlin state of bosons (paired electrons) a paired state at $\nu=1/2$ whose Laughlin quasihole carries charge $e/4$. Extensive numerical calculations 
are more compatible with the MR state than with the Halperin state of pairs \cite{Morf-1998}.
We will see shortly that the Halperin state is an abelian relative of the MR state.

Moore and Read derived the wave function of Eq.\eqref{eq:MR-wf} by considering a CFT with two sectors: a chiral boson CFT with compactification radius $R=\sqrt{n}$, and a chiral Majorana fermion CFT, 
i.e. the chiral part of the 2D Ising CFT, discussed in section \ref{sec:TFIM}. 
The chiral Ising CFT has central charge $c=1/2$ and several primary fields: 
the identity $I$, the twist field $\sigma$, and the (chiral) Majorana fermion $\chi$ with (chiral) scaling dimensions $0$, $1/16$, and $1/2$, respectively. 
The propagator of the Majorana fermion, which is a free field, is 
\begin{equation}
\langle \chi(z)\; \chi(\varw)\rangle=\frac{1}{z-\varw}
\label{eq:Majorana-propagator}
\end{equation}
The primary field $\sigma$, the ``twist field'', is non-local with respect to the Majorana fermion $\chi$, and changes its boundary conditions from periodic to antiperiodic. 

The fusion rules of the chiral Ising CFT are
\begin{equation}
\chi \star \chi=I, \qquad \sigma \star \sigma=I+\chi, \qquad \sigma \star \chi=\chi
\label{eq:chiral-ising-CFT}
\end{equation}
What will be important below is that the $\sigma$ field has two fusion channels.
As in the analysis of the Laughlin states, the compactified boson $n$ primaries (anyons), the vertex operators $V_p(x)=\exp(i p \phi(x)/n)$, with $n=0,1, \ldots n$ and there are $n$ types of anyons, 
and a (charge) current $J=\frac{i}{\sqrt{n}} \partial_z \phi(z)$.

The first task is to identify the electron operator which has to be a fermion and has to carry electric charge. 
This means that it is a composite of an operators (with Fermi statistics) the (neutral) chiral ising CFT and a vertex operator whose charge is an integer multiple of $e$. 
The desired electron operator is $\psi_e(z)=\chi(z) \; \exp(i\sqrt{n} \phi(z))$ whose correlator is
$\langle \chi(z) V_n(z) \chi(\varw) V_{-n}(z)\rangle=1/(z-\varw)^{1+n}$,  which is a fermion with  charge $e$.

The $n$-point function of the chiral Majorana fermions is
\begin{equation}
\langle \chi(z_1) \ldots \chi(z_N)\rangle=\textrm{Pf}\left(\frac{1}{z_i-z_j}\right)
\label{eq:majorana-N-point}
\end{equation}
which follows fro applying Wick's theorem.
Then we see that the MR wave function is the expectation value
\begin{equation}
\Psi_{MR}(z_1,\ldots,z_N)=\langle \chi(z_1) \ldots \chi(z_N)\rangle \times \Big<\left(\prod_{i=1}^N \exp\left(i \sqrt{n} \phi(z_i)\right)\right)\; \exp\left(-i \int d^2z' \sqrt{n} \;\rho_0\; \phi(z')\right)\Big>
\label{eq:MR-CFT}
\end{equation}

Next we need to identify the primary fields of the tensor product of the chiral Ising CFT, $\mathbb{Z}_2$, and the chiral U(1) CFT with compactification radius $R=\sqrt{n}$. 
The allowed primary field must be{\it local} with respect to the electron operator, $\psi_e$. 
This leaves us with four primary fields: 
a) the identity $I$,
b) the non-abelian vortex $\sigma (z) \times \exp\left(\frac{i}{2\sqrt{n}} \phi(z)\right)$ with charge $e/(2n)$ and non-abelian statistics, 
c) the charge-neutral Majorana fermion $\chi$, and 
d) the abelian vortex (the Laughlin quasihole) $\exp\left(\frac{i}{\sqrt{n}}\phi(z) \right)$, with charge $e/n$ and abelian statics $\delta=\pi/n$.

The he new feature here is the non-abelian fractional statistics of the non-abelian vortex (sometimes called a ``half-vortex''.
 Its non-abelian character is a consequence of the fact that the fusions of two twist fields has two channels, see Eq.\eqref{eq:chiral-ising-CFT}. 
 This implies that the wave function of four non-abelian vortices can be expressed as a linear combination of two so-called {\it conformal blocks}. 
 In other words, this wave function belongs to a degenerate two-dimensional Hilbert space, each component  labeled by a conformal block \cite{difrancesco-1997}. 
 A  braiding operation between two non-abelian vortices is a monodromy of the wave function that induces a unitary transformation $U$ in this Hilbert space
\begin{equation}
U=\frac{1}{\sqrt{2}} \exp\left[i \pi \left(\frac{1}{8}+\frac{1}{4n}\right)\right]
\times
\begin{pmatrix}
1, & 1\\
-1 &1
\end{pmatrix}
\label{eq:non-abelian}
\end{equation}
Therefore, the degenerate Hilbert space of four quasiholes provides a two-dimensional representation of the Braid Group. 
This observation is the basis for regarding a state of four non-abelian quasiholes as a qubit. 
Furthermore, Nayak and Wilczek showed that a state of $2p$ quasiholes spans a $2^{p-1}$-dimensional degenerate Hilbert space \cite{nayak-1996}. 
This means that, for large $p$, the degeneracy {\it per quasihole} is $\sqrt{2}$. 
In other words, this degeneracy is not due to a local degree of freedom attached to each quasihole but that it is shared in a non-local fashion!

In what sense are the Moore-Read states paired? In an insightful paper Read and Green \cite{read-2000} used a BCS theory approach to investigate the pairing properties of a  
condensate  of composite fermions in the $p_x+ip_y$ channel. The mean-field BCS Hamiltonian is
\begin{equation}
H_F=\int d^2k \left[\left(\varepsilon(\bm k)-\mu\right) \psi^\dagger(\bm k) \psi(\bm k)+\frac{1}{2} \left(\Delta^* (\bm k) \psi(-\bm k) \psi(\bm k)+\Delta(\bm k) \psi^\dagger(\bm k) \psi^\dagger(-\bm k)\right)\right]
\label{eq:BCS}
\end{equation}
where $\psi(\bm k)$ is the composite fermions field, $\varepsilon(\bm k)=\frac{\bm k^2}{2M}$, and $\Delta(\bm k)$ is the gap function. For a $p_x+ip_y$ condensate the gap function the gap function transforms under rotations as an $\ell=-1$ angular momentum eigenstate, and for $\bm k \to 0$ it behaves as $\Delta(\bm k) = (k_x-ik_y) \Delta$, where $\Delta$ is a constant pairing amplitude.

The BCS ground state has the form
\begin{equation}
|G\rangle=\prod_{\bm k} \left(u(\bm k)+\varv(\bm k) \psi^\dagger(\bm k)\psi^\dagger(-\bm k)\right)|0\rangle
\label{eq:BCS-wf}
\end{equation}
where $|u(\bm k)|^2=|\varv(\bm k)|^2=1$. The amplitudes $u(\bm k)$ and $\varv(\bm k)|$ satisfy the Bogoliubov-de Gennes Equation (BdG)
\begin{equation}
\begin{pmatrix}
\xi(\bm k) & -\Delta^*(\bm k)\\
-\Delta(\bm k) & -\xi^*(\bm k)
\end{pmatrix}
\begin{pmatrix}
u(\bm k)\\
\varv(\bm k)
\end{pmatrix}
\equiv E(\bm k) \hat {\bm n}(\bm k)\cdot \bm \sigma
\begin{pmatrix}
u(\bm k)\\
\varv(\bm k)
\end{pmatrix}
=E(\bm k) 
\begin{pmatrix}
u(\bm k)\\
\varv(\bm k)
\end{pmatrix}
\label{eq:BdG}
\end{equation}
where $\xi(\bm k)=\varepsilon(\bm k)-\mu$, $\bm \sigma=(\sigma_x, \sigma_y, \sigma_z)$ are the Pauli matrices, and $\hat {\bm n}(\bm k)=(-\textrm{Re} \Delta(\bm k), \textrm{Im}\Delta(\bm k), \xi(\bm k))$ is a unit vector defined for every $\bm k$. The eigenvalues $E(\bm k)$ and the eigenvectors $(u(\bm k), \varv(\bm k)$ are
\begin{equation}
E(\bm k)=\sqrt{\xi^2(\bm k)+|\Delta(\bm k)|^2}, \qquad \frac{\varv(\bm k)}{u(\bm k)}=-\frac{E(\bm k)-\xi(\bm k)}{\Delta^*(\bm k)}
\label{eq:ev-BdG}
\end{equation}
The spinor amplitudes are $|u(\bm k)|^2=\frac{1}{2} \left(1+\frac{\xi(\bm k)}{E(\bm k)}\right)$ and $\varv(\bm k)=\frac{1}{2} \left(1-\frac{\xi(\bm k)}{E(\bm k)}\right)$.

In the low momentum limit and in real space the BdG Equations become 
\begin{align}
i \partial_t u=& -\mu u+\Delta^*i (\partial_x+i\partial_y)\varv\nonumber\\
i\partial_t \varv=&\mu \varv+\Delta i(\partial_x-i\partial_y)u
\label{eq:BdG-cont}
\end{align}
which is just the Dirac Equation in 2+1 dimensions, with $\mu$ playing the ole of the mass, and with the restriction that the spinor $(u, \varv)$ obeys the Majorana condition
\begin{equation}
(u, \varv) 
\begin{pmatrix}
0 & 1\\
1 & 0
\end{pmatrix}
=
\begin{pmatrix}
u\\
\varv
\end{pmatrix}
\label{eq:majorana-condition}
\end{equation}
The Majorana condition is obeyed in all superconductors and reflects that fact that in these condensates only the fermion parity $(-1)^{N_F}$ (with $N_F$ being the number of fermions) is conserved, 
while the fermion number $N_F$ is not.

This is a good BCS state in that it is fully gapped and it is chiral. Assuming that the pair fields in the $p_x$ and $p_y$ 
channels are equal, they showed that the BCS ground state $|G\rangle$ has the form
\begin{equation}
|G\rangle\sim \exp\left(\frac{1}{2} \sum_{\bm k} g(\bm k) \psi^\dagger(\bm k) \psi^\dagger(-\bm k)\right)|0\rangle
\label{eq:paired-wf}
\end{equation}
Projected onto a state with $N$ fermions with real space coordinates $\bm x_i$ the wave function is a Pfaffian
\begin{equation}
\Psi(\bm x_1, \dots,\bm x_N)=\langle \bm x_1, \ldots, \bm x_N |G\rangle=\textrm{Pf}(g(\bm x_i - \bm x_j))
\label{eq:BCS-wf-pfaffian}
\end{equation}
The long distance behavior of this wave function depends on whether the chemical potential $\mu >0$ (this is the weak-pairing or BCS regime), or $\mu<0$ (which is the strong pairing BEC-like regime). 
While in the case of an s-wave superconductor these two regimes are smoothly connected by a crossover, in the $p_x+ip_y$ case they are separated by a quantum phase transition at $\mu=0$.
In the weak pairing regime the function $g(\bm k)$ has the asymptotic long distance form, as $\bm k \to 0$,
\begin{equation}
g(\bm k) \simeq -\frac{2\mu}{(k_x+ik_y)\Delta^*}
\label{eq:gofk}
\end{equation}
where the pair field behaves as  $\Delta(\bm k) \simeq (k_x-ik_y) \Delta$. In real space (in complex coordinates) the function $g(z)$ becomes
\begin{equation}
g(z)=\left(\frac{i\mu}{\pi \Delta^*}\right)\frac{1}{z}
\label{eq:gofz}
\end{equation}
which is the form used in the Pfaffian wave function. On the other hand, in the strong pairing regime, $\mu <0$, $g(\bm k)$ has the asymptotic behavior
\begin{equation}
g(\bm k)\simeq -A \frac{(k_x-ik_y)}{a_0^{-2}+\bm k^2}
\label{eq:gofk-strong}
\end{equation}
where $A$ and $a_0$ are functions of $\Delta$ and $\mu$,
\begin{equation}
A=\frac{2|\mu| M \Delta}{2|\mu|+m |\Delta|^2}, \qquad a_0=\frac{1}{2|\mu|} \sqrt{\frac{2|\mu|}{M}+|\Delta|^2}
\label{eq:A-a0}
\end{equation}
 In real space, at distances $|\bm x|\gg a_0$ the Fourier transform of $g(\bm k)$ decays exponentially, and as  $a_0 \to \infty$,  $\mu \to 0$ and the Fourier transform $g(z)$ has the power-law behavior
 \begin{equation}
 g(z) \simeq \left(\frac{i|\Delta|}{2\pi \Delta^*}\right)\; \frac{1}{z|z|}
 \label{eq:critical-gz}
 \end{equation}
 This means that $\mu=0$ is a quantum critical point that separates the weak pairing phase from the strong pairing phase, which have distinct properties.

Read and Green then used a topological argument, originally formulated by Grisha Volovik \cite{volovik-1988} in the context of superfluid $^3$He films. 
To see this we notice that the three component unit vector $\hat {\bm n}(\bm k)$ takes values on the surface of a sphere $S_2$. 
In addition, since $\varv(\bm k) \to 0$ for $\bm k \to \infty$(and, hence $u(\bm k) \to 1$), we can wrap the momentum space $\bm k$ onto a sphere $S_2$. Thus $\hat {\bm n}(\bm k)$ 
are smooth functions of $S_2 \mapsto S_2$. such maps are classified into the homotopy classes of $\pi_2(S_2)\simeq \mathbb{Z}$ given in terms of the integer-valued topological invariant
\begin{equation}
Q=\frac{1}{4\pi}\int d^2k \; \hat {\bm n}(\bm k)\cdot \partial_{k_x} \hat {\bm n}(\bm k) \times \partial_{k_y}\hat {\bm n}(\bm k)
\label{eq:instanton-number}
\end{equation}
In the strong pairing phase $\mu<0$ and $\xi(\bm k)>0$ and $\hat {\bm n}(\bm k)$ takes values only on the northern hemisphere of the target space $S_2$. 
Such maps can be deformed continuously to the North Pole and, hence, are topologically trivial, $Q=0$. On the other hand, in the weak pairing phase, $\mu>0$, $\xi(\bm k)$
takes both positive and negative values. In this phase $\hat {\bm n}(\bm k)$ is topologically non-trivial and the topological invariant $Q$ takes  non-trivial values $\pm 1$. 

The upshot of this analysis is that, in the weak pairing phase, the paired FQH state is, at this mean field level, a two-dimensional {\it topological superconductor}. 
Read and Green \cite{read-2000} further showed that this pairing state has vortices with an interesting fermionic spectrum. A vortex is a state win which at long distances the (complex) amplitude of the 
$p_x+ip_y$ condensate behaves as $\Delta \exp(i\varphi)$, where $\varphi$ is the azimuthal angle, which winds by $2\pi$ on a circumference of large radius $R$. 
The BdG equation, Eq.\eqref{eq:BdG-cont}, has zero mode solutions. In polar coordinates $(r, \varphi)$ the BdG Equation the zero modes satisfy
\begin{align}
\Delta i e^{i\varphi}\left(\partial_r+\frac{i}{r} \partial_\varphi\right)\varv=&\mu u\nonumber\\
\Delta i e^{-i\varphi} \left(\partial_r-\frac{i}{r} \partial_\varphi\right)u=&-\mu \varv
\label{eq:dirac-vortex}
\end{align}
The normalizable zero mode spinor solution is
\begin{equation}
\begin{pmatrix}
u(r,\varphi)\\
\varv(r,\varphi)
\end{pmatrix}
=
\frac{f(r)}{\sqrt{r}} 
\begin{pmatrix}
\frac{1}{\sqrt{i}}e^{i \varphi/2}\\
\frac{1}{\sqrt{-i}} e^{-i \varphi/2}
\end{pmatrix}
\label{eq:vortex-fermion-zero-mode}
\end{equation}
where $f(r)$ is given by
\begin{equation}
f(r) \sim \exp\left(-\int_0^r dr' \frac{\mu(r')}{|\Delta|}\right) \sim \exp(-\mu r/|\Delta|)
\label{eq:zero-mode-f(r)}
\end{equation}
Under a $2\pi$ rotation this spinor solution is double-valued
\begin{equation}
(u(r, \varphi+2\pi), \varv(r, \varphi+2\pi))=-(u(r,\varphi), \varv(r, \varphi))
\label{eq:double-valued}
\end{equation}
which follows form the global phase invariance. I fact, in all paired states, topological or not, under a global transformation of the pair field by a phase $\theta$, 
the (composite) fermion must transform with a phase of $\theta/2$, 
\begin{equation}
\Delta(\bm x) \mapsto e^{i\theta} \Delta(\bm x), \qquad \psi(\bm x) \mapsto e^{i\theta/2} \psi(\bm x), \qquad \psi^\dagger(\bm x) \mapsto e^{-i \theta/2} \psi^\dagger(\bm x)
\label{eq:U(1)-transformation-spinor}
\end{equation}
In other words, in a paired state the fermion is non-local to the vortex.
Hence, under a $2\pi$ phase transformation of the pair field, the fermions must change sign, and the spinor  zero mode solution must be double-valued. This means that this state has a brach cut.

The structure of the vortex and of the fermion zero mode is closely related to the problem of solitons with fractional charge that we discussed in section \ref{sec:soliton-fractional-charge}. In fact, Roman Jackiw and Paolo Rossi \cite{jackiw-1981} investigated a closely related problem of a theory of Dirac field in 2+1 dimensions coupled to a charged scalar field through a Yukawa coupling with the form of a Majorana mass. They showed that such a theory admits vortex solutions with fermion zero modes. In a subsequent paper Erik Weinberg \cite{weinberg-1981} showed that these zero modes are counted by an index theorem which relates the number of zero modes to the vorticity. More recently, Nishida, Santos and Chamon \cite{nishida-2010} that the relativistic theory of Jackiw and Rossi in the non-relativistic approximation reduces to the theory of a $p_x+ip_y$ superconductor.

There is, however, a subtle but profound difference between the fermion zero modes of the one-dimensional fractionally charged solitons and the fermion zero modes of a superconductor. In the case of teh 1D solitons, the zero modes can be either occupied or empty rendering a soliton charge of $-e/2$ or $+e/2$. However, in the case of the superconductor, the BdG fermions are charge-neutral since their charge has gone into the condensate. Thus, in a superconductor fermion number is not conserved, and only the fermion parity is conserved. This means that the field operator associated with the vortex zero mode, which we will denote by $\gamma$, must be a self-adjoint operator, $\gamma^\dagger=\gamma$. 

Ivanov \cite{ivanov-2001} showed that this behavior is the origin of the non-abelian fractional statistics in this system. 
Indeed, if one considers a configuration with $2n$ vortices, each will carry a  Majorana zero mode $\gamma_i$. 
Since these are fermion operators they satisfy the usual anticommutator algebra, $\{ \gamma_i, \gamma_j \}=2\delta_{ij}$. 
We can (arbitrarily) group the $2n$ Majorana fermion operators into $n$ pairs. For each {\it pair} we can define a complex Dirac operator $\psi_j=(\gamma_{2j}+i\gamma_{2j+1})/2$ (and its adjoint),
which satisfies the usual fermionic algebra, $\{ \psi_j, \psi^\dagger_k\}=\delta_{jk}$. 
Each Dirac fermion can be either in an empty state or in an occupied state. Thus a system of $2n$ vortices supports a degenerate Hilbert space of dimension $2^{n-1}$, 
which agrees with the results of Nayak and Wilczek \cite{nayak-1996}. 
However, the assignment of Majorana operators to specific pairs is actually arbitrary, which amounts to a particular definition of the branch cuts.  
Changing the assignment of the operators into pairs is then equivalent to a rearrangement of the cuts.  
In 2006 Michael Stone and Suk Bum Chung \cite{stone-2006}  showed that the these vortices obey the braiding and fusion properties of {\it Ising} anyons. 
These properties follow from the branch cut configurations which affect the monodromies of the vortices. 
It is important to stress that the vortices with zero modes are the non-abelions. Majorana fermions are fermions and, as such, are abelian. 

In addition to the Moore-read Pfaffian state, the CFT approach has led to the formulation of new non-trivial FQH states. 
Read and Rezayi \cite{read-1999} proposed a series of so-called {\it cluster} states, which generalize the concept of paired states. 
They investigated a particular type of cluster states in which the Pfaffian factor of the MR state is replaced by a correlator of {\it parafermion} primary fields of a $\mathbb{Z}_k$ 
CFT of Zamolodchikov and Fateev \cite{zamolodchikov-1985}. Parafermions were originally introduced by Fradkin and Kadanoff \cite{fradkin-1980} (see also Ref.\cite{alcaraz-1981}) 
as a generalization of the fermions of the 2D Ising model to the $\mathbb{Z}_k$ clock model. 
In this model one can define several types of parafermions each consisting of the fusion of a charge operator and a magnetic (disorder) operator. 
These operators obey an algebra of the type $ A B = \exp(i p 2\pi/k) BA$, where the integer $p$ depends on the electric and magnetic charges of the parafermions, which is reminiscent of fractional statistics. 
In close analogy with the Majorana fermions of the 2D Ising model, $\mathbb{Z}_k$ CFT describes the behavior of these systems their self-dual points.

The $\mathbb{Z}_k$ CFT is actually much richer and has more primary fields than the $\mathbb{Z}_2$ case of the Ising model. 
As a CFT, the $\mathbb{Z}_k$ theory is the same as the CFT on the coset SU(2)$-k$/U(1). 
A coset means that a U(1) sector has been projected out of the spectrum. Here we will focus on a special case of the $\mathbb{Z}_3$ CFT. This CFT has a 
parafermion primary field, which we will call $\psi$, and a non-abelian primary that we will  call $\tau$. Read and Rezayi \cite{read-1999} proposed to replace the Pfaffian factor of the MR state with a correlator 
of $N$ parafermions of a $\mathbb{Z}_3$ CFT and a Laughlin factor with an exponent of $n+2/3$. Such states require that the number of electrons $N$ be a multiple of 3. Thus, these states can be viewed as 
sates in which the electrons cluster in groups of 3. They also considered the more general case of the $\mathbb{Z}_k$ CFT in which case $N$ is a multiple of $k$. 
The resulting filling fractions are $\nu=k/(mk+2)$ (with $m \geq 0$). 
The Read-Rezayi $k=3$ parafermion state is a leading candidate for the observed FQH plateau at $\nu=12/5=2+2/5$ \cite{pan-2003,pan-2004} (the compering state being the 2/5  Jain state). 
There is strong numerical support for the 12/5 state being a $\mathbb{Z}_3$ parafermion \cite{Rezayi-2009}.

The quasiholes obtained by inserting the primary fields $\tau$ in the correlator of parafermions have interesting properties which stem from their basic fusion rule
\begin{equation}
\tau \star \tau=I+\tau
\label{eq:Fib}
\end{equation}
Read and Rezayi \cite{read-1999} showed that the number of conformal blocks  for $3n$ quasiholes, i.e. the number of degenerate states, in the parafermion theory is the Fibonacci number $F_{3n-2}$, where 
$F1=1$, $F_2=2$, $F_3=3$, $F_4=5$. In general $F_m=F_{m-1}+F_{m-2}$.  For $m \to \infty$, the rate of increase approaches the limit $\lim_{m \to \infty} F_m/F_{m-1}=(1+\sqrt{5})/2$ which is known as the 
Golden Mean. In other words, for large $m$, the dimension of the Hilbert space increases exponentially as $((1+\sqrt{5})/2)^m$. Aside from these interesting mathematical curiosities, the significance of this 
fusion rule is that these states, regarded as``qubits'', define unitary transformations that cover uniformly the Bloch sphere which is required for universal quantum computation \cite{freedman-2002}.

At the level of topological quantum field theory the Moore-Read and Read-Rezayi states are related to the non-abelian Chern-Simons gauge theory with gauge group SU(2) at Chern-Simons level $k$. In the 
more general case of the gauge group SU(N) on a manifold $\mathcal{M}$ action is given by
\begin{equation}
S_{\rm CS}[A_\mu]=\frac{k}{4\pi} \int_\mathcal{M}d^3x\; \epsilon_{\mu \nu \lambda}\left(A^\mu_a \partial^\nu A^\lambda_a+\frac{2}{3} f_{abc} A^\mu_aA^\nu_bA^\lambda_c\right)
\label{eq:CS-SU(N)k}
\end{equation}
where $k \in \mathbb{Z}$ is the level, the gauge fields $A^\mu=A^\mu_a t_a$ take values in the algebra of SU(N), with $\{ t_a \}$ being the $N^2-1$ generators, and $f_{abc}$ the structure constants of SU(N). 
The expectation values of Wilson loop operators of this theory compute the Jones polynomials which are topological invariants of knot theory \cite{witten-1989}.
The connection with SU2)$_k$ Chern-Simons gauge theory has been essential  in formulating effective low energy quantum field theories for the non-abelian states \cite{fradkin-1998,fradkin-1999,barkeshli-2010,goldman-2019,goldman-2020,goldman-2021}.  

\subsubsection{Edge States and Chiral Conformal Field Theory}
\label{sec:edge}

We will now discuss the edge states of the FQH states. As we saw the FQH states are incompressible in the bulk and all bulk excitations are gapped. 
The edge of the region occupied by the fluid (in many cases that edge of the physical sample) is where the bulk gap collapses and hence where the system has low energy excitations.

The role of edges and their nature can be seen already in the simple case of the integer Hall effect treated as a system of free fermions in the lowest Landau level. In this state all single particle states are occupied, and the bulk ground state wave function is a Slater state which takes the form of the Vandermonde determinant of Eq.\eqref{eq:vandermonde}. The potential that keeps the electrons inside the sample increases monotonously near the edge. As we discussed in section \ref{sec:LL}, in this region the electrons experience an electric field $E$ that pushes them into the sample, and in the presence of the perpendicular magnetic field $B$ the electrons move {\it along the edge} at the drift velocity $\varv=c |E|/|B|$. At some spatial location, corresponding to the locus of a single particle Landau state,  the potential crosses the Fermi energy and in that region potential is essentially a linear function of the coordinate and hence of momentum (or angular momentum depending on the gauge that is being used). In other words, the low energy states are one branch of a one-dimensional chiral excitation, such as in the right-moving states of our discussion of the chiral anomaly in section \ref{sec:global-chiral-symmetry}

In section \ref{sec:CS-quantization} we showed that a Chern-Simons gauge theory on a manifold with a boundary projects onto a chiral CFT on that boundary. In that section we showed that an abelian Chern-Simons theory U(1)$_m$ integrates to the boundary, the 1+1-dimensional Minkowski spacetime  $S^1\times \mathbb{R}$, as a chiral U(1)$_m$ CFT for a compactified boson with compactification radius $R=1$ whose action is given by
\begin{equation}
S[\varphi]=\int_{S^1\times \mathbb{R}}  d^2x\; \frac{1}{4\pi}\; \Big[ \partial_0 \phi \partial_1\phi- \varv (\partial_1 \phi^2)\Big]
\label{eq:chiral-U(1)-level-m-CFT}
\end{equation}
where we rescaled the compactified scalar by a factor of $\sqrt{m}$ to that its compactification radius $R=\sqrt{m}$ as in section \ref{sec:FQH+CFT}. The only difference between the theory of Eq.\eqref{eq:chiral-U(1)-level-k-CFT} and what we did in section \ref{sec:FQH+CFT} is that this theory is in a 1+1-dimensional Minkowski spacetime $S_1\times \mathbb{R}$ while before we were in Euclidean spacetime.In the case of the integer quantum Hall effect $\nu=1$ (and hence $m=1$) and the (time-ordered) correlator of the vertex operator, $V_1(x, t)=\exp(i \phi(x, t))$ is
\begin{equation}
\langle T \exp(i \phi(x, t)) \exp(-i \phi(0, 0))\rangle=\frac{1}{x-\varv t-i\varepsilon}
\label{eq:chiral-fermion-propagator}
\end{equation}
which is, indeed, the propagator of a chiral free fermion.

On the other hand, for the Laughlin states at filling fraction $\nu=1/m$ (and compactification radius $R=\sqrt{m}$) the propagator of the electron $\psi_e\sim \exp(i \sqrt{m} \phi)$ is
\begin{equation}
\langle T \exp(i \sqrt{m} \phi(x, t)) \exp(-i \sqrt{m} \phi(0,0))\rangle\sim \frac{1}{(x-\varv t-i\varepsilon)^m}
\label{eq:fermion-laughlin}
\end{equation}
while the propagator of the quasihole $\exp\left(\frac{i}{\sqrt{m}} \phi\right)$ is
\begin{equation}
\Big<T \exp\left(\frac{i}{\sqrt{m}} \phi(x, t)\right) \exp\left(-\frac{i}{\sqrt{m}} \phi(0,0)\right)\Big> \sim \frac{1}{(x-\varv t-i\varepsilon)^{1/m}}
\label{eq:qh-propagator}
\end{equation}
Therefore, the CFT of the edge is the same as the CFT of the ideal wavefunction with the only difference that the former is in Minkowski spacetime while the latter is in Euclidean spacetime.

In this sense there is a one-to-one correspondence between the bilk and the edge. We can also see how that works by considering a fundamental Wilson arc in the bulk along a path $\gamma(x,y)$ where $x$ and $y$ are at the boundary
\begin{equation}
\langle W[\gamma]\rangle=\langle \exp(i \int_{\gamma(x,y)} dx_\mu \mathscr{A}^\mu)\equiv \langle \exp(\frac{i}{\sqrt{m}} \phi(x)) \exp(-\frac{i}{\sqrt{m}} \phi(y))\rangle
\label{eq:bulk-boundary-identification}
\end{equation}
where on the right hand side we have rescaled the field by$\sqrt{m}$, as before. The scaling dimensions of the electron, the quasihole and the current are $\Delta_e=\frac{m}{2}$, $\Delta_{qh}=\frac{1}{2m}$ and $\Delta_{\rm current}=1$. These results will be important shortly.

The same structure applies to the multi-component abelian FQH states. The only difference is that in multi-component FQH states the edge consists, in general, of a charge field which couples to the electromagnetic field and one or more neutral edge states. An example is the theory of the non-abelian Pfaffian states at filling fraction $\nu=1/n$. In this case the edge theory consists of a compactified chiral boson $\phi$ of radius $R=\sqrt{n}$ and a chiral Majorana (neutral) fermion $\chi$. At a formal level the Lagrangian for the edge state(s) is a sum of two terms
\begin{equation}
\mathcal{L}=\frac{1}{4\pi}( \partial_x \phi \partial_t \phi-\varv_c (\partial_x\phi )^2)+\chi i (\partial_t-\varv_n \partial_x)\chi
\label{eq:L-edge-MR}
\end{equation}
where $\varv_c$ and $\varv_n$ are the velocities of the (charged) compactified chiral scalar $\phi$ and of the neutral Majorana field $\chi$. Numerically (and experimentally) it is known that the charge mode is (substantially) faster than the neutral mode(s), $\varv_c > \varv_n$, and often by significant factors.

A superficial reading of this Lagrangian suggests that these degrees of freedom are decoupled. However this is not correct. 
Only operators from both sectors which are local with respect to the electron $\psi_e \sim \chi \exp(i \sqrt{n} \phi)$ are physically allowed. 
Here an operator is local with respect to the electron means that the operator braids trivially with the electron. 
This condition restricts the allowed observables. In the case of the fermionic MR state, with $n=2$, the allowed operators are the non-abelion $\sigma \exp(i\phi/2\sqrt{2})$ 
with scaling dimension is $\Delta=1/8$ and  electric charge is $Q=1/4$, the Majorana fermion with scaling dimension $\Delta=1/2$ and no electric charge $Q=0$, 
and the Laughlin quasiparticle with scaling dimension $\Delta=1/4$ and charge $Q=1/2$. Its electron operator has scaling dimension $\Delta=3/2$ and charge $Q=1$.

Conformal field theory has an important defining universal quantity called the central charge which is closely related to the energy-momentum tensor of the theory \cite{difrancesco-1997}. 
For pedagogical introductions to CFT see Refs.\cite{affleck-1988,cardy-1996} and chapter 21 of Ref.\cite{fradkin-2021}. 
The energy-momentum tensor $T_{\mu \nu}$ is a locally conserved current, $\partial^\mu T_{\mu \nu}=0$. 
Its local conservation implies the global conservation of energy and momentum. As such, the energy-momentum tensor  is a fundamental observable of any quantum field theory. 
In a {\it conformal} field theory, the energy-momentum is also the generator of local scale and conformal transformations. 
In the special case of 1+1-dimensional systems, such as the edge states of the FQH fluids, the energy momentum tensor has special and crucially important properties. 
In addition of being locally conserved, conformal invariance requires that $T_{\mu \nu}$ must be traceless, $T^\mu_\nu=0$, since its trace is the generator of dilations. 
In this case, if the theory is chiral, the energy momentum tensor has only one (right-moving) component $T=(T_{00}-T_{11})/2$. In complex coordinates of the Euclidean metric, 
the correlator of the energy momentum tensor $T=T_{zz}$ is \cite{belavin-1984}
\begin{equation}
\langle T(z) T(\varw)\rangle=\frac{c/2}{(z-\varw)^4}
\label{eq:TT}
\end{equation}
where $c$ is a universal quantity known as the {it central charge} of the CFT. In the case of the compactified boson the central charge is $c=1$ whereas for a massless 
Majorana fermion $c=1/2$.

The central charge of the CFT enters in many observables of fundamental physical importance. For example, for a 1+1-dimensional  system of length $L$, 
such as the edge state of a FQH droplet,  the ground state energy $E_{\rm gnd}$ for large $L$ has the behavior \cite{affleck-1986,blote-1986}  
\begin{equation}
E_{\rm gnd}=\varepsilon_0 L-c \frac{\pi \varv}{6L} +O(1/L^2)
\label{eq:casimir}
\end{equation}
where $\varepsilon_0$ is the ground state energy density, which is a non-universal quantity, $c$ is the central charge of the CFT and $\varv$ is the velocity of the massless modes. 
The second term in Eq.\eqref{eq:casimir} is known as the Casimir energy. Similarly, the free energy density $f(T)$ of a CFT has the low temperature Stephan-Boltzmann behavior (in one dimension)
\begin{equation}
f(T)=\varepsilon_0+c \frac{\pi T^2}{6\varv} +O(T^3)
\label{eq:stephan-boltzmann}
\end{equation}
where we set the Boltzmann constant to unity. The low-temperature specific heat $c(T)$ is
\begin{equation}
c(T)=c \frac{\pi T}{3\varv} +O(T^2)
\label{eq:low-T-heat}
\end{equation}
In a chiral CFT, such as the edge states of the FQH fluids, the central charge enters in the low temperature behavior of the (Hall) thermal conductivity $\kappa_{xy}$, \cite{read-2000}
\begin{equation}
\kappa_{xy}=c \frac{\pi T}{6\hbar}
\label{eq:kappa-xy}
\end{equation}

Much of what is known about FQH states comes form experiments involving their edge states. In a series of exquisite experiments of \cite{granger-2009} measured $\kappa_{xy}$ in the edge states of a $\nu=1$ quantum Hall fluid and showed that the heat transport is indeed chiral, i.e. 
the edge state behaves as a heat ``diode''. This effect was confirmed in the FQH states by \cite{bid-2010} who, 
in addition, where used this effect to detect the neutral modes in several  Jain states at filling fractions $2/3$ and $2/5$ and in the non-abelian state at $\nu=5/2$. These experiments are a key confirmation that the edge states of fractional quantum Hall states are indeed chiral Luttinger liquids.

\subsection{Point contact tunneling and QH chiral edge states}
\label{eq:qp-contact}

The simplest experimental probes are quantum point contacts and typically are of two types: inter-edge tunneling inside a FQH liquid or tunneling of electrons into the edge state of a 
FQH liquid. Since the bulk is gapped, only the edge states participate in these tunneling processes. In the general case we will have two edges and a tunnel process at a point contact. 
Here the two edges can be either the two edges of the same FQH state, in which case the process happens inside the FQH liquid and involves tunneling of quasiparticles, or the 
edges of different liquids, in which case this process is external and involves tunneling of electrons. Problems of these types  were first investigated by  \cite{kane-1992} which led to a considerable amount of work on these problems.

Let us consider first the case of tunneling into the edge state of a $\nu=1/m$ Laughlin state from a Fermi liquid, which we will take to be a QH fluid in the $\nu=1$ state. 
The point contact is at $x=0$ The total Lagrangian is 
\begin{equation}
\mathcal{L}=\mathcal{L}_{\rm edge}+\mathcal{L}_{\rm FL}-\Gamma \; e^{i \omega_0 t}\; \psi^\dagger_{e, {\rm edge}}(0,t)\; \psi_{e,{\rm FL}}(0,t)+\textrm{h.c.}
\label{eq:tunnel1}
\end{equation}
where $\omega_0=eV/\hbar$, and $V$ is the bias voltage between the two fluids, and $\Gamma$ is a tunneling matrix element. 
The local electron spectral density (the density of states) $N(\omega)$ at energy $\omega$ is
\begin{equation}
N(\omega)=\textrm{Im} \lim_{x \to 0^+} \int_{-\infty}^\infty dt\; G_e(x, t)\; e^{i\omega t}=\textrm{const.}\;  |\omega|^{m-1}
\label{eq:electron-DOS}
\end{equation} 
where $G_e(x, t)=\langle T \psi_e^\dagger(x, t) \psi_e(0, 0)\rangle$ is the electron correlator in the FQH edge, shown in Eq.\eqref{eq:chiral-fermion-propagator}. 
This result, combined with Fermi's Golden Rule for a point contact with voltage bias $V$, predicts a tunneling current from a  Fermi liquid (FL) into the chiral Luttinger liquid (CLL) 
of the edge state to be (\cite{chamon-1997b})
\begin{equation}
I(V)=2\pi \frac{e}{\hbar} |\Gamma|^2 \int_{-eV}^0\; dE\; N_{\rm CLL} (E, T) N_{\rm FL}(E+eV, T) \propto V^m
\label{eq:I-V-curve-tunnel}
\end{equation}
and a tunneling differential conductance $G(V)$
\begin{equation}
G(V)=\frac{dI}{dV} = 2\pi \frac{e}{\hbar} |\Gamma|^2 N_{\rm FL}(0) N_{\rm CLL}(V, T)\propto V^{m-1}
\label{eq:G(V)-into}
\end{equation}
which is non-Ohmic and vanishes as $V\to 0$.

Early experiments by \cite{chang-1996} (reviewed by \cite{chang-2003}), on a geometry that (most likely) had many point contacts, confirmed the predicted CLL behavior, 
although there were discrepancies in the measured exponents. 
The behavior seen in more recent experiments by  \cite{hashisaka-2021} and \cite{cohen-2022} on a point contact in graphene, are consistent with the theoretical predictions of \cite{chamon-1997b}.  
Interestingly, these newer class of experiments  also show evidence of an analog of Andreev reflection predicted to exist near the strong coupling fixed point 
of the theory  of  \cite{sandler-1999} as a consequence of electron fractionalization.

Most experiments are done on a geometry call a Hall bar in which the QH fluid occupies a rectangular region with its length $L$ larger than its width $W$. 
In this geometry, there are chiral edge states at opposite sides of the Hall bar propagating in opposite directions. 
One type of point contact consists in creating a constriction in the quantum Hall fluid by applying a gate, a bias potential on a narrow strip accross the Hall bar. 
This gate repels the electrons in the QH fluid, forcing the opposite edges to approach each other in the proximity of the gate. 
In the absence of the gate, momentum is conserved on each edge which  forbids tunneling accross the Hall bar since the edge states have opposite Fermi momentum. 
However, the gate breaks translation invariance on both edges and tunneling between them  is now allowed. 
In other words, the gate creates a point contact between the opposite propagating edge states leading to a tunneling process between the edges of the QH liquid. 
Thus, this is internal tunneling as oppose the process of tunneling between two different liquids that we discussed above. 

The tunneling Lagrangian for this system has the same form as in Eq.\eqref{eq:tunnel1} except that now the two edges are identical. The theory Kane and Fisher shows that in the factional case the most relevant process is the tunneling of FQH quasiparticles. The Fermi Golden Rule argument used in Eq.\eqref{eq:G(V)-into} to compute the differential conductance also applies in the present case except that now the two densities of states are the densities of states of the (Laughlin) quasiparticles (instead of electrons). The local quasiparticle density of states is $N_{qp}(\omega)\sim |\omega|^{\frac{2}{m}-1}$. Then, the differential tunneling conductance $G(V)$ in this case is
\begin{equation}
G(v)\sim V^{2(\frac{1}{m}-1)}
\label{eq:tunnel2}
\end{equation}
This behavior is also non-Ohmic but, unlike the case of electron tunneling of Eq.\eqref{eq:tunnel1}, the differential tunneling conductance now {\it diverges} as $V \to 0$. This behavior means that the $\Gamma \to 0$ fixed point is {\it unstable} and that the point contact flow to a {\it strong coupling fixed point} at $\Gamma \to \infty$. Early experiments by \cite{milliken-1996} showed indications of CLL behavior. The predicted behavior was confirmed in the experiments of  \cite{roddaro-2004,roddaro-2004b}.

In the RG language, we can define a dimensionless tunneling amplitude $g$ by $\Gamma=a^{\Delta-1} g$, where $\Delta$ is the scaling dimension of the tunneling operator and $a$ is a UV cutoff. The ``tree-level'' beta function is readily found to be
\begin{equation}
\beta(g)=a\frac{\partial g}{\partial a}=(1-\Delta) \; g + O(g^2)
\label{eq:tree-level-beta-function}
\end{equation}
which shows that if the scaling dimension of the operator of the tunneling particle is $\Delta < 1$, then this tunneling process is relevant, while is $\Delta>1$ it is irrelevant.

Kane and Fisher argued  that there should be a crossover from the IR unstable weak-coupling fixed point governed by quasiparticle tunneling 
to an IR stable strong coupling fixed point governed by electron tunneling. 
This crossover is reminiscent of the the crossover in quantum impurity problems such as the Kondo problem of a magnetic impurity in a metal (\cite{anderson-1970,wilson-1975,andrei-1980,wiegmann-1980,read-1983}). 
The main difference is that the impurity coupling is marginal and the crossover scale, the Kondo temperature $T_K$, which is related to the coupling constant $g$ as $T_K \sim \exp(-1/g)$, 
while in the FQH constriction the crossover scale is $T_K(g) \sim g^{1/(1-\Delta)}$. Kane and Fisher argued that this crossover can be viewed as a process that interpolated
 between a FQH fluid with a weak constriction to a regime in which the Hall bar splits into two pieces with weak electron tunneling between them.

It turns out that, after some manipulations, the model of the constriction can be mapped into a one-dimensional compactified boson $\varphi$ on a semi-infinite line, $x\geq 0$, with a vertex operator acting at the boundary. The action of this system, known as boundary sine-Gordon is
\begin{equation}
S=\frac{1}{8\pi} \int_{-\infty}^\infty dt \int_0^\infty dx\; (\partial_\mu \varphi)^2+\Gamma_{qp} \int_{-\infty}^\infty dt\; \cos\left(\sqrt{\frac{\nu}{2}} \varphi(0, t)\right)
\label{eq:BSG}
\end{equation}
This theory has two fixed points: a) the IR unstable fixed point at $\Gamma_{qp}=0$ where the field Neumann boundary conditions at $x=0$, $\partial_x \varphi=0$, and b) an IR stable fixed point at $\Gamma_{qp} \to \infty$ where the field has Dirichlet boundary conditions, $\varphi=2\pi \sqrt{2/\nu} \; n$ (with $n\in \mathbb{Z}$). It turns out that this is an integrable field theory. Fendley, Ludwig and Saleur \cite{fendley-1995a} used the thermodynamic Bethe Ansatz to calculate the differential tunneling conductance for the Laughlin state at $\nu=1/3$ as a function of voltage $V$ and temperature $T$ and obtained the full weak to strong crossover RG flow. Detailed predictions of this theory have been verified experimentally by the work of  \cite{roddaro-2004,roddaro-2004b}. Point contact experiments were performed in the more challenging $\nu=5/2$ FQH state by  \cite{miller-2007} and  \cite{radu-2008}. They found that the MR state is the the one that best fits their experimental results.

\subsubsection{Experimental evidence of fractional charge}
\label{sec:expt-charge}

The charge of a quasiparticle can, in principle, be found directly by measuring the noise of a weak current. This process is known as shot noise. In the case of the constriction, the quasiparticle current operator is $I_{qp}=2e\nu \Gamma_{qp} \sin\left(\sqrt{\nu/2}\phi+\omega_0^*t\right)$, where $\omega_0^*=e\nu V/\hbar$. The noise spectrum, $S(\omega)$, of the tunneling current $I_{qp}$  is (\cite{kane-1994})
\begin{equation}
S(\omega)=\int_{-\infty}^\infty dt\; \langle \{ I_{qp}(t), I_{qp}(0)\}\rangle \; e^{i \omega t}
\label{eq:Sofomega}
\end{equation}
Using the expression of the quasiparticle correlator of Eq.\eqref{eq:qh-propagator}, one readily finds that, to leading order in $\Gamma_{qp}$, the noise spectrum is
\begin{equation}
S(\omega)=e\nu \langle I_{qp} \rangle \left[\left(1-\frac{\omega}{\omega_0^*}\right)^{2\nu-1}+\left(1+\frac{\omega}{\omega_0^*}\right)^{2\nu-1}\right]
\label{eq:Soomega-leading}
\end{equation}
In the limit of zero frequency the noise spectrum takes the shot noise form
\begin{equation}
\lim_{\omega \to 0}S(\omega)=2e^*\langle I_{qp}\rangle
\label{eq:shot-noise}
\end{equation}
where the expectation value of the tunneling current is
\begin{equation}
\langle I_{qp}\rangle=\frac{2\pi}{\Gamma(2\nu)}e\nu |\Gamma_{qp}|^2\; {\omega_0^*}^{2\nu-1}
\label{eq:avcurrent}
\end{equation}
Chamon, Freed and Wen \cite{chamon-1996} calculated the exact noise spectrum for the case of a $\nu=1/2$ bosonic FQH state and were able to investigate the crossover as well, and Fendley, Ludwig and Saleur \cite{fendley-1995b} used the Bethe Ansatz to construct a soliton basis to compute the DC shot noise. 

These theoretical predictions were tested in tour-de-force experiments by  \cite{de-picciotto-1997} and by \cite{samindayar-1997} who were able to measure the fractional charge of $e/3$ in the $\nu=1/3$ state and, with some caveats, $e/5$ in the $\nu=2/5$ state. Subsequent experiments by \cite{dolev-2008} went on to measure the charge from noise experiments in the nonabelian state at $\nu=5/2$ and obtained results consistent with $e^*=e/4$ as predicted by theory.
A separate experiment by \cite{yacoby-2011} used a local electrometer device to measure the electric charge in localized quasiparticles in puddles of the 2DEG at filling fractions $\nu=5/2$ and $\nu=7/3$. A statistical analysis of the data showed that the the quasiparticle charges are consistent with the theoretical predictions of $e^*=e/4$ for $\nu=5/2$ and $e^*=e/3$ for $\nu=7/3$.

\subsubsection{Quantum interferometers and fractional statistics}
\label{sec:interferometers}

We will now consider experiments on Hall bars with two quantum point contacts created by two narrow gates transversal to the bar and separated at some distance $d$ each other. Quantum devices of this type are Fabry-P\'erot quantum interferometers. The way they operate is as follows. A current is injected in the bottom edge.  At the first quantum point contact (QPC) part if the current $I_1$ tunnels to the opposite edge and the other part $I_2$ goes on and tunnels at the second QPC. The FQH fluid is assumed to occupy uniformly the region between the two QPC's. There is some magnetic  flux $\Phi$ is that region which also contains  a number of localized vortices (quasiparticles). When both currents rejoin at the top edge they interfere and the interference has information on the charge of the particles that tunnels through the Aharonov-Bohm effect with the flux $\Phi$. The interference also has information on the fractional statistics of the quasiparticles that tunneled through their braiding with the static vortices. Chamon, Freed, Kivelson, Sondhi and Wen proposed a setup of this type to measure the fractional statistics of the quasiparticles for the abelian states (\cite{chamon-1997}). They showed that 
the total current $I_t=I_1+I_2$ is given by
\begin{equation}
I_t=e^*|\Gamma_{\rm eff}|^2 \frac{2\pi}{\Gamma(2\nu)}\; |\omega_0|^{2\nu-1}\; \textrm{sign}(\omega_0^*)
\label{eq:total-current}
\end{equation}
where $\Gamma_{\rm eff}$ is give by
\begin{equation}
|\Gamma_{\rm eff}|^2=|\Gamma_1|^2+|\Gamma_2|^2+(\Gamma_1 \Gamma_2^*+\Gamma_1^*\Gamma_2) \; F_\nu\left(\frac{\omega_0^*d}{\varv}\right)
\label{eq:interference-qpcs}
\end{equation}
Here $\Gamma_1$ and $\Gamma_2$ are the tunneling amplitudes at the two QPCs, $\nu=1/m$ is the filling fraction, $\varv$ is the velocity of the edge modes, $d$ is the distance between the two QPCs and
\begin{equation}
F_\nu(x)=\sqrt{\pi} \frac{\Gamma(2\nu)}{\Gamma(\nu)} \frac{J_{\nu-1/2}(x)}{(2x)^{\nu-1/2}}
\label{eq:Fnu}
\end{equation}
where $\Gamma(z)$ is the Euler Gamma function and $J_{\nu-1/2}(z)$ is the Bessel function of the first kind. In the presence of $N_q$ localized quasiparticles in the area between the two QPCs, the contribution of the phases of the tunneling matrix elements get shifted to
\begin{equation}
\Gamma_1^*\Gamma_2=\bar \Gamma_1^*\bar \Gamma_2\; \exp\left[-2\pi i \left(\nu \frac{\Phi}{\phi_0}-\nu N_q\right)\right]
\label{eq:interference}
\end{equation}
where $\phi_0$ is the flux quantum. In Eq.\eqref{eq:interference} the first term in the phase shift is the Aharonov-Bohm effect of the tunneling quasiparticles, while the  second term in the phase shift $2\pi \nu N_q$  is the contribution of the fractional statistics of the tunneling quasiparticle as its worldline braids with the $N_q$ localized quasiparticles. This means that there is an interference contribution to the tunneling current (and also to the transmitted current) which is sensitive to both the charge and to the fractional statistics of the quasiparticles. 
This is  this effect that is being measured in experiments. 

Early attempts at doing this experiment were made by Camino and coworkers \cite{camino-2005} but were difficult to interpret partly due to subtle reasons related to the difficulty in controlling the actual area of the region comprised between the two QPCs \cite{halperin-2011}. Recent theoretical work by \cite{halperin-2011} showed that the size of the 2DEG in the region comprised by the interferometer plays an important role in the interpretation of these  experiments. Thus, if the 2DEG in th region between the two  constrictions is small, the device is in a Coulomb blockade regime. A properly working interferometer requires that the interferometer be large enough (and the 2DEG be uniform enough) not to be in this regime. These authors also provided a tell-tale pattern of the oscillations which reveal the detection of fractional statistics. Technical advances in the fabrication and design of these quantum interferometers led in 2020 to the first successful measurement of the fractional statistics of the quasiholes of the Laughlin state at $\nu=1/3$ (and also at the Jain state at $\nu=2/5$) by Nakamura, Liang, Gardner and Manfra \cite{nakamura-2020}. 

The nonabelian case is more subtle both theoretically and experimentally. The theory of the abelian interferometer of \cite{chamon-1997} was generalized  to the nonabelian FQH states by Fradkin, Nayak, Tsvelik and Wilczek in 1998 \cite{fradkin-1998}. The structure of the interferometer is the same as in the abelian case but the interference effects are different. In addition, in the case of the MR state, aside from the nonabelian quasiparticle $\sigma \sim \chi \exp(i\phi/2\sqrt{2})$ has charge $e/4$ and nonabelian fractional statistics, the MR state has  two more abelian anyons, the charge neutral Majorana fermion $\chi$ and the charge $e/2$ Laughlin quasiparticle. The number of anyons and their properties are very different in different nonabelian FQH states.

Ignoring for now the existence of more quasiparticles, let us focus now on the fundamental anyon which is nonabelian. Fradkin, Nayak, Tsvelik and Wilczek  showed \cite{fradkin-1998} considered a nonabelian quasihole that is injected to the bottom edge, tunnels at the first QPC and arrives at the left end of the top edge in state $|\psi\rangle$. If a second such quasiparticle is now injected but now tunnels to the top edge at the second QPC., arriving at the left end of the top edge in the state $e^{i\alpha} B_{N_q} |\psi\rangle$, where $\alpha$ is the Aharonov-Bohm phase determined by the flux $\Phi$ piercing the interferometer and $B_{N_q}$ si the braiding operator os the second quasiparticle that is circling around the $N_q$ localized quasiparticles in the interferometer. 
Then the tunneling conductance measured at the left exit of the top edge has an interference contribution
\begin{equation}
\sigma_{xx}\propto |\Gamma_1|^2+|\Gamma_2|^2+\textrm{Re}\left[\Gamma_1^*\Gamma_2 e^{i\alpha} \langle \psi | B_{N_q} |\psi\rangle\right]
\label{eq:nonabelian-interference}
\end{equation}
The matrix element $\langle \psi | B_{N_q} |\psi\rangle$ is given in terms of the expectation value of the Wilson loop operators of the tunneling quasiparticles braided with the Wilson loops of the $N_q$ localized quasiparticles in the area of the interferometer. In 1989 Edward Witten \cite{witten-1989} showed that this expectation value, which is computed in the nonabelian Chern-Simons gauge theory,  is equal to a topological invariant of the braid known as the  Jones polynomial $V_{N_q}(e^{1\pi/4})$.  Ref. \cite{fradkin-1998} showed that the Verlinde algebra yields a general algorithm for the computation of this matrix element. Therefore, the oscillatory component of the tunneling current (and of the conductance) measures a topological invariant!

However, in the particular case of the MR states, the non-abelian gauge theory associated with these states is the SU(2)$_2$ Chern-Simons gauge theory. In this case, an explicit calculation Bonderson, Kitaev and Shtengel \cite{bonderson-2006b} and by Stern and Halperin \cite{stern-2006} leads to the result
\begin{align}
\sigma_{xx} \propto & |\Gamma_1|^2+|\Gamma_2|^2, & \textrm{for}\; N_q \; \textrm{odd}\nonumber\\
\sigma_{xx} \propto & |\Gamma_1|^2+|\Gamma_2|^2+2\; |\Gamma_1| |\Gamma_2| (-1)^{N_\psi}\; \cos\left(\alpha+\arg\left(\frac{\Gamma_2}{\Gamma_1}\right)+\frac{\pi}{4} N_q\right), &\textrm{for}\; N_q\; \textrm{even}
\label{eq:SU2-level2}
\end{align}
Here, $N_\psi=1$ when the $N_q$ quasiparticles fuse into the state $\psi$ and $N_\psi=0$ otherwise. The interference effect is absent for $N_q$ odd since an odd number of $\sigma$ quasiparticles cannot fuse into the identity state $I$. This simple even-odd effect is special for SU(1)$_2$. In the general case and, in particular in the SU2)$_3$ case which applies to the $k=3$ Read-Rezayi (``Fibonacci'') state, the expressions are more complex.

At any rate, even in the MR state the situation is more complicated for two reasons. One is that the charge $e/2$ abelian Laughlin quasiparticle is always able to tunnel thus spoiling the even-odd effect. the other complication is that the nonabelian FQH occurs in the $N=1$ Landau level and the edge states of the nonabelian FQH state is surrounded by an abelian $\nu=2$ state, which makes accessing the interesting edge states more difficult. Robert Willett has pioneered the fabrication and operation of the interferometer for the nonabelian state at $\nu=5/2$ \cite{willett-2009,willett-2013,willett-2019}. With some significant caveats these experiments provide solid evidence  of nonabelian braiding, although more work remains to be done on this system.

A variant of the Fabry-Perot interferometer is the Mach-Zehnder well known from quantum optics. Much like the Fabry-P\'erot interferometer, this setup has a source and a drain and two ``beam splitters''. However, many features of the Fabry-P\'erot interferometer which make its operation complex, such as the role of Coulomb interactions and the control of the area of the interferometer, do not play an essential role in the Mach-Zehnder system. In a recent paper \cite{kundu-2023} report experiments in an electronic Mach-Zehnder interferometer, developed originally by \cite{ji-2003}, done in fractional quantum Hall states at $\nu=1/3$ and $\nu=2/5$, and found values of the fractional statistics consistent with theoretical predictions.

Quantum Mechanics predicts that in  collision experiments of identical particles bosons tend to have enhanced correlations localized in space, effect known as ``bunching'', while fermions exclude each other , due to the exclusion principle, an effect known as ``anti-bunching''. The bunching effect of photons was first confirmed by the celebrated experiments of Hanbury Brown and Twiss \cite{hanbury-brown1956}. In 2003 S. Vishveshwara proposed a collider-type setup for anyons in a quantum Hall bar with a ``beam'' of anyons created by two sources colliding at a ``beam splitter', and detected at two sinks \cite{vishveshwara-2003}. In that setup fractional statistics is manifest in the cross correlations of the currents detected at the two sinks. Such correlations are present in the noise (quantum fluctuations) of the detected currents. There were several subsequent proposals for detecting the role of fractional statistics in current noise, for example in quantum point contacts at a tri-junction \cite{kim-2005}. Further theoretical work on the collider setup by \cite{vishveshwara-2010} and by \cite{rosenow-2016} refined and clarified the role of fractional statistics in anyon collider setups. An experiment by \cite{bartolomei-2020} confirmed the theoretical prediction that, compared to fermions,  anyons of a fractional quantum Hall state with filling fraction $\nu=1/3$ have a bunching tendency (compared to fermions) at a beam splitter consistent with the predicted  fractional statistics $\pi/3$.

\section{Particle-Vortex Dualities in 2+1 dimensions}
\label{sec:pv-2+1}

\subsection{Electromagnetic Duality}
\label{sec:EM-duality}

 The oldest form of duality in Physics is, perhaps, Dirac's observation that in the absence of electric charges and currents Maxwell's equations are invariant under the exchange of electric and magnetic fields, ${\bm E} \to {\bm B}$ and $\bm B \to - \bm E$ \cite{dirac-1931}. This observation led him to conjecture the existence of magnetic monopoles. In a relativistic invariant formulation, Maxwell's equations in free space can be expressed in terms of  the electromagnetic field tensor $F_{\mu \nu}$ and the dual field tensor $F_{\mu \nu}^*$
 \begin{equation}
 \partial^\mu F_{\mu \nu}=0, \qquad \partial^\mu F_{\mu \nu}^*=0
 \label{eq:maxwell-selfdual}
 \end{equation}
 where $F_{\mu \nu}^*=\frac{1}{2} \epsilon_{\mu \nu \lambda \rho} F^{\lambda \rho}$, and where $ \epsilon_{\mu \nu \lambda \rho}$ is the  fourth-rank antisymmetric  Levi-Civita tensor,
 The first Maxwell equation in Eq.\eqref{eq:maxwell-selfdual} is just the wave equation in free Minkowski spacetime. The second equation in Eq.\eqref{eq:maxwell-selfdual} is  known as the Bianchi identity. The Bianchi identity is a {\it constraint} which implies that there are no magnetic monopoles and that the second rank antisymmetric  field strength tensor  can be expressed in terms of the vector potential $A_\mu$ as $F_{\mu \nu}=\partial_\mu A_\nu-\partial_\nu A_\mu$. 
 
This is an example of what in differential geometry is called {\it Hodge duality}, which relates a vector or, more generally a tensor field to its Hodge dual. In general in $D$ dimensions the Hodge dual of a (fully antisymmetric) tensor of rank $p$ is an antisymmetric tensor of rank $D-p$. When contracted with the oriented infinitesimal element of a $p$-dimensional hypersurface, $dx_1 \wedge dx_2 \wedge \ldots \wedge dx_p$, an  antisymmetric tensor of rank $p$, $F_{\mu_1\mu_2 \ldots \mu_p}$, defines a $p$-differential form, called a $p$-form. Differential forms embody the physically intuitive notions of circulation of a vector field, flux of a second rank tensor, etc. We will see below that duality transformations are closely related to these geometric notions of duality.

\subsection{Particle-Vortex Duality in 2+1 dimensions}
\label{sec:PV-duality}

In this section we will extend the particle-vortex duality discussed in 2D in section \ref{sec:2d-vortices} (see Eq.\eqref{eq:particle-vortex-duality-2D}) to 3D. In the field theory interpretation, 2D is a 1+1-dimensional Minkowski spacetime and the XY model is a representation of a complex scalar field of unit modulus. In the duality, the particles are the particle-like excitations of the complex scalar field. The high temperature phase of the XY model is viewed as a partition function of a set of oriented loops that carry  charge. the loops are the worldlines of the particles of the XY model. 

\subsubsection{The 3D XY Model}
\label{sec:3DXY}

This picture can be seen as follows. Consider an XY model on a D-dimensional hypercubic lattice whose sites are labelled by $\{\bm r \}$. At each site there is a periodic variable $\theta(\bm r) \in [0, 2\pi)$. The partition function is
\begin{equation}
Z_{\rm XY}=\prod_{\bm r} \int_0^{2\pi} \frac{d\theta(\bm r)}{2\pi} \; \exp\left(\frac{1}{T} \sum_{\bm r} \sum_{j=1, \ldots, D} \left(\cos \Delta_j \theta(\bm r)\right)\right)
\label{eq:PF-XY-D}
\end{equation}
where $\Delta_j \theta(\bm r)=\theta(\bm r + \bm e_j)-\theta(\bm r)$, with $j=1,\ldots, D$, is the lattice difference, and $T$ is the temperature (in the classical statistical mechanical picture). Since the interaction on each link is a periodic function of the phase difference $\Delta_j \theta(\bm r)$, we can expand the Gibbs weight (for each link!) in a Fourier series or, what is the same, in the integer-valued representations of the group U(1) (which is the global symmetry of the XY model) to obtain
\begin{equation}
Z_{\rm XY}=\prod_{\bm r} \int_0^{2\pi} \frac{d\theta(\bm r)}{2\pi} \sum_{\ell_j(\bm r)=-\infty}^\infty  \exp\left(-\sum_{\bm r, j} \frac{T}{2} \ell_j^2(\bm r)+i \sum_{\bm r} \theta(\bm r) \Delta_j \ell_j(\bm r)\right)
\label{eq:XY-Fourier}
\end{equation}
where we used a Gaussian  approximation  for  the modified Bessel function  $I_n(z)$,
and $\Delta_j \ell_j(\bm r)=\sum_{j=1}^D(\ell_j(\bm r)-\ell_j(\bm r-\bm e_j))$ (where $\bm e_j$ is the lattice unit vector along the direction $j$) denotes the lattice divergence. Integrating-out the phase variables $\theta(\bm r)$ we obtain
\begin{equation}
Z_{\rm XY}=\prod_{\bm r} \sum_{\ell_j(\bm r)=-\infty}^\infty\; \prod_{\bm r} \delta(\Delta_j \ell_j(\bm r)) \exp\left(-\sum_{\bm r} \sum_{j=1}^D \frac{T}{2} \ell^2_j(\bm r)\right)
\label{eq:XY-highT}
\end{equation}
Therefore, the partition function is given by a sum over loops of  conserved currents $\ell_j(\bm r)$ defined on the links of the lattice, with a weight on each link  $\propto \exp(-\ell_j^2/2\beta)$. These loops are the (lattice) worldlines of the particles of the complex scalar field. In the phase where this representation is convergent, the complex scalar field is massive, the  XY model is gapped,  and these particles have short range interactions. This picture is true in all dimensions, 3D included. 

On the other hand, in 2D the vortices are point-like events in Euclidean spacetime which interact with each other through long range, logarithmic, interactions.  In the field theory language, in 2D the point-like vortices are {\it instantons} which govern the low temperature phase of the XY model.
In spacetime dimensions $D>2$ the vortices  become extended objects: vortex loops (or strings) in 3D, closed vortex surfaces in 4D, etc.  In  3D the vortex loops are magnetic flux tubes which interact with each other through a Biot-Savart type interaction, namely  bits of vortices interact with each other with a  3D Coulomb interaction much in the same way as with loops of current in magnetostatics,
\begin{equation}
Z_{\rm 3D XY}=\sum_{\{ m_j(\tilde{\bm r}\} } \prod_{\tilde{\bm r}} \delta(\Delta_j m_j(\tilde{\bm r})) \exp\left(-\frac{2\pi^2}{T} \sum_{\tilde{\bm r}, \tilde{\bm r}'} \sum_{j=1}^3 m_j(\tilde{\bm r}) G_0(\tilde {\bm r}- \tilde{\bm r}') m_j(\tilde{\bm r}')-\alpha \sum_{\tilde{\bm r}, j} m_j^2(\tilde{\bm r})\right)
\label{eq:vortex-loops}
\end{equation}
where $\{ \tilde{\bm r} \}$ labels the sites of the dual (cubic) lattice, the variables $m_j(\tilde{\bm r})$ take values on the integers, $\alpha$ is a (short-distance) vortex core energy, and $G_0(\tilde {\bm r}- \tilde{\bm r}')$ is the 3D lattice propagator (Green function)  which at long distances has the standard form
\begin{equation}
G_0(\bm x- \bm x')=\frac{1}{4\pi |\bm x - \bm x'|}
\label{eq:3D-coulomb}
\end{equation}
In Eq.\eqref{eq:vortex-loops} the vortex loops are represented by the integer-valued conserved currents $m_j(\tilde{\bm r})$, which are naturally interpreted as the world lines of magnetic charges.

We could have also reached the same result by following the line of reasoning that we used in section \ref{sec:2d-vortices}. Indeed, let $\theta(x)$ be the phase field of a complex scalar field $\phi(x)$ deep in its broken symmetry state where the amplitude of the field $\phi(x)$ can be taken to be approximately fixed. The partition function in this phase reduces to
\begin{equation}
Z[a_\mu]=\int \mathcal{D} \theta \; \exp\left(-\frac{1}{2g} \int d^3x  \left(\partial_\mu \theta -a_\mu\right)^2\right)
\label{eq:3D-eff-action-XY}
\end{equation}
where $g$ is a coupling constant which in the XY model is proportional to the temperature. Here, much as in Eq.\eqref{eq:Z[A]}, the gauge field $a_\mu$ represents the vortices. Indeed, in 3D the vorticity is represented by a locally conserved vector field $\omega_\mu$
\begin{equation}
\omega_\mu=\epsilon_{\mu \nu \lambda} \partial_\nu a_\lambda=2\pi \sum_k m_\mu^k \delta(x-x_k)
\label{eq:3D-vorticity}
\end{equation}
where $m_\mu^k$ are the integer-valued vortex (magnetic) currents we used above. As before, the periodic nature of the phase field $\theta(x)$ implies that the vortex currents must be quantized and be integer-valued.

Here too we can perform a Hubbard-Stratonovich transformation in terms of  a vector field $b_\mu$ to find a dual theory which now is
\begin{equation}
Z(a_\mu)=\int \mathcal{D}b_\mu \mathcal{D} \theta \exp\left(-\frac{g}{2} \int d^3x \; b_\mu^2-i \int d^3x \, b_\mu (\partial_\mu \theta-a_\mu)\right)
\label{eq:HS-3D}
\end{equation}
Upon integrating out the phase field $\theta$, which acts as a Lagrange multiplier which imposes the constraint $\partial_\mu b_\mu=0$. This constraint implies that we can  write the field $b_\mu$ in terms of a dual gauge field $\vartheta_\mu$
\begin{equation}
b_\mu=\epsilon_{\mu \nu \lambda} \partial_\nu \vartheta_\lambda
\label{eq:vartheta-3D}
\end{equation}
Therefore, we can rewrite the partition function as
\begin{equation}
Z(a_\mu)=\int \mathcal{D} \vartheta \; \exp \left(-\frac{g}{4}\int d^3x \; f^2_{\mu \nu}+i \int d^3x \; \omega_\mu \vartheta_\mu\right)
\label{eq:dual-3D}
\end{equation}
where $f_{\mu \nu}=\partial_\mu \vartheta_\nu-\partial_\nu \vartheta_\nu$ is the field strength of the dual gauge field $\vartheta_\mu$. Notice that the vortex current $\omega_\mu$ is minimally coupled to the dual gauge field $\vartheta_\mu$. If we now integrate-out the gauge field $\vartheta_\mu$ we obtain an expression for the weight in the path integral for a configuration of vortices $m_\mu^k$ which is identical to the result of Eq.\eqref{eq:vortex-loops}.

What we have shown above is that in 3D the Goldstone phase of a complex scalar field with coupling constant $g$ is the dual of a Maxwell gauge theory. with coupling constant $1/g$.  Moreover, the periodicity of the phase field implies that the dual gauge field $\vartheta_\mu$ is compact in the sense that its fluxes must obey flux quantization. It is straightforward to show that a charge operator $V_n=\exp(i n \theta)$ with electric charge $n$ in the scalar field theory is represented in the dual gauge theory by a magnetic monopole of magnetic charge $n$.

In summary we showed that the 3D XY model (equivalent to the theory of the complex scalar field) can be written in terms of two different models of loop configurations. We saw that the XY model is a theory of particle (electric) loops in the high temperature phase and of vortex (magnetic) loops in the low temperature phase. However the particle loops have short range interactions while the vortex loops have long range interactions. Thus, the 3D XY model is not self-dual. In spite of having a representation in terms of vortex loops, the 3D XY model is in reality has a very different behavior that the 2D system. The Kosterlitz-Thouless theory describes the phase transition in terms of a vortex-antivortex unbinding transition upon which the vortices proliferate. In addition, in 2D the Goldstone phase is actually a line of fixed points, there is never true long range order but, instead, power-law correlations of the physical observables. In the Kosterlitz-Thouless theory the disordered phase arises the strong fluctuations of the phase field due to the proliferation of vortices while, at the local level, the order parameter still has a finite magnitude. A  consequence of this behavior is that the superfluid density has a universal jump at the Kosterlitz-Thouless transition \cite{nelson-1977}. 

In contrast in the 3D XY model the Goldstone phase is a true phase with long range order. The theory has a continuous phase transition (the thermodynamic superfluid transition) at which the superfluid density vanishes continuously. This means that the phase transition of the 3D XY is better described by the Wilson-Fisher fixed point of a complex scalar field \cite{wilson-1972, amit-book}, rather than by a vortex proliferation picture of the Kosterlitz-Thouless theory \cite{kosterlitz-1973, kosterlitz-1974}. In particular, since this theory does not have magnetic monopoles, the vortex loops cannot proliferate as they do in 2D. Instead, as the phase transition is approached, the vortex loops grow large in size but also become fractal-like objects whose effective core size diverges as the transition is approached. In fact, qualitative vortex proliferation arguments readily lead to the incorrect conclusion that the transition should be (strongly) first order \cite{thomas-1978}.

In spite of these differences, particle-vortex duality still plays an important role in 3D by relating the 3D XY model to another theory which is its dual  under electromagnetic duality. Here electromagnetic duality is understood as the exchange of the electric worldlines (electric loops) and the magnetic loops (vortex loops) while exchanging strong and weak coupling, $T \leftrightarrow 1/T$,  in this case between different theories.  This duality was investigated by Michael Peskin \cite{peskin-1978}, by  Paul Thomas and Michael Stone \cite{thomas-1978}, and by Chandan Dasgupta and Bertrand Halperin \cite{dasgupta-1981}.

\subsubsection{Scalar QED in 3D}
\label{eq:scalar-QED}

These authors considered a theory of a superconductor represented by a complex scalar field $\varphi(x)$, coupled to a fluctuating electromagnetic (Maxwell) field, also known as the abelian Higgs model, or scalar electrodynamics (scalar QED) $a_\mu$, with $\mu=1,2,3$ The partition function of this theory is
\begin{equation}
Z_{\rm SC}=\int \mathcal{D} \varphi  \mathcal{D} \varphi^* \mathcal{D}a_\mu \exp\left(-\int d^3x \; \mathcal{L}[\varphi, \varphi^*,a_\mu]\right)
\label{eq:3D-scalar-QED}
\end{equation}
where
\begin{equation}
\mathcal{L}=|D_\mu(a) \varphi |^2 +m^2 |\varphi |^2+u |\varphi |^4+\frac{1}{4e^2} f^2_{\mu \nu}
\label{eq:L-scalar-electrodynamics}
\end{equation}
where the covariant derivative is $D_\mu(a)=\partial_\mu+iqa_\mu$, with the integer $q$ being the charge of the scalar field (in units of the coupling constant $e$), and $f_{\mu \nu}=\partial_\mu a_\nu-\partial_\nu a_\mu$ is the field strength. This theory is known as (Euclidean) scalar electrodynamics (or, the abelian Higgs model). 

The scalar electrodynamics  was extensively studied using the perturbative renormalization group within the epsilon expansion (near 4 dimensions) which leads to the conclusion that it has a weakly (fluctuation-induced) first order phase transition \cite{coleman-1973, halperin-1974}. Scalar QED of an $N$-component scalar field coupled to a Maxwell gauge field has a continuous phase transition with a non-trivial fixed point for $N \geq N_c \simeq 183$  (!) \cite{moshe-2003}. Since these results rely on perturbation theory (or in the large-$N$ limit) it was long  suspected that the physics may be different in $D=3$. We will see that particle-vortex duality provides the answer to this question \cite{dasgupta-1981}.

We begin by writing the lattice version of Eq.\eqref{eq:3D-scalar-QED} which is obtained by coupling the XY model of Eq.\eqref{eq:PF-XY-D} to a dynamical (Euclidean) Maxwell field
\begin{equation}
Z_{\rm SC}=\prod_{\bm r} \int_0^{2\pi} \frac{d\theta(\bm r)}{2\pi}  \int_{-\infty}^\infty \frac{da_\mu(\bm r)}{2\pi}\; 
\exp\left(-S(\theta, a_\mu)\right)
\label{eq:3D-lattice-scalar-QED}
\end{equation}
where the action $S(\theta,a_\mu)$ is
\begin{equation}
S(\theta,a_\mu)=-\frac{1}{T} \sum_{\bm r, \mu}\cos \left( \Delta_\mu\theta(\bm r)-qa_\mu(\bm r)\right)
+\frac{1}{4e^2} \sum_{{\bm r}, \mu, \nu} \left(\Delta_\mu a_\nu(\bm r)-\Delta_\nu a_\mu(\bm r)\right)^2
\label{eq:3D-lattice-scalar-QED-action}
\end{equation}
In this action we  assumed that the abelian gauge fields do not have monopole configurations.

The partition function of Eq.\eqref{eq:3D-scalar-QED} admits a representation as a sum over loops. We can then used the same line of reasoning that led to Eq.\eqref{eq:XY-highT} and write the partition function as a sum over loops of the worldlines of the particles (charges) of the complex scalar field
\begin{equation}
\frac{Z_{\rm SC}}{Z_0^{\rm gauge}}= \prod_{\bm r} \sum_{\ell_\mu(\bm r)=-\infty}^\infty\; \delta(\Delta_\mu \ell_\mu(\bm r)) \exp\left(-\sum_{\bm r, \mu} \frac{T}{2} \ell^2_\mu(\bm r)\right)\Big< \exp \left(i q \sum_{\bm r, \mu} \ell_\mu(\bm r) a_\mu(\bm r)\right)\Big>_{a}
\label{eq:SC-highT}
\end{equation}
where $\langle \mathcal{O}[a] \rangle_a$ is the expectation value of the operator $\mathcal{O}[a]$ over the gauge fields $a_\mu$, and $Z_0^{\rm gauge}$ is the partition function of the free gauge fields. After computing this free-field expectation value we obtain the result
\begin{equation}
\frac{Z_{\rm SC}}{Z_0^{\rm gauge}}=  \sum_{\{ \ell_\mu(\bm r)\}=-\infty}^\infty\;  \delta(\Delta_\mu \ell_\mu(\bm r)) \exp\left(-\sum_{\bm r, \mu}  \frac{T}{2} \ell^2_\mu(\bm r)-\frac{q^2 e^2}{2}\sum_{\bm r, \mu} \sum_{\bm r', \nu} \ell_\mu(\bm r) \; G_{\mu \nu}(\bm r-\bm r')\; \ell_\nu(\bm r')\right)\
\label{eq:SC-highT-loops}
\end{equation}
where
\begin{equation}
G_{\mu \nu}(\bm r-\bm r')=\langle a_\mu(\bm r) a_\nu(\bm r')\rangle
\label{eq:A-propagator}
\end{equation}
is the Euclidean propagator of the free gauge fields. Since the configurations of the worldlines, the currents represented by $\ell_\mu(\bm r)$, are conserved and  satisfy the local constraint $\Delta_\mu \ell_\mu=0$, the second term in the exponent of Eq.\eqref{eq:SC-highT-loops} is gauge invariant. Hence, we can use the propagator in the Feynman gauge  
\begin{equation}
G_{\mu \nu}(\bm r-\bm r')=\delta_{\mu \nu} G_0(\bm r-\bm r')
\label{eq:propagator-feynman-gauge}
\end{equation}
where $G_0(\bm r-\bm r')$ is the 3D lattice propagator which has the same Coulomb form at long distances already given in Eq.\eqref{eq:3D-coulomb}.

Therefore we can write the partition function of the scalar QED model (the superconductor) as a sum over worldline loop configurations of the particles of the scalar field in the equivalent form
\begin{equation}
\frac{Z_{\rm SC}}{Z_0^{\rm gauge}}= \sum_{\{ \ell_\mu(\bm r)\} =-\infty}^\infty\; \delta(\Delta_\mu \ell_\mu(\bm r)) \exp\left(-\sum_{\bm r, \mu} \frac{T}{2} \ell^2_\mu(\bm r)-\frac{q^2 e^2}{2}\sum_{\bm r, \bm r', \mu} \ell_\mu(\bm r) \; G_0(\bm r-\bm r')\; \ell_\mu(\bm r')\right)\
\label{eq:SC-highT-loops2}
\end{equation}
which is the same as the partition function of the 3D XY model in the broken symmetry state, given in Eq.\eqref{eq:vortex-loops}, with the identification of the particle loops of the abelian Higgs model with the vortices of the 3D XY model and a relation between the coupling constants. Hence we obtain the duality between the broken symmetry phase of the 3DXY model (``low T'') and the symmetric phase of the abelian Higgs model (``high T'')
\begin{equation}
\ell_\mu \leftrightarrow m_\mu, \qquad q^2e^2 \leftrightarrow \frac{2\pi^2}{T}, \qquad \frac{T}{2}\leftrightarrow \alpha
\label{eq:3D-pv-duality1}
\end{equation}
where the quantities on the left hand side of the identifications refer to the 3D abelian Higgs model and the quantities on the right hand side to the 3D XY model.

We can now repeat the same analysis but in the broken symmetry of the abelian Higgs model. In this phase the gauge field $a_\mu$ becomes massive by the Higgs mechanism (or, what is the same, by the Meissner effect of the superconductor), and in this phase the vortex loops $m_\mu$ of the complex scalar field have short range interactions. This phase then is mapped onto the unbroken phase of the 3DXY model with the screened vortex loops of the abelian Higgs model identified with the particle loops of the 3D XY model, and with the same identifications between the coupling constants of the two theories given in Eq.\eqref{eq:3D-pv-duality1}.
The identification between the two partition functions implies that the phase transitions must be the same. In other words, the duality implies that the 3D abelian Higgs model has a continuous phase transition with the same fixed point as that of the complex scalar field in 3D. The only difference is that the phases are reversed and, for this reason, this is sometimes called an ``inverted'' 3D XY transition \cite{dasgupta-1981}. 

\subsubsection{The duality mapping}
\label{eq:pv-duality-mapping}

We can summarize these results with the following identification of two Lagrangians (in real time, Minkowski signature) \cite{seiberg-2016}
\begin{equation}
|\partial_\mu(B) \phi|^2-m^2 |\phi|^2-u|\phi|^4 \leftrightarrow |\partial_\mu(a) \varphi|^2+m^2 |\varphi|^2-u|\varphi|^4-\frac{1}{4e^2} f_{\mu \nu}^2+\frac{1}{2\pi} \epsilon_{\mu\nu\lambda} a^\mu \partial^\nu B^\lambda
\label{eq:pv-duality}
\end{equation}
The left hand side of this equivalence is the Lagrangian of the 3D complex scalar field $\phi$ and $B_\mu$ is a background electromagnetic gauge field. The right hand side is the Lagrangian of the abelian Higgs model with a scalar field $\varphi$ and a U(1) gauge field $a_\mu$. Notice that the mass terms of the two sides are inverted reflecting the reverse order of their two phases. This identification also shows that the current of the scalar field $j_\mu=-\frac{i}{2} (\phi^* \partial_\mu \phi-\phi \partial_\mu \phi^*)$ is identified as
\begin{equation}
j_\mu \leftrightarrow \frac{1}{2\pi} \epsilon_{\mu \nu \lambda} \partial_\nu a_\lambda
\label{eq:duality-current}
\end{equation}
The field operator of the complex scalar field  creates worldlines (as opposed to loops) of charged particles. In the dual abelian Higgs theory this operator is identified with an operator that creates a magnetic monopole  of the  U(1) gauge field $a_\mu$ with unit magnetic charge. It is easy to see that in the broken symmetry phase of the abelian Higgs model a monopole-antimonopole pair creation operator decays exponentially with distance due to the flux expulsion effect (the Meissner effect of a superconductor). This means that  the monopoles are confined with a linear potential. This is the same behavior of the complex scalar field $\phi$ in its symmetric phase where the propagator of the complex scalar field decays exponentially with distance. Conversely, in the broken symmetry phase of the complex scalar field theory, the field operator $\phi$ condenses and its propagator approaches the value $|\langle \phi \rangle|^2$ at long distances. This is the same behavior one readily finds for the monopole operator of the abelian Higgs model in the symmetric phase .

\subsection{Bosonization  in 2+1 dimensions}
\label{sec:fermionization}

The particle-vortex duality that we discussed in section \ref{sec:PV-duality} essentially has been understood since the 1980s. In section  \ref{sec:bosonization-rules} we discussed in detail the problem of bosonization  in 1+1 dimensions whic is a mapping between a massless Dirac fermion and a massless compactified scalar field. As we noted, this mapping is a powerful theoretical tool to investigate non-perturbatively the structure of many theories in 1+1 dimensions of great physical interest. This success has motivated a sustained effort to extend these concepts to higher dimensions where the problem in many ways is more subtle.

Fermi systems in dimensions higher differ from the 1+1-dimensional case in two significant ways. Due to the kinematic restriction of one space dimension a massless relativistic fermion and a theory of fermions (relativistic or not) at finite density are mostly  equivalent to each other. The reason is quite simple. Consider a free massless Dirac theory in 1+1 dimensions. As we saw, it is equivalent to a theory of two chiral fermions, the right and left moving component of the spinor. If we add a chemical potential $\mu$, the right and left moving states will be filled up to $\mu$, which is the Fermi energy. While in space dimensions $d>1$ a finite chemical potential leads to a finite fermi surface, in $d=1$ space dimensions the Fermi ``surface" is just two points, for the right and left moving fermion modes respectively. The Dirac Hamiltonian at finite chemical potential is
\begin{equation}
\mathcal{H}=\psi^\dagger (-i) \alpha \partial_x \psi-\mu \psi^\dagger \psi
\label{eq:H-dirac-chempot}
\end{equation}
Under a chiral transformation with angle $\theta(x)$, the field operators change  as $\psi_R \to e^{i\theta} \psi_R$ and $\psi_L \to e^{-i\theta} \psi_L$ and the Hamiltonian becomes
\begin{equation}
\mathcal{H}=\psi^\dagger (-i) \alpha \partial_x \psi-(\mu-\partial_x \theta)\; \psi^\dagger \psi
\label{eq:H-dirac-chempot-transformed}
\end{equation}
Clearly if we choose $\partial_x \theta=\mu$ (or, what is the same, $\theta=\mu x$) the explicit dependence on the chemical potential has been cancelled and the transformed Dirac Hamiltonian is the same as the Hamiltonian at zero chemical potential. However under this transformation the mass operators transform as $\bar \psi \psi \to \cos (2\mu x) \; \bar \psi \psi$ and $i\bar \psi \gamma_5 \psi \to \sin(2\mu x)\; i\bar \psi \gamma_5$. In the non-relativistic context this change amounts to  a modulation of the charge density with wave vector $Q=2p_F(\mu)$.

In contrast, in space dimensions $d>1$ the massless Dirac system and a system of fermions at finite density (relativistic or not) are no longer  equivalent to each other as the latter system has a Fermi surface (of co-dimension $d-1$) wile the former system does not. This difference has profound effects on their dynamics: the system of fermions at finite density is, at least in the weak coupling regime, is expected to be a Fermi liquid (except for an instability to a superconducting state even for infinitesimal attractive interactions) while the massless Dirac theory is stable as all local interactions are irrelevant.

\subsubsection{Bosonization of the Dirac theory in  2+1 dimensions}
\label{sec:bosonization-dirac-2+1}

We will now turn to the problem of Bose-Fermi mappings in relativistic systems in 2+1 dimensions. 
This problem is of great interest both in Condensed Matter Physics and in High Energy Physics.

Bosonization in 2+1 dimensions is a more subtle problem than in 1+1 dimensions that we discussed in section \ref{sec:bosonization-rules}.
There we saw, in 1+1 dimensions bosonization consists of a set of operator identities relating two free field theories, 
a theory of massless Dirac fermions and a compactified massless free scalar field $\phi(x)$. The success of that program is largely based on the fact  that both theories 
are massless (and hence are scale and conformally invariant), on  the chiral anomaly, 
and on the relation between the gauge current of the Dirac theory with  the topological current of the compactified boson.
 
The physics in 2+1 dimensions (and in general) is very different. 
For instance, instead of the chiral anomaly in 2+1 dimensions theories of relativistic fermions have a parity anomaly, discussed in section \ref{sec:parity-anomaly}.
We will also see that although in 2+1 dimensions there are theories of free massless Dirac fermions and free massless compactified bosons they are no longer dual to each other.
 For these and other reasons it has been difficult to extend the bosonization program to 2+1 dimensions. 
 
 One of the first bosonization constructions  for a  theory of massive  (relativistic) Dirac fields was put forth by Alexander Polyakov in 1988 \cite{polyakov-1988,grundberg-1990}. 
 For reasons that no longer relevant, Polyakov considered a theory of $CP^1$  complex scalar fields (in its unbroken phase) coupled to a U(1)$_1$ Chern-Simons gauge theory. 
 Nevertheless, his main argument, with some corrections associated with the parity anomaly, still holds correct.
 
 Here we will follow the work of Hart Goldman and myself \cite{goldman-2018} and consider instead a theory of a single massive complex field  coupled to a U(1)$_1$ Chern-Simons gauge theory whose Lagrangian density is
 \begin{equation}
 \mathcal{L}=|D_\mu(\mathscr{A}_\mu) \phi|^2-m^2 |\phi|^2-\lambda |\phi|^4+\frac{1}{4\pi} \epsilon_{\mu \nu\lambda}\mathscr{A}^\mu \partial^\nu \mathscr{A}^\lambda
 \label{eq:csf+cs}
 \end{equation}
 In essence, this theory is a relativistic version of what we did in section \ref{sec:CB-QFT-FQH} where we mapped non-relativistic fermions to (also non-relativistic) bosons whereas here the theory is relativistic and we are mapping bosons to fermions. In his work Polyakov actually considered the problem of the propagator of a free massive scalar field coupled to the Chern-Simons gauge field. In this limit, the propagator $G(x, x')$ is the transition amplitude from $x$ to $x'$ (in Minkowski spacetime). 
In the case in which $\phi(x)$ is a free massive field this propagator can be expressed in the for of a Feynman path-integral as a sum over all open  oriented  paths $\{ P_{x, x'} \}$ with endpoints at $x$ and $x'$ which in Euclidean spacetime is 
 \begin{equation}
 G_{\gamma(x, x')}=\sum_{\{ P_{x, x'}\}} \exp(-mL(P_{x, x'}))\; \Big< \exp\left(i \int_{P_{x,x'}} dz_\mu \mathscr{A}^\mu(z)\right)\Big>_{U(1)_1}
 \label{eq:sum-paths-prop-CS}
 \end{equation}
 where the expectation value is over the Chern-Simons gauge fields $\mathscr{A}_\mu$ of Eq.\eqref{eq:csf+cs}.
 
Let us consider now the partition function of the bosons which is a sum over all closed paths $\{ P \}$, which we will assume to be non-intersecting (which requires a short distance repulsion)
\begin{equation}
Z=\sum_{\{ P \} } \exp(-m L(P))\; \Big<\exp(i \oint_{P} dz_\mu \mathscr{A}^\mu(z))\Big>_{U(1)_1}
\label{eq:closed-paths-boson}
\end{equation}
The expectation value is given by
\begin{equation}
\Big<\exp\left(i \int_{P} dz_\mu \mathscr{A}^\mu(z)\right)\Big>_{U(1)_1}=\exp\left(-\frac{1}{2} \oint_P \oint_P dx_\mu dy_\nu \langle \mathscr{A}_\mu(x) \mathscr{A}_\nu(y)\right)
\equiv \exp(i \pi W(P))
\label{eq:writhe}
\end{equation}
where $W(P)$ is the {\it writhe} of the closed path $P$. The computation of the writhe requires a regularization. One possible regularization is to thicken the path which is equivalent 
to including a Maxwell term for the gauge field in the Lagrangian. In general, the writhe of the path $P$ is
\begin{equation}
W(P)=SL(P)-T(P)
\label{eq:writhe2}
\end{equation}
where $SL(P)$ is an integer-valued  topological invariant called the self-linking number and $T(P)$ is the twist (or torsion) of the path. The twist $T(P)$ is a Berry phase which depends on the coordinates of the space in which the path $P$ is embedded, and it is not a topological invariant. 

We will now compute the Berry phase $T(P)$. Let $\hat {\bm e}(s)$ be a unit vector tangent to the path $P$ and where we parametrized the closed path $P$ by a coordinate $s\in [0, L]$. 
The closed path $P$ is the boundary of a surface $\Sigma$ we can take to be a disk. 
We will write the Berry phase by extending $\hat {\bm e}(s)$ smoothly  to the interior of $\Sigma$ as $\hat {\bm e}(s, u)$, 
where $0\leq u \leq 1$ and $\hat {\bm e}(s, u=1)=\hat {\bm e}(s)$ and $\hat {\bm e}(s, u=0)=\hat {\bm e}_0\equiv$ constant. With these definition the Berry phase is
\begin{equation}
T(P)\equiv W(\hat {\bm e})=\frac{1}{2\pi} \int_0^L ds \int_0^1du\; \hat {\bm e}\cdot \partial_s \hat {\bm e} \times \partial_u \hat {\bm e}
\label{eq:berry-twist}
\end{equation}
which is defined modulo an integer.

In this formulation,  in the bosonic theory the amplitude for a path of length $L$ with tangent vector $\hat { e}(s)$ with endpoints at $x$ and $x'$ is
\begin{equation}
G(x-x')=\int_0^\infty dL \int \; \mathcal{D}\; \hat {\bm e}(s)\; \delta\left(1-|\hat {\bm e}|^2\right)\; \delta\left(x-x'-\int_0^L ds\; \hat {e}(s)\right)\; \exp(-|m|L\pm i \pi W(\hat {\bm e}))
\label{eq:propagator-sum}
\end{equation}
In momentum space we find
\begin{equation}
G(p)=\int_0^\infty dL \int \mathcal{D} \hat e \; \delta(1-|\hat e|^2)\; \exp(-|m|L \pm i \pi \; W(\hat e))\; \exp(i p^\mu \int_0^L ds\; \hat e_\mu(s))
\label{eq:G(p)}
\end{equation}
By inspection of Eq.\eqref{eq:berry-twist} we see that this is the same expression that we looked in the theory of the path integral for spin  in section \ref{sec:spin-S}.
for  spin $S=1/2$ particle in a magnetic field $b_\mu=\pm 2p_\mu$. 

The equation of motion for $\hat e$ is 
\begin{equation}
\partial_s \hat e_\mu=\pm 2 \epsilon_{\mu \nu \lambda}\hat e^\lambda=i[H, \hat e_\mu]
\end{equation}
 where $H=\mp p_\mu \hat e^\mu$ is the Hamiltonian. Upon quantization, $\hat e_\mu$ satisfies the commutation relations 
\begin{equation}
 [\hat e_\mu, \hat e_\nu]=2i \epsilon_{\mu \nu \lambda} \hat e^\lambda
 \label{eq:cr-spin}
 \end{equation}
  This means that we should make the identification $\hat e_\mu \to \sigma_\mu$ where $\sigma_\mu$ are the three $2 \times 2$ Pauli matrices. Up to rescaling of the mass $m \to M$, upon performing the path integral of Eq.\eqref{eq:G(p)} we find that $G(P)$ is the (Euclidean) Dirac propagator \cite{polyakov-1988}
\begin{equation}
G(p)=\frac{1}{i p_\mu \sigma^\mu-M}
\label{eq:dirac-propagator}
\end{equation}

Using these results the partition function of Eq.\eqref{eq:closed-paths-boson} becomes
\begin{equation}
Z_{\rm fermion}=\textrm{det} [i\slashed{\partial}-M]=\int \mathcal{D}J \; \delta(\partial_\mu J^\mu)\; \exp\left(-|m|L[J]-i \textrm{sgn}(M) \pi \Phi[J]\right)
\label{eq:polyakov-duality}
\end{equation}
where $\Phi[J]=W[J]$, defined in Eq.\eqref{eq:writhe2}. 

We will return to this construction shortly below when we include the effects of the parity anomaly.
 
 Early attempts at deriving a mapping for a theory of Dirac fermions in 2+1 dimensions were based on their behavior in the presence of gauge fields and the associated parity anomaly,
 discussed in section \ref{sec:parity-anomaly}. These early theories are essentially a hydrodynamic description of the Dirac theory deep in the massive phase. 
 With minor differences, the same results were  derived independently (and simultaneously) by two groups, Fidel Schaposnik and myself \cite{fradkin-1994} 
 and Cliff Burgess and Francisco Quevedo \cite{burgess-1994}. This approach was reexamined more recently by A. Chan, T. Hughes, S. Ryu and myself in 2016 in a derivation of a hydrodynamic effective field theory for topological insulators in different dimensions \cite{chan-2013,chan-2016}. These derivations are similar in spirit to what we discussed in section \ref{sec:QFT-FQH-hydro} for the fractional quantum Hall effect. We will follow the approach of Ref.\cite{chan-2013} for the special case of a Chern insulator 2+1 dimenxions.

 The Dirac theory, in any dimension, is globally gauge invariant. By Noether's theorem this means that it has a locally conserved current, $j_\mu=\bar \psi \gamma_\mu \psi$ which thus satisfies $\partial_\mu j^\mu=0$. Here too, this means that we can write $j_\mu= \epsilon_{\mu \nu \lambda}\partial^\nu b^\lambda$, where $b_\mu$ is a gauge field. We expect that the effective action of the gauge field $b_\mu$ should be local, gauge invariant and, in this case, relativistic invariant. In 2+1 dimensions the effective action should break time reversal invariance and hence it should have a Chern-Simons term.

 We will consider the free massive Dirac theory coupled to a background probe gauge field $A_\mu$. The Lagrangian is $\mathcal{L}=\bar \psi (i \slashed(D)-M) \psi$, where $D_\mu=\partial_\mu+iA_\mu$. As we saw in section \ref{sec:parity-anomaly}, gauge invariance and locality require that we have an even number of Dirac fermions. Here we will assume that one of the Dirac fermions is massive and acts as a regulator. The partition function 
 \begin{equation}
 Z[A_\mu]=\int \mathcal{D} \bar \psi \mathcal{D} \psi\; \exp(i S_F[\bar \psi, \psi, A_\mu])
 \label{eq:pf-dirac-A}
 \end{equation}
 By definition the expectation value of a product of current operators $j_\mu$ can be obtained by functional differentiation of the partition function  with respect to $A_\mu$.
 
 The partition function is gauge invariant, i.e.
 \begin{equation}
 Z[A_\mu=Z[A_\mu +a_\mu]
 \label{eq:pf-gauge-inv}
 \end{equation}
 where the vector field $a_\mu$ is a pure gauge, $a_\mu=\partial_\mu \phi$ and $f_{\mu nu}=\partial_\mu a-\nu-\partial_\mu a_\mu=0$. Hence, $a_\mu$ is said to be flat. Therefore, up to a normalization
 \begin{align}
 Z[A_\mu]=&\int \mathcal{D}[a_\mu]_{\textrm{pure}} \; Z[A_\mu+a_\mu] \nonumber\\
 =&\int \mathcal{D} a_\mu \; \delta(f_{\mu \nu}=0) \; Z[A_\mu+a_\mu] \nonumber\\
 =&\int \mathcal{D} a_\mu \mathcal{D}b_\mu \; Z[A_\mu+a_\mu]\; \exp\left(-\frac{i}{2}\int d^3x \;  \epsilon^{\mu \nu \lambda} \; b_\mu \; f_{\nu \lambda} \right)
 \end{align}
 where in the last equality we introduced a representation of the delta function and the vector field $b_\mu$ plays the role of a Lagrange multiplier field. using the invariance of the integration measure under $a_\mu \to a_\mu-A_\mu$ we find
 \begin{equation}
 Z[A_\mu]=\int \mathcal{D} a_\mu \mathcal{D}b_\mu Z[a_\mu] \; \; \exp\left(-\frac{i}{2}\int d^3x \;  \epsilon^{\mu \nu\lambda}\; b_\mu \; (f_{\nu \lambda}-F^{\nu \lambda})\right)
 \label{eq:Z-A-a}
 \end{equation}
 where $F_{\mu \nu}$ is the field strength of the external field $A_\mu$.
 From these identities and by differentiation of Eq.\eqref{eq:Z-A-a}   it follows that a general expectation value of products of currents is
 \begin{equation}
 \langle j_\mu(x)\;  j_\nu(y) \ldots\rangle=\frac{\delta}{\delta A_\mu(x)} \frac{\delta}{\delta A_\nu(y)} \ldots Z[A_\mu]=\langle \epsilon_{\mu \alpha \beta} \partial^\alpha b^\beta \;  \epsilon_{\nu \gamma \delta}\partial^\gamma b^\delta \ldots\rangle
 \label{eq:currents-fields}
 \end{equation}
 which means that, at the operator level, we can identify \cite{fradkin-1994}
 \begin{equation}
 j_\mu(x) \Leftrightarrow \epsilon_{\mu \nu \lambda}\partial^\nu b^\lambda
 \label{eq:bosonization-current}
 \end{equation}
 This result is actually general. The only difference is the nature of the field $b$ isn different dimensions: in 1+1 is a pseudo-scalar, in 2+1 is a vector (gauge) field, in 3+1 is an anti-symmetric 
 (Kalb-Ramond) tensor field, etc. 
 
Returning to the partition function of Eq.\eqref{eq:Z-A-a} we find, using the result of Eq.\eqref{eq:eff-action-doubler} applied the partition function  $Z[a_\mu]$, that $Z[A_\mu]$ is given by 
\begin{equation}
Z[A_mu]=\int \mathcal{D}a_\mu \mathcal{D}b_\mu \exp(i \int d^3x \; \mathcal{L}_{\rm eff}[a_\mu, b_\mu, A_\mu])
\end{equation}
where the effective Lagrangian is
\begin{equation}
\mathcal{L}_{\rm eff}=- \epsilon^{\mu \nu\lambda}\; b_\mu\; \partial_\nu a_\lambda +\frac{s}{4\pi} \epsilon^{\mu \nu\lambda} a_\mu  \partial_\nu a_\lambda -\frac{1}{4g_{\rm eff}} f_{\mu \nu}f^{\mu \nu}+A_\mu \epsilon^{\mu\nu \lambda} \partial_\nu b_\lambda+\ldots
\label{eq:Leff-2+1}
\end{equation}
where $f_{\mu \nu}=\partial_\mu a_\nu-\partial_\nu a_\mu$. Here $s= 1$ in the Chern insulator, $s=0$ in the trivial insulator and $s=1/2$ at the quantum critical point, 
and $g_{\rm eff}$ an effective coupling constant (with dimensions of length$^{-1}$). Here we included the effects of the parity anomaly. We recognize that the first term of the effective 
Lagrangian of Eq.\eqref{eq:Leff-2+1} is a BF term.

In Ref.\cite{chan-2013} a similar result is also derived for a 3+1-dimensional topological insulator. The main differences are that there is no parity anomaly but an axial anomaly and that the Lagrange multiplier field is a Kalb-Ramond field,
\begin{equation}
\mathcal{L}_{\rm eff}[a_\mu, b_{\mu\nu}, A_\mu]=\epsilon^{\mu\nu \lambda\rho}\; b_{\mu \nu}\; \partial_\lambda a_\rho+ \frac{\theta}{8\pi^2} \epsilon^{\mu\nu\lambda\rho}\partial_\mu a_\nu \partial_\lambda a_\rho-\frac{1}{4g^2} f_{\mu \nu} f^{\mu \nu}+A_\mu \epsilon^{\mu\nu\lambda\rho} \partial_\nu b_{\lambda \rho} +\ldots
\label{eq:Leff+3+1}
\end{equation}
where $\theta=\pi$ in the topological insulator and $\theta=0$ in the trivial insulator.
 
 Although these results are correct, they do not give a full bosonization mapping. 
 In particular, these results do not identify a fixed point for the dual bosonic theory and, along with it, a full mapping of the observables. 
 Progress on this problem has only been achieved in the past few years both in the high energy literature 
 \cite{giombi-2012,jain-2013,aharony-2012,aharony-2012b,aharony-2016,witten-2016,seiberg-2016b,karch-2016}, 
 and in the condensed matter physics literature \cite{son-2015,metlitski-2016,wang-2015,chen-2017,mross-2016}. 
 Much of that new insight was presented in a 2016 insightful paper by Nathaniel Seiberg, T. Senthil, Chong Wang and Edward Witten \cite{seiberg-2016}. 
 Many of the dualities are actually conjectures supported by strong consistency checks. 
A derivation of the basic bosonization duality based on loop models was constructed by Hart Goldman and myself \cite{goldman-2018}.

The basic conjectured bosonization duality is a mapping of a free Dirac fermion to a gauge complex scalar field with a Chern-Simons term \cite{aharony-2016,seiberg-2016}. 
We can think of the Dirac theory as either being defined entirely in 2+1 dimensions or as a being defined at the boundary of a non-trivial 3+1 dimensional topological insulator. 
In the first scenario we need to take into account the contribution of the fermionic doublers (this is what happens in the case of a lattice theory) or as the Pauli-Villars heavy fermionic regulators. 
In both cases, the additional heavy degrees of freedom cancel the anomaly of the 2+1-dimensional Dirac fermion, see section \ref{sec:parity-anomaly}. 
In the second scenario, there is only one Dirac fermion at the boundary and the  bulk $\theta$-term cancels the anomaly of the boundary theory, see section \ref{sec:CH-effect}. 
With these provisos, the Lagrangian $\mathcal{L}_A$ of the free massive (or massless) Dirac fermion in 2+1 dimensions coupled to the electromagnetic gauge field $A_\mu$ is
\begin{equation}
\mathcal{L}_A=\bar \psi (i \slashed{D}(A_\mu)-M)\psi
-\frac{1}{8\pi} \epsilon_{\mu \nu \lambda} A^\mu \partial^\nu A^\lambda
\label{eq:theory-A}
\end{equation}
where $M$ is the Dirac mass. 
We will call this Theory $A$. Here  $\slashed{D}_\mu(A)=\gamma^\mu(\partial_\mu-i A_\mu)$  and $A_\mu$ is a background (non-dynamical) gauge field. 
The last term in Eq.\eqref{eq:theory-A} is the Chern-Simons term with the 1/2-quantized coefficient, the usual short-hand for the $\eta$-invariant term of Eq.\eqref{eq:eta-1/2} 
needed to cancel the parity anomaly.

From the effective field theories we discussed above we know that the fermionic current maps to the curl of a gauge field, see Eq\eqref{eq:bosonization-current}. 
Then, the conjectured bosonic dual is given  by the Lagrangian of Eq\eqref{eq:csf+cs} with an extra term for the (dual) coupling to the electromagnetic gauge filed. 
We will call this Theory $B$ whose Lagrangian $\mathcal{L}_B$  is
\begin{equation}
 \mathcal{L}_B=|D_\mu(\mathscr{A}_\mu) \phi|^2-m^2 |\phi|^2-\lambda |\phi|^4+\frac{1}{4\pi} \epsilon_{\mu \nu\lambda}\mathscr{A}^\mu \partial^\nu \mathscr{A}^\lambda
 +\frac{1}{2\pi} \epsilon_{\mu \nu\lambda}A^\mu \partial^\nu \mathscr{A}^\lambda
 \label{eq:theory-B}
 \end{equation}
To check the consistence we first observe that both theories are anomaly free. 
By functional differentiation of the two partition functions we check that we get the correct mapping for the fermionic current, $j_\mu \leftrightarrow \epsilon_{\mu \nu \lambda} \partial^\nu \mathscr{A}^\lambda$.
This mapping implies that the charge density of Theory $A$ maps onto the gauge field flux of Theory $B$, which is electromagnetic duality. 
We see that this bosonization is a relativistic version of flux attachment.

Let us now identify the mapping of the phases of both theories. If the Dirac mass $M<0$, Theory $A$ describes an anomalous quantum Hall insulator. 
Integrating out the massive fermions we fund that the effective action for the electromagnetic gauge field $A_\mu$ is a U(1)$_1$ Chern-Simons theory. 
This means that in this phase $\sigma_{xy}=-1/(2\pi)$ (in units in which $e=\hbar=c=1$).
 Looking now at Theory $B$ we see that if $m^2>0$, the scalar field is in the unbroken phase, and it is massive. In this phase we set $\phi=0$ and find that the 
 low energy resulting theory if just a U(1)$_1$ Chern-Simons gauge theory coupled to the curl of the electromagnetic field. Upon integrating out the $\mathscr{A}_\mu$ gauge field we find that the 
 effective action of  $A_\mu$ is also a U(1)$_1$ Chern-Simons term, which implies that $\sigma_{xy}=-1/(2\pi)$. 
 
 Conversely, for  $M>0$ the effective electromagnetic action of  the fermionic theory is a Maxwell term (the Chern-Simons term canceled). This phase is a trivial insulator. 
 Looking now at Theory $B$ we see that for $m^2<0$ this theory is in its Higgs phase where $\langle \phi \rangle \neq 0$ and the gauge field $\mathscr{A}_\mu$ now has a mass term, 
 $\propto \mathscr{A}_\mu^2$. In the low energy limit $\mathscr{A}_\mu \to 0$, and the Hall conductivity vanishes, consistent to what is expected from Theory $A$.
 
 On the other hand, for $M=0$ Theory $A$ is at a quantum critical point with  a very simple CFT structure but a non-vanishing Hall conductivity $\sigma_{xy}=1/(4\pi)$. 
 It is natural to map this CFT to a Wilson-Fisher (WF) fixed point of Theory $B$. At the WF one sets the (renormalized!) mass $m^2_R=0$ (not the {\it bare} mass). 
 In the absence of the Chern-Simons gauge field this WF fixed point in well understood form high quality (five loop!) epsilon expansion calculations (enhanced with Borel resummation) \cite{zinn-book}, 
 and by more recent numerical Conformal Bootstrap methods \cite{simmons_duffin-2017}. 
 However, not much is known of the {\it gauged} version of the CFT of Theory $B$ since it cannot be accessed by either methods. 
 If the mapping to Theory $A$ is correct it should have a Hall conductivity of $\sigma_{xy}=1/(4\pi)$. This conjectured value hast not yet been confirmed. 

 Consider now a monopole operator of the gauge field $\mathscr{A}_\mu$ of Theory $B$. We will denote this operator $\mathcal{M}_{\mathscr{A}}$. 
 The Chern-Simons term is not gauge invariant in a monopole background. To put it differently it has charge $1$ under the Chern-Simons gauge field $\mathscr{A}_\mu$. 
 It also has charge $1$ under the electromagnetic field $A_\mu$. Likewise the scalar field $\phi$ has charge $1$ under the Chern-Simons gauge field $\mathscr{A}_\mu$. 
 The operator $\phi^\dagger \mathcal{M}_{\mathscr{A}}$ is gauge-invariant under the Chern-Simons gauge field $\mathscr{A}_\mu$ (i.e. it has charge 0). 
 This composite operator has electromagnetic charge $1$ (which it inherited from the monopole). Moreover, it has spin-1/2 required by the Wu-Yang construction \cite{wu-1976}. 
 In other words, the operator $\phi^\dagger \mathcal{M}_{\mathscr{A}}$ has the {\it same} quantum numbers as the Dirac fermion and are identified by this conjecture. 
 Notice, however, that the free massless Dirac field has scaling dimension $\Delta_\psi=1$ in 2+1 dimensions. The conjecture implies the composite operator $\phi^\dagger \mathcal{M}_{\mathscr{A}}$ 
 should also have scaling dimensions $1$ at the (gauged) WF fixed point of Theory $B$. 
 Many of these conjectures have been verified in non-abelian versions of Theory $A$ and Theory $B$: a theory of free massless Dirac fermions coupled to  a Chern-Simons gauge theory 
 with gauge group  SU(N)$_k$ and a theory of complex scalars coupled to a non-abelian Chern-Simons gauge theory with gauge group SU(k)$_N$, both in the limits $N \to \infty$ and $k\to \infty$ 
 with $N/k$ fixed \cite{giombi-2012,jain-2013,aharony-2012,aharony-2012b,aharony-2016}.
 
The particle-vortex duality discussed in section  \ref{sec:PV-duality} combined with the fermion-vortex duality we sketched here provide a web of dualities. These identifications constitute a powerful non-perturbative tool which has led to many significant developments. For instance, Goldman and myself \cite{goldman-2018b} used the web of dualities to explain the experimentally observed self-duality at fractional quantum Hall plateau transitions, which was a long standing puzzle. It has also provided a powerful new tool to derive effective field theories of non-abelian fractional quantum Hall states \cite{goldman-2019} and even to propose novel states with a single (Fibonacci) anyon \cite{goldman-2021}. 


\subsubsection{Bosonization of the Fermi Surface}
\label{sec:bosonization-FS}

A form of bosonization has been developed for  the case of systems of fermions with a Fermi surface. Dense Fermi systems  at sufficiently weak coupling are well described by the Landau theory of the Fermi liquid \cite{baym-1991,abrikosov-1964,nozieres-1966,polchinski-tasi,shankar-1994}. In this regime, the system of fermions has a collective mode, a bound state of particles and holes, which at long wavelengths is well described by the random phase approximation (RPA) \cite{bohm-1953}. However, in space dimensions $d>1$ there is no kinematical restriction and quasi-particles and quasi-holes may move in their separate ways. The result is that, in addition to the collective modes, there is a low-energy spectrum of renormalized but essentially free quasi-particles. Because of the existence of this quasi-particle spectrum the collective modes generally (but not always) decay into particle-hole  pairs resulting in a finite lifetime of the collective modes. 

Superficially the particle-hole collective modes are similar to the scalar field of the bosonized theory in $d=1$. To an extent it has been possible to ``bosonize the Fermi surface'' and to treat it as a quantum mechanical object \cite{haldane-1994,houghton-1993,castro_neto-1993,castro_neto-1994,houghton-2000,delacretaz-2022}. In this approach one essentially regards each direction normal to the Fermi surface as a one-dimensional chiral fermion, including the current algebra structure, subject to global constraints that cancels the anomalies. This point of view ensures that the total fermion number in the Fermi sea is conserved. Here I will only present a short summary of the main ideas. 

As in the one-dimensional case one defines a filled Fermi sea state $|\textrm{FS}\rangle$. This is a state of non-interacting fermions filling up all one-particle states up to the Fermi energy $E_F$. For concreteness we consider a system in two space dimensions. In this case, neglecting lattice effects, the Fermi surface is a circumference of radius $p_F$, the Fermi momentum. At fixed fermion number, the excitations are particle-hole pairs. The  operator  $n_{\bm k}(\bm q)=c^\dagger({\bm k}+\frac{{\bm q}}{2})\; c({\bm k}-\frac{{\bm q}}{2})$ creates a particle-hole pair with relative momentum $\bm q$ with total  momentum $\bm k$. In the low energy regime $\bm k$ is a point on the Fermi surface (with $|\bm k |=p_F$) and $\bm q$ is small momentum compared with $p_F$. In what follows we will label the point $\bm k$ on the Fermi surface by the angle $\theta$ of the arc spanned by $\bm k$ and an  arbitrary origin on the Fermi surface. 

We will normal-order the particle-hole creation operator  with respect to the filled Fermi sea and define $\delta (\bm q, \theta)=:n_{\bm k}(\bm q):=n_{\bm k}(\bm q)-\langle \textrm{FS}| n_{\bm k}(\bm q)|\textrm{FS}\rangle$. Haldane \cite{haldane-1994}, Houghton and Marston \cite{houghton-1993,houghton-2000}, and Castro Neto and Fradkin \cite{castro_neto-1993,castro_neto-1994} showed that in real space that these normal ordered density operators obey the equal-time commutation relations
\begin{equation}
[\delta n(\bm x, \theta), \delta n(\bm x',\theta')]=-\frac{1}{2\pi} \bm k_F(\theta) \cdot \bm \bigtriangledown \delta^2(\bm x-\bm x') \delta(\theta-\theta')
\label{eq:2D-FS-schwinger}
\end{equation}
where $\bm k_F(\theta)$ is a unit vector normal to the Fermi surface at angle $\theta$. This is the algebra of the quantum fluctuations of the Fermi surface. We can further define at each point $\theta$ on the Fermi surface a Bose field $\varphi(\bm x, \theta)$ such that
\begin{equation}
\delta n(\bm x, \theta)=N(0) \bm \varv_F(\theta) \cdot \bm \bigtriangledown \varphi(\bm x, \theta)
\label{eq:bosonization-FS}
\end{equation}
where $N(0)$ is the density of one-particle states at the Fermi surface and $\bm \varv_F(\theta)$ is the Fermi velocity at the location $\theta$. The scalar field $\varphi(\bm x, \theta)$ is a chiral boson at $\theta$ which parametrizes the quantum fluctuations of the Fermi surface. The quantum dynamics of the chiral bosons $\varphi(\bm x, \theta)$ is governed by the action
\begin{align}
S=&\frac{1}{2} N(0) \int_0^{2\pi} \frac{d\theta}{2\pi} \int d^2x \; dt\; 
\left(-\partial_t \varphi(x, \theta)\; \bm \varv_F(\theta) \cdot \bm \bigtriangledown \varphi(x,\theta)
-(\bm \varv_F(\theta)\cdot \bm \bigtriangledown \varphi(x,\theta))^2\right)\nonumber\\
+&\frac{1}{2} N(0) \int_0^{2\pi} \frac{d\theta}{2\pi} \int_0^{2\pi} \frac{d\theta'}{2\pi} \int d^2x\; d^2x' \; dt \;
F(\bm x-\bm x'; \theta-\theta')\;  \bm \varv_F(\theta)\cdot \bm \bigtriangledown \varphi(x,\theta)\; \bm \varv_F(\theta')\cdot \bm \bigtriangledown \varphi(x,\theta')
\label{eq:2D-bosonized-action}
\end{align}
where $x=(\bm x, t)$, and $F(\bm x-\bm x'; \theta-\theta')$ are the Fermi liquid parameters that parametrize the effective forward scattering interactions among the fermion quasiparticles on the Fermi surface \cite{baym-1991}. The equation of motion predicted by this (quadratic) action is equivalent to the linearized quantum Boltzmann equation familiar from the Landau theory of the Fermi liquid. A non-linear extension has been introduced recently by Delacr\'etaz and coworkers \cite{delacretaz-2022}. Recent work on the role of quantum anomalies in dense Fermi systems has yielded new insights on these problems \cite{shi-2022}.

In the regime, where the Landau theory of the Fermi liquid is expected to work \cite{baym-1991},  bosonization of the Fermi surface approach has reproduced the previously known results. There remain many open problems in this approach (and others)  in the vicinity of quantum phase transitions. In spite of intense research using many different approaches, quantum phase transitions in metallic systems are not yet fully understood beyond perturbation theory \cite{hertz-1976,millis-1993}. Non-perturbative  ideas such as deconfined quantum criticality have been proposed \cite{senthil-2004} and significant work has been done using large-$N$ methods \cite{fitzpatrick-2013,metlitski-2015}. A particularly important metallic quantum phase transition (and perhaps the simplest) occurs
near the Pomeranchuk instability of the fermi liquid. A quantum phase transition to an electron nematic state \cite{oganesyan-2001} has been predicted and studied within the Hertz-Millis approach. It has also been studied using higher dimensional bosonization \cite{oganesyan-2001,lawler-2006,lawler-2007}. Quantum Monte Carlo simulations  by \cite{lederer-2015} have shown that the vicinity of a nematic quantum critical point can trigger a superconducting state.  Nematic Fermi fluids have been found in many physical systems of interest ranging from high temperature superconductors (such as cuprates and iron superconductors), to electron gases in large magnetic fields \cite{kivelson-1998,fradkin-2010,fradkin-2015,fernandes-2019}.




\section{Conclusions}
\label{sec:conclusions}

In this chapter I presented the deep and broad role that Quantum Field Theory has in modern Condensed Matter Physics. In spite of the length of this chapter, I have not done justice to its role in many important areas that I did not cover.  In particular, I have not touched on the the theory of quantum spin liquids, fractons, and particularly topological phases in three space dimensions, among others. An area in which the ideas and concepts of QFT have a huge impact, both conceptually and as a tool, is in the problem of quantum entanglement in Condensed Matter. The concept of quantum entanglement as applied to condensed matter systems, both in equilibrium and out of equilibrium, is a major area of current research and  has become a crucial tool for characterizing systems at quantum criticality as well as topological phases of matter. 

\section*{Acknowledgments}
This work was supported in part by the US National Science Foundation through grant No. DMR 1725401 and DMR 225920 at the University of Illinois.






\end{document}